\documentclass[aps]{revtex4}
\usepackage{eurosym}
\usepackage{amsfonts}
\usepackage{amsmath}
\usepackage{amssymb,epsf}
\usepackage{color}
\usepackage{xcolor}
\usepackage{graphicx}
\usepackage{epstopdf}
\usepackage{float}
\usepackage{caption}
\usepackage{subfig}
\usepackage{hyperref}
\usepackage{mathtools}
\usepackage{hyperref}
\usepackage{booktabs}
\usepackage{orcidlink}

\begin{document}
	\title{Testing the AdS/CFT Correspondence Through Thermodynamic Geometry of Nonlinear Electrodynamics AdS Black Holes with Generalized Entropies}
	
	\author{Abhishek Baruah\,\orcidlink{0009-0006-2069-0872}}
	\affiliation{Department of Physics, Patkai Christian College, Ch\"umoukedima, 797103, Nagaland, India.}
	\affiliation{Department of Physics, Dibrugarh University, Dibrugarh 786004, Assam, India}
	\author{Amijit Bhattacharjee\,\orcidlink{0009-0002-8947-2356}}
	\affiliation{Department of Physics, Dibrugarh University, Dibrugarh 786004, Assam, India}
	\author{Prabwal Jyoti Phukon\,\orcidlink{0000-0002-4465-7974}}
	\affiliation{Department of Physics, Dibrugarh University, Dibrugarh 786004, Assam, India}
	\affiliation{Theoretical Physics Division, Centre for Atmospheric Studies,
		Dibrugarh University, Dibrugarh, 786004, Assam, India}

\begin{abstract}
	We investigate the thermodynamics and thermodynamic geometry of several Anti--de Sitter black hole solutions arising from nonlinear electromagnetic theories, namely the ModMax, nonlinear electrodynamics (NED), and Euler--Heisenberg AdS black holes, together with their holographically dual conformal field theory (CFT) descriptions. The analysis is carried out within three entropy frameworks: the standard Bekenstein--Hawking entropy and the generalized R\'enyi and Kaniadakis entropies. For each system, we analyze the phase structure through the behavior of temperature, specific heat, and the scalar curvature obtained from geometrothermodynamics (GTD).
	We find that thermodynamic critical points correspond to extrema in the temperature--entropy relation and coincide with divergences of the specific heat. These locations are reproduced by singularities in the Legendre--invariant GTD curvature, demonstrating a consistent geometric interpretation of the phase transitions. A comparison between the bulk black hole systems and their dual CFT counterparts shows that the number and structure of critical points are preserved under the holographic correspondence.
	Our results further reveal that the Euler--Heisenberg AdS black hole exhibits a more intricate phase structure compared with the ModMax and NED cases, while the Kaniadakis entropy consistently generates an additional critical point across all systems considered. These findings highlight the combined influence of nonlinear electromagnetic dynamics and generalized entropy formalisms on the critical behavior of AdS black holes and their dual CFTs.
\end{abstract}
	
	\maketitle
	\tableofcontents

\section{Introduction}
\label{sec:1}
\noindent Black hole thermodynamics represents one of the most remarkable and profoundly revealing connections
between gravitatiZon, quantum theory, and statistical physics. The rather seminal works of Bekenstein and
Hawking \cite{1,2,3,4,5,6,7,8} revealed that black holes are not merely classical gravitational objects but behave as authentic
thermodynamic systems, possessing well-defined temperature and entropy as described by the relations
\begin{equation}
T=\frac{\kappa}{2\pi}, \quad S_{BH}=\frac{A}{4}
\end{equation}
where $\kappa$ and $A$ are the surface gravity and the area of the black hole respectively. These results suggest that
gravitational systems possess an underlying microscopic description and have motivated extensive investigations into the thermodynamic behaviour of black holes.\\
A major breakthrough in understanding the connection between gravity and quantum field theory came
with the proposal of the AdS/CFT correspondence \cite{54}.  This conjecture states that a gravitational theory
in a $(d+1)$-dimensional Anti-de Sitter (AdS) spacetime is dual to a conformal field theory (CFT) living on its
$d$-dimensional boundary. Within this framework, thermodynamic properties of AdS black holes correspond to
thermodynamic behaviour in the dual field theory. A well-known example is the Hawking-Page phase transition \cite{HawkingPage1983}, which is interpreted holographically as the confinement-deconfinement transition in the boundary gauge
theory.\\
In recent years, black hole solutions arising from nonlinear electrodynamics have attracted considerable
attention. Such theories introduce nonlinear corrections to Maxwell electrodynamics that become significant
in strong electromagnetic fields. Examples include ModMax electrodynamics \cite{modmax1,modmax2,modmax3,modmax4,modmax5,modmax6,modmax7,modmax8}, nonlinear electrodynamics (NED) models \cite{ned1,ned2,ned3,ned4,ned5,ned6,ned7,ned8}, and Euler–Heisenberg electrodynamics \cite{eu1,eu2,eu3,eu4,eu5,eu6}. Black holes constructed within these
frameworks in AdS space-time provide a useful setting for exploring the influence of nonlinear electromagnetic
interactions on black hole thermodynamics and holography.\\
Another important direction in black hole thermodynamics involves generalized entropy formalisms. In
addition to the standard Bekenstein–Hawking entropy, generalized statistical frameworks such as R\'enyi \cite{renyi1,renyi2,renyi3,renyi4,renyi5}
and Kaniadakis entropy \cite{kani1,kani2,kani3,kani4} have been proposed to capture possible non-extensive effects and quantum
corrections to the thermodynamic description. These entropy formalisms introduce deformation parameters
that can significantly modify the phase structure and stability properties of black hole systems whereas in
appropriate limits they can reduce back to the Bekenstein-Hawking entropy formalism.\\
Thermodynamic geometry is a widely accepted approach that applies the methods of differential geometry to study the thermodynamic systems in order to reveal their underlying interaction structure and critical behaviour. In this framework, one defines a Riemannian metric  on the space of equilibrium states using thermodynamic potentials such as entropy or internal energy. Early formulations were put forward by Frank Weinhold \cite{weinhold1975} and later developed by George Ruppeiner \cite{ruppeiner1979,ruppeiner1995review}, where geometric quantities such as the thermodynamic curvature provide information about microscopic interactions and phase transitions of the system. But later it was seen that since these metrics were not legendre-invariant a different choice of potential would give slightly different results, so in order to deal with this problem  the geometrothermodynamic (GTD) \cite{Quevedo2007} framework was proposed by Hernando Quevedo, which constructs a Legendre-invariant geometric structure on the thermodynamic phase space. The GTD formalism ensures that the geometric description of thermodynamics remains invariant under Legendre transformations between different thermodynamic potentials, thereby providing a consistent and unified geometric description of thermodynamic systems particularly, black holes.\\
Geometrothermodynamics (GTD) \cite{Quevedo2007} presents itself as a powerful geometric formulation of thermodynamics in which phase transitions, critical behaviour and microscopic interactions in black holes are intertwined in the geometric curvature of an aptly constructed thermodynamic manifold. The formalism, quite successfully, has been widely applied to diverse gravitational systems which has inturn revealed how different entropy models, modified gravity effects, and non-linear matter sources influence the thermodynamic geometry of a particular system. Early explorations included GTD analysis of non-linear sources \cite{sanchez2021geometrothermodynamics}, STU black holes \cite{sekhmani2023thermodynamic}, and perturbatively corrected $f(R)$ black holes \cite{upadhyay2021perturbed}. Further studies extended the GTD framework to pure Lovelock black holes \cite{khuzani2022thermodynamic}, AdS black holes with nonlinear electrodynamics \cite{kruglov2025thermodynamic}, RN-AdS black holes with non-local observables probing thermodynamic curvature \cite{wang2024thermodynamic}. The framework has also been used to systems with global monopoles \cite{zafar2025thermodynamic}, SU(N) non-linear sigma model black holes \cite{chaudhary2025analysis} and regular black holes such as that of the Bardeen solution where the conventional and modified GTD metrics highlighted metric-dependence effects \cite{mahanta2025thermodynamic}. More recent works of this GTD thermodynamic geometric formulation emphasizes the role of generalized entropies such as that of Kaniadakis entropy \cite{luciano2023p}, Barrow entropy \cite{rani2023impact}, and Tsallis entropy in both standard AdS black holes \cite{luciano2023black} and modified gravity such as that of the Einstein-Gauss-Bonnet theories \cite{jawad2025geometrothermodynamics}. Geometrothermodynamics (GTD) has also enabled the study of geodesic flows \cite{gogoi2023geodesics}, certain topological signatures of criticality in regularized Maxwell black holes \cite{sekhmani2025topological}, and various microstructural phenomena such as shadows and phase transitions in RN-AdS spacetimes \cite{ladino2024phase}. Beyond the regime of black holes, GTD has also found diverse applications in ordinary thermodynamic systems which includes nonlinear real  \cite{quevedo2023geometrothermodynamic} and ideal quantum gases \cite{zaldivar2023ideal} by further enhancing the universality and versatility of this geometric thermodynamic approach of ours.\\
In geometrothermodynamics (GTD), the thermodynamic behaviour of a system is well described through the methods of differential geometry particularly with the aim of capturing phase transitions as geometric features \cite{Quevedo2007,Soroushfar2016,GarciaPelaez:2014}. The framework is premised upon a thermodynamic phase space $\mathcal{T}$ of dimension $2n+1$, which is further endowed with coordinates $Z^A=\{\Phi,E^a,I^a\}$, where $\Phi$ is a thermodynamic potential, $E^a$ being the extensive variables and $I^a$ is regarded to be their conjugate intensive variables $(a=1,\dots,n)$. This space is endowed with the contact 1-form which is given by: 
\begin{equation}
	\Theta = d\Phi - I_a\, dE^a ,
\end{equation}
which further encodes the first law of thermodynamics. A Legendre invariant metric $G$ is then introduced on $\mathcal{T}$, for example, 
\begin{equation}
	G = (d\Phi - I_a dE^a)^2 + (\delta_{ab} E^a I^b)\,\eta_{cd}\, dE^c dI^d ,
\end{equation}
with $\eta_{cd}=\text{diag}(-1,1,\dots,1)$ ensures invariance under Legendre transformations \cite{Quevedo:2010curvature,Quevedo:2017homogeneity}. The space of equilibrium states $\mathcal{E}$ is then defined through the smooth embedding $\varphi:\mathcal{E}\to\mathcal{T}$ given as: 
\begin{equation}
	\varphi: \{E^a\} \mapsto \{\Phi(E^a),E^a,I^a=\partial \Phi/\partial E^a\},
\end{equation}
such that $\varphi^*(\Theta)=0$, which further corresponds to the Gibbs relation. The induced metric on $\mathcal{E}$ is given by: 
\begin{equation}
	g = \varphi^*(G),
\end{equation}
then captures the thermodynamic interaction: flat geometry ($R[g]=0$) is said to correspond to having no thermodynamic interaction (ideal gas), while curvature singularities are meant to signal critical points and phase transitions \cite{QGas:2023,GTD:BH2021,Luongo:2023,Beissen:2023}. Therefore, the Geometrothermodynamic (GTD) framework provides a Legendre-invariant geometric criterion which can be further used to study features such as: thermodynamic stability and critical phenomena.\\
The GTD metric is therefore a Legendre invariant metric and thus would not depend on any specific choice of thermodynamic potential. The phase transitions as obtained from the specific heat capacity of the black hole system are meant to be  properly encoded in the scalar curvature of the GTD metric, such as that a curvature singularity in the GTD scalar `$R_{GTD}$'  is meant to imply the occurrence of a phase transition. The general form of the type II GTD metric is derived from the following \cite{{Soroushfar2016}} :
\begin{equation}
	\label{eq2}
	g = \left(E^{c} \frac{\partial{\varphi}}{\partial{E^{c}}} \right)
	\left(\eta_{ab} \delta^{bc} \frac{\partial^2 \varphi}{\partial E^{c} \partial E^{d}} dE^{a} dE^{d}\right)
\end{equation}     
where `$\varphi$' is the thermodynamic potential and `$E^a$' is an extensive thermodynamic variable with $a=1,2,3....$..
The thermodynamic scalar curvature, computed within GTD, is therefore meant to diverge exactly at the points where the specific heat diverges, thereby establishing a precise correspondence between geometric and thermodynamic signals of phase transitions. This similarity is therefore central to interpreting the GTD framework as a powerful diagnostic tool for black hole criticality and phase transitions.\\
The behavior of different black hole solutions near critical points has been widely studied in the literature. In particular, within the framework of the AdS/CFT correspondence, the phase structure of charged AdS black holes has been investigated extensively~\cite{12,13}. Furthermore, the thermodynamic stability and phase transition properties of several charged black hole configurations have also been examined by placing the system inside a finite cavity, which effectively mimics thermal equilibrium conditions~\cite{14,15,16}. \\
Another powerful approach to the study of black hole critical phenomena is provided by thermodynamic geometry, where the curvature of the thermodynamic parameter space is interpreted as a probe of phase transitions and critical behavior~\cite{17,18,19,20,21,22}. In parallel, the recently developed framework of thermodynamic topology offers a complementary perspective by characterizing phase structures through topological invariants such as the Euler characteristic, thereby providing a global description of critical phenomena and phase transitions~\cite{23,24,25,26,27,28}. These geometric and topological methods have significantly broadened the understanding of black hole phase transitions beyond conventional thermodynamic analysis.\\
Motivated by the formulation introduced in~\cite{29}, known as Extended Phase Space Thermodynamics (EPST), the cosmological constant is treated as a thermodynamic pressure while its conjugate quantity is interpreted as the thermodynamic volume. This viewpoint establishes a direct analogy between AdS black holes and ordinary liquid--gas systems, enabling the exploration of $P$--$V$ criticality and associated phase behavior for a broad class of AdS black hole solutions~\cite{30,31,32,33,34,35,36,37,38,39,40,41,42}. Despite its considerable success, the EPST framework faces certain conceptual issues, including ambiguities related to the choice of ensemble and the lack of homogeneity in the thermodynamic energy function, which complicate the physical interpretation of the thermodynamic variables.\\
To address the aforementioned limitations, the framework of Restricted Phase Space Thermodynamics (RPST) was introduced~\cite{43}. Within this formulation, the cosmological constant is treated as a fixed parameter, and the thermodynamic description is reconstructed in terms of quantities defined on the boundary conformal field theory. In particular, variables such as the central charge $C$ and the CFT volume $V$, together with their conjugate potentials, play the role of fundamental thermodynamic parameters. This approach restores the homogeneity of the thermodynamic relations, removes ambiguities associated with ensemble dependence, and enables a consistent holographic interpretation of phase transitions occurring in AdS black holes~\cite{44,45,46,47,48,49,50,51,52,53}. A recurring result emerging from these studies is that the phase transition at the critical point generally exhibits characteristics of a second–order transition.
\\
The conceptual basis for interpreting these thermodynamic features from a field–theoretic perspective arises from the AdS/CFT correspondence, originally proposed by Maldacena. This correspondence establishes a remarkable duality between gravitational dynamics in asymptotically Anti--de Sitter (AdS) spacetimes and conformal field theories (CFTs) defined on the boundary of the spacetime. Through this holographic relationship, gravitational phenomena associated with black holes acquire a natural interpretation in terms of thermodynamic properties of the dual quantum field theory. In this context, fundamental black hole parameters such as the mass $M$, electric charge $Q$, temperature $T$, and entropy $S$ are mapped to corresponding observables in the boundary CFT, including the energy $E$, electric potential $\Phi$, and the central charge $C$~\cite{55,56,57,58}. Consequently, the thermodynamic behavior of AdS black holes can be interpreted holographically as the thermodynamics of the dual conformal field theory.
Within this holographic framework, the thermodynamic first law describing charged AdS black holes can be recast in terms of boundary CFT variables as
\begin{equation}
	dE=\mathcal{T}d\mathcal{S}+\varphi d\mathcal{Q}-pd\mathcal{V}+\mu C
\end{equation}
with the Smarr relation written as
\begin{equation}
E=\mathcal{T}\mathcal{S}+\varphi \mathcal{Q}+\mu C 
\end{equation}
Here, $\mu$ denotes the chemical potential associated with the central charge $C$, while $p$ and $V$ correspond to the pressure and volume of the boundary CFT, respectively. Within the holographic description, these thermodynamic variables are related through a conformal scaling factor $\omega$, which provides the appropriate mapping between the bulk and boundary quantities as
\begin{equation}
E=\frac{M}{\omega}, \quad \mathcal{T}=\frac{T}{\omega}, \quad \varphi=\frac{\Phi \sqrt{G}}{\omega l},\quad \mathcal{Q}=\frac{Q l}{\sqrt{G}}
\end{equation}
where the CFT boundary geometry is defined as $\omega=R/l$ \cite{55,56,57,58,59,60,61,62,63}.\\
Motivated by these developments, in this work we perform a systematic thermodynamic and geometric investigation of three nonlinear electrodynamic black hole systems in the Anti-de Sitter spacetime: the ModMaxAdS black hole, the NED-AdS black hole, and the Euler-Heisenberg-AdS black hole. The
analysis is carried out within three different entropy formalisms, namely the standard Bekenstein–Hawking
entropy, the Kaniadakis entropy, and the R´enyi entropy. Using the framework of geometrothermodynamics, we
compute the thermodynamic scalar curvature and analyze its behavior with temperature in order to identify
the phase transition structure of these systems. Testing the AdS/CFT correspondence for different classes of
black hole solutions is essential for assessing the robustness and universality of the conjecture. In particular,
black holes arising from nonlinear electrodynamics theories, provide an important arena for such investigations since the nonlinear corrections to the electromagnetic sector can significantly modify the thermodynamic
behaviour and phase structure of the system. Analysing these black holes therefore allows one to explore
how strong-field electromagnetic effects influence the holographic correspondence. Furthermore, it is equally
important to examine these systems under different entropy formalisms. While the Bekenstein–Hawking entropy represents the standard thermodynamic description of black holes, generalized entropy measures such as
the R´enyi and Kaniadakis entropies arise from non-extensive statistical frameworks and may encode possible
quantum or statistical corrections to the conventional area law. Studying the thermodynamic and holographic
behaviour of these black holes within both the standard and generalized entropy frameworks enables a broader
comparison of the phase structures obtained on the AdS and CFT sides. Such a systematic analysis provides
a more comprehensive and complete picture of the thermodynamic properties of these systems and allows one
to examine the extent to which the AdS/CFT correspondence may remain consistent.\\
The primary objective of this study is to examine the holographic consistency predicted by the AdS/CFT
correspondence by comparing the thermodynamic behaviour of the bulk AdS black holes with that of their
corresponding conformal field theory counterparts. In particular, we investigate whether the number and
structure of phase transitions obtained from thermodynamic geometry remain consistent on both the AdS and
CFT sides under different entropy formalisms.\\
The main contributions of this work includes, providing a unified thermodynamic geometric analysis of several nonlinear electrodynamic AdS black holes within multiple entropy frameworks. Secondly, we investigate
whether and how generalized entropy formalisms such as R´enyi and Kaniadakis entropies modify the thermodynamic phase structure of these systems. Finally, we use thermodynamic geometry as a tool to compare the
phase transition behaviour of the bulk AdS black holes with their boundary CFT counterparts, thereby offering
a geometric perspective on testing the AdS/CFT correspondence in the presence of nonlinear electrodynamics
and non-extensive entropy corrections.\\
The structure of this paper is organized as follows. In Section~\ref{sec:1}, we present a concise overview of the thermodynamics and geometrothermodynamics (GTD) of AdS black holes together with a brief discussion of the AdS/CFT correspondence. Sections~\ref{sec:2} and \ref{sec:3} are devoted to the analysis of the thermodynamic properties and GTD structure of ModMax-AdS black holes and their holographically dual CFT description, considered within three different entropy frameworks, namely the Bekenstein--Hawking entropy, R\'enyi entropy, and Kaniadakis statistics.\\
Subsequently, in Sections~\ref{sec:4} and \ref{sec:5}, we extend this investigation to nonlinear electrodynamics (NED) AdS black holes, examining both their thermodynamic behavior and the corresponding GTD geometry, along with the associated CFT dual under the same three entropy formalisms. In Sections~\ref{sec:6} and \ref{sec:7}, a similar analysis is carried out for Euler--Heisenberg AdS black holes and their dual CFT counterparts within the Bekenstein--Hawking, R\'enyi, and Kaniadakis entropy descriptions. Finally, Section~\ref{sec:8} summarizes the main findings of this work and discusses possible future directions for further investigation.
\section{  ModMax-AdS Black Hole Solution}
\label{sec:2}
The action describing the coupling of Einstein's gravity with the
cosmological constant in the presence of the Mod(A)Max electrodynamic fields
is expressed as 
\begin{equation}
I=\frac{1}{16\pi }\int_{\partial \mathcal{M}}d^{4}x\sqrt{-g}\left[
R-2\Lambda -4\eta \mathcal{L}\right] ,  \label{Action}
\end{equation}%

where $g=\text{det}(g_{\mu \nu })$ is the determinant of  $%
g_{\mu \nu }$, the metric tensor. Also, $\Lambda $ and $R$ are  the cosmological
constant and the Ricci scalar, respectively. In addition, $\eta =+1$, and $%
\eta =-1$ are related to Maxwell and anti-Maxwell (phantom) cases,
respectively. It is notable that, Mod(A)Max devotes to both ModMax and
ModAMax fields. In other words, Mod(A)Max is related to ModMax field when $%
\eta =+1$ or the ModAMax field when $\eta =-1$. In the action \eqref{Action}%
, $\mathcal{L}$ refers to ModMax's Lagrangian and is defined as \cite
{ModMaxI,ModMaxII} 
\begin{equation}
\mathcal{L}=\mathcal{S}\cosh \gamma -\sqrt{\mathcal{S}^{2}+\mathcal{P}^{2}}%
\sinh \gamma ,  \label{ModMaxL}
\end{equation}%
where $\gamma $ is known as ModMax's parameter,. This parameter is a
dimensionless quantity. Furthermore, $\mathcal{S}$ and $\mathcal{P}$
represent a true scalar and a pseudoscalar, respectively, in the following
expressions 
\begin{equation}
\mathcal{S}=\frac{\mathcal{F}}{4},~~~\&~~~\mathcal{P}=\frac{\widetilde{%
\mathcal{F}}}{4},
\end{equation}%
in the equations above, the term $\mathcal{F}=F_{\mu \nu }F^{\mu \nu }$ is
referred to as the Maxwell invariant. Moreover, $F_{\mu \nu }$ represents
the electromagnetic tensor field and is defined as $F_{\mu \nu }=\partial
_{\mu }A_{\nu }-\partial _{\nu }A_{\mu }$, with $A_{\mu }$ being the gauge
potential. Also, $\widetilde{\mathcal{F}}=$ $F_{\mu \nu }\widetilde{F}^{\mu
\nu }$, where $\widetilde{F}^{\mu \nu }=\frac{1}{2}\epsilon _{\mu \nu
}^{~~~\rho \lambda }F_{\rho \lambda }$. By considering $\gamma =0$, the
Lagrangian of ModMax Eq. \eqref{ModMaxL} reduces to the Maxwell theory as $%
\mathcal{L}=\frac{\mathcal{F}}{4}$.

In order to get electrically charged black holes solutions, we set $\mathcal{%
P}=0$ in the Lagrangian of ModMax (Eq. (\ref{ModMaxL})). So, the generalized
Einstein-$\Lambda $ equations in the presence of the modification Maxwell
(ModMax) and the modification anti-Maxwell (ModAMax) fields write in the
following form \cite{MM1} 
\begin{eqnarray}
G_{\mu \nu }+\Lambda g_{\mu \nu } &=&8\pi \mathrm{T}_{\mu \nu },  \label{eq1}
\\
&&  \notag \\
\partial _{\mu }\left( \sqrt{-g}\widetilde{E}^{\mu \nu }\right) &=&0,
\label{eq2}
\end{eqnarray}%
where $\mathrm{T}_{\mu \nu }$ is the energy-momentum tensor in the presence
of Mod(A)Max field which is given by 
\begin{equation}
8\pi \mathrm{T}^{\mu \nu }=2\eta \left( F^{\mu \sigma }F_{~~\sigma }^{\nu
}e^{-\gamma }\right) -2\eta e^{-\gamma }\mathcal{S}g^{\mu \nu },  \label{eq3}
\end{equation}%
and $\widetilde{E}_{\mu \nu }$ is defined as 
\begin{equation}
\widetilde{E}_{\mu \nu }=\frac{\partial \mathcal{L}}{\partial F^{\mu \nu }}%
=2\left( \mathcal{L}_{\mathcal{S}}F_{\mu \nu }\right) ,  \label{eq4}
\end{equation}%
where $\mathcal{L}_{\mathcal{S}}=\frac{\partial \mathcal{L}}{\partial 
\mathcal{S}}$.

For charged case, the equation (\ref{eq2}), turns to 
\begin{equation}
\partial _{\mu }\left( \sqrt{-g}e^{-\gamma }F^{\mu \nu }\right) =0.
\label{Maxwell Equation}
\end{equation}

We consider a topological four-dimensional static metric as 
\begin{equation}
ds^{2}=-f\left( r\right) dt^{2}+\frac{dr^{2}}{f(r)}+r^{2}d\Omega _{k}^{2},
\label{Metric}
\end{equation}%
where $f(r)$ is the metric function, which we are going to extract it. In
addition, $d\Omega _{k}^{2}$ is given by 
\begin{equation}
d\Omega _{k}^{2}=\left\{ 
\begin{array}{ccc}
d\theta ^{2}+\sin ^{2}\theta d\varphi ^{2} &  & k=1 \\ 
d\theta ^{2}+d\varphi ^{2} &  & k=0 \\ 
d\theta ^{2}+\sinh ^{2}\theta d\varphi ^{2} &  & k=-1%
\end{array}%
\right. .
\end{equation}

Applying the following gauge potential, we can have a radial electric field 
\begin{equation}
A_{\mu }=h(r)\delta _{\mu }^{t},  \label{gauge
potential}
\end{equation}%
and using the metric (\ref{Metric}) and the equation (\ref{Maxwell Equation}%
), we find the following relation 
\begin{equation}
2h^{\prime }(r)+rh^{\prime \prime }(r)=0,  \label{heq}
\end{equation}%
where the prime and double prime are, respectively, the first and second
derivatives with respect to $r$. By solving the equation (\ref{heq}), we
obtain 
\begin{equation}
h(r)=-\frac{Q}{r},  \label{h(r)}
\end{equation}%
where $q$ is an integration constant related to the electric charge.

To extract the metric function, $f(r)$, we investigate the equations (\ref%
{eq1}), (\ref{eq3}), (\ref{Metric}), and (\ref{h(r)}). We obtain the
following differential equations 
\begin{eqnarray}
&&eq_{tt}=eq_{rr}=rf^{\prime }(r)+f\left( r\right) +\Lambda r^{2}-k+\frac{%
\eta Q^{2}e^{-\gamma }}{r^{2}}=0,  \label{eqENMax1} \\
&&  \notag \\
&&eq_{\theta \theta }=eq_{\varphi \varphi }=f^{\prime \prime }(r)+\frac{2}{r}%
f^{\prime }(r)+2\Lambda -\frac{2\eta Q^{2}}{r^{4}}e^{-\gamma }=0,
\label{eqENMax2}
\end{eqnarray}%
which $eq_{tt}$, $eq_{rr}$, $eq_{\theta \theta }$, and $eq_{\varphi \varphi
} $ are representative $tt$, $rr$, $\theta \theta $, and $\varphi \varphi $
components of Eq. (\ref{eq1}), respectively. Applying the differential
equations (\ref{eqENMax1}) and (\ref{eqENMax2}), we can obtain an exact
solution for the metric function in the following form 
\begin{equation}
f(r)=k-\frac{m}{r}-\frac{\Lambda r^{2}}{3}+\frac{\eta G Q^{2}e^{-\gamma }}{%
r^{2}}=\left\{ 
\begin{array}{ccc}
k-\frac{m}{r}-\frac{\Lambda r^{2}}{3}+\frac{G Q^{2}e^{-\gamma }}{r^{2}}, &  & 
\text{ModMax} \\ 
&  &  \\ 
k-\frac{m}{r}-\frac{\Lambda r^{2}}{3}-\frac{G Q^{2}e^{-\gamma }}{r^{2}}, &  & 
\text{ModAMax}%
\end{array}%
\right. ,  \label{f(r)}
\end{equation}%
where $m$ is an integration constant that is related to the geometrical mass
of the black hole. Also, the metric function (\ref{f(r)}) satisfies all the
components of the field equation (\ref{eq1}), simultaneously. It is notable
that, $\eta =+1$ belongs to the ModMax solution and  i.e. 
\begin{equation}
f(r)= k-\frac{m}{r}-\frac{\Lambda r^{2}}{3}+\frac{q^{2}e^{-\gamma }}{r^{2}}.
\end{equation}%
Solving the lapse function for k=1 we obtain the mass $M$ of the black hole which is given as:
\begin{equation}
M=\frac{r}{2G} + \frac{r^3}{2 G l^2} + \frac{q^{2}e^{-\gamma }}{2r}.
\label{eqENM}
\end{equation}
 where, the integration constant, $m$  is replaced by $2G M$ and where $M$ is the mass of the ModMax black hole.
 \subsection{Bekenstein-Hawking Entropy}
 \label{subsec:2a}
The Bekenstein-Hawking entropy of the ModMax AdS black hole is given as:
\begin{eqnarray}
&&S= \frac{\pi r^2}{G}  \\
&&  \notag \\
&&r = \sqrt{\frac{GS}{\pi}},
\label{eqENr}
\end{eqnarray}%
Now substituting the value of r from eqtn.\ref{eqENr} to eqtn.\ref{eqENM} we get the mass of the ModMax AdS black hole, which is given as:

\begin{equation}
M=\frac{e^{-\gamma}\big(e^{\gamma} G S^{2}+l^{2}\pi(\pi Q^{2}+e^{\gamma} S)\big)}{2\sqrt{G}\,l^{2}\pi^{3/2}\sqrt{S}}
\label{eq:massadsbh}
\end{equation}

The temperature of the black hole is therefore given as:

\begin{equation}
T= \frac{d M}{d S}= \frac{e^{-\gamma}\big(3 e^{\gamma} G S^{2}-l^{2}\pi(\pi Q^{2}-e^{\gamma} S)\big)}{4\sqrt{G}\,l^{2}\pi^{3/2} S^{3/2}}
\end{equation}

and the specific heat capacity of the black hole is given by:

\begin{equation}
\mathcal{C}= T \frac{ d S}{d T}= \frac{6 e^{\gamma} G S^{3}-2 l^{2}\pi S(\pi Q^{2}-e^{\gamma} S)}{3 e^{\gamma} G S^{2}+l^{2}\pi(3\pi Q^{2}-e^{\gamma} S)}
\end{equation}

\begin{figure}[ht]
\begin{center}
\includegraphics[width=.52\textwidth]{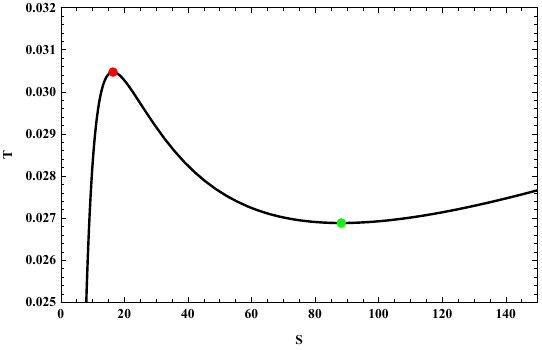}\vspace{3pt}
\caption{$T-S$ curve for the ModMax AdS black hole for $Q=2$, $\gamma=1$ and $l=10$.}\label{fig1}
\end{center}
\end{figure}

\begin{figure}[ht]
\begin{center}
\includegraphics[width=.52\textwidth]{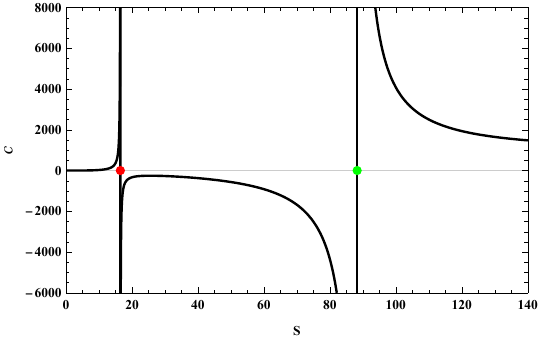}\vspace{3pt}
\caption{$\mathcal{C}-S$ curve for the ModMax AdS black hole for $Q=2$, $\gamma=1$ and $l=10$.}\label{fig2}
\end{center}
\end{figure}

We plot the temperature, T and specific heat capcity, C versus the Bekenstein-Hawking entropy, S for the ModMax AdS black hole as can be seen from Fig.\ref{fig1} and Fig.\ref{fig2} respectively, where we see for the T-S curve in Fig.\ref{fig1} there are visible peaks at $S=16.454$ and $S=88.265$  respectively, whereas for the C-S curve in Fig.\ref{fig2} we see potential discontinuities exactly at the same points as that of the peaks observed in the temperature curve of the black hole.

The GTD metric for the ModMax AdS black hole for the Bekenstein-Hawking entropy can be written as:
\begin{equation}
g  =S \left(\frac{\partial M}{\partial S}\right)\left(- \frac{\partial^2 M}{\partial S^2} dS^2  + \frac{\partial^2 M}{\partial \tilde{Q}^2} d\tilde{Q}^2 \right)
\end{equation}

\begin{figure}[ht]
\begin{center}
\includegraphics[width=.52\textwidth]{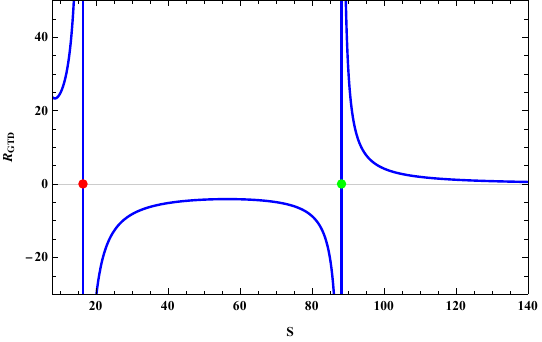}\vspace{3pt}
\caption{$R_{GTD}-S$ curve for the ModMax AdS black hole for $Q=2$, $\gamma=1$ and $l=10$.}\label{fig3}
\end{center}
\end{figure}

Now by substituting the value of M from eqtn.\ref{eqENM} we get the GTD metric from which we can easily calculate the GTD scalar $R_{GTD}$ given in appendix \ref{ap:1} and therefore plot it against the Bekenstein-Hawking entropy for the ModMax AdS black hole as can be seen from Fig.\ref{fig3} where we see that in accordance with the peak and trough observed in the T-S and the discontinuities observed in the C-S curves the scalar $R_{GTD}$ versus entropy, S curve produces discontinuities exactly at the same points i.e. $S=16.454$ and $S=88.265$. 

 \subsection{Renyi Entropy}
 \label{subsec:2b}
The Renyi entropy of the ModMax AdS black hole is given as:
\begin{eqnarray}
&&S_R=\frac{1}{\lambda} \ln\left[1 + \lambda \left(\frac{\pi r^2}{G}\right)\right]  \\
&&  \notag \\
&&r = \sqrt{G\left( \frac{e^{\lambda S_R} -1}{\lambda \pi}\right)},
\label{eqENRr}
\end{eqnarray}%
where $\lambda$ is the R\'enyi parameter and after making series expansion for small $\lambda$ and then substituting the value of r from eqtn.\ref{eqENRr} to eqtn.\ref{eqENM} we get the mass of the ModMax AdS black hole for R\'enyi entropy, which is given as:

\begin{equation}
M=\frac{e^{-\gamma}\big(e^{\gamma} G S_R^{2}(4+3 S \lambda)-l^{2}\pi\big(\pi Q^{2}(-4+S_R \lambda)-e^{\gamma} S_R(4+S_R \lambda)\big)\big)}{8\,l^{2}\pi^{3/2}\sqrt{G S_R}}
\label{eq:massadsrenyi}
\end{equation}

The temperature of the black hole is therefore given as:

\begin{equation}
T=\frac{e^{-\gamma} G\big(3 e^{\gamma} G S_R^{2}(4+5 S \lambda)-l^{2}\pi\big(\pi Q^{2}(4+S_R \lambda)-e^{\gamma} S_R(4+3 S_R \lambda)\big)\big)}{16\,l^{2}\pi^{3/2}(G S_R)^{3/2}}
\end{equation}

and the specific heat capacity of the black hole is given by:

\begin{equation}
\mathcal{C}=\frac{6 e^{\gamma} G S_R^{3}(4+5 S_R \lambda)-2 l^{2}\pi S\big(\pi Q^{2}(4+S_R \lambda)-e^{\gamma} S_R(4+3 S_R \lambda)\big)}{3 e^{\gamma} G S_R^{2}(4+15 S_R \lambda)+l^{2}\pi\big(\pi Q^{2}(12+S_R \lambda)+e^{\gamma} S_R(-4+3 S_R \lambda)\big)}
\end{equation}

\begin{figure}[ht]
\begin{center}
\includegraphics[width=.52\textwidth]{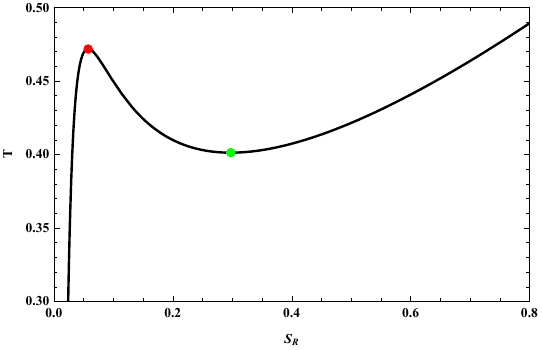}\vspace{3pt}
\caption{$T-S_R$ curve for the ModMax AdS black hole for R\'enyi entropy for $Q=0.2$, $\gamma=2$, $l=1$ and $\lambda =0.05$.}\label{fig4}
\end{center}
\end{figure}

\begin{figure}[ht]
\begin{center}
\includegraphics[width=.52\textwidth]{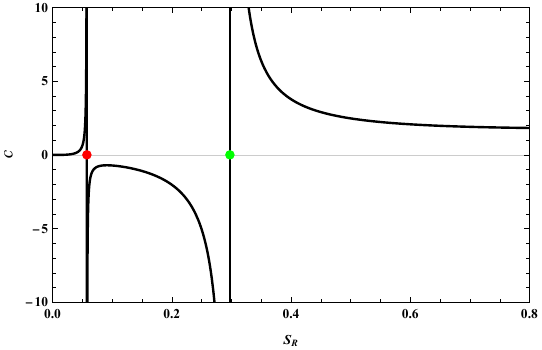}\vspace{3pt}
\caption{$\mathcal{C}-S_R$ curve for the ModMax AdS black hole for R\'enyi entropy for $Q=0.2$, $\gamma=2$, $l=1$ and $\lambda =0.05$.}\label{fig5}
\end{center}
\end{figure}

We plot the temperature, $T$ and specific heat capacity, $\mathcal{C}$ versus the R\'enyi entropy, $S_R$ for the ModMax AdS black hole as can be seen from Fig.\ref{fig4} and Fig.\ref{fig5} respectively, where we see for the $T-S_R$ curve in Fig.\ref{fig4} there are visible peaks at $S_R=0.054$ and $S_R=0.823$  respectively, whereas for the $\mathcal{C}-S_R$ curve in Fig.\ref{fig5} we see potential discontinuities exactly at the same points as that of the peaks observed in the temperature curve of the black hole.

The GTD metric for the ModMax AdS black hole for the Renyi entropy can be written as:
\begin{equation}
g  =S_R \left(\frac{\partial M}{\partial S_R}\right)\left(- \frac{\partial^2 M}{\partial S_R^2} dS_R^2  + \frac{\partial^2 M}{\partial \tilde{Q}^2} d\tilde{Q}^2 \right)
\end{equation}

\begin{figure}[ht]
\begin{center}
\includegraphics[width=.52\textwidth]{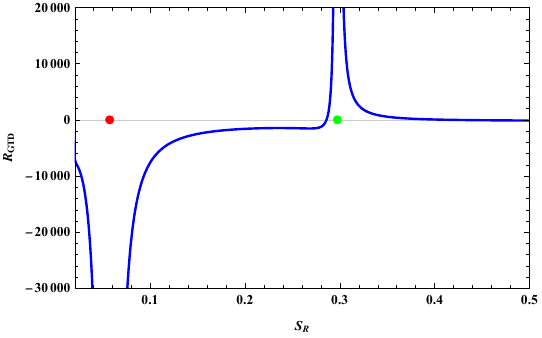}\vspace{3pt}
\caption{$R_{GTD}-S_R$ curve for the ModMax AdS black hole for R\'enyi entropy for $Q=0.2$, $\gamma=2$, $l=1$ and $\lambda =0.05$.}\label{fig6}
\end{center}
\end{figure}

Now by substituting the value of $M$ from \ref{eqENM} we get the GTD metric from which we can easily calculate the GTD scalar $R_{GTD}$ which is given in appendix \ref{ap:2} and therefore plot it against the R\'enyi entropy for the ModMax AdS black hole as can be seen from Fig.\ref{fig6} where we see that in accordance with the peak and trough observed in the $T-S_R$ and the discontinuities observed in the $\mathcal{C}-S_R$ curves the scalar $R_{GTD}$ versus entropy, $S_R$ curve produces discontinuities exactly at the same points i.e. $S_R=0.054$ and $S_R=0.823$. 

\subsection{Kaniadakis Entropy}
\label{subsec:2c}
The Kaniadakis entropy of the ModMax AdS black hole is given as:
\begin{eqnarray}
&&S_K=\frac{1}{\kappa} Sinh\left(\kappa \left(\frac{\pi r^2}{G}\right)\right)  \\
&&  \notag \\
&&r = \sqrt{G\left( \frac{ArcSinh(\kappa S_K)}{\kappa \pi}\right)},
\label{eqENKr}
\end{eqnarray}%
where $\kappa$ is the Kaniadakis parameter and after making series expansion for small $\kappa$ and then substituting the value of $r$ from eqtn.\ref{eqENKr} to eqtn.\ref{eqENM} we get the mass of the ModMax AdS black hole for Kaniadakis entropy, which is given as:

\begin{equation}
M=\frac{e^{-\gamma}\big(-3 e^{\gamma} G S_K^{2}(-4+S_K^{2}\kappa^{2})+l^{2}\pi\big(-e^{\gamma} S_K(-12+S_K^{2}\kappa^{2})+\pi Q^{2}(12+S_K^{2}\kappa^{2})\big)\big)}{24\,l^{2}\pi^{3/2}\sqrt{G S_K}}
\label{eq:masskania}
\end{equation}

The temperature of the black hole is therefore given as:

\begin{equation}
T=\frac{e^{-\gamma} G\big(-3 e^{\gamma} G S_K^{2}(-12+7 S_K^{2}\kappa^{2})+l^{2}\pi\big(3\pi Q^{2}(-4+S_K^{2}\kappa^{2})-e^{\gamma} S_K(-12+5 S_K^{2}\kappa^{2})\big)\big)}{48\,l^{2}\pi^{3/2}(G S_K)^{3/2}}
\end{equation}

and the specific heat capacity of the black hole is given by:

\begin{equation}
C=\frac{6 e^{\gamma} G S_K^{3}(-12+7 S_K^{2}\kappa^{2})-2 l^{2}\pi S_K\big(3\pi Q^{2}(-4+S_K^{2}\kappa^{2})-e^{\gamma} S_K(-12+5 S_K^{2}\kappa^{2})\big)}{3 e^{\gamma} G S_K^{2}(-12+35 S_K^{2}\kappa^{2})-3 l^{2}\pi\big(\pi Q^{2}(12+S_K^{2}\kappa^{2})-e^{\gamma} S_K(4+5 S_K^{2}\kappa^{2})\big)}
\end{equation}

\begin{figure}[ht]
\begin{center}
\includegraphics[width=.82\textwidth]{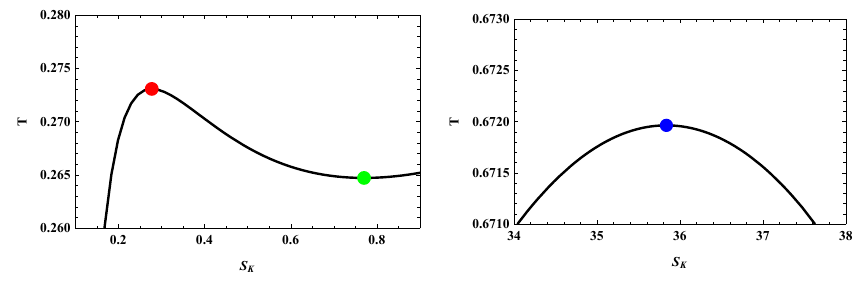}\vspace{3pt}
\caption{$T-S_K$ curve for the ModMax AdS black hole for Kaniadakis entropy for $Q=0.4$, $\gamma=2$, $l=1$ and $\kappa =0.016$.}\label{fig7}
\end{center}
\end{figure}

\begin{figure}[ht]
\begin{center}
\includegraphics[width=.82\textwidth]{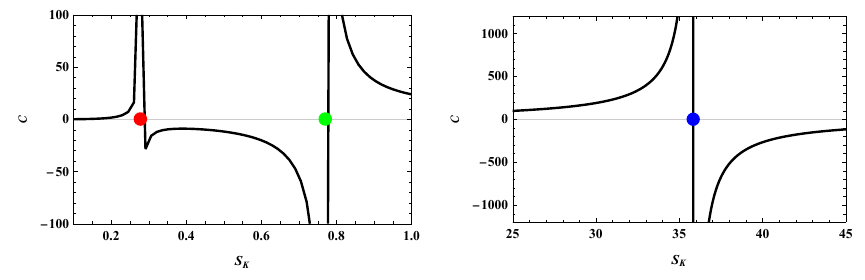}\vspace{3pt}
\caption{$\mathcal{C}-S_K$ curve for the ModMax AdS black hole for Kaniadakis entropy for $Q=0.4$, $\gamma=2$, $l=1$ and $\kappa =0.016$.}\label{fig8}
\end{center}
\end{figure}

We plot the temperature, $T$ and specific heat capcity, $\mathcal{C}$ versus the Kaniadakis entropy, $S_K$ for the ModMax AdS black hole as can be seen from Fig.\ref{fig7} and Fig.\ref{fig8} respectively, where we see for the $T-S_K$ curve in Fig.\ref{fig7} there are visible peaks at $S_K=0.278, 0.770$ and $S_K=35.837$  respectively, whereas for the $\mathcal{C}-S_K$ curve in Fig.\ref{fig8} we see potential discontinuities exactly at the same points as that of the peaks observed in the temperature curve of the black hole.

The GTD metric for the ModMax AdS black hole for the Kaniadakis entropy can be written as:
\begin{equation}
g  =S_K \left(\frac{\partial M}{\partial S_K}\right)\left(- \frac{\partial^2 M}{\partial S_K^2} dS_K^2  + \frac{\partial^2 M}{\partial \tilde{Q}^2} d\tilde{Q}^2 \right)
\end{equation}

\begin{figure}[ht]
\begin{center}
\includegraphics[width=.82\textwidth]{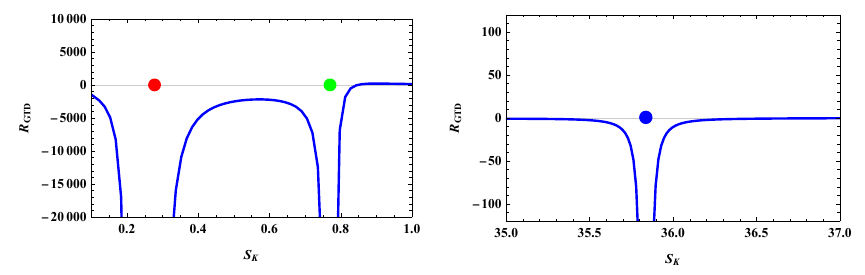}\vspace{3pt}
\caption{$R_{GTD}-S_K$ curve for the ModMax AdS black hole for Kaniadakis entropy for $Q=0.4$, $\gamma=2$, $l=1$ and $\kappa =0.016$.}\label{fig9}
\end{center}
\end{figure}

Now by substituting the value of $M$ from eqtn.\ref{eqENM} we get the GTD metric from which we can easily calculate the GTD scalar $R_{GTD}$ and is given in appendix \ref{ap:3} and therefore plot it against the Kaniadakis entropy for the ModMax AdS black hole as can be seen from Fig.\ref{fig9} where we see that in accordance with the peak and trough observed in the $T-S_K$ and the discontinuities observed in the $\mathcal{C}-S_K$ curves the scalar $R_{GTD}$ versus entropy, $S_K$ curve produces discontinuities exactly at the same points i.e. $S_K=0.278, 0.770$ and $S_K=35.837$. 

\section{CFT Thermodynamic Geometry fo ModMax AdS Black Holes}
\label{sec:3}
To incorporate the boundary central charge into the thermodynamic first law, we make use of the holographic duality that relates the AdS length scale $l$, Newton’s constant $G$, and the central charge $C$. This relation provides a natural framework for extending the thermodynamic phase space to include variations of the central charge and is given as
\begin{equation}
	C=\frac{\Omega_2 l^2}{16\pi G}
	\label{eq:centralcharge}
\end{equation}
In the context of CFT thermodynamics, we analyze a charged AdS black hole governed by ModMax nonlinear electrodynamics, with the mass identified as the internal energy of the system. The associated background spacetime is specified by the metric given below:

\begin{equation}
ds^2=\omega^2(-dt^2+L^2d\Omega_2^2)
\end{equation}
The line element of the two-sphere is written as $d\Omega_2^2$, and $\omega$ denotes a dimensionless conformal factor. In certain treatments, this factor is chosen as $\omega = R/l$, where $R$ characterizes the curvature radius. This parametrization enables a convenient implementation of the holographic first law, in which variations of the cosmological constant are permitted while Newton’s constant is kept fixed. Consequently, the spatial volume of the boundary sphere can be expressed as
\begin{equation}
\mathcal{V}=\Omega_2(\omega l)^2
\label{eq:volume}
\end{equation}
Within the CFT thermodynamic description, the definition of a volume $\mathcal{V}$ naturally allows for the introduction of an associated pressure $p$, giving rise to a work contribution of the form $-p d\mathcal{V}$. Employing the holographic correspondence, bulk thermodynamic variables including the mass, entropy, temperature, electric potential and charge can be systematically mapped to their corresponding quantities in the dual field theory as follows:
\begin{equation}
E=\frac{M}{\omega}, \quad \mathcal{T}=\frac{T}{\omega},\quad \mathcal{S}=S,\quad \varphi=\frac{\Phi \sqrt{G}}{\omega l},\quad \mathcal{Q}=\frac{Q l}{\sqrt{G}}
\label{eq:holographic}
\end{equation}
\subsection{Bekenstein-Hawking Entropy}
\begin{figure}[ht]
		\begin{center}
			\includegraphics[width=.52\textwidth]{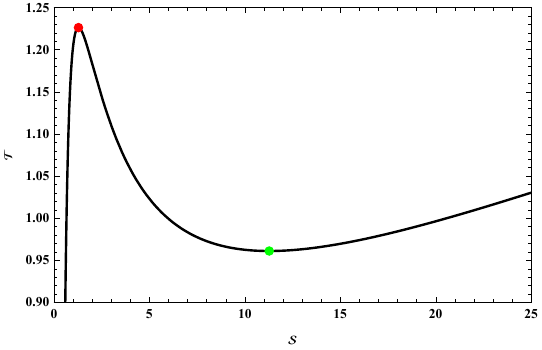}\vspace{3pt}
			\caption{$\mathcal{T}$--$\mathcal{S}$ curve for the dual CFT at fixed parameters $\mathcal{Q}=2$, $\gamma=1$, $C=3$ and $\mathcal{V}=1$
			}\label{fig:10}
		\end{center}
	\end{figure}

\begin{figure}[ht]
		\begin{center}
			\includegraphics[width=.52\textwidth]{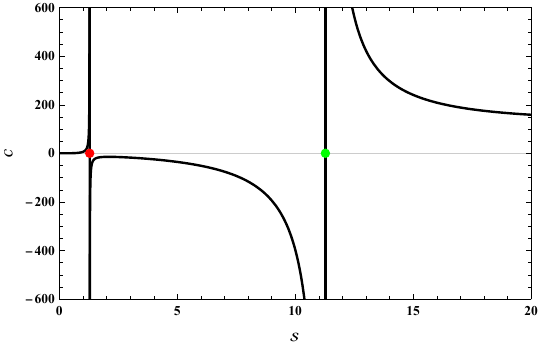}\vspace{3pt}
			\caption{$\mathcal{C}$--$\mathcal{S}$ curve for the dual CFT at fixed parameters $\mathcal{Q}=2$, $\gamma=1$, $C=3$ and $\mathcal{V}=1$}\label{fig:11}
		\end{center}
	\end{figure}

\begin{figure}[ht]
		\begin{center}
			\includegraphics[width=.52\textwidth]{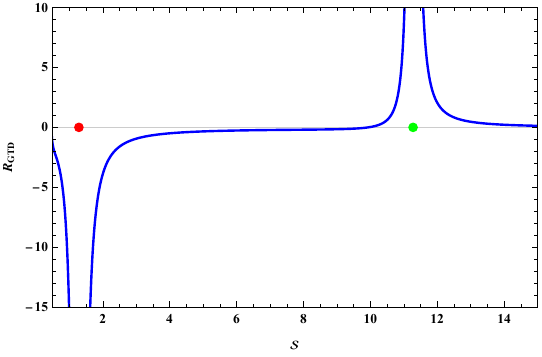}\vspace{3pt}
			\caption{$R_{\text{GTD}}$--$\mathcal{S}$ profile for the dual CFT at fixed parameters $\mathcal{Q}=2$, $\gamma=1$, $C=3$ and $\mathcal{V}=1$}
			\label{fig:12}
		\end{center}
	\end{figure}

By combining Eqs.~\eqref{eqENr}, \eqref{eq:massadsbh} \eqref{eq:centralcharge}, \eqref{eq:volume}, and \eqref{eq:holographic}, the energy of the dual CFT in Bekenstein-Hawking entropy can be expressed in the form
\begin{equation}
E=	\frac{e^{-\gamma } \left(4 \pi  C \mathcal{S} e^{\gamma }+\mathcal{S}^2 e^{\gamma }+\pi ^2 \mathcal{Q}^2\right)}{2 \pi  \sqrt{\mathcal{S}} \sqrt{C \mathcal{V}}}
\label{eq:energycft}
	\end{equation}
	The thermodynamic first law is related to the corresponding Smarr formula through the following relation:
	\begin{equation}
		\begin{split}
&dE=\mathcal{T}d\mathcal{S}+\varphi d\mathcal{Q}-p d\mathcal{V}+\mu dC\\
&E=\mathcal{T} \mathcal{S}+\varphi \mathcal{Q}+\mu dC
\end{split}
	\end{equation}
	We calculate the CFT temperature and is given as
	\begin{equation}
\mathcal{T}=\left(\frac{\partial E}{\partial \mathcal{S}}\right)=\frac{e^{-\gamma } \left(4 \pi  C \mathcal{S} e^{\gamma }+3 \mathcal{S}^2 e^{\gamma }-\pi ^2 \mathcal{Q}^2\right)}{4 \pi  \mathcal{S}^{3/2} \sqrt{C \mathcal{V}}}
	\end{equation}
	The heat capacity can be expressed in the following expression
	\begin{equation}
		\mathcal{C}=\frac{\partial E}{\partial\mathcal{T}}=\mathcal{T}\frac{\partial \mathcal{S}}{\partial \mathcal{T}}=\frac{2 \mathcal{S} \left(4 \pi  C \mathcal{S} e^{\gamma }+3 \mathcal{S}^2 e^{\gamma }-\pi ^2 \mathcal{Q}^2\right)}{-4 \pi  C \mathcal{S} e^{\gamma }+3 \mathcal{S}^2 e^{\gamma }+3 \pi ^2 \mathcal{Q}^2}
	\end{equation}
	The temperature $\mathcal{T}$ and specific heat capacity $\mathcal{C}$ of the CFT analogous to the ModMax AdS black hole are plotted as functions of the Bekenstein-Hawking entropy $\mathcal{S}$, as illustrated in Figures \ref{fig:10} and \ref{fig:11}, respectively. The $\mathcal{T}$--$\mathcal{S}$ profile shown in Figure \ref{fig:10} exhibits two pronounced extrema located at $\mathcal{S}=1.287$ and $S=11.278$. Notably, the $\mathcal{C}$--$\mathcal{S}$ curve displayed in Figure \ref{fig:11} develops clear discontinuities precisely at these same entropy values, coinciding with the extrema observed in the temperature curve. This correspondence signals the presence of critical thermodynamic behavior at these points.\\
	Within the CFT thermodynamic framework, the geometrothermodynamic (GTD) metric associated with the Bekenstein--Hawking entropy in the ModMax setup can be expressed as
	\begin{equation}
	\tilde{g}=\mathcal{S}\left(\frac{\partial E}{\partial \mathcal{S}}\right)\left(-\frac{\partial^2E}{\partial \mathcal{S}^2}d\mathcal{S}^2+\frac{\partial^2E}{\partial \mathcal{Q}^2}d\mathcal{Q}^2\right)
	\end{equation}
	By substituting the expression for $E$ from Eq. \ref{eq:energycft} into the geometrothermodynamic metric, we obtain the explicit GTD line element, from which the corresponding GTD scalar curvature $R_{GTD}$ can be straightforwardly evaluated and is given in appendix \ref{ap:4}. We then examine the behavior of this scalar as a function of the Bekenstein--Hawking entropy $\mathcal{S}$ within the CFT framework. As illustrated in Figure \ref{fig:12}, the $R_{\text{GTD}}$--$\mathcal{S}$ profile develops distinct singular features at $\mathcal{S}=1.287$ and $S=11.278$. Remarkably, these locations coincide precisely with the extrema observed in the $\mathcal{T}$--$\mathcal{S}$ diagram and the discontinuities present in the $\mathcal{C}$--$\mathcal{S}$ curve, thereby reinforcing the consistency of the thermodynamic and geometric descriptions of the phase structure.
	\subsection{R\'enyi Entropy}
	Making use of Equations~\eqref{eqENRr}, \eqref{eq:massadsrenyi} \eqref{eq:centralcharge}, \eqref{eq:volume}, and \eqref{eq:holographic}, the energy of the holographically dual CFT formulated in terms of R\'enyi entropy can be written as
	\begin{equation}
	E=\frac{\sqrt{\frac{\mathcal{S}_R}{C}} e^{-\gamma } \left(4 \pi  C \mathcal{S}_R e^{\gamma } (\lambda  \mathcal{S}_R+4)+\mathcal{S}_R^2 e^{\gamma } (3 \lambda  \mathcal{S}_R+4)+\pi ^2 \mathcal{Q}^2 (4-\lambda  \mathcal{S}_R)\right)}{8 \pi  \mathcal{S}_R \sqrt{\mathcal{V}}}
	\label{eq:energycftrenyi}
	\end{equation}
	The corresponding temperature of the dual CFT incorporating the R\'enyi entropy is thus obtained as
	\begin{equation}
\mathcal{T}=	\frac{e^{-\gamma } \left(4 \pi  C \mathcal{S}_R e^{\gamma } (3 \lambda  \mathcal{S}_R+4)+3 \mathcal{S}_R^2 e^{\gamma } (5 \lambda  \mathcal{S}_R+4)-\pi ^2 \mathcal{Q}^2 (\lambda  \mathcal{S}_R+4)\right)}{16 \pi  C^2 \sqrt{\mathcal{V}} \left(\frac{\mathcal{S}_R}{C}\right)^{3/2}}
	\end{equation}
	The specific heat of the dual CFT is given as
	\begin{equation}
\mathcal{C}=\frac{2 \mathcal{S}_R \left(4 \pi  C \mathcal{S}_R e^{\gamma } (3 \lambda  \mathcal{S}_R+4)+3 \mathcal{S}_R^2 e^{\gamma } (5 \lambda  \mathcal{S}_R+4)-\pi ^2 \mathcal{Q}^2 (\lambda  \mathcal{S}_R+4)\right)}{4 \pi  C \mathcal{S}_R e^{\gamma } (3 \lambda  \mathcal{S}_R-4)+3 \mathcal{S}_R^2 e^{\gamma } (15 \lambda  \mathcal{S}+4)+\pi ^2 \mathcal{Q}^2 (\lambda  \mathcal{S}_R+12)}
	\end{equation}
	\begin{figure}[ht]
		\begin{center}
			\includegraphics[width=.52\textwidth]{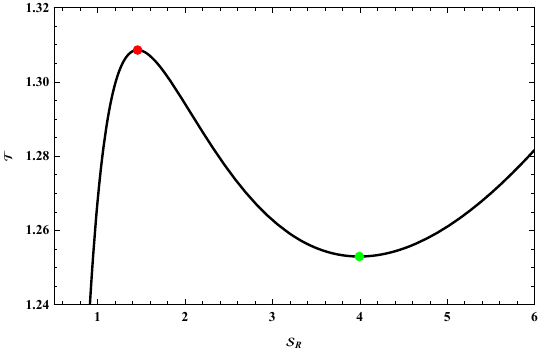}\vspace{3pt}
			\caption{$\mathcal{T}$--$\mathcal{S}_R$ curve for the dual CFT using R\'enyi entropy at fixed parameters $\mathcal{Q}=2$, $\gamma=1$, $C=3$, $\lambda=0.05$ and $\mathcal{V}=1$
			}\label{fig:13}
		\end{center}
	\end{figure}
		The temperature $\mathcal{T}$ and specific heat $\mathcal{C}$ of the dual CFT, formulated using R\'enyi entropy, are examined as functions of the Rényi entropy $\mathcal{S}$, as depicted in Figs.~\ref{fig:13} and \ref{fig:14}, respectively. As seen from the	 $\mathcal{T}$--$\mathcal{S}_R$ curve in Fig.~\ref{fig:13}, the temperature develops two distinct extremal points located at $\mathcal{S}_R=1.457$ and $\mathcal{S}_R=3.995$. Correspondingly, the $\mathcal{C}$--$\mathcal{S}_R$ plot shown in Fig.~\ref{fig:14} exhibits sharp discontinuities precisely at these same entropy values. The exact alignment of the temperature extrema with the divergences in the specific heat provides strong evidence for the emergence of the thermodynamic behavior within the R\'enyi incorporated CFT framework.
	\begin{figure}[ht]
		\begin{center}
			\includegraphics[width=.52\textwidth]{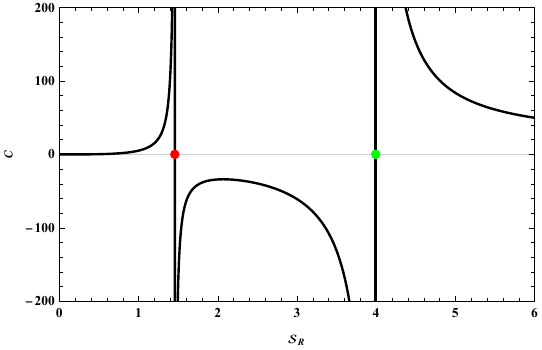}\vspace{3pt}
			\caption{$\mathcal{C}$--$\mathcal{S}_R$ curve for the dual CFT using R\'enyi entropy at fixed parameters $\mathcal{Q}=2$, $\gamma=1$, $C=3$, $\lambda=0.05$ and $\mathcal{V}=1$}\label{fig:14}
		\end{center}
	\end{figure}
	Within the CFT thermodynamic formalism, the geometrothermodynamic (GTD) metric corresponding to the R\'enyi entropy in the ModMax-deformed setup may be written as
	\begin{equation}
		\tilde{g}=\mathcal{S}_{\mathrm{R}}\left(\frac{\partial E}{\partial \mathcal{S}_{\mathrm{R}}}\right)
		\left(
		-\frac{\partial^2 E}{\partial \mathcal{S}_{\mathrm{R}}^2}\, d\mathcal{S}_{\mathrm{R}}^2
		+\frac{\partial^2 E}{\partial \mathcal{Q}^2}\, d\mathcal{Q}^2
		\right).
	\end{equation}
	Substituting the CFT energy $E$ from Equation~\eqref{eq:energycftrenyi} into the above expression yields the explicit form of the GTD metric, from which the associated scalar curvature can be computed directly. The resulting curvature scalar $R_{GTD}$ which is given in \ref{ap:5} is then analyzed as a function of the R\'enyi entropy $\mathcal{S}_R$ within the CFT description. As shown in Figure~\ref{fig:15}, the $R_{\mathrm{GTD}}$--$\mathcal{S}_R$ curve exhibits pronounced divergences at  $\mathcal{S}_R=1.457$ and $\mathcal{S}_R=3.995$. These singular points occur at the same entropy values where the $\mathcal{T}$--$\mathcal{S}_R$ plot displays extremal behavior and the $\mathcal{C}$--$\mathcal{S}_R$ curve develops discontinuities, thereby providing a coherent geometric interpretation of the R\'enyi-deformed CFT phase structure.
		\begin{figure}[ht]
		\begin{center}
			\includegraphics[width=.52\textwidth]{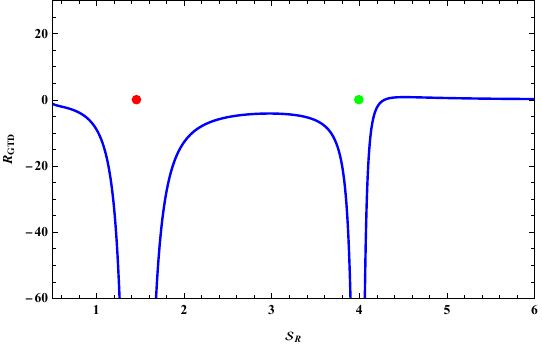}\vspace{3pt}
			\caption{$R_{\text{GTD}}$--$\mathcal{S}_R$ profile for the dual CFT using R\'enyi entropy at fixed parameters $\mathcal{Q}=2$, $\gamma=1$, $C=3$, $\lambda=0.05$ and $\mathcal{V}=1$}
			\label{fig:15}
		\end{center}
	\end{figure}
	
	\subsection{Kaniadakis Entropy}
	Combining Eqs.~\eqref{eqENKr}, \eqref{eq:masskania}, \eqref{eq:centralcharge}, \eqref{eq:volume}, and \eqref{eq:holographic}, the resulting form of the energy in the dual CFT, formulated in terms of Kaniadakis entropy, takes the form
	\begin{equation}
E=\frac{e^{-\gamma } \left(-4 \pi C \mathcal{S}_K e^{\gamma } \left(\kappa ^2 \mathcal{S}_K^2-12\right)-3 \mathcal{S}_K^2 e^{\gamma } \left(\kappa ^2 \mathcal{S}_K^2-4\right)+\pi ^2 \mathcal{Q}^2 \left(\kappa ^2 \mathcal{S}_K^2+12\right)\right)}{24 \pi  C \sqrt{\mathcal{V}} \sqrt{\frac{\mathcal{S}_K}{C}}}
\label{eq:energykaniacft}
	\end{equation}
	The resulting temperature of the dual CFT, formulated within the Kaniadakis entropy framework, is therefore given by
	\begin{equation}
\mathcal{T}=\frac{e^{-\gamma } \left(-4 \pi  C \mathcal{S}_K e^{\gamma } \left(5 \kappa ^2 \mathcal{S}_K^2-12\right)-3 \mathcal{S}_K^2 e^{\gamma } \left(7 \kappa ^2 \mathcal{S}_K^2-12\right)+3 \pi ^2 \mathcal{Q}^2 \left(\kappa ^2 \mathcal{S}_K^2-4\right)\right)}{48 \pi  C^2 \sqrt{\mathcal{V}} \left(\frac{\mathcal{S}_K}{C}\right)^{3/2}}
	\end{equation}
	The Specific Heat is given as
	\begin{equation}
\mathcal{C}=\frac{2 \mathcal{S}_K \left(4 \pi  C \mathcal{S}_K e^{\gamma } \left(5 \kappa ^2\mathcal{S}_K^2-12\right)+3 \mathcal{S}_K^2 e^{\gamma } \left(7 \kappa ^2 \mathcal{S}_K^2-12\right)-3 \pi ^2 \mathcal{Q}^2 \left(\kappa ^2 \mathcal{S}_K^2-4\right)\right)}{12 \pi  C \mathcal{S}_K e^{\gamma } \left(5 \kappa ^2 \mathcal{S}_K^2+4\right)+3 \mathcal{S}_K^2 e^{\gamma } \left(35 \kappa ^2 \mathcal{S}_K^2-12\right)-3 \pi ^2 \mathcal{Q}^2 \left(\kappa ^2 \mathcal{S}_K^2+12\right)}
	\end{equation}
	\begin{figure}[ht]
		\begin{center}
			\includegraphics[width=.82\textwidth]{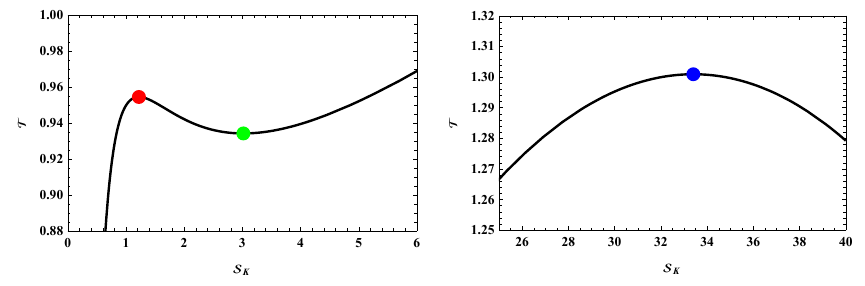}\vspace{3pt}
			\caption{$\mathcal{T}$--$\mathcal{S}_K$ curve for the dual CFT using Kaniadakis entropy at fixed parameters $\mathcal{Q}=1$, $\gamma=1$, $C=1$, $\kappa=0.016$ and $\mathcal{V}=1$
			}\label{fig:16}
		\end{center}
	\end{figure}
	\begin{figure}[ht]
		\begin{center}
			\includegraphics[width=.82\textwidth]{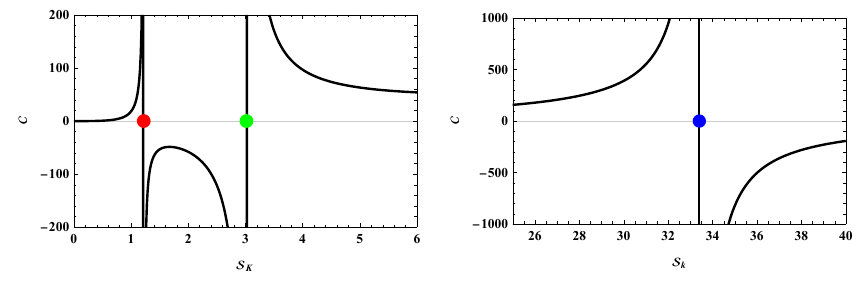}\vspace{3pt}
			\caption{$\mathcal{C}$--$\mathcal{S}_K$ curve for the dual CFT using Kaniadakis entropy at fixed parameters $\mathcal{Q}=1$, $\gamma=1$, $C=1$, $\kappa=0.016$ and $\mathcal{V}=1$
			}\label{fig:17}
		\end{center}
	\end{figure}
	The temperature $\mathcal{T}$ and specific heat capacity $\mathcal{C}$ of the dual CFT, formulated within the Kaniadakis entropy framework, are analyzed as shown in Figs.~\ref{fig:16} and \ref{fig:17}, respectively. The $\mathcal{T}$--$\mathcal{S}$ plot in Fig.~\ref{fig:16} reveals the presence of three distinct extrema occurring at $\mathcal{S}_K = 1.222$, $3.018$, and $33.404$. In parallel, the $\mathcal{C}$--$\mathcal{S}_K$ curve displayed in Fig.~\ref{fig:17} exhibits pronounced discontinuities precisely at these same entropy values. The exact correspondence between the temperature extrema and the divergences in the specific heat indicates the emergence of critical thermodynamic behavior in the Kaniadakis-deformed CFT description.\\
	Within the CFT thermodynamic description, the geometrothermodynamic (GTD) metric corresponding to the Kaniadakis entropy in the ModMax-deformed setup can be expressed as
		\begin{equation}
		\tilde{g}=\mathcal{S}_{\mathrm{K}}\left(\frac{\partial E}{\partial \mathcal{S}_{\mathrm{K}}}\right)
		\left(
		-\frac{\partial^2 E}{\partial \mathcal{S}_{\mathrm{K}}^2}\, d\mathcal{S}_{\mathrm{K}}^2
		+\frac{\partial^2 E}{\partial \mathcal{Q}^2}\, d\mathcal{Q}^2
		\right).
	\end{equation}
		\begin{figure}[ht]
		\begin{center}
			\includegraphics[width=0.82\textwidth]{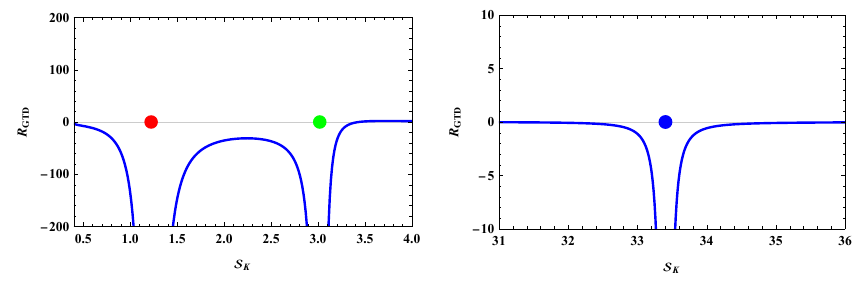}\vspace{3pt}
			\caption{$R_\mathrm{GTD}$--$\mathcal{S}_\mathrm{K}$ curve for the dual CFT using Kaniadakis entropy at fixed parameters $\mathcal{Q}=1$, $\gamma=1$, $C=1$, $\kappa=0.016$ and $\mathcal{V}=1$
			}\label{fig:18}
		\end{center}
	\end{figure}
	By inserting the expression for the energy $E$ from Eq.~\eqref{eq:energykaniacft} into the geometrothermodynamic formalism, the explicit GTD metric of the dual CFT can be constructed. This allows for a direct evaluation of the associated GTD scalar curvature $R_{GTD}$ which is given in appendix \ref{ap:6}, which is subsequently examined as a function of the Kaniadakis entropy $\mathcal{S}_{\mathrm{K}}$. The resulting behavior of the curvature scalar is displayed in Fig.~\ref{fig:18}. One observes that the $R_{\mathrm{GTD}}$--$\mathcal{S}_{\mathrm{K}}$ profile develops pronounced singularities at $\mathcal{S}_{\mathrm{K}}=1.222$, $3.018$, and $33.404$. Notably, these entropy values coincide precisely with the extrema identified in the $\mathcal{T}$--$\mathcal{S}_{\mathrm{K}}$ diagram and the discontinuities appearing in the $\mathcal{C}$--$\mathcal{S}_{\mathrm{K}}$ curve, thereby providing a consistent geometric characterization of the critical structure in the Kaniadakis-deformed CFT thermodynamics.
	
	\section{NED-AdS Black Hole Solution}
	\label{sec:4}
	
	We begin with the action of NED-AdS black holes \cite{FII}
	\begin{equation}
		I=\int d^{4}x\sqrt{-g}\left(\frac{R-2\Lambda}{16\pi G_N}+\mathcal{L}(\mathcal{F}) \right),
		\label{2.1}
	\end{equation}
	where $\Lambda=-3/l^2$ is the negative cosmological constant, $G_N$ is Newton's constant and $l$ is the AdS radius. The NED Lagrangian \cite{Bronnikov} is given by
	\begin{equation}
		{\cal L}(\mathcal{F}) =-\frac{{\cal F}}{4\pi\cosh^2\left(a\sqrt[4]{2|{\cal F}|}\right)},
		\label{2.2}
	\end{equation}
	where $a$ is a coupling constant and ${\cal F}=F^{\mu\nu}F_{\mu\nu}/4=(B^2-E^2)/2$ is the field invariant. From action (\ref{2.1}) one can obtain the gravitation and electromagnetic fields equations
	\begin{equation}
		R_{\mu\nu}-\frac{1}{2}g_{\mu \nu}R+\Lambda g_{\mu \nu} =8\pi G_N T_{\mu \nu},
		\label{2.3}
	\end{equation}
	\begin{equation}
		\partial _{\mu }\left( \sqrt{-g}\mathcal{L}_{\mathcal{F}}F^{\mu \nu}\right)=0.
		\label{2.4}
	\end{equation}
	Here, $R$ and $R_{\mu \nu }$  are the Ricci scalar and the Ricci tensor, respectively. The stress tensor of electromagnetic fields is given as:
	\begin{equation}
		T_{\mu\nu }=F_{\mu\rho }F_{\nu }^{~\rho }\mathcal{L}_{\mathcal{F}}+g_{\mu \nu }\mathcal{L}\left( \mathcal{F}\right),
		\label{2.5}
	\end{equation}
	where $\mathcal{L}_{\mathcal{F}}=\partial \mathcal{L}( \mathcal{F})/\partial \mathcal{F}$.
	We here analyse the space-time with spherical symmetry that has the line element squared, given by:
	\begin{equation}
		ds^{2}=-f(r)dt^{2}+\frac{1}{f(r)}dr^{2}+r^{2}\left( d\theta^{2}+\sin ^{2}\theta d\phi ^{2}\right).
		\label{2.6}
	\end{equation}
	We here consider only magnetized black holes as electrically charged black holes for NED has the Maxwell weak-field limit and therefore leads to singularities \cite{Bronnikov}. Now the tensor $F_{\mu\nu}$ possesses the radial electric field $F_{01}=-F_{10}$ and radial
	magnetic field $F_{23}=-F_{32}=q\sin(\theta)$, where $q$ is the magnetic charge. The stress tensor is seen to be diagonal, $T_{0}^{~0}=T_{r}^{~r}$ and $T_{\theta}^{~\theta}=T_{\phi}^{~\phi}$. One can then find the metric function from the relation \cite{Bronnikov} given as:
	\begin{equation}
		f(r)=1-\frac{2m(r)G_N}{r}.
		\label{2.7}
	\end{equation}
	The mass function is given by
	\begin{equation}
		m(r)=m_0+4\pi\int_{0}^{r}\rho(r)r^{2}dr,
		\label{2.8}
	\end{equation}
	where the integration constant $m_0$ refers to the Schwarzschild mass and $\rho(r)$ is the energy density. It is important to mention here that $\rho(r)$ in \ref{2.8} includes the term which is due to the cosmological constant.
	We here study the static magnetic black holes with the field invariant $\mathcal{F}=Q^2/(2r^4)$. Therefore, the black hole is considered as the magnetic monopole with the magnetic induction field $B=Q/r^2$. Making use of Eq. \ref{2.5}, we find the magnetic energy density including the term corresponding to the negative cosmological constant
	\begin{equation}
		\rho=\frac{Q}{8\pi r^4\cosh^2(b/r)}-\frac{3}{8\pi G_Nl^2},
		\label{2.9}
	\end{equation}
	where for brevity we use parameter $b=a\sqrt{q}$.
	By use of Eqs. \ref{2.8} and \ref{2.9} we obtain the mass function
	\begin{equation}
		m(r)=m_0+\frac{Q^2}{2b}\left[1-\tanh\left(\frac{b}{r}\right)\right]-\frac{r^3}{2G_Nl^2}.
		\label{2.10}
	\end{equation}
	One can then find the finite magnetic mass of black holes
	\begin{equation}
		m_M=\int_0^\infty \frac{Qdr}{2r^2\cosh^2(b/r)}=\frac{Q^2}{2b}.
		\label{2.11}
	\end{equation}
	Then from Eqs. \ref{2.10} and \ref{2.11} we obtain the mass function
	\begin{equation}
		m(r)=M-\frac{Q^2}{2b}\tanh\left(\frac{b}{r}\right)-\frac{r^3}{2G_Nl^2}.
		\label{2.12}
	\end{equation}
	The ADM black hole mass is given by $M=m_0+m_M$. One can observe that at the Maxwell limit, $b=0$, the magnetic energy is found to be infinite.
	With the help of Eqs. \ref{2.7} and \ref{2.10} one finds the metric function
	\begin{equation}
		f(r)=1-\frac{2MG}{r}+\frac{Q^2G}{br}\tanh\left(\frac{b}{r}\right)+\frac{r^2}{l^2}.
		\label{2.13}
	\end{equation}
	Solving the lapse function by equating it to zero i.e.($f(r)=0$) we get the mass of the black hole:
	\begin{equation}
		M=\frac{r}{2 G}+\frac{Q^2}{2 b}\tanh\left(\frac{b}{r}\right)+\frac{r^3}{2 G l^2}.
		\label{2.14}
	\end{equation}
	Now, a series expansion of $\tanh\left(\frac{b}{r}\right)$ for $b < r$ is given as:
	\begin{equation}
		\tanh\left(\frac{b}{r}\right)= \frac{b}{r} - \frac{b^3}{3 r^3} + \frac{2b^5}{15r^5}......
	\end{equation}
	Neglecting terms in the expansion which have order greater than five we get:
	\begin{equation}
		M=\frac{r}{2 G}+\frac{Q^2}{2 b}\left(\frac{b}{r} - \frac{b^3}{3 r^3} + \frac{2b^5}{15r^5}\right)+\frac{r^3}{2 G l^2}.
		\label{2.15}
	\end{equation}
	
	\subsection{Bekenstein-Hawking Entropy}
	Substituting the value of r from eqtn.\ref{eqENr} to eqtn.\ref{2.15} we get the mass of the NED AdS black hole, which is given as:
	
	\begin{equation}
		M =
		\frac{
			2 b^{4} l^{2} \pi^{4} Q^{2}
			- 5 b^{2} G l^{2} \pi^{3} Q^{2} S
			+ 15 G^{2} S^{2} \left( G S^{2} + l^{2} \pi (\pi Q^{2} + S) \right)
		}{
			30 G^{5/2} l^{2} \pi^{3/2} S^{5/2}
		}
		\label{eq:massnedBH}
	\end{equation}

	The temperature of the black hole is therefore given as:
	
	\begin{equation}
		T =
		\frac{
			- 2 b^{4} l^{2} \pi^{4} Q^{2}
			+ 3 b^{2} G l^{2} \pi^{3} Q^{2} S
			+ 3 G^{2} S^{2} \left( 3 G S^{2} + l^{2} \pi (-\pi Q^{2} + S) \right)
		}{
			12 G^{5/2} l^{2} \pi^{3/2} S^{7/2}
		}
	\end{equation}

	and the specific heat capacity of the black hole is given by:
	
	\begin{equation}
		\mathcal{C} =
		\frac{
			2 S \left(
			- 2 b^{4} l^{2} \pi^{4} Q^{2}
			+ 3 b^{2} G l^{2} \pi^{3} Q^{2} S
			+ 3 G^{2} S^{2} \left( 3 G S^{2} + l^{2} \pi (-\pi Q^{2} + S) \right)
			\right)
		}{
			14 b^{4} l^{2} \pi^{4} Q^{2}
			- 15 b^{2} G l^{2} \pi^{3} Q^{2} S
			+ 3 G^{2} S^{2} \left( l^{2} \pi (3 \pi Q^{2} - S) + 3 G S^{2} \right)
		}
	\end{equation}

\begin{figure}[ht]
		\begin{center}
			\includegraphics[width=.52\textwidth]{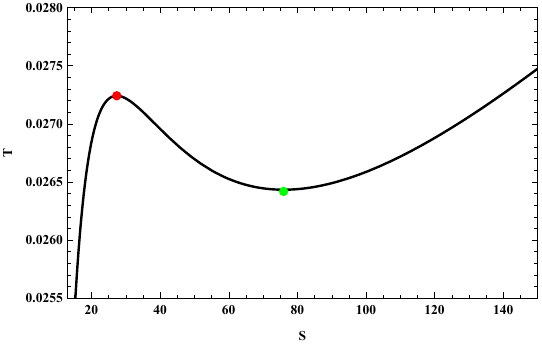}\vspace{3pt}
			\caption{$T-S$ curve for the NED AdS black hole for $Q=1.5$, $b=0.5$ and $l=10$.}\label{fig10}
		\end{center}
	\end{figure}

\begin{figure}[ht]
		\begin{center}
			\includegraphics[width=.52\textwidth]{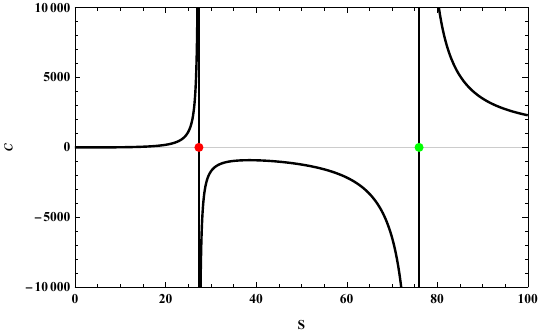}\vspace{3pt}
			\caption{$C-S$ curve for the NED AdS black hole for $Q=1.5$, $b=0.5$ and $l=10$.}\label{fig11}
		\end{center}
	\end{figure}

	We plot the temperature, $T$ and specific heat capcity, $\mathcal{C}$ versus the Bekenstein-Hawking entropy, $S$ for the NED AdS black hole as can be seen from Fig.\ref{fig10} and Fig.\ref{fig11} respectively, where we see for the $T-S$ curve in Fig.\ref{fig10} there are visible peaks at $S=27.351$ and $S=75.998$  respectively, whereas for the $\mathcal{C}-S$ curve in Fig.\ref{fig11} we see potential discontinuities exactly at the same points as that of the peaks observed in the temperature curve of the black hole.
	
	The GTD metric for the NED AdS black hole for the Bekenstein-Hawking entropy can be written as:
	\begin{equation}
		g  =S \left(\frac{\partial M}{\partial S}\right)\left(- \frac{\partial^2 M}{\partial S^2} dS^2  + \frac{\partial^2 M}{\partial \tilde{Q}^2} d\tilde{Q}^2 \right)
	\end{equation}

\begin{figure}[ht]
		\begin{center}
			\includegraphics[width=.52\textwidth]{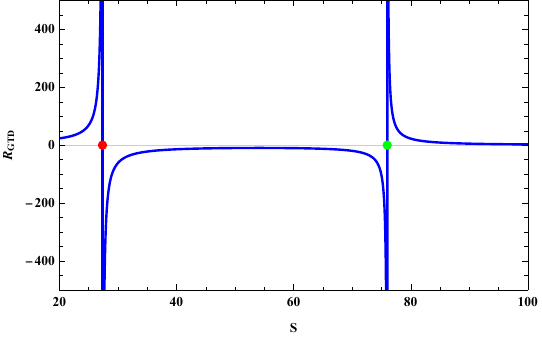}\vspace{3pt}
			\caption{$R_{GTD}-S$ curve for the NED AdS black hole for $Q=1.5$, $b=0.5$ and $l=10$.}\label{fig12}
		\end{center}
	\end{figure}

	Now by substituting the value of M from above we get the GTD metric from which we can easily calculate the GTD scalar $R_{GTD}$ given in appendix \ref{ap:7} and therefore plot it against the Bekenstein-Hawking entropy for the NED AdS black hole as can be seen from Fig.\ref{fig12} where we see that in accordance with the peak and trough observed in the $T-S$ and the discontinuities observed in the $\mathcal{C}-S$ curves the scalar $R_{GTD}$ versus entropy, S curve produces discontinuities exactly at the same points i.e. $S=27.351$ and $S=75.998$ . 
	
	\subsection{R\'enyi Entropy}
	After making the series expansion for small $\lambda$ and then substituting the value of r from eqtn.\ref{eqENRr} to eqtn.\ref{2.15} we get the mass of the NED AdS black hole for R\'enyi entropy, which is given as:
	
	\begin{equation}
		M =
		\frac{
			2 b^{4} l^{2} \pi^{4} Q^{2} (4 - 5 S_R \lambda)
			+ 5 b^{2} G l^{2} \pi^{3} Q^{2} S_R (-4 + 3 S_R \lambda)
			+ 15 G^{2} S_R^{2} \left(
			G S^{2} (4 + 3 S_R \lambda)
			+ \mathcal{A}
			\right)
		}{
			120 l^{2} \pi^{3/2} (G S_R)^{5/2}
		}
		\label{eq:massnedrenyi}
	\end{equation}
	where, $\mathcal{A}=l^{2} \pi \left( \pi Q^{2} (4 -S_R \lambda)
	+ S (4 + S_R \lambda) \right)$. The temperature of the black hole is therefore given as:
	
	\begin{equation}
		T =
		\frac{
			G \left(
			- 3 b^{2} G l^{2} \pi^{3} Q^{2} S_R (-4 + S_R \lambda)
			+ 2 b^{4} l^{2} \pi^{4} Q^{2} (-4 + 3 S_R \lambda)
			+ 3 G^{2} S_R^{2} \left(
			3 G S_R^{2} (4 + 5 S_R \lambda)
			+ \mathcal{B}\right)
			\right)
		}{
			48 l^{2} \pi^{3/2} (G S_R)^{7/2}
		}
	\end{equation}

	where, $\mathcal{B}=l^{2} \pi \left(
	- \pi Q^{2} (4 + S_R \lambda)
	+ S_R (4 + 3S_R \lambda)
	\right)$ and the specific heat capacity of the black hole is given by:
	
	\begin{equation}
		\mathcal{C} =
		\frac{
			2 S_R \left(
			- 3 b^{2} G l^{2} \pi^{3} Q^{2} S_R (-4 + S_R \lambda)
			+ 2 b^{4} l^{2} \pi^{4} Q^{2} (-4 + 3 S_R \lambda)
			+ 3 G^{2} S_R^{2} \left(
			3 G S_R^{2} (4 + 5 S_R \lambda)
			+ \mathcal{B}
			\right)
			\right)
		}{
			2 b^{4} l^{2} \pi^{4} Q^{2} (28 - 15 S_R \lambda)
			+ 3 b^{2} G l^{2} \pi^{3} Q^{2} S_R (-20 + 3 S_R \lambda)
			+ 3 G^{2} S_R^{2} \left(
			3 G S_R^{2} (4 + 15 S_R \lambda)
			+ \mathcal{D}
			\right)
		}
	\end{equation}
	where, $\mathcal{D}=l^{2} \pi \left(
	\pi Q^{2} (12 + S_R \lambda)
	+S_R (-4 + 3 S_R \lambda)
	\right)$

\begin{figure}[ht]
		\begin{center}
			\includegraphics[width=.52\textwidth]{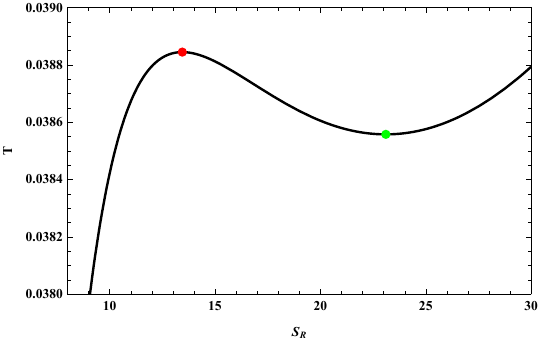}\vspace{3pt}
			\caption{$T-S_R$ curve for the NED AdS black hole for R\'enyi entropy for $Q=1$, $b=1$, $l=10$ and $\lambda =0.01$.}\label{fig13}
		\end{center}
	\end{figure}

\begin{figure}[ht]
		\begin{center}
			\includegraphics[width=.52\textwidth]{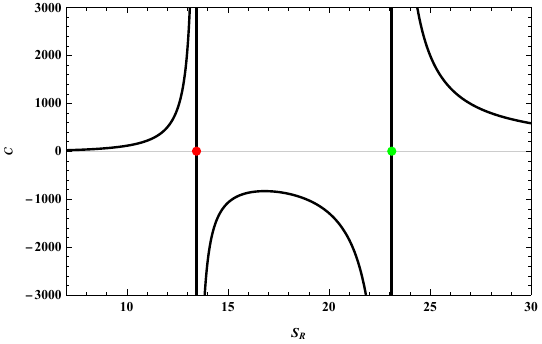}\vspace{3pt}
			\caption{$\mathcal{C}-S_R$ curve for the NED AdS black hole for R\'enyi entropy for $Q=1$, $b=1$, $l=10$ and $\lambda =0.01$.}\label{fig14}
		\end{center}
	\end{figure}

	We plot the temperature, $T$ and specific heat capcity, $\mathcal{C}$ versus the R\'enyi entropy, $S_R$ for the NED AdS black hole as can be seen from Fig.\ref{fig13} and Fig.\ref{fig14} respectively, where we see for the $T-S_R$ curve in Fig.\ref{fig13} there are visible peaks at $S_R=13.450$ and $S_R=23.107$  respectively, whereas for the $\mathcal{C}-S_R$ curve in Fig.\ref{fig14} we see potential discontinuities exactly at the same points as that of the peaks observed in the temperature curve of the black hole.
	
	The GTD metric for the NED AdS black hole for the R\'enyi entropy can be written as
	\begin{equation}
		g  =S_R\left(\frac{\partial M}{\partial S_R}\right)\left(- \frac{\partial^2 M}{\partial S_R^2} dS^2  + \frac{\partial^2 M}{\partial \tilde{Q}^2} d\tilde{Q}^2 \right)
	\end{equation}

\begin{figure}[ht]
		\begin{center}
			\includegraphics[width=.52\textwidth]{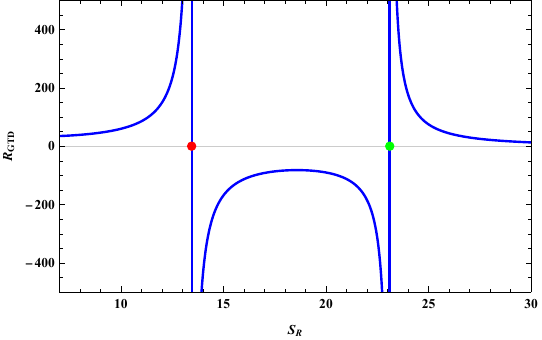}\vspace{3pt}
			\caption{$R_{GTD}-S_R$ curve for the NED AdS black hole for R\'enyi entropy for $Q=1$, $b=1$, $l=10$ and $\lambda =0.01$.}\label{fig15}
		\end{center}
	\end{figure}

	Now by substituting the value of $M$ from above we get the GTD metric from which we can easily calculate the GTD scalar $R_{GTD}$ given in appendix \ref{ap:8} and therefore plot it against the R\'enyi entropy for the NED AdS black hole as can be seen from Fig.\ref{fig15} where we see that in accordance with the peak and trough observed in the $T-S_R$ and the discontinuities observed in the $\mathcal{C}-S_R$ curves the scalar $R_{GTD}$ versus entropy, $S_R$ curve produces discontinuities exactly at the same points i.e. $S_R=13.450$ and $S_R=23.107$. 
	
	\subsection{Kaniadakis Entropy}
\begin{figure}[ht]
		\begin{center}
			\includegraphics[width=.92\textwidth]{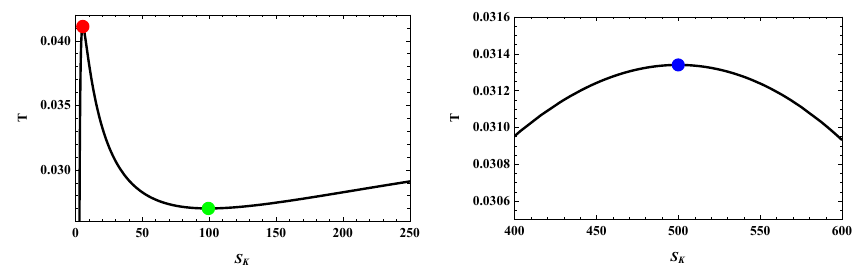}\vspace{3pt}
			\caption{$T-S_K$ curve for the NED AdS black hole for Kaniadakis entropy for $Q=1$, $b=1$, $l=10$ and $\kappa =0.001$.}\label{fig16}
		\end{center}
	\end{figure}

\begin{figure}[ht]
		\begin{center}
			\includegraphics[width=.92\textwidth]{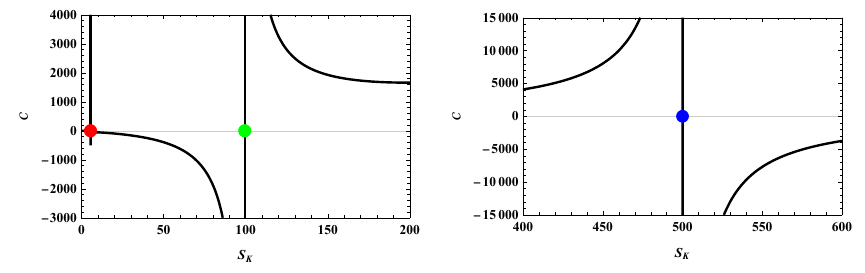}\vspace{3pt}
			\caption{$C-S_K$ curve for the NED AdS black hole for Kaniadakis entropy for $Q=1$, $b=1$, $l=10$ and $\kappa =0.001$.}\label{fig17}
		\end{center}
	\end{figure}

	After making series expansion for small $\kappa$ and then substituting the value of r from eqtn.\ref{eqENKr} to eqtn.\ref{2.15} we get the mass of the NED AdS black hole for Kaniadakis entropy, which is given as:
	
	\begin{equation}
		M =
		\frac{
			- 45 G^{3} S_K^{4} (-4 + S_K^{2} \kappa^{2})
			- 15 b^{2} G l^{2} \pi^{3} Q^{2} S_K (4 + S_K^{2} \kappa^{2})
			+ 2 b^{4} l^{2} \pi^{4} Q^{2} (12 + 5 S_K^{2} \kappa^{2})
			+ \mathcal{E}
		}{
			360 l^{2} \pi^{3/2} (G S_K)^{5/2}
		}
		\label{eq:massnedkania}
	\end{equation}

	where, $\mathcal{E}=15 G^{2} l^{2} \pi S_K^{2}
	\left(
	12 S - S_K^{3} \kappa^{2}
	+ \pi Q^{2} (12 + S_K^{2} \kappa^{2})
	\right)$. The temperature of the black hole is therefore given as:
	
	\begin{equation}
		T =
		\frac{
			G \left(
			- 3 b^{2} G l^{2} \pi^{3} Q^{2} S_K (-12 + S_K^{2} \kappa^{2})
			- 2 b^{4} l^{2} \pi^{4} Q^{2} (12 + S_K^{2} \kappa^{2})
			+ 3 G^{2} S_K^{2} \left(
			3 G S_K^{2} (12 - 7 S_K^{2} \kappa^{2})
			+ \mathcal{F}
			\right)
			\right)
		}{
			144 l^{2} \pi^{3/2} (G S_K)^{7/2}
		}
	\end{equation}

	where, $\mathcal{F}=l^{2} \pi \left(
	12 S - 5 S_K^{3} \kappa^{2}
	+ 3 \pi Q^{2} (-4 + S_K^{2} \kappa^{2})
	\right)$ and the specific heat capacity of the black hole is given by:
	
	\begin{equation}
	\mathcal{C} =
		- \frac{
			2 S_K \left(
			3 b^{2} G l^{2} \pi^{3} Q^{2} S_K (-12 + S_K^{2} \kappa^{2})
			+ 2 b^{4} l^{2} \pi^{4} Q^{2} (12 + S_K^{2} \kappa^{2})
			+ 3 G^{2} S_K^{2} \left(
			3 G S_K^{2} (-12 + 7 S_K^{2} \kappa^{2})
			+ \mathcal{G}
			\right)
			\right)
		}{
			3 b^{2} G l^{2} \pi^{3} Q^{2} S_K (-60 + S_K^{2} \kappa^{2})
			+ 6 b^{4} l^{2} \pi^{4} Q^{2} (28 + S_K^{2} \kappa^{2})
			+ 9 G^{2} S_K^{2} \left(
			G S_K^{2} (12 - 35 S_K^{2} \kappa^{2})
			+ \mathcal{H}
			\right)
		}
	\end{equation}
		where, $\mathcal{G}=l^{2} \pi \left(
	- 3 \pi Q^{2} (-4 + S_K^{2} \kappa^{2})
	+ S_K (-12 + 5 S_K^{2} \kappa^{2})
	\right)$ and $\mathcal{H}=l^{2} \pi \left(
	\pi Q^{2} (12 + S_K^{2} \kappa^{2})
	- S (4 + 5 S_K^{2} \kappa^{2})
	\right)$
	We plot the temperature, $T$ and specific heat capacity, $\mathcal{C}$ versus the Kaniadakis entropy, $S_K$ for the NED AdS black hole as can be seen from Fig.\ref{fig16} and Fig.\ref{fig17} respectively, where we see for the $T-S_K$ curve in Fig.\ref{fig16} there are visible peaks at $S_K=5.524, 99.459$ and $S_K=499.975$  respectively, whereas for the $\mathcal{C}-S_K$ curve in Fig.\ref{fig17} we see potential discontinuities exactly at the same points as that of the peaks observed in the temperature curve of the black hole.
	
	The GTD metric for the NED AdS black hole for the Kaniadakis entropy can be written as:
	\begin{equation}
		g  =S_K \left(\frac{\partial M}{\partial S_K}\right)\left(- \frac{\partial^2 M}{\partial S_K^2} dS_K^2  + \frac{\partial^2 M}{\partial \tilde{Q}^2} d\tilde{Q}^2 \right)
	\end{equation}

\begin{figure}[ht]
		\begin{center}
			\includegraphics[width=.92\textwidth]{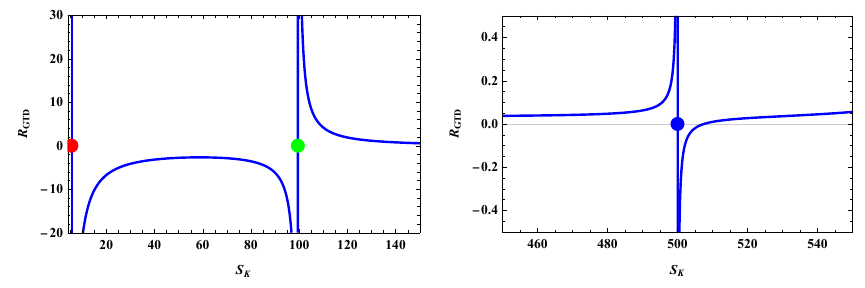}\vspace{3pt}
			\caption{$R_{GTD}-S_K$ curve for the NED AdS black hole for Kaniadakis entropy for $Q=1$, $b=1$, $l=10$ and $\kappa =0.001$.}\label{fig18}
		\end{center}
	\end{figure}

	Now by substituting the value of $M$ from above we get the GTD metric from which we can easily calculate the GTD scalar $R_{GTD}$ given in appendix \ref{ap:9}  and therefore plot it against the Kaniadakis entropy for the NED AdS black hole as can be seen from Fig.\ref{fig18} where we see that in accordance with the peak and trough observed in the $T-S_K$ and the discontinuities observed in the $\mathcal{C}-S_K$ curves the scalar $R_{GTD}$ versus entropy, $S_K$ curve produces discontinuities exactly at the same points i.e.$S_K=5.524, 99.459$ and $S_K=499.975$ . 
	
	\newpage
	\section{CFT Thermodynamic Geometry For NED AdS Black Holse}
\label{sec:5}
	\subsection{Bekenstein-Hawking Entropy}
	By employing Eqs.\eqref{eq:massnedBH}, \eqref{eq:centralcharge}, \eqref{eq:volume}, and \eqref{eq:holographic}, the energy of the holographically dual CFT corresponding to a NED black hole, expressed in terms of Bekenstein--Hawking entropy, can be written as
	\begin{equation}
E=\frac{\sqrt{\frac{1}{C}} \left(32 \pi ^4 \mathit{b}^4 C^2 \mathcal{Q}^2+20 C \left(\pi ^3 \mathit{b}^2 \mathcal{Q}^2 \mathcal{S}+3 \pi  \mathcal{S}^3\right)+15 \left(\pi ^2 \mathcal{Q}^2 \mathcal{S}^2+\mathcal{S}^4\right)\right)}{30 \pi  \mathcal{S}^{5/2} \sqrt{\mathcal{V}}}
\label{eq:cftenergy1}
	\end{equation}
	The temperature of the dual CFT is then evaluated and is given by
	\begin{equation}
\mathcal{T}=-\frac{\sqrt{\frac{1}{C}} \left(32 \pi ^4 \mathit{b}^4 C^2 \mathcal{Q}^2+12 C \left(\pi ^3 \mathit{b}^2 \mathcal{Q}^2 \mathcal{S}-\pi  \mathcal{S}^3\right)+3 \pi ^2 \mathcal{Q}^2 \mathcal{S}^2-9 \mathcal{S}^4\right)}{12 \pi  \mathcal{S}^{7/2} \sqrt{\mathcal{V}}}
	\end{equation}
		The heat capacity can be expressed in the following expression as
		\begin{equation}
\mathcal{C}=\frac{2 \mathcal{S} \left(-32 \pi ^4 \mathit{b}^4 C^2 \mathcal{Q}^2-12 \pi ^3 \mathit{b}^2 C \mathcal{Q}^2 \mathcal{S}+12 \pi  C \mathcal{S}^3-3 \pi ^2 \mathcal{Q}^2 \mathcal{S}^2+9 \mathcal{S}^4\right)}{224 \pi ^4 \mathit{b}^4 C^2 \mathcal{Q}^2+60 \pi ^3 \mathit{b}^2 C \mathcal{Q}^2 \mathcal{S}-12 \pi  c \mathcal{S}^3+9 \pi ^2 \mathcal{Q}^2 \mathcal{S}^2+9 \mathcal{S}^4}
		\end{equation}
	
\begin{figure}[ht]
		\begin{center}
			\includegraphics[width=.52\textwidth]{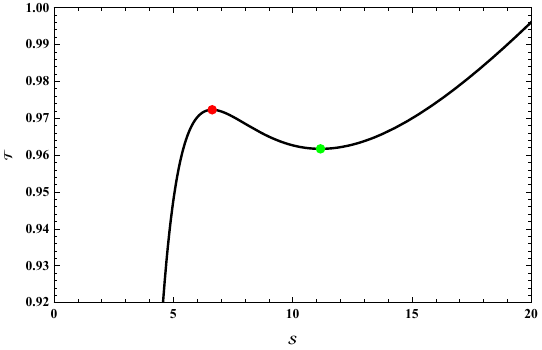}\vspace{3pt}
			\caption{$\mathcal{T}$--$\mathcal{S}$ curve for the dual CFT at fixed parameters $\mathcal{Q}=1.5$, $b=0.5$, $C=3$ and $\mathcal{V}=1$
			}\label{fig:28}
		\end{center}
	\end{figure}

\begin{figure}[ht]
		\begin{center}
			\includegraphics[width=.52\textwidth]{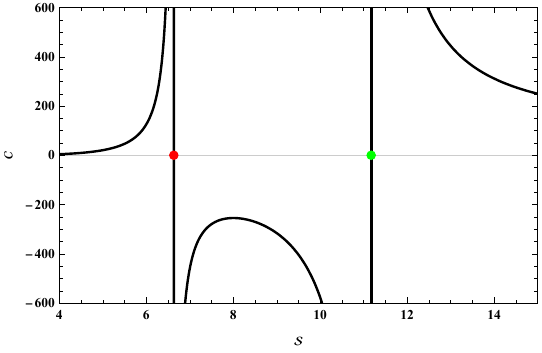}\vspace{3pt}
			\caption{$\mathcal{C}$--$\mathcal{S}$ curve for the dual CFT at fixed parameters $\mathcal{Q}=1.5$, $b=0.5$, $C=3$ and $\mathcal{V}=1$}\label{fig:29}
		\end{center}
	\end{figure}

	The thermodynamic behavior of the dual CFT is examined by plotting the temperature $\mathcal{T}$ and the specific heat $\mathcal{C}$ as functions of the Bekenstein--Hawking entropy $\mathcal{S}$, as shown in Figs.~\ref{fig:28} and \ref{fig:29}, respectively. From the $\mathcal{T}$--$\mathcal{S}$ diagram in Fig.~\ref{fig:28}, the temperature curve clearly develops two distinct extremal points occurring at $\mathcal{S}=6.633$ and $\mathcal{S}=11.173$. A corresponding analysis of the $\mathcal{C}$--$\mathcal{S}$ behavior presented in Fig.~\ref{fig:29} reveals sharp divergences appearing precisely at these same entropy values. The coincidence between the extrema of the temperature profile and the singular features of the specific heat indicates the onset of critical thermodynamic phenomena in the CFT description.
	\\
	In the context of CFT thermodynamics, the geometrothermodynamic (GTD) metric constructed from the Bekenstein--Hawking entropy for the ModMax configuration can be written as
	\begin{equation}
		\tilde{g}=\mathcal{S}\left(\frac{\partial E}{\partial \mathcal{S}}\right)
		\left(-\frac{\partial^{2}E}{\partial \mathcal{S}^{2}}\,d\mathcal{S}^{2}
		+\frac{\partial^{2}E}{\partial \mathcal{Q}^{2}}\,d\mathcal{Q}^{2}\right)
	\end{equation}
	
	Substituting the CFT energy expression $E$ from Eq.~\ref{eq:cftenergy1} into the above relation yields the explicit form of the geometrothermodynamic line element. From this metric, the associated GTD scalar curvature $R_{GTD}$ given in \ref{ap:10} can be readily computed. The resulting curvature scalar is then analyzed as a function of the Bekenstein--Hawking entropy $\mathcal{S}$ within the CFT thermodynamic description. The behavior of this quantity is depicted in Fig.~\ref{fig:12}, where the $R_{\mathrm{GTD}}$--$\mathcal{S}$ curve exhibits clear divergences at $\mathcal{S}=6.633$ and $\mathcal{S}=11.173$. Interestingly, these entropy values coincide exactly with the extremal points observed in the $\mathcal{T}$--$\mathcal{S}$ profile as well as the discontinuities appearing in the $\mathcal{C}$--$\mathcal{S}$ diagram. This agreement highlights the close correspondence between the geometric structure encoded in the GTD curvature and the thermodynamic critical behavior of the system.
	\begin{figure}[ht]
		\begin{center}
			\includegraphics[width=.52\textwidth]{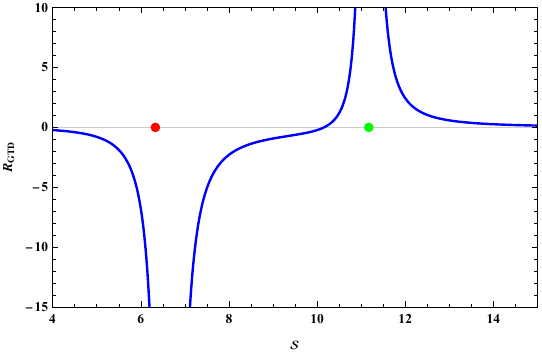}\vspace{3pt}
			\caption{$R_{\text{GTD}}$--$\mathcal{S}$ profile for the dual CFT at fixed parameters $\mathcal{Q}=1.5$, $b=0.5$, $C=3$ and $\mathcal{V}=1$}\label{fig:30}
		\end{center}
	\end{figure}
	
	\subsection{R\'enyi Entropy}
	Employing the relations provided in Eqs.~\eqref{eq:massnedrenyi}, \eqref{eq:centralcharge}, \eqref{eq:volume}, and \eqref{eq:holographic}, one can derive the expression for the energy of the holographically dual CFT when the thermodynamic description is formulated using R\'enyi entropy. The resulting form of the CFT energy is given by
	\begin{equation}
		E=\frac{\sqrt{\frac{\mathcal{S}_R}{C}} \left(\begin{aligned}&-32 \pi ^4 \mathit{b}^4 C^2 \mathcal{Q}^2 (5 \lambda  \mathcal{S}_R-4)+20 C \left(\pi ^3 \mathit{b}^2 \mathcal{Q}^2 \mathcal{S}_R (3 \lambda  \mathcal{S}_R-4)+3 \pi  \mathcal{S}_R^3 (\lambda  \mathcal{S}_R+4)\right)\\&+15 \mathcal{S}_R^2 \left(\pi ^2 \mathcal{Q}^2 (4-\lambda  \mathcal{S}_R)+\mathcal{S}_R^2 (3 \lambda  \mathcal{S}_R+4)\right)\end{aligned}\right)}{120 \pi  \mathcal{S}_R^3 \sqrt{\mathcal{V}}}
		\label{eq:cftenergy2}
	\end{equation}
	The temperature associated with the holographically dual CFT, when described within the R\'enyi entropy framework, is therefore obtained as
	\begin{equation}
		\mathcal{T}=\frac{\sqrt{\frac{\mathcal{S}_R}{C}} \left(\begin{aligned}&32 \pi ^4 \mathit{b}^4 C^2 \mathcal{Q}^2 (3 \lambda  \mathcal{S}_R-4)-12 \pi ^3 \mathit{b}^2 C \mathcal{Q}^2 \mathcal{S}_R (\lambda  \mathcal{S}_R-4)+12 \pi  C \mathcal{S}_R^3 (3 \lambda \mathcal{S}_R+4)\\&-3 \pi ^2 \mathcal{Q}^2 \mathcal{S}_R^2 (\lambda  \mathcal{S}_R+4)+9 \mathcal{S}_R^4 (5 \lambda  \mathcal{S}_R+4)\end{aligned}\right)}{48 \pi  \mathcal{S}_R^4 \sqrt{\mathcal{V}}}
	\end{equation}
	The specific heat is given as
	\begin{equation}
		\mathcal{C}=\frac{2 \mathcal{S}_R \left(\begin{aligned}&32 \pi ^4 \mathit{b}^4 c^2 \mathcal{Q}^2 (3 \lambda  \mathcal{S}_R-4)-12 \pi ^3 \mathit{b}^2 c \mathcal{Q}^2 \mathcal{S}_R (\lambda  \mathcal{S}_R-4)+12 \pi  c \mathcal{S}_R^3 (3 \lambda  \mathcal{S}_R+4)\\&-3 \pi ^2 \mathcal{Q}^2 \mathcal{S}_R^2 (\lambda  \mathcal{S}_R+4)+9 \mathcal{S}_R^4 (5 \lambda  \mathcal{S}_R+4)\end{aligned}\right)}{\begin{aligned}&32 \pi ^4 \mathit{b}^4 c^2 \mathcal{Q}^2 (28-15 \lambda  \mathcal{S}_R)+12 c \left(\pi ^3 \mathit{b}^2 \mathcal{Q}^2 \mathcal{S}_R(3 \lambda  \mathcal{S}_R-20)+\pi  \mathcal{S}_R^3 (3 \lambda  \mathcal{S}_R-4)\right)\\&+3 \pi ^2 \mathcal{Q}^2 \mathcal{S}_R^2 (\lambda  \mathcal{S}_R+12)+9 \mathcal{S}_R^4 (15 \lambda \mathcal{S}_R+4)\end{aligned}}
	\end{equation}
		\begin{figure}[ht]
		\begin{center}
			\includegraphics[width=.52\textwidth]{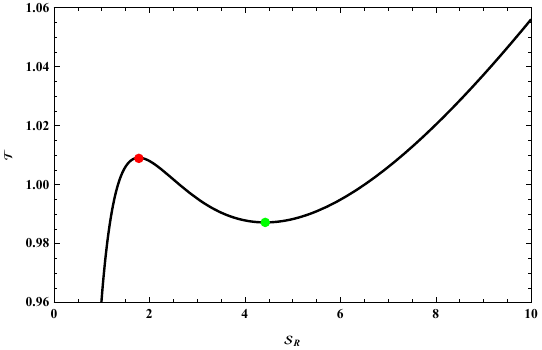}\vspace{3pt}
			\caption{$\mathcal{T}$--$\mathcal{S}_R$ curve for the dual CFT using R\'enyi entropy at fixed parameters $\mathcal{Q}=1.2$, $b=0.1$, $C=2$, $\lambda=0.01$ and $\mathcal{V}=1$
			}\label{fig:31}
		\end{center}
	\end{figure}
	The thermodynamic properties of the dual CFT incorporating R\'enyi entropy are analyzed by plotting the temperature $\mathcal{T}$ and the specific heat $\mathcal{C}$ as functions of the entropy $\mathcal{S}_R$, as illustrated in Figs.~\ref{fig:31} and \ref{fig:32}, respectively. The $\mathcal{T}$--$\mathcal{S}_R$ profile presented in Fig.~\ref{fig:31} reveals two well-defined extremal points occurring at $\mathcal{S}_R=1.780$ and $\mathcal{S}_R=4.428$. A corresponding examination of the $\mathcal{C}$--$\mathcal{S}_R$ curve in Fig.~\ref{fig:32} shows pronounced divergences appearing exactly at these same entropy values. The coincidence between the extrema in the temperature curve and the singular behavior of the specific heat indicates the presence of critical thermodynamic features in the CFT description governed by R\'enyi entropy.\\
		\begin{figure}[ht]
		\begin{center}
			\includegraphics[width=.52\textwidth]{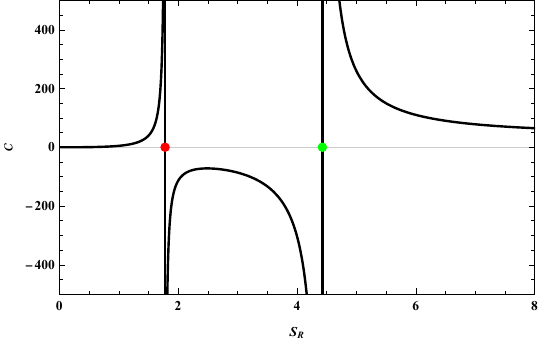}\vspace{3pt}
			\caption{$\mathcal{C}$--$\mathcal{S}_R$ curve for the dual CFT using R\'enyi entropy at fixed parameters $\mathcal{Q}=1.2$, $b=0.1$, $C=2$, $\lambda=0.01$ and $\mathcal{V}=1$}\label{fig:32}
		\end{center}
	\end{figure}
	In the thermodynamic description of the dual CFT, the geometrothermodynamic (GTD) metric associated with the R\'enyi entropy for the NED system can be expressed as
		\begin{equation}
		\tilde{g}=\mathcal{S}_{\mathrm{R}}\left(\frac{\partial E}{\partial \mathcal{S}_{\mathrm{R}}}\right)
		\left(
		-\frac{\partial^2 E}{\partial \mathcal{S}_{\mathrm{R}}^2}\, d\mathcal{S}_{\mathrm{R}}^2
		+\frac{\partial^2 E}{\partial \mathcal{Q}^2}\, d\mathcal{Q}^2
		\right).
	\end{equation}
	By inserting the expression for the CFT energy $E$ from Eq.~\eqref{eq:cftenergy2} into the preceding relation, the explicit form of the geometrothermodynamic metric is obtained. This metric allows the corresponding GTD scalar curvature $R_{GTD}$ given in \ref{ap:11} to be evaluated directly. We then investigate the behavior of this curvature scalar as a function of the R\'enyi entropy $\mathcal{S}_R$ within the CFT thermodynamic framework. The resulting profile is displayed in Fig.~\ref{fig:33}, where the $R_{\mathrm{GTD}}$--$\mathcal{S}_R$ curve develops clear divergences at $\mathcal{S}_R=1.780$ and $\mathcal{S}_R=4.428$. Notably, these entropy values coincide with the extrema observed in the $\mathcal{T}$--$\mathcal{S}_R$ diagram and with the discontinuities appearing in the $\mathcal{C}$--$\mathcal{S}_R$ curve. This correspondence highlights the consistency between the geometric information encoded in the GTD curvature and the critical thermodynamic features of the R\'enyi-modified CFT system.
		\begin{figure}[ht]
		\begin{center}
			\includegraphics[width=.52\textwidth]{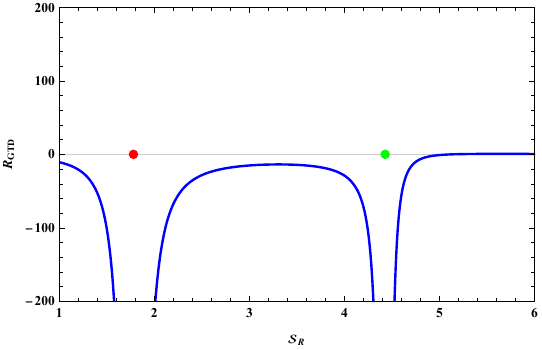}\vspace{3pt}
			\caption{$R_{\text{GTD}}$--$\mathcal{S}_R$ profile for the dual CFT using R\'enyi entropy at fixed parameters $\mathcal{Q}=1.2$, $b=0.1$, $C=2$, $\lambda=0.01$ and $\mathcal{V}=1$}
			\label{fig:33}
		\end{center}
	\end{figure}
	\subsection{Kaniadakis Entropy}
	Employing the relations given in Eqs.~\eqref{eq:massnedkania}, \eqref{eq:centralcharge}, \eqref{eq:volume}, and \eqref{eq:holographic}, one can derive the expression for the energy of the holographically dual CFT when the thermodynamic description is formulated using Kaniadakis entropy. The resulting CFT energy can be written as
	\begin{equation}
	E=\frac{\sqrt{\frac{\mathcal{S}_K}{C}} \left(\begin{aligned}&32 \pi ^4 \mathit{b}^4 C^2 \mathcal{Q}^2 \left(5 \kappa ^2 \mathcal{S}_K^2+12\right)-60 C \left(\pi ^3 \mathit{b}^2 \mathcal{Q}^2 \mathcal{S}_K \left(\kappa ^2 \mathcal{S}_K^2+4\right)+\pi  \mathcal{S}_K^3 \left(\kappa ^2 \mathcal{S}_K^2-12\right)\right)\\&+15 \left(\pi ^2 \mathcal{Q}^2 \mathcal{S}_K^2 \left(\kappa ^2 \mathcal{S}_K^2+12\right)-3 \mathcal{S}_K^4 \left(\kappa ^2 \mathcal{S}_K^2-4\right)\right)\end{aligned}\right)}{360 \pi  \mathcal{S}_K^3 \sqrt{\mathcal{V}}}
	\label{eq:cftenergy3}
	\end{equation}
	Accordingly, the temperature of the holographically dual CFT within the Kaniadakis entropy formulation can be written as
	\begin{equation}
	\mathcal{T}=\frac{\sqrt{\frac{\mathcal{S}_K}{C}} \left(\begin{aligned}&-32 \pi ^4 \mathit{b}^4 C^2 \mathcal{Q}^2 \left(\kappa ^2 \mathcal{S}_K^2+12\right)-12 C \left(\pi ^3 \mathit{b}^2 \mathcal{Q}^2 \mathcal{S}_K \left(\kappa ^2 \mathcal{S}_K^2-12\right)+\pi  \mathcal{S}_K^3 \left(5 \kappa ^2 \mathcal{S}_K^2-12\right)\right)\\&+9 \left(\pi ^2 \mathcal{Q}^2 \mathcal{S}_K^2 \left(\kappa ^2 \mathcal{S}_K^2-4\right)-7 \kappa ^2 \mathcal{S}_K^6+12 \mathcal{S}_K^4\right)\end{aligned}\right)}{144 \pi  \mathcal{S}_K^4 \sqrt{\mathcal{V}}}
	\end{equation}
	Therefore the specific heat is given as
	\begin{equation}
		\mathcal{C}=-\frac{2 \mathcal{S}_K \left(\begin{aligned}&32 \pi ^4 \mathit{b}^4 C^2 \mathcal{Q}^2 \left(\kappa ^2 \mathcal{S}_K^2+12\right)+12 C \left(\pi ^3 \mathit{b}^2 \mathcal{Q}^2 \mathcal{S}_K \left(\kappa ^2 \mathcal{S}_K^2-12\right)+\right.\\&\left.\pi  \mathcal{S}_K^3 \left(5 \kappa ^2 \mathcal{S}_K^2-12\right)\right)-9 \left(\pi ^2 \mathcal{Q}^2 \mathcal{S}_K^2 \left(\kappa ^2 \mathcal{S}_K^2-4\right)-7 \kappa ^2 \mathcal{S}_K^6+12 \mathcal{S}_K^4\right)\end{aligned}\right)}{3 \left(\begin{aligned}&32 \pi ^4 \mathit{b}^4 C^2 \mathcal{Q}^2 \left(\kappa ^2 \mathcal{S}_K^2+28\right)+4 \pi ^3 \mathit{b}^2 C \mathcal{Q}^2 \mathcal{S}_K \left(\kappa ^2 \mathcal{S}_K^2-60\right)\\&-12 \pi  C \mathcal{S}_K^3 \left(5 \kappa ^2 \mathcal{S}_K^2+4\right)+3 \pi ^2 \mathcal{Q}^2 \mathcal{S}_K^2 \left(\kappa ^2 \mathcal{S}_K^2+12\right)-105 \kappa ^2 \mathcal{S}_K^6+36 \mathcal{S}_K^4\end{aligned}\right)}
	\end{equation}
	\begin{figure}[ht]
		\begin{center}
			\includegraphics[width=.82\textwidth]{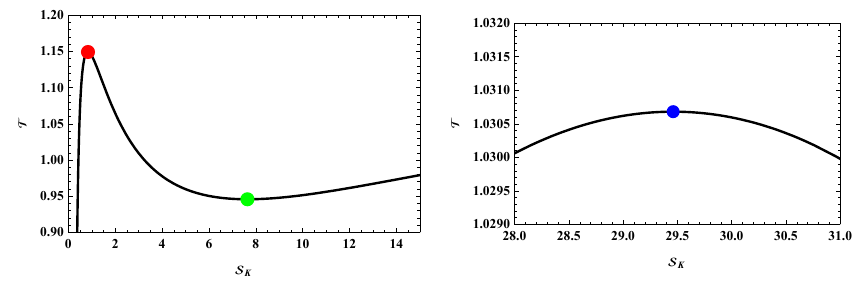}\vspace{3pt}
			\caption{$\mathcal{T}$--$\mathcal{S}$ curve for the dual CFT using Kaniadakis entropy at fixed parameters $\mathcal{Q}=1$, $b=0.1$, $C=2$, $\kappa=0.016$ and $\mathcal{V}=1$}\label{fig:34}
		\end{center}
	\end{figure}
	
		\begin{figure}[ht]
		\begin{center}
			\includegraphics[width=.82\textwidth]{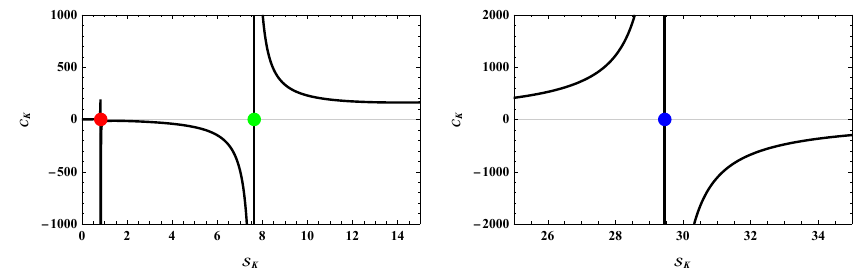}\vspace{3pt}
			\caption{$\mathcal{T}$--$\mathcal{S}$ curve for the dual CFT using Kaniadakis entropy at fixed parameters $\mathcal{Q}=1$, $b=0.1$, $C=2$, $\kappa=0.016$ and $\mathcal{V}=1$	}\label{fig:35}
		\end{center}
	\end{figure}
	The thermodynamic properties of the dual CFT in the Kaniadakis entropy formulation are investigated through the behavior of the temperature $\mathcal{T}$ and the specific heat $\mathcal{C}$, as illustrated in Figs.~\ref{fig:34} and \ref{fig:35}, respectively. The $\mathcal{T}$--$\mathcal{S}_K$ profile presented in Fig.~\ref{fig:34} displays three prominent extremal points located at $\mathcal{S}_K=0.837$, $\mathcal{S}_K=7.643$, and $\mathcal{S}_K=29.461$. A corresponding examination of the $\mathcal{C}$--$\mathcal{S}_K$ diagram in Fig.~\ref{fig:35} reveals clear divergences emerging exactly at these same entropy values. The coincidence of the temperature extrema with the singular behavior of the specific heat signals the presence of critical thermodynamic features within the CFT framework governed by Kaniadakis entropy.\\
	In the thermodynamic formulation of the dual CFT, the geometrothermodynamic (GTD) metric constructed from the Kaniadakis entropy for the ModMax-deformed system can be written as
		\begin{equation}
		\tilde{g}=\mathcal{S}_{\mathrm{K}}\left(\frac{\partial E}{\partial \mathcal{S}_{\mathrm{K}}}\right)
		\left(
		-\frac{\partial^2 E}{\partial \mathcal{S}_{\mathrm{K}}^2}\, d\mathcal{S}_{\mathrm{K}}^2
		+\frac{\partial^2 E}{\partial \mathcal{Q}^2}\, d\mathcal{Q}^2
		\right).
	\end{equation}
		\begin{figure}[ht]
		\begin{center}
			\includegraphics[width=0.82\textwidth]{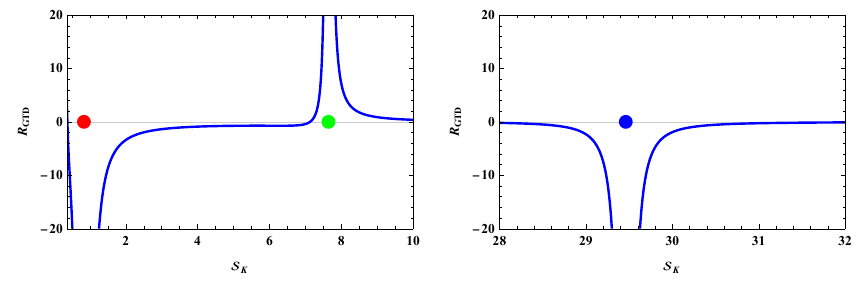}\vspace{3pt}
			\caption{$R_\mathrm{GTD}$--$\mathcal{S}_\mathrm{K}$ curve for the dual CFT using Kaniadakis entropy at fixed parameters  $\mathcal{Q}=1$, $b=0.1$, $C=2$, $\kappa=0.016$ and $\mathcal{V}=1$}\label{fig:36}
		\end{center}
	\end{figure}
	Substituting the CFT energy expression $E$ from Eq.~\eqref{eq:cftenergy3} into the geometrothermodynamic framework yields the explicit form of the GTD metric. From this metric, the corresponding scalar curvature $R_{GTD}$ can be determined straightforwardly and is given in appendix \ref{ap:12}. We then investigate the dependence of this curvature scalar on the Kaniadakis entropy $\mathcal{S}_{\mathrm{K}}$ within the CFT thermodynamic description. The resulting behavior is illustrated in Fig.~\ref{fig:36}, where the $R_{\mathrm{GTD}}$--$\mathcal{S}_{\mathrm{K}}$ curve exhibits clear divergences at $\mathcal{S}_{\mathrm{K}}=0.837$, $7.643$, and $29.461$. Interestingly, these entropy values coincide with the extremal points observed in the $\mathcal{T}$--$\mathcal{S}_{\mathrm{K}}$ profile and with the discontinuities appearing in the $\mathcal{C}$--$\mathcal{S}_{\mathrm{K}}$ diagram. This agreement highlights the close relationship between the geometric information encoded in the GTD curvature and the critical thermodynamic characteristics of the CFT system governed by Kaniadakis entropy.
	\section{Euler-Heisenberg AdS Black hole Solution}
	\label{sec:6}
	We here give an overview of the EH theory being coupled to gravity \cite{FIII}.
	The four-dimensional action of general relativity with cosmological constant
	$\Lambda$ coupled to the nonlinear electrodynamics (NLED) \cite{Plebanski},
	\cite{Salazar} is given as: 
	
	\begin{equation}
		S=\frac{1}{4\pi}\int_{M^4} d^4x \sqrt{-g}\left[\frac{1}{4}(R-2\Lambda)-
		\mathcal{L}(F,G) \right],\label{action}
	\end{equation} 
	where $g$ is the determinant of the metric tensor and $R$ is the Ricci scalar,
	$\mathcal{L}(F,G)$ is the non-linear electrodynamic (NLED) Lagrangian that depends on the electromagnetic
	invariants,    $F=\frac{1}{4}F_{\mu \nu} F^{\mu \nu}$ and $G=\frac{1}{4}F_{\mu \nu}
	{^*F^{\mu \nu}}$ along with $F_{\mu\nu}$ denoting the electromagnetic field strength
	tensor and ${^*F^{\mu \nu}}=\epsilon _{\mu\nu\sigma\rho} F^{\sigma\rho}
	/(2\sqrt{-g})$ being its dual. The antisymmetric  tensor
	$\epsilon_{\mu\nu\sigma\rho}$, satisfies the condition
	$\epsilon_{\mu\nu\sigma\rho}\epsilon^{\mu\nu\sigma\rho}=-4!$\\
	The Lagrangian density for the Euler-Heisenberg NLED \cite{EH} is given as: 
	\begin{equation}
		\mathcal{L}(F,G)=-F+\frac{a}{2}F^2+ \frac{7a}{8} G^2,\label{Lagrangian}
	\end{equation}
	where $a=8 \alpha^2/45 m^4$ is known as the Euler-Heisenberg parameter which regulates the intensity of the
	NLED contribution; $\alpha$ is called
	the fine structure constant and $m$ is the well-known electron mass (here, we take $c=1= \hbar$), such that the EH parameter
	becomes of the order of $\alpha / E_c^2$.
	For $a=0$ we recover the usual Maxwell electrodynamics $\mathcal{L}(F)=-F$.
	
	Regarding NLED
	there are two possible frameworks one being the usual one ($F$ framework) in terms of
	the electromagnetic field tensor $F^{\mu \nu}$.
	Alternatively, there is the lesser known $P$ framework with the tensor $P_{\mu\nu}$ as the main field,
	defined by
	\begin{equation} 
		P_{\mu\nu}= -(\mathcal{L}_F F_{\mu\nu}+ {^*F}_{\mu\nu} \mathcal{L}_G ),
	\end{equation}
	where the subscript $X$ in $\mathcal{L}$ is used to denote the derivative, $  \mathcal{L}_{X}= d
	\mathcal{L} /d X$. In  the E-H (Euler-Heisenberg) theory, $P_{\mu \nu}$ takes the form:
	\begin{equation}
		P_{\mu\nu}=(1-a F)F_{\mu\nu} - {^*F}_{\mu\nu} \frac{7a}{4}G .\label{Pmunu_a1}
	\end{equation}
	This tensor corresponds to the electric induction {\bf D} and the
	magnetic field {\bf H} and  Eqs. (\ref{Pmunu_a1}) are the well established relations
	between {\bf D},  {\bf H} and the magnetic intensity {\bf B} and the electric field {\bf E} in the E-H NLED. \\
	
	The two independent invariants $P$ and $O$ associated to the P framework are given as:
	\begin{equation}
		P=-\frac{1}{4} P_{\mu\nu} P^{\mu\nu}, \hspace{1cm}
		O= -\frac{1}{4} P_{\mu\nu} {^*P^{\mu\nu}},
	\end{equation}
	with $^*P_{\mu\nu}=\frac{1}{2\sqrt{-g}}\epsilon_{\mu\nu\rho\sigma}P^{\sigma\rho}$.
	$\mathcal{L}$'s legendre transformation defines the Hamiltonian or structural
	function $\mathcal{H}$,
	\begin{equation}
		\mathcal{H} (P,O)= -\frac{1}{2}P^{\mu\nu} F_{\mu\nu}-\mathcal{L}.
	\end{equation}
	Neglecting  the  higher order terms (second order and above) in $a$, the structural function for the EH
	theory can take the following form \cite{Ruffini2013} as:
	\begin{equation}
		\mathcal{H}(P,O)= P-\frac{a}{2}P^2-\frac{7a}{8}O^2. \label{HamiltonianEH}
	\end{equation}
	
	The field equations are \cite{Salazar}
	\begin{equation}
		\nabla_\mu P^{\mu\nu}=0, \hspace{.7cm} G_{\mu\nu}+\Lambda g_{\mu\nu} =8\pi
		T_{\mu\nu}.\label{motion}
	\end{equation}
	The energy momentum tensor $T_{\mu\nu}$ for the E-H theory in the $P$ framework is given by:
	\begin{equation}
		T_{\mu\nu}=\frac{1}{4\pi}\left[(1-a P)P_\mu^\beta
		P_{\nu\beta}+g_{\mu\nu}\left(P-\frac{3}{2}a
		P^2-\frac{7a}{8}O^2\right)\right].\label{emtensor}
	\end{equation}
	
	We then present the static spherically symmetric solution of
	the E-H equations with cosmological constant $\Lambda$. We then determine the solution to the field Eqs. (\ref{motion}) for a static, spherically
	symmetric metric of the form
	\begin{equation}
		ds^2= -f(r)dt^2 + f(r)^{-1}dr^2 + r^2(d\theta^2 +\sin^2{\theta} d\phi^2),\label{metric}
	\end{equation}
	with $f(r)=1-\frac{2m(r)}{r}$. Regarding the electromagnetic field by restricting to
	an electric charge $Q$, the symmetry of the space-time allows for the non-vanishing
	components, 
	\begin{equation}
		P_{\mu\nu}= \frac{Q}{r^2}\delta^0_{[\mu}\delta^1_{\nu]}; \label{Pmunu}
	\end{equation}
	and then the electromagnetic invariants become
	\begin{equation}
		P=\frac{Q^2}{2r^4},\hspace{.7cm} O=0  \label{PO} . 
	\end{equation}
	Substituting these into the (0,0) component of the field Eqs. (\ref{motion}),
	we find that
	\begin{equation}
		\frac{dm}{dr} = \frac{Q^2}{2r^2}-\frac{a Q^4}{8 r^6}+\frac{\Lambda r^2}{2}.
	\end{equation}
	By integrating this equation, and putting the Newton's gravitational constant, G in the right place, the metric function for the electric case is given by
	\begin{equation}
		f(r)= 1-\frac{2 G M}{r}+\frac{G Q^2}{r^2}-\frac{\Lambda r^2}{3}-\frac{a G^2 Q^4}{20 r^6},
		\label{gtt}
	\end{equation}
	where $M$ is the mass of the BH, $Q$ its electric charge, $a$ is the EH parameter and $\Lambda$ is the cosmological constant that can be positive or negative. For AdS BH we take $\Lambda= \frac{-3}{l^2}$ and thereby solving for the lapse function $f(r)=0$, we obtain the mass of NED AdS BH as:
	
	\begin{equation}
		M =
		\frac{
			r^3
			\left(
			- a G^{2} l^{2} Q^{4}
			+ 20 \left( G l^{2} Q^{2} r^{4} + l^{2} r^{6} + r^{8} \right)
			\right)
		}{
			40 l^{2} G r^{8}
		}
		\label{2.16}
	\end{equation}

	\subsection{Bekenstein-Hawking Entropy}
	Substituting the value of r from eqtn.\ref{eqENr} to eqtn.\ref{2.16} we get the mass of the EH AdS black hole, which is given as:
	
	\begin{equation}
		M =
		\frac{- a l^{2} \pi^{4} Q^{4}
			+ 20 G S^{2} \left(
			G S^{2}
			+ l^{2} \pi \left( \pi Q^{2} + S \right)
			\right)}{	40 G^{3/2} l^{2} \pi^{3/2} S^{5/2}}
			\label{eq:massehBH}
	\end{equation}

	The temperature of the black hole is therefore given as:
	
	\begin{equation}
		T =
		\frac{
			a l^{2} \pi^{4} Q^{4}
			+ 4 G S^{2} \left(
			3 G S^{2}
			+ l^{2} \pi \left( - \pi Q^{2} + S \right)
			\right)
		}{
			16 G^{3/2} l^{2} \pi^{3/2} S^{7/2}
		}
	\end{equation}

	and the specific heat capacity of the black hole is given by:
	
	\begin{equation}
		\mathcal{C} =
		- \frac{
			2 \left(
			a l^{2} \pi^{4} Q^{4} S
			+ 4 G S^{3} \left(
			3 G S^{2}
			+ l^{2} \pi \left( - 3 \pi Q^{2} + S \right)
			\right)
			\right)
		}{
			7 a l^{2} \pi^{4} Q^{4}
			+ 4 G S^{2} \left(
			- 3 G S^{2}
			+ l^{2} \pi \left( - 3 \pi Q^{2} + S \right)
			\right)
		}
	\end{equation}

\begin{figure}[ht]
		\begin{center}
			\includegraphics[width=.52\textwidth]{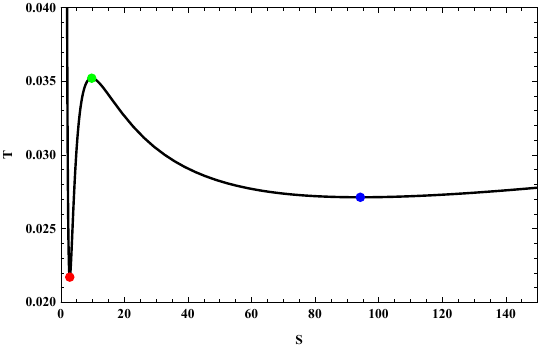}\vspace{3pt}
			\caption{$T-S$ curve for the EH AdS black hole for $Q=1$, $a=1$ and $l=10$.}\label{fig19}
		\end{center}
	\end{figure}

\begin{figure}[ht]
		\begin{center}
			\includegraphics[width=.82\textwidth]{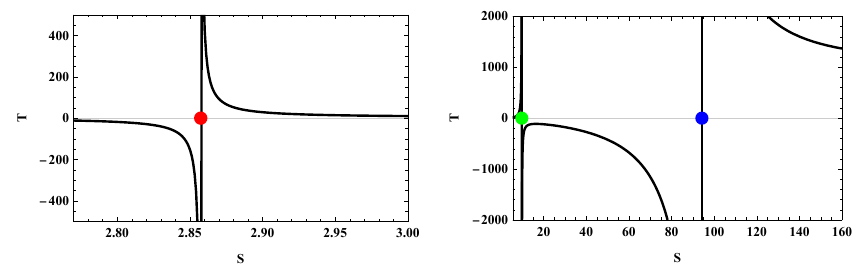}\vspace{3pt}
			\caption{$\mathcal{C}-S$ curve for the EH AdS black hole for $Q=1$, $a=1$ and $l=10$.}\label{fig20}
		\end{center}
	\end{figure}

	We plot the temperature, $T$ and specific heat capcity, $\mathcal{C}$ versus the Bekenstein-Hawking entropy, S for the EH AdS black hole as can be seen from Fig.\ref{fig19} and Fig.\ref{fig20} respectively, where we see for the T-S curve in Fig.\ref{fig19} there are visible peaks at $S=2.857$, $S=9.766$ and $S=94.255$  respectively, whereas for the $\mathcal{C}-S$ curve in Fig.\ref{fig20} we see potential discontinuities exactly at the same points as that of the peaks observed in the temperature curve of the black hole.
	
	The GTD metric for the NED AdS black hole for the Bekenstein-Hawking entropy can be written as:
	\begin{equation}
		g  =S \left(\frac{\partial M}{\partial S}\right)\left(- \frac{\partial^2 M}{\partial S^2} dS^2  + \frac{\partial^2 M}{\partial \tilde{Q}^2} d\tilde{Q}^2 \right)
	\end{equation}

\begin{figure}[ht]
		\begin{center}
			\includegraphics[width=.82\textwidth]{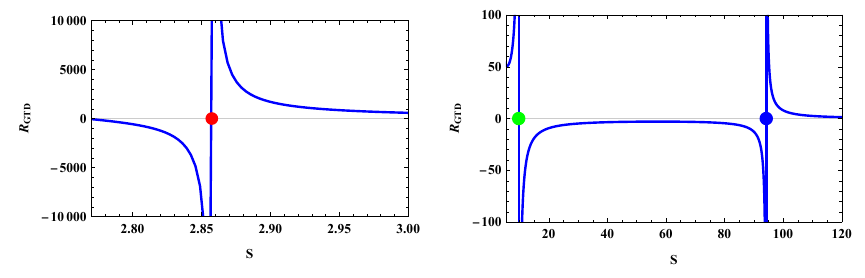}\vspace{3pt}
			\caption{$R_{GTD}-S$ curve for the EH AdS black hole for $Q=1$, $a=1$ and $l=10$.}\label{fig21}
		\end{center}
	\end{figure}

	Now by substituting the value of M from above we get the GTD metric from which we can easily calculate the GTD scalar $R_{GTD}$ and is given in appendix \ref{ap:13} and therefore plot it against the Bekenstein-Hawking entropy for the EH AdS black hole as can be seen from Fig.\ref{fig21} where we see that in accordance with the peak and trough observed in the T-S and the discontinuities observed in the C-S curves the scalar $R_{GTD}$ versus entropy, S curve produces discontinuities exactly at the same points i.e. $S=2.857$, $S=9.766$ and $S=94.255$ respectively. 
	
	\subsection{R\'enyi Entropy}
	After making the series expansion for small $\lambda$ and then substituting the value of r from eqtn.\ref{eqENRr} to eqtn.\ref{2.16} we get the mass of the EH AdS black hole for R\'enyi entropy, which is given as:
	
	\begin{equation}
		M =
		\frac{
			G \left(
			a l^{2} \pi^{4} Q^{4} (-4 + 5 S_R \lambda)
			+ 20 G S_R^{2} \left(
			G S_R^{2} (4 + 3 S_R \lambda)
			+ l^{2} \pi \left(
			\pi Q^{2} (4 - S_R \lambda)
			+ S_R (4 + S \lambda)
			\right)
			\right)
			\right)
		}{
			160 l^{2} \pi^{3/2} (G S_R)^{5/2}
		}
		\label{eq:massehrenyi}
	\end{equation}

	The temperature of the black hole is therefore given as:
	
	\begin{equation}
		T =
		\frac{
			a l^{2} \pi^{4} Q^{4} (4 - 3 S_R \lambda)
			+ 4 G S_R^{2} \left(
			3 G S_R^{2} (4 + 5 S_R \lambda)
			+ l^{2} \pi \left(
			- \pi Q^{2} (4 + S_R \lambda)
			+ S (4 + 3 S_R \lambda)
			\right)
			\right)
		}{
			64 l^{2} \pi^{3/2} S_R^{2} (G S_R)^{3/2}
		}
	\end{equation}

	The specific heat capacity of the black hole is given by:
	
	\begin{equation}
		\mathcal{C} =
		\frac{
			2 a l^{2} \pi^{4} Q^{4} S_R (4 - 3 S_R \lambda)
			+ 8 G S_R^{3} \left(
			3 G S_R^{2} (4 + 5 S_R \lambda)
			+ l^{2} \pi \left(
			- \pi Q^{2} (4 + S_R \lambda)
			+ S (4 + 3 S_R \lambda)
			\right)
			\right)
		}{
			a l^{2} \pi^{4} Q^{4} (-28 + 15 S_R \lambda)
			+ 4 G S_R^{2} \left(
			3 G S_R^{2} (4 + 15 S_R \lambda)
			+ l^{2} \pi \left(
			\pi Q^{2} (12 + S_R \lambda)
			+ S_R (-4 + 3 S_R \lambda)
			\right)
			\right)
		}
	\end{equation}

\begin{figure}[ht]
		\begin{center}
			\includegraphics[width=.52\textwidth]{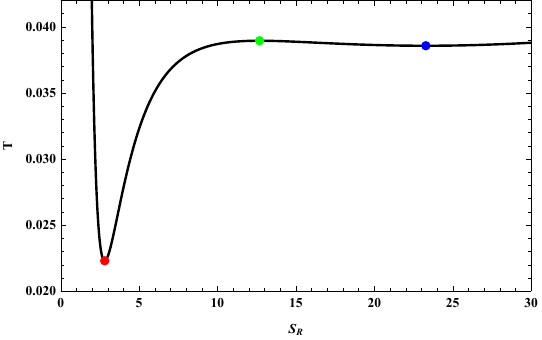}\vspace{3pt}
			\caption{$T-S_R$ curve for the EH AdS black hole for R\'enyi entropy for $Q=1$, $a=1$, $l=10$ and $\lambda =0.01$.}\label{fig22}
		\end{center}
	\end{figure}

\begin{figure}[ht]
		\begin{center}
			\includegraphics[width=.82\textwidth]{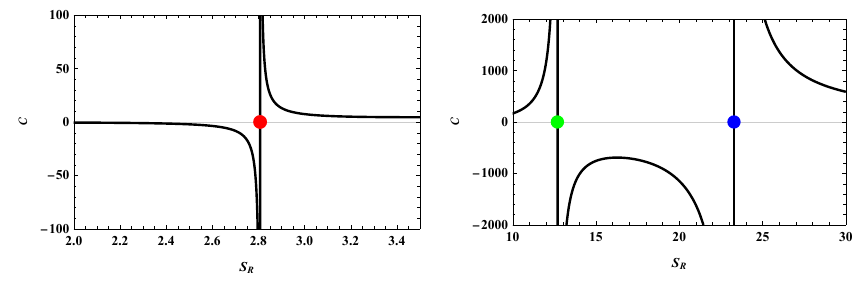}\vspace{3pt}
			\caption{$C-S_R$ curve for the EH AdS black hole for R\'enyi entropy for $Q=1$, $a=1$, $l=10$ and $\lambda =0.01$.}\label{fig23}
		\end{center}
	\end{figure}

	We plot the temperature, T and specific heat capacity, $\mathcal{C}$ versus the R\'enyi entropy, S for the EH AdS black hole as can be seen from Fig.\ref{fig22} and Fig.\ref{fig23} respectively, where we see for the $T-S_R$ curve in Fig.\ref{fig22} there are visible peaks at $S_R=2.807$, $12.683$ and $S_R=23.285$  respectively, whereas for the C-S curve in Fig.\ref{fig23} we see potential discontinuities exactly at the same points as that of the peaks observed in the temperature curve of the black hole.
	
	The GTD metric for the EH AdS black hole for the R\'enyi entropy can be written as:
	\begin{equation}
		g  =S_R \left(\frac{\partial M}{\partial S_R}\right)\left(- \frac{\partial^2 M}{\partial S_R^2} dS_R^2  + \frac{\partial^2 M}{\partial \tilde{Q}^2} d\tilde{Q}^2 \right)
	\end{equation}

\begin{figure}[ht]
		\begin{center}
			\includegraphics[width=.82\textwidth]{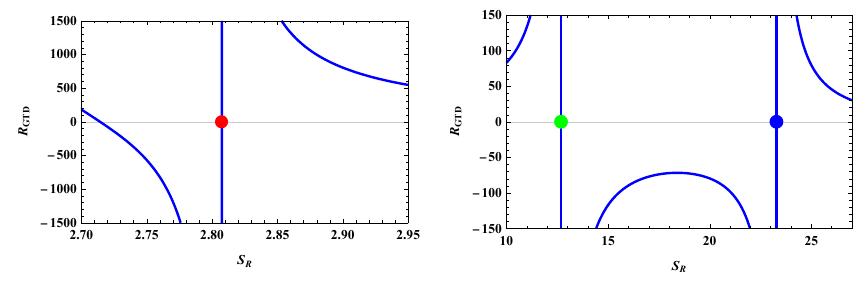}\vspace{3pt}
			\caption{$R_{GTD}-S_R$ curve for the EH AdS black hole for Renyi entropy for $Q=1$, $a=1$, $l=10$ and $\lambda =0.01$.}\label{fig24}
		\end{center}
	\end{figure}

	Now by substituting the value of M from above we get the GTD metric from which we can easily calculate the GTD scalar $R_{GTD}$ given in appendix \ref{ap:14}  and therefore plot it against the R\'enyi entropy for the EH AdS black hole as can be seen from Fig.\ref{fig24} where we see that in accordance with the peak and trough observed in the $T-S_R$ and the discontinuities observed in the $\mathcal{C}-S_R$ curves the scalar $R_{GTD}$ versus entropy, $S_R$ curve produces discontinuities exactly at the same points i.e. $S_R=2.807$, $12.683$ and $S_R=23.285$  respectively. 
	
	\subsection{Kaniadakis Entropy}
	After making series expansion for small $\kappa$ and then substituting the value of r from eqtn.\ref{eqENKr} to eqtn.\ref{2.16} we get the mass of the EH AdS black hole for Kaniadakis entropy, which is given as:
	
	\begin{equation}
		M =
		\frac{
			G \left(
			- 60 G^{2} S_K^{4} (-4 + S_K^{2} \kappa^{2})
			- a l^{2} \pi^{4} Q^{4} (12 + 5 S_K^{2} \kappa^{2})
			+ 20 G l^{2} \pi S_K^{2} \left(
			12 S_K
			- S_K^{3} \kappa^{2}
			+ \pi Q^{2} (12 + S_K^{2} \kappa^{2})
			\right)
			\right)
		}{
			480 l^{2} \pi^{3/2} (G S_K)^{5/2}
		}
		\label{eq:massehkania}
	\end{equation}

	The temperature of the black hole is therefore given as:
	
	\begin{equation}
		T =
		\frac{
			a l^{2} \pi^{4} Q^{4} (12 + S_K^{2} \kappa^{2})
			+ 4 G S_K^{2} \left(
			3 G S_K^{2} (12 - 7 S_K^{2} \kappa^{2})
			+ l^{2} \pi \left(
			12 S
			- 5 S_K^{3} \kappa^{2}
			+ 3 \pi Q^{2} (-4 + S_K^{2} \kappa^{2})
			\right)
			\right)
		}{
			192 l^{2} \pi^{3/2} S_K^{2} (G S_K)^{3/2}
		}
	\end{equation}

	The specific heat capacity of the black hole is given by:
	
	\begin{equation}
		C =
		- \frac{
			2 \left(
			a l^{2} \pi^{4} Q^{4} S_K (12 + S_K^{2} \kappa^{2})
			+ 4 G S_K^{3} \left(
			3 G S_K^{2} (12 - 7 S_K^{2} \kappa^{2})
			+ l^{2} \pi \left(
			12 S_K
			- 5 S_K^{3} \kappa^{2}
			+ 3 \pi Q^{2} (-4 + S_K^{2} \kappa^{2})
			\right)
			\right)
			\right)
		}{
			3 \left(
			a l^{2} \pi^{4} Q^{4} (28 + S_K^{2} \kappa^{2})
			+ 4 G S_K^{2} \left(
			G S_K^{2} (-12 + 35 S_K^{2} \kappa^{2})
			+ l^{2} \pi \left(
			- \pi Q^{2} (12 + S_K^{2} \kappa^{2})
			+ S_K (4 + 5 S_K^{2} \kappa^{2})
			\right)
			\right)
			\right)
		}
	\end{equation}

\begin{figure}[ht]
		\begin{center}
			\includegraphics[width=.82\textwidth]{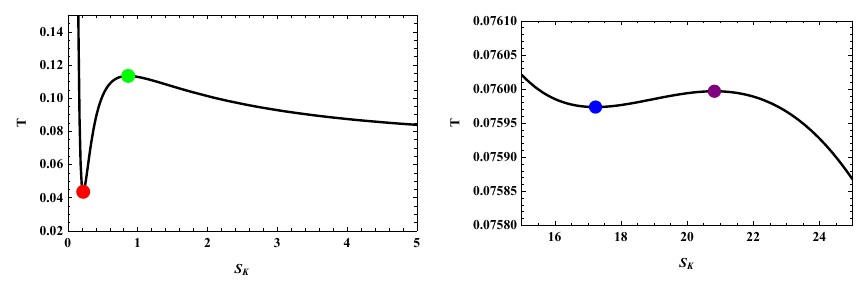}\vspace{3pt}
			\caption{$T-S_K$ curve for the NED AdS black hole for Kaniadakis entropy for $Q=1$, $a=1$, $l=10$ and $\kappa =0.016$.}\label{fig25}
		\end{center}
	\end{figure}

\begin{figure}[ht]
		\begin{center}
			\includegraphics[width=.999\textwidth]{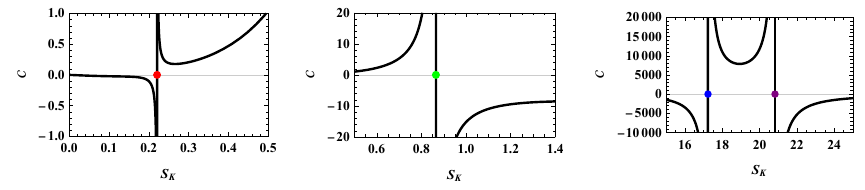}\vspace{3pt}
			\caption{$\mathcal{C}-S_K$ curve for the EH AdS black hole for Kaniadakis entropy for $Q=1$, $a=1$, $l=10$ and $\kappa =0.016$.}\label{fig26}
		\end{center}
	\end{figure}

	We plot the temperature, $T$ and specific heat capacity, $\mathcal{C}$ versus the Kaniadakis entropy, $S_K$ for the EH AdS black hole as can be seen from Fig.\ref{fig25} and Fig.\ref{fig26} respectively, where we see for the $T-S_K$ curve in Fig.\ref{fig25} there are visible peaks at $S_K=0.220, 0.865, 17.324$ and $S_K=20.827$  respectively, whereas for the $\mathcal{C}-S_K$ curve in Fig.\ref{fig26} we see potential discontinuities exactly at the same points as that of the peaks observed in the temperature curve of the black hole.
	
	The GTD metric for the EH AdS black hole for the Kaniadakis entropy can be written as:
	\begin{equation}
		g  =S_K \left(\frac{\partial M}{\partial S_K}\right)\left(- \frac{\partial^2 M}{\partial S_K^2} dS^2  + \frac{\partial^2 M}{\partial \tilde{Q}^2} d\tilde{Q}^2 \right)
	\end{equation}

\begin{figure}[ht]
		\begin{center}
			\includegraphics[width=1\textwidth]{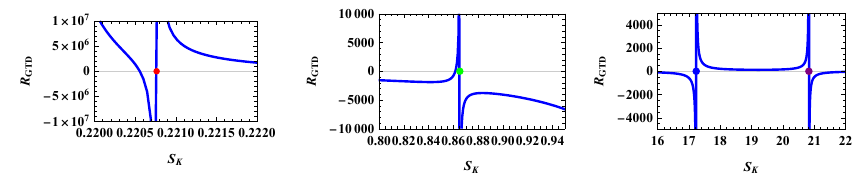}\vspace{3pt}
			\caption{$R_{GTD}-S_K$ curve for the EH AdS black hole for Kaniadakis entropy for $Q=1$, $a=1$, $l=10$ and $\kappa =0.016$.}\label{fig27}
		\end{center}
	\end{figure}

	Now by substituting the value of M from above we get the GTD metric from which we can easily calculate the GTD scalar $R_{GTD}$ given in appendix \ref{ap:15}  and therefore plot it against the Kaniadakis entropy for the EH AdS black hole as can be seen from Fig.\ref{fig27} where we see that in accordance with the peak and trough observed in the $T-S_K$ and the discontinuities observed in the $\mathcal{C}-S_K$ curves the scalar $R_{GTD}$ versus entropy, $S_K$ curve produces discontinuities exactly at the same points i.e. $S_K=0.220, 0.865, 17.324$ and $S_K=20.827$  respectively. 
	
	\section{CFT Thermodynamic geometry for Euler Heisenberg AdS black Holes}
	\label{sec:7}
	\subsection{Bekenstein-Hawking Entropy}
\begin{figure}[ht]
		\begin{center}
			\includegraphics[width=.52\textwidth]{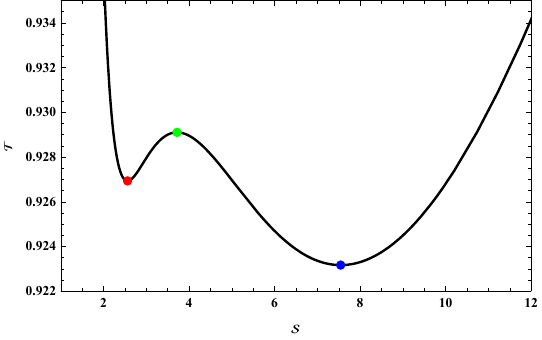}\vspace{3pt}
			\caption{$\mathcal{T}$--$\mathcal{S}$ curve for the dual CFT at fixed parameters $\mathcal{Q}=2$, $\alpha=0.1$, $C=3$ and $\mathcal{V}=1$
			}\label{fig:46}
		\end{center}
	\end{figure}

\begin{figure}[ht]
		\begin{center}
			\includegraphics[width=.52\textwidth]{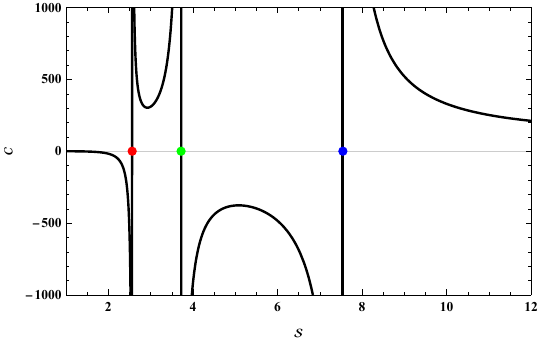}\vspace{3pt}
			\caption{$\mathcal{C}$--$\mathcal{S}$ curve for the dual CFT at fixed parameters $\mathcal{Q}=2$, $\alpha=0.1$, $C=3$ and $\mathcal{V}=1$}\label{fig:47}
		\end{center}
	\end{figure}

\begin{figure}[ht]
		\begin{center}
			\includegraphics[width=.52\textwidth]{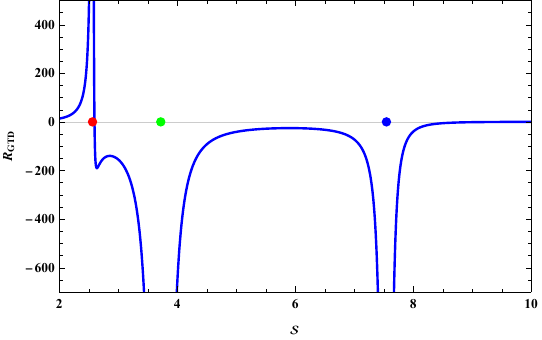}\vspace{3pt}
			\caption{$R_{\text{GTD}}$--$\mathcal{S}$ profile for the dual CFT at fixed parameters $\mathcal{Q}=2$, $\alpha=0.1$, $C=3$ and $\mathcal{V}=1$}\label{fig:48}
		\end{center}
	\end{figure}

	Utilizing the relations provided in Eqs.~\eqref{eq:massehBH}, \eqref{eq:centralcharge}, \eqref{eq:volume}, and \eqref{eq:holographic}, one can obtain the expression for the energy of the holographically dual CFT associated with a nonlinear electrodynamics (NED) black hole. When the thermodynamic description is formulated using the Bekenstein--Hawking entropy, the CFT energy takes the form
	\begin{equation}
		E=\frac{80 \pi  C \mathcal{S}^3-\pi ^4 \alpha  \mathcal{Q}^4+20 \pi ^2 \mathcal{Q}^2 \mathcal{S}^2+20 \mathcal{S}^4}{40 \pi  \mathcal{S}^{5/2} \sqrt{C \mathcal{V}}}
		\label{eq:cftenergy4}
	\end{equation}
	The temperature associated with the holographically dual CFT can therefore be determined as
	\begin{equation}
\mathcal{T}=\frac{16 \pi  C \mathcal{S}^3+\pi ^4 \alpha  \mathcal{Q}^4-4 \pi ^2 \mathcal{Q}^2 \mathcal{S}^2+12 \mathcal{S}^4}{16 \pi  \mathcal{S}^{7/2} \sqrt{C \mathcal{V}}}
	\end{equation}
	The heat capacity can then by expressed as
	\begin{equation}
	\mathcal{C}=\frac{2 \mathcal{S} \left(16 \pi  C \mathcal{S}^3+\pi ^4 \alpha  \mathcal{Q}^4-4 \pi ^2 \mathcal{Q}^2 \mathcal{S}^2+12 \mathcal{S}^4\right)}{-16 \pi  C \mathcal{S}^3-7 \pi ^4 \alpha  \mathcal{Q}^4+12 \pi ^2 \mathcal{Q}^2 \mathcal{S}^2+12 \mathcal{S}^4}
	\end{equation}
	The thermodynamic characteristics of the dual CFT for Euler Heisenberg AdS black holes are investigated by analyzing the temperature $\mathcal{T}$ and specific heat $\mathcal{C}$ as functions of the Bekenstein--Hawking entropy $\mathcal{S}$, as illustrated in Figs.~\ref{fig:46} and \ref{fig:47}, respectively. The $\mathcal{T}$--$\mathcal{S}$ curve displayed in Fig.~\ref{fig:46} exhibits three prominent extremal points located at $\mathcal{S}=2.563$, $\mathcal{S}=3.720$ and $\mathcal{S}=7.546$. A parallel examination of the $\mathcal{C}$--$\mathcal{S}$ profile shown in Fig.~\ref{fig:47} reveals distinct divergences occurring exactly at these same three entropy values. The alignment of the temperature extrema with the singular behavior of the specific heat provides strong evidence for the presence of critical thermodynamic phenomena within the CFT framework.
	Within the framework of CFT thermodynamics, the geometrothermodynamic (GTD) metric derived from the Bekenstein--Hawking entropy for the  dual Euler Heisenberg AdS black hole is expressed as
	\begin{equation}
		\tilde{g}=\mathcal{S}\left(\frac{\partial E}{\partial \mathcal{S}}\right)
		\left(
		-\frac{\partial^{2}E}{\partial \mathcal{S}^{2}}\, d\mathcal{S}^{2}
		+\frac{\partial^{2}E}{\partial \mathcal{Q}^{2}}\, d\mathcal{Q}^{2}
		\right).
	\end{equation}
	By inserting the expression for the CFT energy $E$ from Eq.~\ref{eq:cftenergy4} into the preceding relation, the explicit geometrothermodynamic line element is obtained. This formulation enables a direct evaluation of the corresponding GTD scalar curvature $R_{GTD}$ given in appendix \ref{ap:16}. We subsequently investigate the dependence of this curvature scalar on the Bekenstein--Hawking entropy $\mathcal{S}$ within the CFT thermodynamic framework. The resulting behavior is illustrated in Fig.~\ref{fig:48}, where the $R_{\mathrm{GTD}}$--$\mathcal{S}$ profile develops pronounced singularities at $\mathcal{S}=2.563$, $\mathcal{S}=3.720$ and $\mathcal{S}=7.546$. Notably, these three entropy values coincide with the extremal points identified in the $\mathcal{T}$--$\mathcal{S}$ curve and with the discontinuities appearing in the $\mathcal{C}$--$\mathcal{S}$ diagram. This correspondence emphasizes the strong connection between the geometric information encoded in the GTD curvature and the underlying critical thermodynamic features of the system.
	\subsection{R\'enyi Entropy}
	By utilizing the relations presented in Eqs.~\eqref{eq:massehrenyi}, \eqref{eq:centralcharge}, \eqref{eq:volume}, and \eqref{eq:holographic}, the energy associated with the holographically dual CFT can be determined when the thermodynamic framework is expressed in terms of R\'enyi entropy. The corresponding expression for the CFT energy is therefore written as
\begin{equation}
E=\frac{\sqrt{\frac{\mathcal{S}_R}{C}} \left(80 \pi  C \mathcal{S}_R^3 (\lambda  \mathcal{S}_R+4)+\pi ^4 \alpha  \mathcal{Q}^4 (5 \lambda  \mathcal{S}_R-4)-20 \pi ^2 \mathcal{Q}^2 \mathcal{S}_R^2 (\lambda  \mathcal{S}_R-4)+20 \mathcal{S}_R^4 (3 \lambda  \mathcal{S}_R+4)\right)}{160 \pi  \mathcal{S}_R^3 \sqrt{\mathcal{V}}}
\label{eq:cftenergy5}
\end{equation}
The temperature of the dual CFT, described with the R\'enyi entropy is written as
\begin{equation}
\mathcal{T}=\frac{\sqrt{\frac{\mathcal{S}_R}{C}} \left(16 \pi  C \mathcal{S}_R^3 (3 \lambda  \mathcal{S}_R+4)+\pi ^4 \alpha  \mathcal{Q}^4 (4-3 \lambda  \mathcal{S}_R)-4 \pi ^2 \mathcal{Q}^2 \mathcal{S}_R^2 (\lambda  \mathcal{S}_R+4)+12 \mathcal{S}_R^4 (5 \lambda  \mathcal{S}_R+4)\right)}{64 \pi  \mathcal{S}_R^4 \sqrt{\mathcal{V}}}
\end{equation}
Hence the specific heat in this case is given as
\begin{equation}
\mathcal{C}=\frac{32 \pi  C \mathcal{S}_R^4 (3 \lambda  \mathcal{S}_R+4)+2 \pi ^4 \alpha  \mathcal{Q}^4 \mathcal{S}_R (4-3 \lambda  \mathcal{S}_R)-8 \pi ^2 \mathcal{Q}^2 \mathcal{S}_R^3 (\lambda  \mathcal{S}_R+4)+24 \mathcal{S}_R^5 (5 \lambda  \mathcal{S}_R+4)}{16 \pi  C \mathcal{S}_R^3 (3 \lambda  \mathcal{S}_R-4)+\pi ^4 \alpha  \mathcal{Q}^4 (15 \lambda  \mathcal{S}_R-28)+4 \pi ^2 \mathcal{Q}^2 \mathcal{S}_R^2 (\lambda  \mathcal{S}_R+12)+12 \mathcal{S}_R^4 (15 \lambda  \mathcal{S}_R+4)}
\end{equation}
	\begin{figure}[ht]
	\begin{center}
		\includegraphics[width=.52\textwidth]{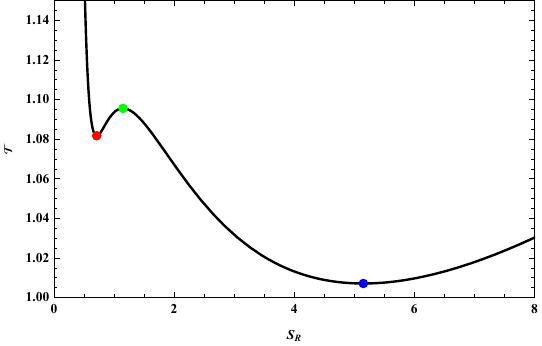}\vspace{3pt}
		\caption{$\mathcal{T}$--$\mathcal{S}_R$ curve for the dual CFT using R\'enyi entropy at fixed parameters $\mathcal{Q}=1$, $\alpha=0.04$, $C=2$, $\lambda=0.01$ and $\mathcal{V}=1$
		}\label{fig:49}
	\end{center}
\end{figure}
The thermodynamic behavior of the dual CFT in the presence of R\'enyi entropy is investigated by examining the variation of the temperature $\mathcal{T}$ and the specific heat $\mathcal{C}$ with respect to the entropy $\mathcal{S}_R$, as displayed in Figs.~\ref{fig:49} and \ref{fig:50}, respectively. The $\mathcal{T}$--$\mathcal{S}_R$ curve shown in Fig.~\ref{fig:49} exhibits three prominent extremal points located at $\mathcal{S}_R=0.712$, $\mathcal{S}_R=1.149$ and $\mathcal{S}_R=5.150$. A parallel inspection of the $\mathcal{C}$--$\mathcal{S}_R$ profile in Fig.~\ref{fig:50} reveals distinct divergences arising precisely at these same entropy values. The alignment of the temperature extrema with the singularities of the specific heat signals the occurrence of critical thermodynamic phenomena within the R\'enyi-modified CFT framework.\\
\begin{figure}[ht]
	\begin{center}
		\includegraphics[width=.72\textwidth]{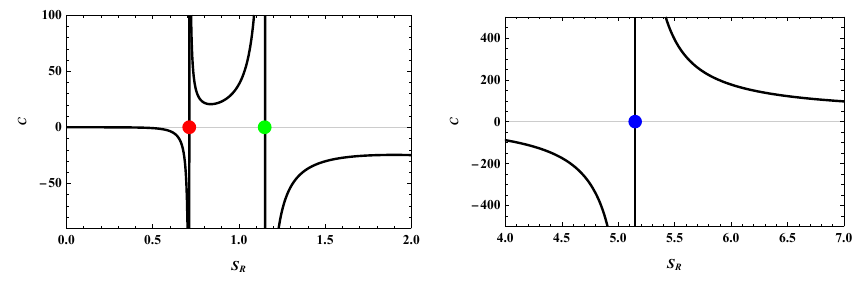}\vspace{3pt}
		\caption{$\mathcal{C}$--$\mathcal{S}_R$ curve for the dual CFT using R\'enyi entropy at fixed parameters  $\mathcal{Q}=1$, $\alpha=0.04$, $C=2$, $\lambda=0.01$ and $\mathcal{V}=1$}\label{fig:50}
	\end{center}
\end{figure}
Within the thermodynamic framework of the holographically dual CFT, the geometrothermodynamic (GTD) metric constructed using the R\'enyi entropy for the nonlinear electrodynamics (NED) system can be written as
\begin{equation}
	\tilde{g}=\mathcal{S}_{\mathrm{R}}\left(\frac{\partial E}{\partial \mathcal{S}_{\mathrm{R}}}\right)
	\left(
	-\frac{\partial^{2}E}{\partial \mathcal{S}_{\mathrm{R}}^{2}}\, d\mathcal{S}_{\mathrm{R}}^{2}
	+\frac{\partial^{2}E}{\partial \mathcal{Q}^{2}}\, d\mathcal{Q}^{2}
	\right).
\end{equation}
	\begin{figure}[ht]
	\begin{center}
		\includegraphics[width=.72\textwidth]{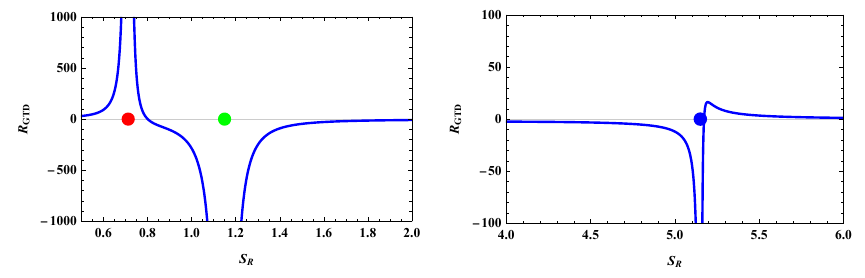}\vspace{3pt}
		\caption{$R_{\text{GTD}}$--$\mathcal{S}_R$ profile for the dual CFT using R\'enyi entropy at fixed parameters $\mathcal{Q}=1$, $\alpha=0.04$, $C=2$, $\lambda=0.01$ and $\mathcal{V}=1$}
		\label{fig:51}
	\end{center}
\end{figure}
Substituting the CFT energy expression $E$ from Eq.~\eqref{eq:cftenergy5} into the above relation yields the explicit geometrothermodynamic metric. From this metric, the associated GTD scalar curvature $R_{GTD}$ given in appendix \ref{ap:17} can be computed in a straightforward manner. We then analyze the dependence of this curvature scalar on the R\'enyi entropy $\mathcal{S}_R$ within the CFT thermodynamic framework. The corresponding behavior is illustrated in Fig.~\ref{fig:51}, where the $R_{\mathrm{GTD}}$--$\mathcal{S}_R$ curve exhibits pronounced singularities at $\mathcal{S}_R=0.712$, $\mathcal{S}_R=1.149$ and $\mathcal{S}_R=5.150$. Interestingly, these entropy values coincide with the extremal points identified in the $\mathcal{T}$--$\mathcal{S}_R$ profile as well as with the discontinuities appearing in the $\mathcal{C}$--$\mathcal{S}_R$ curve. This agreement underscores the close relationship between the geometric structure encoded in the GTD curvature and the critical thermodynamic behavior characterizing the R\'enyi-modified CFT system.
\subsection{Kaniadakis Entropy}
Utilizing the relations presented in Eqs.~\eqref{eq:massehkania}, \eqref{eq:centralcharge}, \eqref{eq:volume}, and \eqref{eq:holographic}, the energy associated with the holographically dual CFT of Euler Heisenberg AdS BH can be obtained when the thermodynamic framework is expressed in terms of Kaniadakis entropy. The corresponding form of the CFT energy is therefore given by
\begin{equation}
E=-\frac{\sqrt{\frac{\mathcal{S}_K}{C}} \left(80 \pi  C \mathcal{S}_K^3 \left(\kappa ^2 \mathcal{S}_K^2-12\right)+\pi ^4 \alpha  \mathcal{Q}^4 \left(5 \kappa ^2 \mathcal{S}_K^2+12\right)-20 \pi ^2 \mathcal{Q}^2 \mathcal{S}_K^2 \left(\kappa ^2 \mathcal{S}_K^2+12\right)+60 \mathcal{S}_K^4 \left(\kappa ^2 \mathcal{S}_K^2-4\right)\right)}{480 \pi  \mathcal{S}_K^3 \sqrt{\mathcal{V}}}
\end{equation}
Therefore the dual CFT temperature of Euler Heisenberg AdS BH can be written as
\begin{equation}
\mathcal{T}=\frac{\sqrt{\frac{\mathcal{S}_K}{C}} \left(16 \pi C \mathcal{S}_K^3 \left(12-5 \kappa ^2 \mathcal{S}_K^2\right)+\pi ^4 \alpha  \mathcal{Q}^4 \left(\kappa ^2 \mathcal{S}_K^2+12\right)+12 \pi ^2 \mathcal{Q}^2 \mathcal{S}_K^2 \left(\kappa ^2 \mathcal{S}_K^2-4\right)-84 \kappa ^2 \mathcal{S}_K^6+144 \mathcal{S}_K^4\right)}{192 \pi  \mathcal{S}_K^4 \sqrt{\mathcal{V}}}
\end{equation}
The specific heat is also written as
\begin{equation}
\mathcal{C}=-\frac{2 \left(16 \pi C \mathcal{S}_K^4 \left(12-5 \kappa ^2 \mathcal{S}_K^2\right)+\pi ^4 \alpha  \mathcal{Q}^4 \mathcal{S}_K \left(\kappa ^2 \mathcal{S}_K^2+12\right)+12 \pi ^2 \mathcal{Q}^2 \mathcal{S}_K^3 \left(\kappa ^2 \mathcal{S}_K^2-4\right)-84 \kappa ^2 \mathcal{S}_K^7+144 \mathcal{S}_K^5\right)}{3 \left(16 \pi  C \mathcal{S}_K^3 \left(5 \kappa ^2 \mathcal{S}_K^2+4\right)+\pi ^4 \alpha  \mathcal{Q}^4 \left(\kappa ^2 \mathcal{S}_K^2+28\right)-4 \pi ^2 \mathcal{Q}^2 \mathcal{S}_K^2 \left(\kappa ^2 \mathcal{S}_K^2+12\right)+4 \mathcal{S}_K^4 \left(35 \kappa ^2 \mathcal{S}_K^2-12\right)\right)}
\end{equation}
	\begin{figure}[ht]
	\begin{center}
		\includegraphics[width=.82\textwidth]{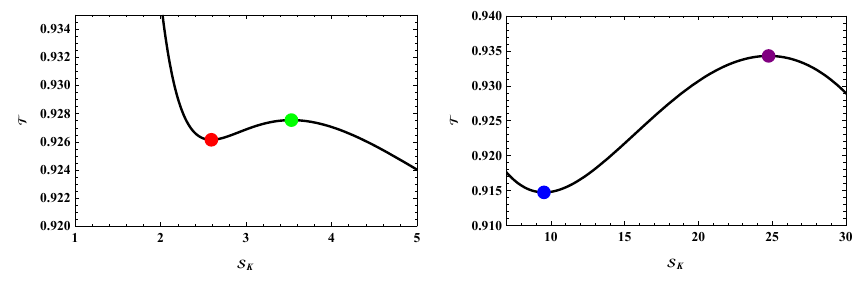}\vspace{3pt}
		\caption{$\mathcal{T}$--$\mathcal{S}_K$ curve for the dual CFT using Kaniadakis entropy at fixed parameters $\mathcal{Q}=2$, $\alpha=0.1$, $C=3$, $\kappa=0.016$ and $\mathcal{V}=1$}\label{fig:52}
	\end{center}
\end{figure}
The thermodynamic characteristics of the dual CFT within the Kaniadakis entropy framework are explored by examining the variation of the temperature $\mathcal{T}$ and the specific heat $\mathcal{C}$ with respect to the entropy $\mathcal{S}_K$, as shown in Figs.~\ref{fig:52} and \ref{fig:52}, respectively. The $\mathcal{T}$--$\mathcal{S}_K$ curve displayed in Fig.~\ref{fig:52} reveals four distinct extremal points occurring at $\mathcal{S}_K=2.594$, $\mathcal{S}_K=3.530$, $\mathcal{S}_K=9.537$ and $\mathcal{S}_K=24.765$. A parallel analysis of the $\mathcal{C}$--$\mathcal{S}_K$ behavior presented in Fig.~\ref{fig:35} indicates pronounced divergences appearing precisely at these same entropy values. The correspondence between the extrema of the temperature profile and the singular features of the specific heat provides strong evidence for the emergence of critical thermodynamic phenomena in the CFT description governed by Kaniadakis entropy.\\
Within the thermodynamic framework of the holographically dual CFT, the geometrothermodynamic (GTD) metric derived from the Kaniadakis entropy for the ModMax-deformed configuration can be expressed as
\begin{equation}
	\tilde{g}=\mathcal{S}_{\mathrm{K}}\left(\frac{\partial E}{\partial \mathcal{S}_{\mathrm{K}}}\right)
	\left(
	-\frac{\partial^{2}E}{\partial \mathcal{S}_{\mathrm{K}}^{2}}\, d\mathcal{S}_{\mathrm{K}}^{2}
	+\frac{\partial^{2}E}{\partial \mathcal{Q}^{2}}\, d\mathcal{Q}^{2}
	\right).
\end{equation}
	\begin{figure}[ht]
	\begin{center}
		\includegraphics[width=.82\textwidth]{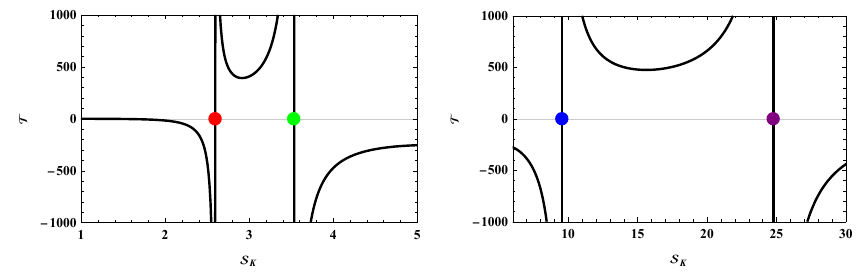}\vspace{3pt}
		\caption{$\mathcal{T}$--$\mathcal{S}_K$ curve for the dual CFT using Kaniadakis entropy at fixed parameters  $\mathcal{Q}=2$, $\alpha=0.1$, $C=3$, $\kappa=0.016$ and $\mathcal{V}=1$	}\label{fig:53}
	\end{center}
\end{figure}
Inserting the CFT energy expression $E$ from Eq.~\eqref{eq:cftenergy3} into the geometrothermodynamic formulation leads to the explicit representation of the GTD metric. This construction allows the associated scalar curvature $R_{GTD}$ given in appendix \ref{ap:18} to be computed directly. The variation of this curvature scalar with respect to the Kaniadakis entropy $\mathcal{S}_{\mathrm{K}}$ is subsequently examined within the CFT thermodynamic framework. The resulting behavior is depicted in Fig.~\ref{fig:36}, where the $R_{\mathrm{GTD}}$--$\mathcal{S}_{\mathrm{K}}$ curve develops pronounced singularities at $\mathcal{S}_{\mathrm{K}}=0.837$, $7.643$, and $29.461$. Notably, these entropy values coincide with the extremal points observed in the $\mathcal{T}$--$\mathcal{S}_{\mathrm{K}}$ profile and with the discontinuities appearing in the $\mathcal{C}$--$\mathcal{S}_{\mathrm{K}}$ curve. This correspondence underscores the strong connection between the geometric structure captured by the GTD curvature and the critical thermodynamic behavior characterizing the CFT system governed by Kaniadakis entropy.
	\begin{figure}[ht]
	\begin{center}
		\includegraphics[width=0.82\textwidth]{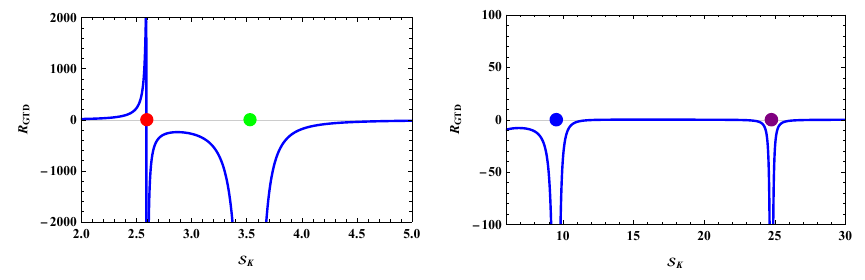}\vspace{3pt}
		\caption{$R_\mathrm{GTD}$--$\mathcal{S}_\mathrm{K}$ curve for the dual CFT using Kaniadakis entropy at fixed parameters  $\mathcal{Q}=2$, $\alpha=0.1$, $C=3$, $\kappa=0.016$ and $\mathcal{V}=1$}\label{fig:54}
	\end{center}
\end{figure}
\section{Conclusion}
\label{sec:8}
We investigated the thermodynamics and thermodynamic geometry of ModMax, NED and Euler Heisenberg AdS BH as well as its dual CFT counterpart using three entropy prescriptions namely the Bekenstein-Hawking, R\'enyi, and Kaniadakis. Across all three formulations, the critical structure is diagnosed consistently: extrema of the temperature profile coincide with divergences of the specific heat, and the thermodynamic curvature reproduces the same critical loci. This establishes a coherent thermodynamic--geometric characterization of boundary phase structure in the presence of nonlinear electromagnetic sectors and nonextensive entropy deformations.\\
In Section~\ref{sec:2}, we investigated the thermodynamic properties and geometrothermodynamic (GTD) structure of the ModMax AdS black hole under different entropy formulations. For each entropy model, the fundamental thermodynamic quantities—including the black hole mass $M$, temperature $T$, specific heat $\mathcal{C}$, and the GTD scalar curvature $R_{\mathrm{GTD}}$ are evaluated and analyzed. 
Within the Bekenstein--Hawking entropy framework, the $T$--$S$ diagram exhibits two distinct extrema located at $S=16.454$ and $S=88.265$. These entropy values coincide precisely with the divergence points observed in both the $\mathcal{C}$--$S$ and $R_{\mathrm{GTD}}$--$S$ curves, signaling the presence of critical thermodynamic behavior. 
When the analysis is performed using the R\'enyi entropy formulation, the critical structure shifts, and the characteristic points appear at $S_R=0.054$ and $S_R=0.823$. These entropy values are consistently reflected across the $T$--$S_R$, $\mathcal{C}$--$S_R$, and $R_{\mathrm{GTD}}$--$S_R$ profiles. 
In the Kaniadakis entropy framework, the thermodynamic phase structure becomes more intricate. The $T$--$S_K$ curve develops three turning points at $S_K=0.278$, $S_K=0.770$, and $S_K=35.837$. These locations are again accompanied by corresponding divergences in the $\mathcal{C}$--$S_K$ and $R_{\mathrm{GTD}}$--$S_K$ plots. This behavior indicates that the deformation of entropy modifies the distribution of critical points while preserving the fundamental thermodynamic criterion, whereby instabilities are associated with divergences of response functions.\\
In Section~\ref{sec:3}, we examined the thermodynamic behavior and geometrothermodynamic (GTD) structure of the CFT that is holographically dual to the ModMax AdS black hole under the three entropy formulations. Within the Bekenstein--Hawking entropy description, the $\mathcal{T}$--$\mathcal{S}$ curve exhibits two well-defined extrema at $\mathcal{S}=1.287$ and $\mathcal{S}=11.278$. These entropy values coincide exactly with the singularities appearing in both the $\mathcal{C}$--$\mathcal{S}$ and $R_{\mathrm{GTD}}$--$\mathcal{S}$ profiles, indicating the presence of critical thermodynamic behavior in the dual CFT.
When the analysis is performed using the R\'enyi entropy framework, the location of the critical points shifts, with the corresponding extrema occurring at $\mathcal{S}_R=1.457$ and $\mathcal{S}_R=3.995$. These values are consistently reproduced across the $\mathcal{T}$--$\mathcal{S}_R$, $\mathcal{C}$--$\mathcal{S}_R$, and $R_{\mathrm{GTD}}$--$\mathcal{S}_R$ diagrams. In the case of Kaniadakis entropy, the thermodynamic phase structure becomes richer, with the $\mathcal{T}$--$\mathcal{S}_K$ curve developing three turning points at $\mathcal{S}_K=1.222$, $\mathcal{S}_K=3.018$, and $\mathcal{S}_K=33.404$. These entropy values correspond precisely to the divergence points observed in the $\mathcal{C}$--$\mathcal{S}_K$ and $R_{\mathrm{GTD}}$--$\mathcal{S}_K$ plots.\\
In Section~\ref{sec:4}, we analyzed the thermodynamic behavior and the corresponding geometrothermodynamic (GTD) structure of the nonlinear electrodynamics (NED) AdS black hole within several entropy formalisms. 
Under the standard Bekenstein--Hawking entropy description, the $T$--$S$ curve exhibits two prominent extrema located at $S=27.351$ and $S=75.998$. These entropy values coincide exactly with the singularities appearing in the $\mathcal{C}$--$S$ and $R_{\mathrm{GTD}}$--$S$ diagrams, indicating the onset of critical thermodynamic phenomena. 
When the analysis is carried out using the R\'enyi entropy framework, the positions of the critical points shift, with the relevant extrema appearing at $S_R=13.450$ and $S_R=23.107$. These entropy values consistently emerge in the $T$--$S_R$, $\mathcal{C}$--$S_R$, and $R_{\mathrm{GTD}}$--$S_R$ profiles, demonstrating a coherent thermodynamic and geometric description of the phase structure.
Within the Kaniadakis entropy formulation, the thermodynamic landscape becomes more complex. The $T$--$S_K$ curve develops three distinct turning points at $S_K=5.524$, $S_K=99.459$, and $S_K=499.975$. Each of these points corresponds to divergences observed in the $\mathcal{C}$--$S_K$ and $R_{\mathrm{GTD}}$--$S_K$ plots, reflecting the emergence of additional critical loci in the phase structure due to entropy deformation. \\
In Section~\ref{sec:5}, we explored the thermodynamic characteristics and geometrothermodynamic (GTD) properties of the conformal field theory that is holographically dual to the NED AdS black hole within three distinct entropy frameworks. 
Under the conventional Bekenstein--Hawking entropy formulation, the $\mathcal{T}$--$\mathcal{S}$ profile displays two pronounced extrema located at $\mathcal{S}=6.633$ and $\mathcal{S}=11.173$. These entropy values coincide with the divergence points observed in both the $\mathcal{C}$--$\mathcal{S}$ and $R_{\mathrm{GTD}}$--$\mathcal{S}$ curves, indicating the emergence of critical thermodynamic behavior in the boundary CFT.
When the analysis is carried out using the R\'enyi entropy description, the critical structure shifts, with the characteristic extrema appearing at $\mathcal{S}_R=1.780$ and $\mathcal{S}_R=4.428$. These entropy values consistently manifest in the $\mathcal{T}$--$\mathcal{S}_R$, $\mathcal{C}$--$\mathcal{S}_R$, and $R_{\mathrm{GTD}}$--$\mathcal{S}_R$ plots, reflecting a coherent correspondence between the thermodynamic and geometric signals of phase transitions.
In the Kaniadakis entropy framework, the phase structure becomes more elaborate. The $\mathcal{T}$--$\mathcal{S}_K$ diagram develops three distinct turning points at $\mathcal{S}_K=0.837$, $\mathcal{S}_K=7.643$, and $\mathcal{S}_K=29.461$. These entropy values are again associated with divergences in the $\mathcal{C}$--$\mathcal{S}_K$ and $R_{\mathrm{GTD}}$--$\mathcal{S}_K$ profiles, demonstrating the appearance of additional critical points within the Kaniadakis-deformed CFT thermodynamics.\\
In Section~\ref{sec:6}, we investigated the thermodynamic characteristics and the associated geometrothermodynamic (GTD) geometry of the Euler--Heisenberg AdS black hole within different entropy formulations. 
Within the conventional Bekenstein--Hawking entropy framework, the $T$--$S$ profile reveals three distinct extrema located at $S=2.857$, $S=9.766$, and $S=94.255$. These entropy values precisely coincide with the divergence points observed in both the $\mathcal{C}$--$S$ and $R_{\mathrm{GTD}}$--$S$ diagrams, signaling the appearance of critical thermodynamic behavior in the system. 
When the analysis is extended to the R\'enyi entropy formulation, the locations of these characteristic points are modified, with the extrema appearing at $S_R=2.807$, $S_R=12.683$, and $S_R=23.285$. These values consistently manifest across the $T$--$S_R$, $\mathcal{C}$--$S_R$, and $R_{\mathrm{GTD}}$--$S_R$ plots, demonstrating a clear correspondence between thermodynamic response functions and the geometric structure encoded in the GTD curvature.
In the case of the Kaniadakis entropy framework, the thermodynamic phase structure becomes significantly richer. The $T$--$S_K$ diagram develops four distinct turning points located at $S_K=0.220$, $S_K=0.865$, $S_K=17.324$, and $S_K=20.827$. Each of these entropy values is accompanied by corresponding divergences in the $\mathcal{C}$--$S_K$ and $R_{\mathrm{GTD}}$--$S_K$ profiles, indicating the emergence of additional critical points in the phase space due to the deformation introduced by the generalized entropy formalism. \\
In Section~\ref{sec:7}, we examined the thermodynamic behavior and the associated geometrothermodynamic (GTD) structure of the conformal field theory that is holographically dual to the Euler--Heisenberg AdS black hole within three different entropy formulations.
Within the standard Bekenstein--Hawking entropy framework, the $\mathcal{T}$--$\mathcal{S}$ curve develops three distinct extrema located at $\mathcal{S}=2.563$, $\mathcal{S}=3.720$, and $\mathcal{S}=7.546$. These entropy values coincide with the divergence points appearing in both the $\mathcal{C}$--$\mathcal{S}$ and $R_{\mathrm{GTD}}$--$\mathcal{S}$ plots, signaling the presence of critical thermodynamic behavior in the boundary CFT description.
When the thermodynamic analysis is performed using the R\'enyi entropy formalism, the positions of the critical points are modified. In this case, the characteristic extrema arise at $\mathcal{S}_R=0.712$, $\mathcal{S}_R=1.149$, and $\mathcal{S}_R=5.150$. These entropy values consistently appear in the $\mathcal{T}$--$\mathcal{S}_R$, $\mathcal{C}$--$\mathcal{S}_R$, and $R_{\mathrm{GTD}}$--$\mathcal{S}_R$ diagrams, illustrating a clear correspondence between the thermodynamic response functions and the geometric structure encoded in the GTD curvature.
In the Kaniadakis entropy formulation, the thermodynamic phase structure becomes considerably richer. The $\mathcal{T}$--$\mathcal{S}_K$ profile exhibits four distinct turning points located at $\mathcal{S}_K=2.594$, $\mathcal{S}_K=3.530$, $\mathcal{S}_K=9.537$, and $\mathcal{S}_K=24.765$. Each of these entropy values corresponds to divergences observed in the $\mathcal{C}$--$\mathcal{S}_K$ and $R_{\mathrm{GTD}}$--$\mathcal{S}_K$ curves, indicating the emergence of additional critical points in the phase structure due to the deformation introduced by the generalized Kaniadakis entropy.\\
A table is shown below where it shows all the result obtained in this study.
\begin{table}[h]
	\centering
	\caption{Summary of critical behavior for different black hole systems and their dual CFTs under various entropy formalisms.}
	\begin{tabular}{|l|l|l|l|}
		\toprule
		\hline
		\textbf{System} & \textbf{Type of Entropy} & \textbf{Number of} & \textbf{Remarks} \\
	 &  & \textbf{Critical Points} &  \\
		\hline
		\midrule
		ModMax AdS BH & Bekenstein--Hawking &  ~~~~~~~~2 &The ModMax AdS BH and its dual CFT  \\
		
		ModMax AdS BH & R\'enyi &~~~~~~~ 2 &counterpart corresponds by having the same  \\
		
		ModMax AdS BH & Kaniadakis & ~~~~~~~~3 &number of critical points across various  \\
		
		Dual CFT (ModMax) & Bekenstein--Hawking & ~~~~~~~~2 &entropy formulation  \\
		
		Dual CFT (ModMax) & R\'enyi & ~~~~~~~~2 &  \\
		
		Dual CFT (ModMax) & Kaniadakis & ~~~~~~~~3 &  \\
		\hline
		NED AdS BH & Bekenstein--Hawking & ~~~~~~~~2 & The NED AdS BH and its dual CFT \\
		
		NED AdS BH & R\'enyi & ~~~~~~~~2 & counterpart corresponds by having the \\
		
		NED AdS BH & Kaniadakis & ~~~~~~~~3 & same number of critical points across \\
		
		Dual CFT (NED) & Bekenstein--Hawking & ~~~~~~~~2 & various entropy formulation \\
		
		Dual CFT (NED) & R\'enyi & ~~~~~~~~2 & \\
		
		Dual CFT (NED) & Kaniadakis & ~~~~~~~~3 & \\
		\hline
		Euler--Heisenberg AdS BH & Bekenstein--Hawking & ~~~~~~~~3 & The Euler Heisenberg AdS BH and its \\
		
		Euler--Heisenberg AdS BH & R\'enyi & ~~~~~~~~3 & dual CFT counterpart corresponds by \\
		
		Euler--Heisenberg AdS BH & Kaniadakis & ~~~~~~~~4 &having the same number of critical \\
		
		Dual CFT (Euler--Heisenberg) & Bekenstein--Hawking & ~~~~~~~~3 & points across various entropy formulations \\
		
		Dual CFT (Euler--Heisenberg) & R\'enyi & ~~~~~~~~3 &  \\
		
		Dual CFT (Euler--Heisenberg) & Kaniadakis & ~~~~~~~~4 &  \\
		\hline
		\bottomrule
	\end{tabular}
	\label{tab:criticalpoints}
\end{table}\\
A comparison between Sections~\ref{sec:2} for ModMax AdS BH and \ref{sec:3} showing dual CFT for the ModMax AdS BH reveals that the number and location of the critical points remain consistent between the ModMax AdS black hole and its holographically dual CFT description. This correspondence also matches for NED AdS BH and its dual CFT interpretation in Section \ref{sec:4}, \ref{sec:5} and also for the Euler heisenberg AdS BH in Section \ref{sec:6}, \ref{sec:7}.  This correspondence further supports the robustness of the holographic interpretation of thermodynamic phase transitions.\\
An important observation emerging from Table~\ref{tab:criticalpoints} is that the inclusion of Euler--Heisenberg nonlinear electrodynamics significantly enriches the thermodynamic phase structure when compared with the ModMax and NED cases. In particular, the Euler--Heisenberg AdS black hole exhibits a larger number of critical points, indicating a more intricate thermodynamic landscape associated with the nonlinear quantum corrections encoded in the Euler--Heisenberg action. This increased complexity is faithfully reflected in the corresponding dual CFT description, thereby reinforcing the consistency of the holographic correspondence across different nonlinear electrodynamics models. Furthermore, the adoption of generalized entropy formalisms introduces an additional layer of structure. While both the Bekenstein--Hawking and R\'enyi entropy frameworks lead to comparable critical behavior, the Kaniadakis entropy consistently generates one additional critical point across all the black hole systems considered in this work. This feature suggests that the deformation parameter inherent in the Kaniadakis statistics effectively enriches the thermodynamic phase space by allowing the emergence of additional critical loci. Consequently, both the choice of nonlinear electromagnetic theory in the bulk and the entropy formalism used to characterize microscopic degrees of freedom play a crucial role in shaping the overall thermodynamic phase structure of AdS black holes and their holographically dual CFTs.
Geometrothermodynamics supplies the geometric completion of these results. The Legendre-invariant GTD scalar curvature $R_{\mathrm{GTD}}$ develops singularities exactly at the entropy values where $\mathcal{C}$ diverges, demonstrating that the equilibrium-state manifold encodes the same critical structure detected by standard response functions. Within the broader AdS/CFT and extended-thermodynamics perspective, this strengthens the interpretation that boundary criticality reflects both the nonlinear bulk matter content and the entropy functional used to parametrize microstate counting, while the GTD curvature provides a robust, potential-independent diagnostic. The main limitations are the dependence of the quoted numerical values on the chosen parameter ranges and ensemble, and (where applicable) the controlled small-deformation expansions used for generalized entropies; nevertheless, the coincidence of thermodynamic and geometric criticality appears robust.\\
In future extensions, we can include analytic determination of critical points (and possible critical exponents), systematic parameter scans over ModMax/NED couplings and nonextensive parameters, generalizations to rotation or higher dimensions, exploration of other generalized entropies, and connections to boundary observables such as entanglement, complexity, and transport.
\newpage
\appendix
\section{Appendix}
In Section \ref{sec:2}, the GTD scalar for ModMax AdS black hole for the Bekenstein-Hawking entropy is given as
\begin{equation}
R_{GTD}=\frac{8 \pi ^3 G l^2 S^2 e^{3 \eta } \left(\frac{l^2 A}{C}+\frac{4 B}{D}\right)}{\pi  l^2 \left(\pi  Q^2-S e^{\eta }\right)-3 G S^2 e^{\eta }}
\label{ap:1}
\end{equation}
where A, B, C, D are given below
\begin{equation}
	\begin{split}
&A=\begin{aligned}\left(-81 G^3 S^5 e^{2 \eta }-9 \pi  G^2 S^4 e^{2 \eta }+\pi ^3 S^2 e^{\eta } \left(2 e^{\eta }+27 G Q^2\right)-\pi ^4 Q^2 S \left(4 e^{\eta }+39 G Q^2\right)+2 \pi ^5 Q^4\right)\end{aligned}\\
&B=\left(-81 G^3 S^5 e^{2 \eta }-36 \pi  G^2 S^4 e^{2 \eta }+16 \pi ^3 S^2 e^{\eta } \left(8 e^{\eta }+27 G Q^2\right)-16 \pi ^4 Q^2 S \left(16 e^{\eta }+39 G Q^2\right)+128 \pi ^5 Q^4\right)\\
&C=\left(3 G S^2 e^{\eta }-\pi  S e^{\eta }+3 \pi ^2 Q^2\right) \left(3 G S^2 e^{\eta }+\pi  S e^{\eta }-\pi ^2 Q^2\right)^2 \left(3 G S^2 e^{\eta }+\pi  l^2 \left(3 \pi  Q^2-S e^{\eta }\right)\right)\\
&D=\left(3 G S^2 e^{\eta }-4 \pi  S e^{\eta }+12 \pi ^2 Q^2\right)^2 \left(3 G S^2 e^{\eta }+4 \pi  S e^{\eta }-4 \pi ^2 Q^2\right)^2
\end{split}
\end{equation}
The GTD scalar for ModMax AdS black hole for the R\'enyi entropy is given as
\begin{equation}
R_{GTD}=\frac{128 e^{2 \eta } G l^2 \pi ^3 S^2}{(S \lambda -4)^2} \left(\frac{4 \left(\begin{aligned}&16 \pi ^4 Q^4 A e^{\eta }+16 \pi ^2 Q^2 S B e^{2 \eta }\\&+C e^{3 \eta }-1024 \pi ^6 Q^6 \lambda\times \\& \left(S^3 \lambda ^3-4 S^2 \lambda ^2-80 S \lambda +64\right)\end{aligned}\right)}{D}-\frac{l^2 \left(\begin{aligned}&\pi ^4 Q^4E e^{\eta }+\pi ^2 Q^2 S F e^{2 \eta }+\\&G e^{3 \eta }-16 \pi ^6 Q^6 \lambda\times \\&  \left(S^3 \lambda ^3-4 S^2 \lambda ^2-80 S \lambda +64\right)\end{aligned}\right)}{H}\right)
\label{ap:2}
\end{equation}
where A, B, C, D, E, F, G, H are given below
\begin{equation}
	\begin{split}
&A=\left(\begin{aligned}&16 \pi  \left(3 S^5 \lambda ^5+70 S^4 \lambda ^4+64 S^3 \lambda ^3-2688 S^2 \lambda ^2+1024 S \lambda -512\right)\\&+3 G S \left(75 S^5 \lambda ^5+836 S^4 \lambda ^4-32 S^3 \lambda ^3-29056 S^2 \lambda ^2+32512 S \lambda +13312\right)\end{aligned}\right)\\
&B=\left(\begin{aligned}&9 G^2 \lambda  \left(75 S^4 \lambda ^4+1040 S^3 \lambda ^3-7072 S^2 \lambda ^2+3072 S \lambda -256\right) S^3\\&+3 G \pi  \left(45 S^5 \lambda ^5+1988 S^4 \lambda ^4-12480 S^3 \lambda ^3+18432 S^2 \lambda ^2-39168 S \lambda -9216\right) S\\&+8 \pi ^2 \left(9 S^5 \lambda ^5+84 S^4 \lambda ^4-976 S^3 \lambda ^3+3264 S^2 \lambda ^2-1536 S \lambda +2048\right)\end{aligned}\right)\\
&C=\left(\begin{aligned}&27 G^3 \left(2625 S^5 \lambda ^5-14900 S^4 \lambda ^4+16480 S^3 \lambda ^3+35968 S^2 \lambda ^2+20736 S \lambda +3072\right) S^5\\&+768 G \pi ^2 \lambda  \left(45 S^4 \lambda ^4-279 S^3 \lambda ^3+592 S^2 \lambda ^2+416 S \lambda +832\right) S^4\\&+36 G^2 \pi  \left(2325 S^5 \lambda ^5-13780 S^4 \lambda ^4+24128 S^3 \lambda ^3+24576 S^2 \lambda ^2+17152 S \lambda +1024\right) S^4\\&+128 \pi ^3 \left(27 S^5 \lambda ^5-216 S^4 \lambda ^4+720 S^3 \lambda ^3-64 S^2 \lambda ^2-1024\right) S^2\end{aligned}\right)\\
&D=\begin{aligned}&\left(4 \pi  S (3 S \lambda +4) e^{\eta }+3 G S^2 (5 S \lambda +4) e^{\eta }-4 \pi ^2 Q^2 (S \lambda +4)\right)^2\times\\& \left(4 \pi  S (3 S \lambda -4) e^{\eta }+3 G S^2 (15 S \lambda +4) e^{\eta }+4 \pi ^2 Q^2 (S \lambda +12)\right)^2\times\\& \left(3 e^{\eta } G S^2 (5 S \lambda +4)-l^2 \pi  \left(\pi  Q^2 (S \lambda +4)-e^{\eta } S (3 S \lambda +4)\right)\right)\end{aligned}\\
&E= \left(\begin{aligned}&4 \pi  \left(3 S^5 \lambda ^5+70 S^4 \lambda ^4+64 S^3 \lambda ^3-2688 S^2 \lambda ^2+1024 S \lambda -512\right)\\&+3 G S \left(75 S^5 \lambda ^5+836 S^4 \lambda ^4-32 S^3 \lambda ^3-29056 S^2 \lambda ^2+32512 S \lambda +13312\right)\end{aligned}\right)\\
&F=\left(\begin{aligned}&36 G^2 \lambda  \left(75 S^4 \lambda ^4+1040 S^3 \lambda ^3-7072 S^2 \lambda ^2+3072 S \lambda -256\right) S^3\\&+3 G \pi  \left(45 S^5 \lambda ^5+1988 S^4 \lambda ^4-12480 S^3 \lambda ^3+18432 S^2 \lambda ^2-39168 S \lambda -9216\right) S\\&+2 \pi ^2 \left(9 S^5 \lambda ^5+84 S^4 \lambda ^4-976 S^3 \lambda ^3+3264 S^2 \lambda ^2-1536 S \lambda +2048\right)\end{aligned}\right)\\
&G=\left(\begin{aligned}&27 G^3 \left(2625 S^5 \lambda ^5-14900 S^4 \lambda ^4+16480 S^3 \lambda ^3+35968 S^2 \lambda ^2+20736 S \lambda +3072\right) S^5\\&+48 G \pi ^2 \lambda  \left(45 S^4 \lambda ^4-279 S^3 \lambda ^3+592 S^2 \lambda ^2+416 S \lambda +832\right) S^4\\&+9 G^2 \pi  \left(2325 S^5 \lambda ^5-13780 S^4 \lambda ^4+24128 S^3 \lambda ^3+24576 S^2 \lambda ^2+17152 S \lambda +1024\right) S^4\\&+2 \pi ^3 \left(27 S^5 \lambda ^5-216 S^4 \lambda ^4+720 S^3 \lambda ^3-64 S^2 \lambda ^2-1024\right) S^2\end{aligned}\right)\\
\end{split}
\end{equation}
\begin{equation*}
H=\left(\begin{aligned}&\left(\pi  S (3 S \lambda +4) e^{\eta }+3 G S^2 (5 S \lambda +4) e^{\eta }-\pi ^2 Q^2 (S \lambda +4)\right)^2 \times\\&\left(\pi  S (3 S \lambda -4) e^{\eta }+3 G S^2 (15 S \lambda +4) e^{\eta }+\pi ^2 Q^2 (S \lambda +12)\right)\times\\& \left(3 G S^2 (15 S \lambda +4) e^{\eta }+l^2 \pi  \left(S (3 S \lambda -4) e^{\eta }+\pi  Q^2 (S \lambda +12)\right)\right)\times\\& \left(l^2 \pi  \left(\pi  Q^2 (S \lambda +4)-e^{\eta } S (3 S \lambda +4)\right)-3 e^{\eta } G S^2 (5 S \lambda +4)\right)\end{aligned}\right)
\end{equation*}
The GTD scalar for ModMax AdS BH for the Kaniadakis entropy is
\begin{equation}
R_{GTD}=\frac{384 e^{2 \eta } G l^2 \pi ^3 S^2 \left(\frac{l^2 \left(\begin{aligned}&3 \pi ^4 Q^4 A e^{\eta }-3 \pi ^2 Q^2 S B e^{2 \eta }\\&+I e^{3 \eta }+576 \pi ^6 Q^6 S \kappa ^2 \left(S^6 \kappa ^6\right.\\&\left.-12 S^4 \kappa ^4-336 S^2 \kappa ^2-576\right)\end{aligned}\right)}{D}-\frac{4 \left(\begin{aligned}&48 \pi ^4 Q^4 E e^{\eta }-48 \pi ^2 Q^2 S F e^{2 \eta }\\&+G e^{3 \eta }+36864 \pi ^6 Q^6 S \kappa ^2 \left(S^6 \kappa ^6\right.\\&\left.-12 S^4 \kappa ^4-336 S^2 \kappa ^2-576\right)\end{aligned}\right)}{H}\right)}{\begin{aligned}&\left(S^2 \kappa ^2+12\right)^2 \left(l^2 \pi  \left(3 \pi  Q^2 \left(S^2 \kappa ^2-4\right)\right.\right.\\& \left.\left.-e^{\eta } S \left(5 S^2 \kappa ^2-12\right)\right)-3 e^{\eta } G S^2 \left(7 S^2 \kappa ^2-12\right)\right)\end{aligned}}
\label{ap:3}
\end{equation}
where A, B, I, D, E, F, G, H are given below
\begin{equation}
	\begin{split}
&	A=\left(\begin{aligned}&\pi  \left(50 S^{10} \kappa ^{10}+72 S^8 \kappa ^8+28032 S^6 \kappa ^6+268800 S^4 \kappa ^4+345600 S^2 \kappa ^2+55296\right)\\&+3 G S \left(175 S^{10} \kappa ^{10}+692 S^8 \kappa ^8+52192 S^6 \kappa ^6+452736 S^4 \kappa ^4+762624 S^2 \kappa ^2-359424\right)\end{aligned}\right)\\
&B= \left(\begin{aligned}&24 G^2 \kappa ^2 \left(245 S^8 \kappa ^8+7700 S^6 \kappa ^6+123504 S^4 \kappa ^4+137664 S^2 \kappa ^2+55296\right) S^4+\\&G \pi  \left(2555 S^{10} \kappa ^{10}+58268 S^8 \kappa ^8+950304 S^6 \kappa ^6+1650816 S^4 \kappa ^4+3158784 S^2 \kappa ^2-746496\right) S\\&+8 \pi ^2 \left(25 S^{10} \kappa ^{10}+540 S^8 \kappa ^8+10088 S^6 \kappa ^6+25440 S^4 \kappa ^4+42624 S^2 \kappa ^2+13824\right)\end{aligned}\right)\\
&I=\left(\begin{aligned}&8 G \pi ^2 \kappa ^2 \left(2975 S^8 \kappa ^8+40200 S^6 \kappa ^6+243936 S^4 \kappa ^4+41472 S^2 \kappa ^2+518400\right) S^5+\\&9 G^3 \left(32585 S^{10} \kappa ^{10}+470988 S^8 \kappa ^8+1989792 S^6 \kappa ^6-3259008 S^4 \kappa ^4+3587328 S^2 \kappa ^2-746496\right) S^5\\&+3 G^2 \pi  \left(48265 S^{10} \kappa ^{10}+669732 S^8 \kappa ^8+3424032 S^6 \kappa ^6-1662336 S^4 \kappa ^4+4416768 S^2 \kappa ^2-248832\right) S^4\\&+2 \pi ^3 \left(625 S^{10} \kappa ^{10}+8500 S^8 \kappa ^8+62880 S^6 \kappa ^6+42624 S^4 \kappa ^4+158976 S^2 \kappa ^2+82944\right) S^2\end{aligned}\right)\\
&D=\left(\begin{aligned}&\left(\pi  S \left(5 S^2 \kappa ^2-12\right) e^{\eta }+3 G S^2 \left(7 S^2 \kappa ^2-12\right) e^{\eta }-3 \pi ^2 Q^2 \left(S^2 \kappa ^2-4\right)\right)^2\times\\& \left(\pi  S \left(5 S^2 \kappa ^2+4\right) e^{\eta }+G S^2 \left(35 S^2 \kappa ^2-12\right) e^{\eta }-\pi ^2 Q^2 \left(S^2 \kappa ^2+12\right)\right)\times \\& \left(l^2 \pi  \left(\pi  Q^2 \left(S^2 \kappa ^2+12\right)-e^{\eta } S \left(5 S^2 \kappa ^2+4\right)\right)-e^{\eta } G S^2 \left(35 S^2 \kappa ^2-12\right)\right)\end{aligned}\right)\\
& E=\left(\begin{aligned}&8 \pi  \left(25 S^{10} \kappa ^{10}+36 S^8 \kappa ^8+14016 S^6 \kappa ^6+134400 S^4 \kappa ^4+172800 S^2 \kappa ^2+27648\right)\\&+3 G S \left(175 S^{10} \kappa ^{10}+692 S^8 \kappa ^8+52192 S^6 \kappa ^6+452736 S^4 \kappa ^4+762624 S^2 \kappa ^2-359424\right)\end{aligned}\right)\\
&F=\left(\begin{aligned}&6 G^2 \kappa ^2 \left(245 S^8 \kappa ^8+7700 S^6 \kappa ^6+123504 S^4 \kappa ^4+137664 S^2 \kappa ^2+55296\right) S^4\\&+G \pi  \left(2555 S^{10} \kappa ^{10}+58268 S^8 \kappa ^8+950304 S^6 \kappa ^6+1650816 S^4 \kappa ^4+3158784 S^2 \kappa ^2-746496\right) S\\&+32 \pi ^2 \left(25 S^{10} \kappa ^{10}+540 S^8 \kappa ^8+10088 S^6 \kappa ^6+25440 S^4 \kappa ^4+42624 S^2 \kappa ^2+13824\right)\end{aligned}\right)\\
&G=\left(\begin{aligned}&128 G \pi ^2 \kappa ^2 \left(2975 S^8 \kappa ^8+40200 S^6 \kappa ^6+243936 S^4 \kappa ^4+41472 S^2 \kappa ^2+518400\right) S^5\\&+9 G^3 \left(32585 S^{10} \kappa ^{10}+470988 S^8 \kappa ^8+1989792 S^6 \kappa ^6-3259008 S^4 \kappa ^4+3587328 S^2 \kappa ^2-746496\right) S^5+\\&12 G^2 \pi  \left(48265 S^{10} \kappa ^{10}+669732 S^8 \kappa ^8+3424032 S^6 \kappa ^6-1662336 S^4 \kappa ^4+4416768 S^2 \kappa ^2-248832\right) S^4\\&+128 \pi ^3 \left(625 S^{10} \kappa ^{10}+8500 S^8 \kappa ^8+62880 S^6 \kappa ^6+42624 S^4 \kappa ^4+158976 S^2 \kappa ^2+82944\right) S^2\end{aligned}\right)\\
&H=\left(\begin{aligned}&\left(4 \pi  S \left(5 S^2 \kappa ^2-12\right) e^{\eta }+3 G S^2 \left(7 S^2 \kappa ^2-12\right) e^{\eta }-12 \pi ^2 Q^2 \left(S^2 \kappa ^2-4\right)\right)^2\times \\& \left(4 \pi  S \left(5 S^2 \kappa ^2+4\right) e^{\eta }+G S^2 \left(35 S^2 \kappa ^2-12\right) e^{\eta }-4 \pi ^2 Q^2 \left(S^2 \kappa ^2+12\right)\right)^2\end{aligned}\right)
\end{split}
\end{equation}
In section \ref{sec:3}, the GTD scalar for the dual CFT of ModMax in Bekensetein Hawking entropy is given as
\begin{equation}
R_{GTD}=-\frac{\begin{aligned}&16 \pi ^2 c \mathcal{S}^2 \mathcal{V} e^{3 \eta } \left(128 \pi ^3 c^3 \mathcal{S}^2 e^{2 \eta }-64 \pi ^4 c^2 \mathcal{Q}^2 \mathcal{S} e^{\eta }+4 \pi  c \left(-9 \mathcal{S}^4 e^{2 \eta }+27 \pi ^2 \mathcal{Q}^2 \mathcal{S}^2 e^{\eta }+2 \pi ^4 \mathcal{Q}^4\right)\right.\\&\left.-81 \mathcal{S}^5 e^{2 \eta }-39 \pi ^4 \mathcal{Q}^4 \mathcal{S}\right)\end{aligned}}{\left(-4 \pi  c \mathcal{S} e^{\eta }+3 \mathcal{S}^2 e^{\eta }+3 \pi ^2 \mathcal{Q}^2\right)^2 \left(4 \pi  c \mathcal{S} e^{\eta }+3 \mathcal{S}^2 e^{\eta }-\pi ^2 \mathcal{Q}^2\right)^3}
\label{ap:4}
\end{equation}
The GTD scalar for the dual CFT of ModMax in R\'enyi entropy is given as
\begin{equation}
R_{GTD}=\frac{256 \pi ^2 C \mathcal{S}_R^2 \mathcal{V} e^{2 \eta } \left(\begin{aligned}&128 \pi ^3 C^3 \mathcal{S}_R^2 e^{3 \eta } A+32 \pi ^2 C^2 \mathcal{S}_R e^{2 \eta }B+4 \pi  C e^{\eta } I+36 \pi ^2 \lambda  \mathcal{Q}^2 \mathcal{S}_R^4 e^{2 \eta }D\\&+27 \mathcal{S}_R^5 e^{3 \eta } E+3 \pi ^4 \mathcal{Q}^4 \mathcal{S}_R e^{\eta } F-16 \pi ^6 \lambda  \mathcal{Q}^6G\end{aligned}\right)}{(\lambda \mathcal{S}_R-4)^2 H^3 \left(4 \pi C \mathcal{S}_R e^{\eta } (3 \lambda  \mathcal{S}_R-4)+3 \mathcal{S}_R^2 e^{\eta } (15 \lambda  \mathcal{S}_R+4)+\pi ^2 \mathcal{Q}^2 (\lambda  \mathcal{S}_R+12)\right)^2}
\label{ap:5}
\end{equation}
where A, B, I, D, E, F, G, H are given below
\begin{equation}
\begin{split}
&A=\left(27 \lambda ^5 \mathcal{S}^5-216 \lambda ^4 \mathcal{S}^4+720 \lambda ^3 \mathcal{S}^3-64 \lambda ^2 \mathcal{S}^2-1024\right)\\
&B= \left(\begin{aligned}&24 \lambda  \mathcal{S}^3 e^{\eta } \left(45 \lambda ^4 \mathcal{S}^4-279 \lambda ^3 \mathcal{S}^3+592 \lambda ^2 \mathcal{S}^2+416 \lambda  \mathcal{S}+832\right)+\\&\pi ^2 \mathcal{Q}^2 \left(9 \lambda ^5 \mathcal{S}^5+84 \lambda ^4 \mathcal{S}^4-976 \lambda ^3 \mathcal{S}^3+3264 \lambda ^2 \mathcal{S}^2-1536 \lambda  \mathcal{S}+2048\right)\end{aligned}\right)\\
&I=\left(\begin{aligned}&9 \mathcal{S}^4 e^{2 \eta } \left(2325 \lambda ^5 \mathcal{S}^5-13780 \lambda ^4 \mathcal{S}^4+24128 \lambda ^3 \mathcal{S}^3+24576 \lambda ^2 \mathcal{S}^2+17152 \lambda  \mathcal{S}+1024\right)\\&+3 \pi ^2 \mathcal{Q}^2 \mathcal{S}^2 e^{\eta } \left(45 \lambda ^5 \mathcal{S}^5+1988 \lambda ^4 \mathcal{S}^4-12480 \lambda ^3 \mathcal{S}^3+18432 \lambda ^2 \mathcal{S}^2-39168 \lambda  \mathcal{S}-9216\right)\\&+4 \pi ^4 \mathcal{Q}^4 \left(3 \lambda ^5 \mathcal{S}^5+70 \lambda ^4 \mathcal{S}^4+64 \lambda ^3 \mathcal{S}^3-2688 \lambda ^2 \mathcal{S}^2+1024 \lambda  \mathcal{S}-512\right)\end{aligned}\right)\\
&D= \left(75 \lambda ^4 \mathcal{S}^4+1040 \lambda ^3 \mathcal{S}^3-7072 \lambda ^2 \mathcal{S}^2+3072 \lambda  \mathcal{S}-256\right)\\
&E=\left(2625 \lambda ^5 \mathcal{S}^5-14900 \lambda ^4 \mathcal{S}^4+16480 \lambda ^3 \mathcal{S}^3+35968 \lambda ^2 \mathcal{S}^2+20736 \lambda  \mathcal{S}+3072\right)\\
&F=\left(75 \lambda ^5 \mathcal{S}^5+836 \lambda ^4 \mathcal{S}^4-32 \lambda ^3 \mathcal{S}^3-29056 \lambda ^2 \mathcal{S}^2+32512 \lambda  \mathcal{S}+13312\right)\\
&G= \left(\lambda ^3 \mathcal{S}^3-4 \lambda ^2 \mathcal{S}^2-80 \lambda  \mathcal{S}+64\right)\\
&H=\left(4 \pi  c \mathcal{S} e^{\eta } (3 \lambda  \mathcal{S}+4)+3 \mathcal{S}^2 e^{\eta } (5 \lambda  \mathcal{S}+4)-\pi ^2 \mathcal{Q}^2 (\lambda  \mathcal{S}+4)\right)
\end{split}
\end{equation}
The GTD scalar for the dual CFT of ModMax BH in Kaniadakis entropy is given as
\begin{equation}
R_{GTD}=\frac{768 \pi ^2 c \mathcal{S}^2 \mathcal{V} e^{2 \eta } \left(128 \pi ^3 c^3 \mathcal{S}^2 e^{3 \eta } A+128 \pi ^2 c^2 \mathcal{S} e^{2 \eta }B+12 \pi  c e^{\eta } D+9 \mathcal{S} E\right)}{\left(\kappa ^2 \mathcal{S}^2+12\right)^2 F^3 \left(4 \pi  c \mathcal{S} e^{\eta } \left(5 \kappa ^2 \mathcal{S}^2+4\right)+\mathcal{S}^2 e^{\eta } \left(35 \kappa ^2 \mathcal{S}^2-12\right)-\pi ^2 \mathcal{Q}^2 \left(\kappa ^2 \mathcal{S}^2+12\right)\right)^2}
\label{ap:6}
\end{equation}
where A, B, D, E, F are given below
\begin{equation}
\begin{split}
&A=\left(625 \kappa ^{10} \mathcal{S}^{10}+8500 \kappa ^8 \mathcal{S}^8+62880 \kappa ^6 \mathcal{S}^6+42624 \kappa ^4 \mathcal{S}^4+158976 \kappa ^2 \mathcal{S}^2+82944\right)\\
&B= \left(\begin{aligned}&\kappa ^2 \mathcal{S}^4 e^{\eta } \left(2975 \kappa ^8 \mathcal{S}^8+40200 \kappa ^6 \mathcal{S}^6+243936 \kappa ^4 \mathcal{S}^4+41472 \kappa ^2 \mathcal{S}^2+518400\right)\\&-3 \pi ^2 \mathcal{Q}^2 \left(25 \kappa ^{10} \mathcal{S}^{10}+540 \kappa ^8 \mathcal{S}^8+10088 \kappa ^6 \mathcal{S}^6+25440 \kappa ^4 \mathcal{S}^4+42624 \kappa ^2 \mathcal{S}^2+13824\right)\end{aligned}\right)\\
&D=\left(\begin{aligned}&\mathcal{S}^4 e^{2 \eta } \left(48265 \kappa ^{10} \mathcal{S}^{10}+669732 \kappa ^8 \mathcal{S}^8+3424032 \kappa ^6 \mathcal{S}^6-1662336 \kappa ^4 \mathcal{S}^4+4416768 \kappa ^2 \mathcal{S}^2-248832\right)\\&-\pi ^2 \mathcal{Q}^2 \mathcal{S}^2 e^{\eta } \left(2555 \kappa ^{10} \mathcal{S}^{10}+58268 \kappa ^8 \mathcal{S}^8+950304 \kappa ^6 \mathcal{S}^6+1650816 \kappa ^4 \mathcal{S}^4+3158784 \kappa ^2 \mathcal{S}^2-746496\right)\\&+2 \pi ^4 \mathcal{Q}^4 \left(25 \kappa ^{10} \mathcal{S}^{10}+36 \kappa ^8 \mathcal{S}^8+14016 \kappa ^6 \mathcal{S}^6+134400 \kappa ^4 \mathcal{S}^4+172800 \kappa ^2 \mathcal{S}^2+27648\right)\end{aligned}\right)\\
&E=\left(\begin{aligned}&-8 \pi ^2 \kappa ^2 \mathcal{Q}^2 \mathcal{S}^4 e^{2 \eta } \left(245 \kappa ^8 \mathcal{S}^8+7700 \kappa ^6 \mathcal{S}^6+123504 \kappa ^4 \mathcal{S}^4+137664 \kappa ^2 \mathcal{S}^2+55296\right)+\\&\mathcal{S}^4 e^{3 \eta } \left(32585 \kappa ^{10} \mathcal{S}^{10}+470988 \kappa ^8 \mathcal{S}^8+1989792 \kappa ^6 \mathcal{S}^6-3259008 \kappa ^4 \mathcal{S}^4+3587328 \kappa ^2 \mathcal{S}^2-746496\right)+\\&\pi ^4 \mathcal{Q}^4 e^{\eta } \left(175 \kappa ^{10} \mathcal{S}^{10}+692 \kappa ^8 \mathcal{S}^8+52192 \kappa ^6 \mathcal{S}^6+452736 \kappa ^4 \mathcal{S}^4+762624 \kappa ^2 \mathcal{S}^2-359424\right)+\\&64 \pi ^6 \kappa ^2 \mathcal{Q}^6 \left(\kappa ^6 \mathcal{S}^6-12 \kappa ^4 \mathcal{S}^4-336 \kappa ^2 \mathcal{S}^2-576\right)\end{aligned}\right)\\
&F=\left(4 \pi  c \mathcal{S} e^{\eta } \left(5 \kappa ^2 \mathcal{S}^2-12\right)+3 \mathcal{S}^2 e^{\eta } \left(7 \kappa ^2 \mathcal{S}^2-12\right)-3 \pi ^2 \mathcal{Q}^2 \left(\kappa ^2 \mathcal{S}^2-4\right)\right)
\end{split}
\end{equation}
In Section \ref{sec:4}, the GTD scalar for NED AdS black hole for the Bekenstein-Hawking entropy is given as
\begin{equation}
R_{GTD}=\frac{72 G^6 l^2 \pi ^3 S^6 \left(\frac{4 \left(\begin{aligned}&12288 \pi ^{15} Q^6 b^{18}-256 G \pi ^{12} Q^4 S A b^6\\&+270 G^7 \pi ^2 S^7 B\end{aligned}\right)}{I}+\frac{l^2 \left(\begin{aligned}&192 \pi ^{15} Q^6 b^{18}-16 G \pi ^{12} Q^4 S D b^6\\&+135 G^7 \pi ^2 S^7 E\end{aligned}\right)}{F}\right)}{\begin{aligned}&\left(2 \pi ^2 b^4-5 G \pi  S b^2+15 G^2 S^2\right)^2\times\\& \left(-2 l^2 \pi ^4 Q^2 b^4+3 G l^2 \pi ^3 Q^2 S b^2+3 G^2 S^2 \left(\pi  \left(S-\pi  Q^2\right) l^2+3 G S^2\right)\right)\end{aligned}}
\label{ap:7}
\end{equation}
where A, B, I, D, E, F, are given below
\begin{equation}
\begin{split}
&A=\left(\begin{aligned}&\left(472 \pi ^2 Q^2+369 G S^2+1248 \pi  S\right) b^{16}+384 G^2 \pi ^{11} Q^4 S^2 \left(1208 \pi ^2 Q^2+513 G S^2+4420 \pi  S\right) b^{14}\\&-192 G^3 \pi ^8 Q^2 S^3 \left(2792 \pi ^4 Q^4+20100 \pi ^3 S Q^2+459 G^2 S^4+234 G \pi  S^3-3 \pi ^2 \left(4609 G Q^2+368\right) S^2\right) b^{12}\\&+144 G^4 \pi ^7 Q^2 S^4 \left(19264 \pi ^4 Q^4+28296 \pi ^3 S Q^2+3069 G^2 S^4+3036 G \pi  S^3-22 \pi ^2 \left(4197 G Q^2+296\right) S^2\right) b^{10}\\&+36 G^5 \pi ^4 S^5 \left(-239936 \pi ^6 Q^6+120080 \pi ^5 S Q^4+24 \pi ^4 \left(44521 G Q^2+2028\right) S^2 Q^2+891 G^3 S^6+972 G^2 \pi  S^5\right.\\&\left.-18 G \pi ^2 \left(2587 G Q^2+96\right) S^4-24 \pi ^3 \left(4187 G Q^2+112\right) S^3\right) b^8-2160 G^6 \pi ^3 S^6 \left(-4848 \pi ^6 Q^6+4628 \pi ^5 S Q^4\right.\\&\left.+7 \pi ^4 \left(3423 G Q^2+160\right) S^2 Q^2+54 G^3 S^6+54 G^2 \pi  S^5-6 G \pi ^2 \left(183 G Q^2+16\right) S^4-\pi ^3 \left(3681 G Q^2+160\right) S^3\right)\end{aligned}\right)\\
&B=\left(\begin{aligned}&\left(-24960 \pi ^6 Q^6+32032 \pi ^5 S Q^4+8 \pi ^4 \left(19377 G Q^2+380\right) S^2 Q^2+1215 G^3 S^6+1008 G^2 \pi  S^5\right.\\&\left.-6 G \pi ^2 \left(843 G Q^2+292\right) S^4-8 \pi ^3 \left(6225 G Q^2+428\right) S^3\right) b^4-8100 G^8 \pi  S^8 \left(-192 \pi ^6 Q^6+144 \pi ^5 S Q^4\right.\\&\left.+12 \pi ^4 \left(183 G Q^2+20\right) S^2 Q^2+54 G^3 S^6+36 G^2 \pi  S^5-3 G \pi ^2 \left(15 G Q^2+16\right) S^4-4 \pi ^3 \left(255 G Q^2+32\right) S^3\right) b^2\\&+6075 G^9 S^{10} \left(81 G^3 S^5+36 G^2 \pi  S^4-16 \pi ^3 \left(27 G Q^2+8\right) S^2+16 \pi ^4 Q^2 \left(39 G Q^2+16\right) S-128 \pi ^5 Q^4\right)\end{aligned}\right)\\
&I=\begin{aligned}&\left(56 \pi ^4 Q^2 b^4-60 G \pi ^3 Q^2 S b^2+3 G^2 S^2 \left(12 \pi ^2 Q^2+3 G S^2-4 \pi  S\right)\right)^2\times\\& \left(-8 \pi ^4 Q^2 b^4+12 G \pi ^3 Q^2 S b^2+3 G^2 S^2 \left(-4 \pi ^2 Q^2+3 G S^2+4 \pi  S\right)\right)^2\end{aligned}\\
&D=\left(\begin{aligned}&\left(118 \pi ^2 Q^2+369 G S^2+312 \pi  S\right) b^{16}+24 G^2 \pi ^{11} Q^4 S^2 \left(302 \pi ^2 Q^2+513 G S^2+1105 \pi  S\right) b^{14}\\&-12 G^3 \pi ^8 Q^2 S^3 \left(698 \pi ^4 Q^4+5025 \pi ^3 S Q^2+1836 G^2 S^4+234 G \pi  S^3-3 \pi ^2 \left(4609 G Q^2+92\right) S^2\right) b^{12}\\&+18 G^4 \pi ^7 Q^2 S^4 \left(2408 \pi ^4 Q^4+3537 \pi ^3 S Q^2+6138 G^2 S^4+1518 G \pi  S^3-11 \pi ^2 \left(4197 G Q^2+74\right) S^2\right) b^{10}\\&+9 G^5 \pi ^4 S^5 \left(-14996 \pi ^6 Q^6+7505 \pi ^5 S Q^4+6 \pi ^4 \left(44521 G Q^2+507\right) S^2 Q^2+3564 G^3 S^6+972 G^2 \pi  S^5\right.\\&\left.-18 G \pi ^2 \left(2587 G Q^2+24\right) S^4-6 \pi ^3 \left(4187 G Q^2+28\right) S^3\right) b^8-135 G^6 \pi ^3 S^6 \left(-1212 \pi ^6 Q^6+1157 \pi ^5 S Q^4\right.\\&\left.+7 \pi ^4 \left(3423 G Q^2+40\right) S^2 Q^2+864 G^3 S^6+216 G^2 \pi  S^5-24 G \pi ^2 \left(183 G Q^2+4\right) S^4-\pi ^3 \left(3681 G Q^2+40\right) S^3\right)\end{aligned}\right)\\
&E=\left(\begin{aligned}&\left(-780 \pi ^6 Q^6+1001 \pi ^5 S Q^4+\pi ^4 \left(19377 G Q^2+95\right) S^2 Q^2+2430 G^3 S^6+504 G^2 \pi  S^5-3 G \pi ^2 \left(843 G Q^2+73\right) S^4\right.\\&\left.-\pi ^3 \left(6225 G Q^2+107\right) S^3\right) b^4-2025 G^8 \pi  S^8 \left(-12 \pi ^6 Q^6+9 \pi ^5 S Q^4+3 \pi ^4 \left(183 G Q^2+5\right) S^2 Q^2+\right.\\&\left.216 G^3 S^6+36 G^2 \pi  S^5-3 G \pi ^2 \left(15 G Q^2+4\right) S^4-\pi ^3 \left(255 G Q^2+8\right) S^3\right) b^2\\&+6075 G^9 S^{10} \left(81 G^3 S^5+9 G^2 \pi  S^4-\pi ^3 \left(27 G Q^2+2\right) S^2+\pi ^4 Q^2 \left(39 G Q^2+4\right) S-2 \pi ^5 Q^4\right)\end{aligned}\right)\\
&F=\begin{aligned}&\left(14 l^2 \pi ^4 Q^2 b^4-15 G l^2 \pi ^3 Q^2 S b^2+3 G^2 S^2 \left(\pi  \left(3 \pi  Q^2-S\right) l^2+3 G S^2\right)\right) \times\\&\left(14 \pi ^4 Q^2 b^4-15 G \pi ^3 Q^2 S b^2+3 G^2 S^2 \left(3 \pi ^2 Q^2+3 G S^2-\pi  S\right)\right)\times\\& \left(-2 \pi ^4 Q^2 b^4+3 G \pi ^3 Q^2 S b^2+3 G^2 S^2 \left(-\pi ^2 Q^2+3 G S^2+\pi  S\right)\right)^2\end{aligned}\\
&
\end{split}
\end{equation}
The GTD scalar for the NED AdS black hole for the R\'enyi entropy is given as
\begin{equation}
R_{GTD}=\frac{1152 G^5 l^2 \pi ^3 S^5}{A8} \left(\frac{l^2 \left(\begin{aligned}&10 \pi ^2 A B I Q^2-D^2\\&+5 E+10 S H F J K\\&+\frac{L M}{N}-\frac{O}{P}\end{aligned}\right)}{R}+\frac{4 S T \left(\begin{aligned}&32768 \pi ^{16} Q^6 \lambda  U b^{16}+384 G^3 \pi ^{11} Q^4 S^2 V b^{12}\\&-144 G^5 \pi ^7 Q^2 S^4 W b^{10}-36 G^6 \pi ^4 S^5 X b^8\\&+1080 G^7 \pi ^3 S^6 \left(-32 \pi ^6 Y+36 G^2 \pi  S^5 Z\right) b^6\\&-270 G^8 \pi ^2 S^7 \left(128 \pi ^6 A1+27 G^3 S^6 A2\right) b^4\\&+8100 G^9 \pi  S^8 \left(64 \pi ^6 A3-4 \pi ^3 S^3 A4\right) b^2\\&-6075 G^{10} S^{10} \left(-1024 \pi ^6 \lambda  A5+48 G \pi ^2 S^4 \lambda  A6\right)\end{aligned}\right)}{A7}\right)
\label{ap:8}
\end{equation}
where A, B, I, D, E, H, F, J, K, L, M, N, O, P, R, T, U, V, W, X, Y, Z, A1, A2, A3, A4, A5, A6, A7, A8 are given below
\begin{equation}
	\begin{split}
		&A=\left(\begin{aligned}&\left(2 \pi ^2 (5 S \lambda -4) b^4-5 G \pi  S (3 S \lambda -4) b^2+15 G^2 S^2 (S \lambda -4)\right) \left(4 \pi ^4 \left(45 S^2 \lambda ^2-144 S \lambda +112\right) b^8\right.\\&\left.-48 G \pi ^3 S \left(3 S^2 \lambda ^2-20 S \lambda +24\right) b^6-3 G^2 \pi ^2 S^2 \left(27 S^2 \lambda ^2+224 S \lambda -560\right) b^4\right.\\&\left.+36 G^3 \pi  S^3 \left(S^2 \lambda ^2-32\right) b^2+9 G^4 S^4 \left(S^2 \lambda ^2+16 S \lambda +48\right)\right)\end{aligned}\right)\\
		&B=\left(\begin{aligned}&\left(2 \pi ^4 Q^2 (3 S \lambda -4) b^4-3 G \pi ^3 Q^2 S (S \lambda -4) b^2+3 G^2 S^2 \left(-\pi ^2 (S \lambda +4) Q^2+\pi  S (3 S \lambda +4)+3 G S^2 (5 S \lambda +4)\right)\right) Q^2\\&-5 \pi ^2 \left(2 \pi ^2 (3 S \lambda -4) b^4-3 G \pi  S (S \lambda -4) b^2-3 G^2 S^2 (S \lambda +4)\right)\\& \left(2 \pi ^2 (5 S \lambda -4) b^4-5 G \pi  S (3 S \lambda -4) b^2+15 G^2 S^2 (S \lambda -4)\right)\end{aligned}\right)\\
		&I=\left(\begin{aligned}&4 \pi ^6 Q^2 \left(45 S^2 \lambda ^2-144 S \lambda +112\right) b^8-48 G \pi ^5 Q^2 S \left(3 S^2 \lambda ^2-20 S \lambda +24\right) b^6\\&+3 G^2 \pi ^2 S^2 \left(\pi ^2 \left(-27 S^2 \lambda ^2-224 S \lambda +560\right) Q^2+4 \pi  S \left(9 S^2 \lambda ^2-32\right)+6 G S^2 \left(15 S^2 \lambda ^2-16 S \lambda -48\right)\right) b^4\\&+9 G^3 \pi  S^3 \left(4 \pi ^2 \left(S^2 \lambda ^2-32\right) Q^2+48 G S^2 (S \lambda +2)+\pi  S \left(-3 S^2 \lambda ^2+16 S \lambda +48\right)\right) b^2\\&+9 G^4 S^4 \left(\pi ^2 \left(S^2 \lambda ^2+16 S \lambda +48\right) Q^2-16 \pi  S (S \lambda +2)+3 G S^2 \left(5 S^2 \lambda ^2-16\right)\right)\end{aligned}\right)\\
		&D=\left(\begin{aligned}&-4 \pi ^6 Q^2 \left(45 S^2 \lambda ^2-128 S \lambda +80\right) b^8+16 G \pi ^5 Q^2 S \left(15 S^2 \lambda ^2-72 S \lambda +64\right) b^6\\&+3 G^2 \pi ^2 S^2 \left(\pi ^2 \left(-35 S^2 \lambda ^2+544 S \lambda -816\right) Q^2-16 \pi  S (S \lambda -4)+6 G S^2 \left(25 S^2 \lambda ^2+16\right)\right) b^4\\&-15 G^3 \pi  S^3 \left(6 G \lambda  (15 S \lambda -4) S^3+\pi  \left(9 S^2 \lambda ^2+16\right) S+32 \pi ^2 Q^2 (S \lambda -4)\right) b^2\\&+45 G^4 S^4 \left(-\pi ^2 \left(S^2 \lambda ^2+16\right) Q^2+2 \pi  S^2 \lambda  (3 S \lambda -4)+3 G S^2 \left(15 S^2 \lambda ^2-32 S \lambda -16\right)\right)\end{aligned}\right)\\
		&E=\left(\begin{aligned}&\left(2 \pi ^2 (5 S \lambda -4) b^4-5 G \pi  S (3 S \lambda -4) b^2+15 G^2 S^2 (S \lambda -4)\right) \left(2 \pi ^4 Q^2 (3 S \lambda -4) b^4-3 G \pi ^3 Q^2 S (S \lambda -4) b^2\right.\\&\left.+3 G^2 S^2 \left(-\pi ^2 (S \lambda +4) Q^2+\pi  S (3 S \lambda +4)+3 G S^2 (5 S \lambda +4)\right)\right) \left(4 \pi ^6 Q^2 \left(45 S^2 \lambda ^2-144 S \lambda +112\right) b^8\right.\\&\left.-48 G \pi ^5 Q^2 S \left(3 S^2 \lambda ^2-20 S \lambda +24\right) b^6+3 G^2 \pi ^2 S^2 \left(\pi ^2 \left(-27 S^2 \lambda ^2-224 S \lambda +560\right) Q^2\right.\right.\\&\left.\left.+4 \pi  S \left(9 S^2 \lambda ^2-32\right)+6 G S^2 \left(15 S^2 \lambda ^2-16 S \lambda -48\right)\right) b^4+9 G^3 \pi  S^3 \left(4 \pi ^2 \left(S^2 \lambda ^2-32\right) Q^2\right.\right.\\&\left.\left.+48 G S^2 (S \lambda +2)+\pi  S \left(-3 S^2 \lambda ^2+16 S \lambda +48\right)\right) b^2+9 G^4 S^4 \left(\pi ^2 \left(S^2 \lambda ^2+16 S \lambda +48\right) Q^2\right.\right.\\&\left.\left.-16 \pi  S (S \lambda +2)+3 G S^2 \left(5 S^2 \lambda ^2-16\right)\right)\right)\end{aligned}\right)\\
		&H=\left(\begin{aligned}&\left(2 \pi ^2 \lambda  b^4+2 G \pi  (2-3 S \lambda ) b^2+3 G^2 S (3 S \lambda -8)\right) \left(2 \pi ^4 Q^2 (3 S \lambda -4) b^4-3 G \pi ^3 Q^2 S (S \lambda -4) b^2\right.\\&\left.+3 G^2 S^2 \left(-\pi ^2 (S \lambda +4) Q^2+\pi  S (3 S \lambda +4)+3 G S^2 (5 S \lambda +4)\right)\right) \left(-4 \pi ^6 Q^2 \left(45 S^2 \lambda ^2-128 S \lambda +80\right) b^8\right.\\&\left.+16 G \pi ^5 Q^2 S \left(15 S^2 \lambda ^2-72 S \lambda +64\right) b^6+3 G^2 \pi ^2 S^2 \left(\pi ^2 \left(-35 S^2 \lambda ^2+544 S \lambda -816\right) Q^2\right.\right.\\&\left.\left.-16 \pi  S (S \lambda -4)+6 G S^2 \left(25 S^2 \lambda ^2+16\right)\right) b^4-15 G^3 \pi  S^3 \left(6 G \lambda  (15 S \lambda -4) S^3+\pi  \left(9 S^2 \lambda ^2+16\right) S\right.\right.\\&\left.\left.+32 \pi ^2 Q^2 (S \lambda -4)\right) b^2+45 G^4 S^4 \left(-\pi ^2 \left(S^2 \lambda ^2+16\right) Q^2+2 \pi  S^2 \lambda  (3 S \lambda -4)+3 G S^2 \left(15 S^2 \lambda ^2-32 S \lambda -16\right)\right)\right)\\&+2 \left(2 \pi ^2 (5 S \lambda -4) b^4-5 G \pi  S (3 S \lambda -4) b^2+15 G^2 S^2 (S \lambda -4)\right)\end{aligned}\right)
	\end{split}
\end{equation}
\begin{equation}
	\begin{split}
				&F=\left(\begin{aligned}&\left(2 \pi ^4 Q^2 (3 S \lambda -4) b^4-3 G \pi ^3 Q^2 S (S \lambda -4) b^2+3 G^2 S^2 \left(-\pi ^2 (S \lambda +4) Q^2+\pi  S (3 S \lambda +4)+3 G S^2 (5 S \lambda +4)\right)\right)\\& \left(-4 \pi ^6 Q^2 \left(45 S^2 \lambda ^2-128 S \lambda +80\right) b^8+16 G \pi ^5 Q^2 S \left(15 S^2 \lambda ^2-72 S \lambda +64\right) b^6\right.\\&\left.+3 G^2 \pi ^2 S^2 \left(\pi ^2 \left(-35 S^2 \lambda ^2+544 S \lambda -816\right) Q^2-16 \pi  S (S \lambda -4)+6 G S^2 \left(25 S^2 \lambda ^2+16\right)\right) b^4\right.\\&\left.-15 G^3 \pi  S^3 \left(6 G \lambda  (15 S \lambda -4) S^3+\pi  \left(9 S^2 \lambda ^2+16\right) S+32 \pi ^2 Q^2 (S \lambda -4)\right) b^2\right.\\&\left.+45 G^4 S^4 \left(-\pi ^2 \left(S^2 \lambda ^2+16\right) Q^2+2 \pi  S^2 \lambda  (3 S \lambda -4)+3 G S^2 \left(15 S^2 \lambda ^2-32 S \lambda -16\right)\right)\right)\\&+2 S \left(2 \pi ^2 (5 S \lambda -4) b^4-5 G \pi  S (3 S \lambda -4) b^2+15 G^2 S^2 (S \lambda -4)\right)\\& \left(6 \pi ^4 Q^2 \lambda  b^4-6 G \pi ^3 Q^2 (S \lambda -2) b^2+3 G^2 S \left(-\pi ^2 (3 S \lambda +8) Q^2+12 \pi  S (S \lambda +1)+3 G S^2 (25 S \lambda +16)\right)\right)\end{aligned}\right)\\
		&J=\left(\begin{aligned}&\left(-4 \pi ^6 Q^2 \left(45 S^2 \lambda ^2-128 S \lambda +80\right) b^8+16 G \pi ^5 Q^2 S \left(15 S^2 \lambda ^2-72 S \lambda +64\right) b^6\right.\\&\left.+3 G^2 \pi ^2 S^2 \left(\pi ^2 \left(-35 S^2 \lambda ^2+544 S \lambda -816\right) Q^2-16 \pi  S (S \lambda -4)+6 G S^2 \left(25 S^2 \lambda ^2+16\right)\right) b^4\right.\\&\left.-15 G^3 \pi  S^3 \left(6 G \lambda  (15 S \lambda -4) S^3+\pi  \left(9 S^2 \lambda ^2+16\right) S+32 \pi ^2 Q^2 (S \lambda -4)\right) b^2+\right.\\&\left.45 G^4 S^4 \left(-\pi ^2 \left(S^2 \lambda ^2+16\right) Q^2+2 \pi  S^2 \lambda  (3 S \lambda -4)+3 G S^2 \left(15 S^2 \lambda ^2-32 S \lambda -16\right)\right)\right)\\&-4 S \left(2 \pi ^2 (5 S \lambda -4) b^4-5 G \pi  S (3 S \lambda -4) b^2+15 G^2 S^2 (S \lambda -4)\right) \left(2 \pi ^4 Q^2 (3 S \lambda -4) b^4\right.\\&\left.-3 G \pi ^3 Q^2 S (S \lambda -4) b^2+3 G^2 S^2 \left(-\pi ^2 (S \lambda +4) Q^2+\pi  S (3 S \lambda +4)+3 G S^2 (5 S \lambda +4)\right)\right)\end{aligned}\right)\\
		&K=\left(\begin{aligned}&-4 \pi ^6 Q^2 \lambda  (45 S \lambda -64) b^8+8 G \pi ^5 Q^2 \left(45 S^2 \lambda ^2-144 S \lambda +64\right) b^6+6 G^2 \pi ^2 S \left(\pi ^2 \left(-35 S^2 \lambda ^2+408 S \lambda -408\right) Q^2\right.\\&\left.-16 \pi  S (S \lambda -3)+3 G S^2 \left(75 S^2 \lambda ^2+32\right)\right) b^4-15 G^3 \pi  S^2 \left(9 G \lambda  (35 S \lambda -8) S^3+\pi  \left(27 S^2 \lambda ^2+32\right) S\right.\\&\left.+64 \pi ^2 Q^2 (S \lambda -3)\right) b^2+45 G^4 S^3 \left(-\pi ^2 \left(3 S^2 \lambda ^2+32\right) Q^2+3 \pi  S^2 \lambda  (7 S \lambda -8)+12 G S^2 \left(15 S^2 \lambda ^2-28 S \lambda -12\right)\right)\end{aligned}\right)\\
		&L=\left(\begin{aligned}&\left(2 \pi ^2 (5 S \lambda -4) b^4-5 G \pi  S (3 S \lambda -4) b^2+15 G^2 S^2 (S \lambda -4)\right) \left(-4 \pi ^6 Q^2 \left(45 S^2 \lambda ^2-128 S \lambda +80\right) b^8\right.\\&\left.+16 G \pi ^5 Q^2 S \left(15 S^2 \lambda ^2-72 S \lambda +64\right) b^6+3 G^2 \pi ^2 S^2 \left(\pi ^2 \left(-35 S^2 \lambda ^2+544 S \lambda -816\right) Q^2-16 \pi  S (S \lambda -4)\right.\right.\\&\left.\left.+6 G S^2 \left(25 S^2 \lambda ^2+16\right)\right) b^4-15 G^3 \pi  S^3 \left(6 G \lambda  (15 S \lambda -4) S^3+\pi  \left(9 S^2 \lambda ^2+16\right) S+32 \pi ^2 Q^2 (S \lambda -4)\right) b^2\right.\\&\left.+45 G^4 S^4 \left(-\pi ^2 \left(S^2 \lambda ^2+16\right) Q^2+2 \pi  S^2 \lambda  (3 S \lambda -4)+3 G S^2 \left(15 S^2 \lambda ^2-32 S \lambda -16\right)\right)\right)\end{aligned}\right)
	\end{split}
\end{equation}
\begin{equation}
	\begin{split}
&M=\left(\begin{aligned}&4 \pi ^8 Q^4 \left(225 S^2 \lambda ^2-864 S \lambda +784\right) b^8-192 G \pi ^7 Q^4 S \left(3 S^2 \lambda ^2-25 S \lambda +36\right) b^6\\&+3 G^2 \pi ^4 Q^2 S^2 \left(\pi ^2 \left(-81 S^2 \lambda ^2-896 S \lambda +2800\right) Q^2+16 \pi  S \left(9 S^2 \lambda ^2-64\right)\right.\\&\left.+12 G S^2 \left(15 S^2 \lambda ^2-32 S \lambda -144\right)\right) b^4+18 G^3 \pi ^3 Q^2 S^3 \left(4 \pi ^2 \left(S^2 \lambda ^2-64\right) Q^2+48 G S^2 (S \lambda +4)\right.\\&\left.+\pi  S \left(-3 S^2 \lambda ^2+32 S \lambda +144\right)\right) b^2+9 G^4 S^4 \left(\pi ^4 \left(S^2 \lambda ^2+32 S \lambda +144\right) Q^4-32 \pi ^3 S (S \lambda +4) Q^2\right.\\&\left.+24 G \pi  S^4 \lambda  (15 S \lambda +8)+9 G^2 S^4 \left(225 S^2 \lambda ^2+160 S \lambda +16\right)+\pi ^2 S^2 \left(-6 G \left(5 S^2 \lambda ^2+16\right) Q^2+9 S^2 \lambda ^2+16\right)\right)\end{aligned}\right)\\
&N=2 \pi ^4 Q^2 (28-15 S \lambda ) b^4+3 G \pi ^3 Q^2 S (3 S \lambda -20) b^2+3 G^2 S^2 \left(\pi ^2 (S \lambda +12) Q^2+\pi  S (3 S \lambda -4)+3 G S^2 (15 S \lambda +4)\right)\\
&O=\left(\begin{aligned}&10 \pi ^2 Q^2 \left(2 \pi ^2 (5 S \lambda -4) b^4-5 G \pi  S (3 S \lambda -4) b^2+15 G^2 S^2 (S \lambda -4)\right) \left(4 \pi ^6 Q^2 \left(45 S^2 \lambda ^2-144 S \lambda +112\right) b^8\right.\\&\left.-48 G \pi ^5 Q^2 S \left(3 S^2 \lambda ^2-20 S \lambda +24\right) b^6+3 G^2 \pi ^2 S^2 \left(\pi ^2 \left(-27 S^2 \lambda ^2-224 S \lambda +560\right) Q^2\right.\right.\\&\left.\left.+4 \pi  S \left(9 S^2 \lambda ^2-32\right)+6 G S^2 \left(15 S^2 \lambda ^2-16 S \lambda -48\right)\right) b^4+9 G^3 \pi  S^3 \left(4 \pi ^2 \left(S^2 \lambda ^2-32\right) Q^2\right.\right.\\&\left.\left.+48 G S^2 (S \lambda +2)+\pi  S \left(-3 S^2 \lambda ^2+16 S \lambda +48\right)\right) b^2+9 G^4 S^4 \left(\pi ^2 \left(S^2 \lambda ^2+16 S \lambda +48\right) Q^2\right.\right.\\&\left.\left.-16 \pi  S (S \lambda +2)+3 G S^2 \left(5 S^2 \lambda ^2-16\right)\right)\right)^2\end{aligned}\right)\\
&P=2 \pi ^4 Q^2 (15 S \lambda -28) b^4+3 G \pi ^3 Q^2 S (20-3 S \lambda ) b^2-3 G^2 S^2 \left(\pi ^2 (S \lambda +12) Q^2+\pi  S (3 S \lambda -4)+3 G S^2 (15 S \lambda +4)\right)\\
&R=\begin{aligned}&\left(2 \pi ^4 Q^2 (3 S \lambda -4) b^4-3 G \pi ^3 Q^2 S (S \lambda -4) b^2+3 G^2 S^2 \left(-\pi ^2 (S \lambda +4) Q^2+\pi  S (3 S \lambda +4)+3 G S^2 (5 S \lambda +4)\right)\right)^2\\& \left(2 l^2 \pi ^4 Q^2 (28-15 S \lambda ) b^4+3 G l^2 \pi ^3 Q^2 S (3 S \lambda -20) b^2+3 G^2 S^2 \left(\pi  \left(\pi  (S \lambda +12) Q^2+S (3 S \lambda -4)\right)\right.\right.\\&\left.\left. l^2+3 G S^2 (15 S \lambda +4)\right)\right)\end{aligned}\\
&T=\left(2 \pi ^2 (5 S \lambda -4) b^4-5 G \pi  S (3 S \lambda -4) b^2+15 G^2 S^2 (S \lambda -4)\right)\\
&U=\left(\begin{aligned}&\left(225 S^3 \lambda ^3-660 S^2 \lambda ^2+560 S \lambda -192\right) b^{20}+4096 G \pi ^{15} Q^6 \left(675 S^5 \lambda ^5-11160 S^4 \lambda ^4+35952 S^3 \lambda ^3\right.\\&\left.-40896 S^2 \lambda ^2+20992 S \lambda -3072\right) b^{18}-256 G^2 \pi ^{12} Q^4 S \left(-8 \pi ^2 \left(3375 S^5 \lambda ^5+13500 S^4 \lambda ^4-163872 S^3 \lambda ^3\right.\right.\\&\left.\left.+320640 S^2 \lambda ^2-223488 S \lambda +60416\right) Q^2+24 \pi  S \left(2025 S^5 \lambda ^5-24480 S^4 \lambda ^4+77664 S^3 \lambda ^3-125568 S^2 \lambda ^2\right.\right.\\&\left.\left.+125696 S \lambda -53248\right)+9 G S^2 \left(55125 S^5 \lambda ^5-280500 S^4 \lambda ^4+457120 S^3 \lambda ^3-260736 S^2 \lambda ^2+82176 S \lambda -41984\right)\right)\end{aligned}\right)\\
&V=\left(\begin{aligned}&\left(-8 \pi ^2 \left(25425 S^5 \lambda ^5-139740 S^4 \lambda ^4+131328 S^3 \lambda ^3+243456 S^2 \lambda ^2-384256 S \lambda +154624\right) Q^2\right.\\&\left.+4 \pi  S \left(141075 S^5 \lambda ^5-846180 S^4 \lambda ^4+1858848 S^3 \lambda ^3-2254464 S^2 \lambda ^2+2215680 S \lambda -1131520\right)\right.\\&\left.+3 G S^2 \left(624375 S^5 \lambda ^5-3531900 S^4 \lambda ^4+6351840 S^3 \lambda ^3-3675776 S^2 \lambda ^2+399104 S \lambda -175104\right)\right) b^{14}\\&-192 G^4 \pi ^8 Q^2 S^3 \left(-8 \pi ^4 \left(146025 S^5 \lambda ^5-1160640 S^4 \lambda ^4+2843376 S^3 \lambda ^3-1972032 S^2 \lambda ^2\right.\right.\\&\left.\left.-390144 S \lambda +357376\right) Q^4+12 \pi ^3 S \left(388575 S^5 \lambda ^5-2521500 S^4 \lambda ^4+5675616 S^3 \lambda ^3-5383808 S^2 \lambda ^2\right.\right.\\&\left.\left.+3149568 S \lambda -1715200\right) Q^2+27 G^2 S^4 \left(5625 S^5 \lambda ^5-3500 S^4 \lambda ^4-32000 S^3 \lambda ^3+103680 S^2 \lambda ^2\right.\right.\\&\left.\left.-35072 S \lambda -17408\right)+18 G \pi  S^3 \left(46125 S^5 \lambda ^5-60300 S^4 \lambda ^4-192320 S^3 \lambda ^3+402944 S^2 \lambda ^2\right.\right.\\&\left.\left.-143616 S \lambda -13312\right)+3 \pi ^2 S^2 \left(G \left(3700125 S^5 \lambda ^5-26058300 S^4 \lambda ^4+58201440 S^3 \lambda ^3-41997696 S^2 \lambda ^2\right.\right.\right.\\&\left.\left.\left.-945920 S \lambda +4719616\right) Q^2+32 \left(2025 S^5 \lambda ^5-6120 S^4 \lambda ^4-4584 S^3 \lambda ^3+28128 S^2 \lambda ^2-28544 S \lambda +11776\right)\right)\right)\end{aligned}\right)\\
&W=\left(\begin{aligned}&64 \pi ^4 \left(24375 S^5 \lambda ^5-275265 S^4 \lambda ^4+904812 S^3 \lambda ^3-881392 S^2 \lambda ^2-170432 S \lambda +308224\right) Q^4\\&-8 \pi ^3 S \left(1149075 S^5 \lambda ^5-10041540 S^4 \lambda ^4+27167904 S^3 \lambda ^3-26592384 S^2 \lambda ^2+8898304 S \lambda -3621888\right) Q^2\\&-12 G \pi  S^3 \left(124875 S^5 \lambda ^5-152700 S^4 \lambda ^4-1603200 S^3 \lambda ^3+3560704 S^2 \lambda ^2-1432320 S \lambda -259072\right)\\&+9 G^2 S^4 \left(196875 S^5 \lambda ^5-544500 S^4 \lambda ^4+635200 S^3 \lambda ^3-1520640 S^2 \lambda ^2+760064 S \lambda +349184\right)\\&-2 \pi ^2 S^2 \left(3 G \left(3363375 S^5 \lambda ^5-31655100 S^4 \lambda ^4+90183520 S^3 \lambda ^3-83978368 S^2 \lambda ^2+1259264 S \lambda\right.\right.\\&\left.\left. +15758336\right) Q^2+8 \left(33075 S^5 \lambda ^5-93060 S^4 \lambda ^4-214176 S^3 \lambda ^3+840576 S^2 \lambda ^2-890112 S \lambda +416768\right)\right)\end{aligned}\right)\\
&X=\left(\begin{aligned}&-64 \pi ^6 \left(46725 S^5 \lambda ^5-973680 S^4 \lambda ^4+4501488 S^3 \lambda ^3-6076992 S^2 \lambda ^2-1024000 S \lambda +3838976\right) Q^6\\&+16 \pi ^5 S \left(1693125 S^5 \lambda ^5-24046380 S^4 \lambda ^4+86945568 S^3 \lambda ^3-104446848 S^2 \lambda ^2+24559872 S \lambda +7685120\right) Q^4\\&+48 \pi ^4 S^2 \left(G \left(1225875 S^5 \lambda ^5-17811450 S^4 \lambda ^4+68946000 S^3 \lambda ^3-86702144 S^2 \lambda ^2+11210496 S \lambda\right.\right.\\&\left.\left. +22794752\right) Q^2+2 \left(20475 S^5 \lambda ^5-8940 S^4 \lambda ^4-652992 S^3 \lambda ^3+1789952 S^2 \lambda ^2-1572608 S \lambda +519168\right)\right) Q^2\\&+81 G^3 S^6 \left(65625 S^5 \lambda ^5-22500 S^4 \lambda ^4-76000 S^3 \lambda ^3-3200 S^2 \lambda ^2+42240 S \lambda +11264\right)\\&+108 G^2 \pi  S^5 \left(73125 S^5 \lambda ^5-42500 S^4 \lambda ^4-86400 S^3 \lambda ^3-6400 S^2 \lambda ^2+71424 S \lambda +9216\right)\\&-24 \pi ^3 S^3 \left(G Q^2 \left(248625 S^5 \lambda ^5-1243500 S^4 \lambda ^4+11029440 S^3 \lambda ^3-23867904 S^2 \lambda ^2+10865408 S \lambda\right.\right.\\&\left.\left. +4287488\right)-16 \left(2025 S^5 \lambda ^5-2520 S^4 \lambda ^4-4944 S^3 \lambda ^3+4672 S^2 \lambda ^2+8192 S \lambda -7168\right)\right)\\&-18 G \pi ^2 S^4 \left(G Q^2 \left(1888125 S^5 \lambda ^5-7078500 S^4 \lambda ^4+8421600 S^3 \lambda ^3-7141760 S^2 \lambda ^2+4999424 S \lambda +2649088\right)\right.\\&\left.-32 \left(7875 S^5 \lambda ^5-3900 S^4 \lambda ^4-11120 S^3 \lambda ^3-1088 S^2 \lambda ^2+13056 S \lambda -3072\right)\right)\end{aligned}\right)
	\end{split}
\end{equation}
\begin{equation}
\begin{split}
	&Y=\left(\begin{aligned}&\left(225 S^5 \lambda ^5-26476 S^4 \lambda ^4+205504 S^3 \lambda ^3-412672 S^2 \lambda ^2+40704 S \lambda +310272\right) Q^6+\\&8 \pi ^5 S \left(19845 S^5 \lambda ^5-854316 S^4 \lambda ^4+4875744 S^3 \lambda ^3-7801216 S^2 \lambda ^2+2357504 S \lambda +1184768\right) Q^4\\&+2 \pi ^4 S^2 \left(3 G Q^2 \left(59925 S^5 \lambda ^5-2479940 S^4 \lambda ^4+14938272 S^3 \lambda ^3-26167168 S^2 \lambda ^2+7443712 S \lambda\right.\right.\\&\left.\left. +8178688\right)-16 \left(1755 S^5 \lambda ^5-18576 S^4 \lambda ^4+128256 S^3 \lambda ^3-292224 S^2 \lambda ^2+235776 S \lambda -71680\right)\right) Q^2\\&+27 G^3 S^6 \left(16875 S^5 \lambda ^5-17250 S^4 \lambda ^4-22200 S^3 \lambda ^3+6880 S^2 \lambda ^2+17280 S \lambda +4096\right)\end{aligned}\right)\\
	&Z=\left(\begin{aligned}&\left(16875 S^5 \lambda ^5-25650 S^4 \lambda ^4-18040 S^3 \lambda ^3+8480 S^2 \lambda ^2+25984 S \lambda +3072\right)\\&-2 \pi ^3 S^3 \left(3 G Q^2 \left(131175 S^5 \lambda ^5-721020 S^4 \lambda ^4+2564096 S^3 \lambda ^3-5375232 S^2 \lambda ^2\right.\right.\\&\left.\left.+2965760 S \lambda +1256448\right)-64 \left(405 S^5 \lambda ^5-945 S^4 \lambda ^4-540 S^3 \lambda ^3+1680 S^2 \lambda ^2+1856 S \lambda -2560\right)\right)\\&-3 G \pi ^2 S^4 \left(3 G Q^2 \left(192375 S^5 \lambda ^5-1060300 S^4 \lambda ^4+1476080 S^3 \lambda ^3-666048 S^2 \lambda ^2+420864 S \lambda\right.\right.\\&\left.\left. +249856\right)-32 \left(3375 S^5 \lambda ^5-5085 S^4 \lambda ^4-4140 S^3 \lambda ^3+1072 S^2 \lambda ^2+8512 S \lambda -2048\right)\right)\end{aligned}\right)\\
	&A1=\left(\begin{aligned}&\left(405 S^5 \lambda ^5+1620 S^4 \lambda ^4-79264 S^3 \lambda ^3+256640 S^2 \lambda ^2-116480 S \lambda -199680\right) Q^6\\&-32 \pi ^5 S \left(15525 S^5 \lambda ^5+141612 S^4 \lambda ^4-2393568 S^3 \lambda ^3+5581440 S^2 \lambda ^2-2288384 S \lambda -1025024\right) Q^4\\&-8 \pi ^4 S^2 \left(3 G \left(44625 S^5 \lambda ^5+414900 S^4 \lambda ^4-6968864 S^3 \lambda ^3+18363008 S^2 \lambda ^2-8641280 S \lambda -6614016\right) Q^2\right.\\&\left.+4 \left(11205 S^5 \lambda ^5-119268 S^4 \lambda ^4+719808 S^3 \lambda ^3-1605120 S^2 \lambda ^2+1097472 S \lambda -97280\right)\right) Q^2\\&+144 G^2 \pi  S^5 \left(23625 S^5 \lambda ^5-92550 S^4 \lambda ^4-12960 S^3 \lambda ^3+41152 S^2 \lambda ^2+68480 S \lambda +7168\right)\end{aligned}\right)\\
	&A2=\left(\begin{aligned}&\left(106875 S^5 \lambda ^5-315000 S^4 \lambda ^4-198000 S^3 \lambda ^3+164160 S^2 \lambda ^2+217088 S \lambda +46080\right)\\&-8 \pi ^3 S^3 \left(3 G Q^2 \left(125325 S^5 \lambda ^5-1090380 S^4 \lambda ^4+3465792 S^3 \lambda ^3-7205888 S^2 \lambda ^2+4384512 S \lambda +2124800\right)\right.\\&\left.-4 \left(6885 S^5 \lambda ^5-39852 S^4 \lambda ^4+15840 S^3 \lambda ^3+57216 S^2 \lambda ^2+42240 S \lambda -109568\right)\right)\\&-6 G \pi ^2 S^4 \left(3 G Q^2 \left(288375 S^5 \lambda ^5-2646300 S^4 \lambda ^4+5337440 S^3 \lambda ^3-2675072 S^2 \lambda ^2-399616 S \lambda +287744\right)\right.\\&\left.-4 \left(66825 S^5 \lambda ^5-289260 S^4 \lambda ^4-42048 S^3 \lambda ^3+61440 S^2 \lambda ^2+356096 S \lambda -74752\right)\right)\end{aligned}\right)\\
	&A3=\left(\begin{aligned}&\left(3 S^5 \lambda ^5-72 S^4 \lambda ^4-720 S^3 \lambda ^3+5440 S^2 \lambda ^2-4608 S \lambda -3072\right) Q^6\\&-16 \pi ^5 S \left(207 S^5 \lambda ^5-1860 S^4 \lambda ^4-32480 S^3 \lambda ^3+145792 S^2 \lambda ^2-72960 S \lambda -9216\right) Q^4\\&-4 \pi ^4 S^2 \left(3 G \left(615 S^5 \lambda ^5-6124 S^4 \lambda ^4-89248 S^3 \lambda ^3+459648 S^2 \lambda ^2-340224 S \lambda -187392\right) Q^2\right.\\&\left.+4 \left(189 S^5 \lambda ^5-3420 S^4 \lambda ^4+20640 S^3 \lambda ^3-59264 S^2 \lambda ^2+42240 S \lambda -15360\right)\right) Q^2\\&+36 G^2 \pi  S^5 \left(2925 S^5 \lambda ^5-13830 S^4 \lambda ^4+6392 S^3 \lambda ^3+11872 S^2 \lambda ^2+11904 S \lambda +1024\right)\\&+27 G^3 S^6 \left(3375 S^5 \lambda ^5-13350 S^4 \lambda ^4-1320 S^3 \lambda ^3+14112 S^2 \lambda ^2+11392 S \lambda +2048\right)\end{aligned}\right)\\
	&A4=\left(\begin{aligned}&\left(3 G Q^2 \left(1845 S^5 \lambda ^5-30756 S^4 \lambda ^4+110464 S^3 \lambda ^3-250112 S^2 \lambda ^2+248576 S \lambda +87040\right)\right.\\&\left.-16 \left(81 S^5 \lambda ^5-594 S^4 \lambda ^4+792 S^3 \lambda ^3+480 S^2 \lambda ^2+128 S \lambda -2048\right)\right)\\&-3 G \pi ^2 S^4 \left(3 G Q^2 \left(3075 S^5 \lambda ^5-71540 S^4 \lambda ^4+218176 S^3 \lambda ^3-135168 S^2 \lambda ^2-20224 S \lambda +5120\right)\right.\\&\left.-16 \left(945 S^5 \lambda ^5-4986 S^4 \lambda ^4+2904 S^3 \lambda ^3+2592 S^2 \lambda ^2+7808 S \lambda -1024\right)\right)\end{aligned}\right)\\
	&A5=\left(\begin{aligned}&\left(S^3 \lambda ^3-4 S^2 \lambda ^2-80 S \lambda +64\right) Q^6+256 \pi ^5 \left(3 S^5 \lambda ^5+70 S^4 \lambda ^4+64 S^3 \lambda ^3-2688 S^2 \lambda ^2+1024 S \lambda -512\right) Q^4\\&+16 \pi ^4 S \left(3 G \left(75 S^5 \lambda ^5+836 S^4 \lambda ^4-32 S^3 \lambda ^3-29056 S^2 \lambda ^2+32512 S \lambda +13312\right) Q^2\right.\\&\left.+8 \left(9 S^5 \lambda ^5+84 S^4 \lambda ^4-976 S^3 \lambda ^3+3264 S^2 \lambda ^2-1536 S \lambda +2048\right)\right) Q^2\\&+36 G^2 \pi  S^4 \left(2325 S^5 \lambda ^5-13780 S^4 \lambda ^4+24128 S^3 \lambda ^3+24576 S^2 \lambda ^2+17152 S \lambda +1024\right)\\&+27 G^3 S^5 \left(2625 S^5 \lambda ^5-14900 S^4 \lambda ^4+16480 S^3 \lambda ^3+35968 S^2 \lambda ^2+20736 S \lambda +3072\right)\end{aligned}\right)\\
	&A6=\left(\begin{aligned}&\left(3 G \left(75 S^4 \lambda ^4+1040 S^3 \lambda ^3-7072 S^2 \lambda ^2+3072 S \lambda -256\right) Q^2\right.\\&\left.+16 \left(45 S^4 \lambda ^4-279 S^3 \lambda ^3+592 S^2 \lambda ^2+416 S \lambda +832\right)\right)\\&+16 \pi ^3 S^2 \left(3 G \left(45 S^5 \lambda ^5+1988 S^4 \lambda ^4-12480 S^3 \lambda ^3+18432 S^2 \lambda ^2-39168 S \lambda -9216\right) Q^2\right.\\&\left.+8 \left(27 S^5 \lambda ^5-216 S^4 \lambda ^4+720 S^3 \lambda ^3-64 S^2 \lambda ^2-1024\right)\right)\end{aligned}\right)\\
	&A7=\left(\begin{aligned}&\left(2 \pi ^2 (4-5 S \lambda ) b^4+5 G \pi  S (3 S \lambda -4) b^2-15 G^2 S^2 (S \lambda -4)\right)\\& \left(8 \pi ^4 Q^2 (3 S \lambda -4) b^4-12 G \pi ^3 Q^2 S (S \lambda -4) b^2+3 G^2 S^2 \left(-4 \pi ^2 (S \lambda +4) Q^2+4 \pi  S (3 S \lambda +4)+3 G S^2 (5 S \lambda +4)\right)\right)^2\\& \left(8 \pi ^4 Q^2 (15 S \lambda -28) b^4+12 G \pi ^3 Q^2 S (20-3 S \lambda ) b^2-\right.\\&\left.3 G^2 S^2 \left(4 \pi ^2 (S \lambda +12) Q^2+4 \pi  S (3 S \lambda -4)+3 G S^2 (15 S \lambda +4)\right)\right)^2\end{aligned}\right)\\
	&A8=\left(\begin{aligned}&\left(2 \pi ^2 (5 S \lambda -4) b^4-5 G \pi  S (3 S \lambda -4) b^2+15 G^2 S^2 (S \lambda -4)\right)^2\\& \left(2 l^2 \pi ^4 Q^2 (3 S \lambda -4) b^4-3 G l^2 \pi ^3 Q^2 S (S \lambda -4) b^2+3 G^2 S^2 \left(\pi  \left(S (3 S \lambda +4)-\pi  Q^2 (S \lambda +4)\right) l^2+3 G S^2 (5 S \lambda +4)\right)\right)\end{aligned}\right)
\end{split}
\end{equation}
The GTD scalar for the NED AdS black hole for the Kaniadakis entropy is given as
\begin{equation}
R_{GTD}=\frac{3456 G^5 l^2 \pi ^3 S^6}{A3} \left(\frac{4 \left(\begin{aligned}&131072 \pi ^{16} Q^6 S \kappa ^2 A b^{16}-384 G^3 \pi ^{11} Q^4 S^2 B b^{14}\\&-576 G^4 \pi ^8 Q^2 S^3 D b^{12}+144 G^5 \pi ^7 Q^2 S^4 E b^{10}\\&+108 G^6 \pi ^4 S^5 \left(64 \pi ^6F+12 G^2 \pi  S^5 H\right) b^8-\\&2160 G^7 \pi ^3 S^6 \left(48 \pi ^6 I-3 G \pi ^2 S^4 J\right) b^6+\\&810 G^8 \pi ^2 S^7 \left(384 \pi ^6 K+9 G^3 S^6 L\right) b^4-\\&8100 G^9 \pi  S^8 \left(-576 \pi ^6 M+3 G \pi ^2 S^4 N\right) b^2\\&+6075 G^{10} S^{10} \left(36864 \pi ^6 S \kappa ^2 O\right.\\&\left.-32 G \pi ^2 S^5 \kappa ^2 P\right)\end{aligned}\right)}{R}+\frac{l^2 \left(\begin{aligned}&2048 \pi ^{16} Q^6 S \kappa ^2 T b^{14}\\&-36 G^4 \pi ^8 Q^2 S^3 U b^{12}+\\&18 G^5 \pi ^7 Q^2 S^4 V b^{10}+\\&27 G^6 \pi ^4 S^5 W b^8-\\&135 G^7 \pi ^3 S^6 X b^6+\\&405 G^8 \pi ^2 S^7 Y b^4-\\&2025 G^9 \pi  S^8 Z b^2+\\&6075 G^{10} S^{10} A1\end{aligned}\right)}{A2}\right)
\label{ap:9}
\end{equation}
where A, B, D, E, F, H, I, J, K, L, M, N, O, P, R, T, U, V, W, X, Y, Z, A1, A2, A3 are given below
\begin{equation}
	\begin{split}
			&A=\left(\begin{aligned}&\left(25 S^6 \kappa ^6+516 S^4 \kappa ^4-1872 S^2 \kappa ^2+1728\right) b^{20}+4096 G \pi ^{15} Q^6 \left(25 S^{10} \kappa ^{10}+2100 S^8 \kappa ^8+38880 S^6 \kappa ^6\right.\\&\left.+897408 S^4 \kappa ^4-145152 S^2 \kappa ^2-248832\right) b^{18}+768 G^2 \pi ^{12} Q^4 S \left(-8 \pi ^2 \left(325 S^{10} \kappa ^{10}+29820 S^8 \kappa ^8\right.\right.\\&\left.\left.+657888 S^6 \kappa ^6+4149888 S^4 \kappa ^4+2518272 S^2 \kappa ^2-1631232\right) Q^2+48 \pi  S \left(125 S^{10} \kappa ^{10}+8100 S^8 \kappa ^8\right.\right.\\&\left.\left.+146320 S^6 \kappa ^6+333120 S^4 \kappa ^4+142848 S^2 \kappa ^2+718848\right)+G S^2 \left(11725 S^{10} \kappa ^{10}+657180 S^8 \kappa ^8\right.\right.\\&\left.\left.+11183904 S^6 \kappa ^6+25370496 S^4 \kappa ^4+1347840 S^2 \kappa ^2+10202112\right)\right)\end{aligned}\right)\\
		&B=\left(\begin{aligned}&-24 \pi ^2 \left(775 S^{10} \kappa ^{10}+80460 S^8 \kappa ^8+2137632 S^6 \kappa ^6+12573312 S^4 \kappa ^4+11904768 S^2 \kappa ^2-4174848\right) Q^2\\&+4 \pi  S \left(11375 S^{10} \kappa ^{10}+1002900 S^8 \kappa ^8+21592800 S^6 \kappa ^6+59393664 S^4 \kappa ^4+14549760 S^2 \kappa ^2+91653120\right)\\&+3 G S^2 \left(23975 S^{10} \kappa ^{10}+2128620 S^8 \kappa ^8+44788896 S^6 \kappa ^6+126117504 S^4 \kappa ^4+13803264 S^2 \kappa ^2+14183424\right)\end{aligned}\right)\\
		&D=\left(\begin{aligned}&72 \pi ^4 \left(275 S^{10} \kappa ^{10}+38180 S^8 \kappa ^8+1307296 S^6 \kappa ^6+8815488 S^4 \kappa ^4+9962240 S^2 \kappa ^2-1072128\right) Q^4\\&-12 \pi ^3 S \left(6625 S^{10} \kappa ^{10}+694100 S^8 \kappa ^8+17063840 S^6 \kappa ^6+66961536 S^4 \kappa ^4+23291136 S^2 \kappa ^2+46310400\right) Q^2\\&+3 G^2 S^4 \left(42875 S^{10} \kappa ^{10}+1717380 S^8 \kappa ^8+3130848 S^6 \kappa ^6+3687552 S^4 \kappa ^4+10098432 S^2 \kappa ^2-4230144\right)\\&+2 G \pi  S^3 \left(58625 S^{10} \kappa ^{10}+2837700 S^8 \kappa ^8+7308000 S^6 \kappa ^6+22232448 S^4 \kappa ^4+35520768 S^2 \kappa ^2-3234816\right)\\&-3 \pi ^2 S^2 \left(G Q^2 \left(36575 S^{10} \kappa ^{10}+4271780 S^8 \kappa ^8+103819744 S^6 \kappa ^6+417627264 S^4 \kappa ^4+144458496 S^2 \kappa ^2\right.\right.\\&\left.\left.-127429632\right)-16 \left(625 S^{10} \kappa ^{10}+33500 S^8 \kappa ^8+107200 S^6 \kappa ^6+456960 S^4 \kappa ^4+744192 S^2 \kappa ^2+635904\right)\right)\end{aligned}\right)
	\end{split}
\end{equation}
\begin{equation}
	\begin{split}
		&E=\left(\begin{aligned}&-64 \pi ^4 \left(125 S^{10} \kappa ^{10}-155040 S^8 \kappa ^8-9940176 S^6 \kappa ^6-82216512 S^4 \kappa ^4-110343168 S^2 \kappa ^2+24966144\right) Q^4\\&-24 \pi ^3 S \left(10125 S^{10} \kappa ^{10}+1996300 S^8 \kappa ^8+70974560 S^6 \kappa ^6+368767104 S^4 \kappa ^4+192478464 S^2 \kappa ^2\right.\\&\left.+97790976\right) Q^2+27 G^2 S^4 \left(84525 S^{10} \kappa ^{10}+3700900 S^8 \kappa ^8+10162144 S^6 \kappa ^6+7202688 S^4 \kappa ^4\right.\\&\left.+21300480 S^2 \kappa ^2-9427968\right)+12 G \pi  S^3 \left(216125 S^{10} \kappa ^{10}+9450300 S^8 \kappa ^8+32879520 S^6 \kappa ^6+\right.\\&\left.71561088 S^4 \kappa ^4+122390784 S^2 \kappa ^2-20984832\right)+2 \pi ^2 S^2 \left(8 \left(49375 S^{10} \kappa ^{10}+2089500 S^8 \kappa ^8\right.\right.\\&\left.\left.+8762400 S^6 \kappa ^6+26847360 S^4 \kappa ^4+42225408 S^2 \kappa ^2+33758208\right)-3 G Q^2 \left(43575 S^{10} \kappa ^{10}+12099100 S^8 \kappa ^8+\right.\right.\\&\left.\left.423351648 S^6 \kappa ^6+2249453952 S^4 \kappa ^4+1349948160 S^2 \kappa ^2-1276425216\right)\right)\end{aligned}\right)\\
		&F= \left(\begin{aligned}&\left(1225 S^{10} \kappa ^{10}-1380 S^8 \kappa ^8-15176736 S^6 \kappa ^6-159933312 S^4 \kappa ^4-266340096 S^2 \kappa ^2+103652352\right) Q^6\\&+16 \pi ^5 S \left(35625 S^{10} \kappa ^{10}+3335300 S^8 \kappa ^8+205961760 S^6 \kappa ^6+1380249216 S^4 \kappa ^4+1221297408 S^2 \kappa ^2\right.\\&\left.-207498240\right) Q^4+8 \pi ^4 S^2 \left(G Q^2 \left(103775 S^{10} \kappa ^{10}+10018260 S^8 \kappa ^8+605066688 S^6 \kappa ^6\right.\right.\\&\left.\left.+4090823424 S^4 \kappa ^4+4199337216 S^2 \kappa ^2-3692749824\right)-4 \left(105625 S^{10} \kappa ^{10}+3907500 S^8 \kappa ^8\right.\right.\\&\left.\left.+26536800 S^6 \kappa ^6+69033600 S^4 \kappa ^4+111165696 S^2 \kappa ^2+42052608\right)\right) Q^2+9 G^3 S^6 \left(471625 S^{10} \kappa ^{10}\right.\\&\left.+1190700 S^8 \kappa ^8+876960 S^6 \kappa ^6-2706048 S^4 \kappa ^4+7236864 S^2 \kappa ^2-2737152\right)\end{aligned}\right)\\
		&H=\left(\begin{aligned}&\left(667625 S^{10} \kappa ^{10}+2295300 S^8 \kappa ^8+3970080 S^6 \kappa ^6-3487104 S^4 \kappa ^4+14038272 S^2 \kappa ^2-2239488\right)\\&-8 \pi ^3 S^3 \left(3 G Q^2 \left(445375 S^{10} \kappa ^{10}+17414700 S^8 \kappa ^8+102854240 S^6 \kappa ^6+186039168 S^4 \kappa ^4\right.\right.\\&\left.\left.+319322880 S^2 \kappa ^2-115762176\right)-16 \left(9375 S^{10} \kappa ^{10}+47500 S^8 \kappa ^8+108000 S^6 \kappa ^6-28800 S^4 \kappa ^4\right.\right.\\&\left.\left.+449280 S^2 \kappa ^2+580608\right)\right)-2 G \pi ^2 S^4 \left(9 G Q^2 \left(471625 S^{10} \kappa ^{10}+20205500 S^8 \kappa ^8+103013728 S^6 \kappa ^6\right.\right.\\&\left.\left.+67128960 S^4 \kappa ^4+124915968 S^2 \kappa ^2-71525376\right)-32 \left(83125 S^{10} \kappa ^{10}+346500 S^8 \kappa ^8\right.\right.\\&\left.\left.+756000 S^6 \kappa ^6-224640 S^4 \kappa ^4+2757888 S^2 \kappa ^2+746496\right)\right)\end{aligned}\right)\\
		&I=\left(\begin{aligned}&\left(105 S^{10} \kappa ^{10}+16852 S^8 \kappa ^8-695328 S^6 \kappa ^6-10456704 S^4 \kappa ^4-20051712 S^2 \kappa ^2+8377344\right) Q^6+\\&4 \pi ^5 S \left(8825 S^{10} \kappa ^{10}+77580 S^8 \kappa ^8+36898272 S^6 \kappa ^6+328708224 S^4 \kappa ^4+365140224 S^2 \kappa ^2-95966208\right) Q^4\\&+3 \pi ^4 S^2 \left(G Q^2 \left(27125 S^{10} \kappa ^{10}+114948 S^8 \kappa ^8+71384736 S^6 \kappa ^6+629665920 S^4 \kappa ^4+830013696 S^2 \kappa ^2\right.\right.\\&\left.\left.-662473728\right)-64 \left(1125 S^{10} \kappa ^{10}+51000 S^8 \kappa ^8+453320 S^6 \kappa ^6+1268448 S^4 \kappa ^4+1761408 S^2 \kappa ^2+483840\right)\right) Q^2\\&+27 G^3 S^6 \left(25725 S^{10} \kappa ^{10}+90160 S^8 \kappa ^8+86688 S^6 \kappa ^6-203904 S^4 \kappa ^4+463104 S^2 \kappa ^2-165888\right)\\&+18 G^2 \pi  S^5 \left(72275 S^{10} \kappa ^{10}+309260 S^8 \kappa ^8+653184 S^6 \kappa ^6-405504 S^4 \kappa ^4+1651968 S^2 \kappa ^2-248832\right)\end{aligned}\right)\\
		&J=\left(\begin{aligned}&\left(3 G Q^2 \left(53655 S^{10} \kappa ^{10}+3056060 S^8 \kappa ^8+22143408 S^6 \kappa ^6+21196224 S^4 \kappa ^4+14031360 S^2 \kappa ^2-10119168\right)\right.\\&\left.-64 \left(4375 S^{10} \kappa ^{10}+21975 S^8 \kappa ^8+61020 S^6 \kappa ^6-4752 S^4 \kappa ^4+164160 S^2 \kappa ^2+41472\right)\right)\\&-3 \pi ^3 S^3 \left(G Q^2 \left(226275 S^{10} \kappa ^{10}+10731700 S^8 \kappa ^8+85631136 S^6 \kappa ^6+170854272 S^4 \kappa ^4\right.\right.\\&\left.\left.+243475200 S^2 \kappa ^2-101772288\right)-32 \left(1875 S^{10} \kappa ^{10}+11500 S^8 \kappa ^8+36000 S^6 \kappa ^6+\right.\right.\\&\left.\left.1920 S^4 \kappa ^4+99072 S^2 \kappa ^2+138240\right)\right)\end{aligned}\right)
	\end{split}
\end{equation}
\begin{equation}
	\begin{split}
		&K=\left(\begin{aligned}&\left(5 S^{10} \kappa ^{10}+5260 S^8 \kappa ^8-100128 S^6 \kappa ^6-2413440 S^4 \kappa ^4-4817664 S^2 \kappa ^2+1797120\right) Q^6\\&+32 \pi ^5 S \left(4925 S^{10} \kappa ^{10}-13020 S^8 \kappa ^8+7751712 S^6 \kappa ^6+92506752 S^4 \kappa ^4+111829248 S^2 \kappa ^2-27675648\right) Q^4\\&+8 \pi ^4 S^2 \left(3 G Q^2 \left(17885 S^{10} \kappa ^{10}-36340 S^8 \kappa ^8+14857632 S^6 \kappa ^6+171531648 S^4 \kappa ^4+245221632 S^2 \kappa ^2\right.\right.\\&\left.\left.-178578432\right)-4 \left(16875 S^{10} \kappa ^{10}+884500 S^8 \kappa ^8+9588000 S^6 \kappa ^6+35402112 S^4 \kappa ^4+42603264 S^2 \kappa ^2\right.\right.\\&\left.\left.+2626560\right)\right) Q^2+192 G^2 \pi  S^5 \left(35525 S^{10} \kappa ^{10}+281085 S^8 \kappa ^8+554376 S^6 \kappa ^6-96192 S^4 \kappa ^4\right.\\&\left.+1085184 S^2 \kappa ^2-145152\right)\end{aligned}\right)\\
		&L=\left(\begin{aligned}&\left(403025 S^{10} \kappa ^{10}+2966460 S^8 \kappa ^8+2573088 S^6 \kappa ^6-5175936 S^4 \kappa ^4+11162880 S^2 \kappa ^2-3732480\right)\\&-6 G \pi ^2 S^4 \left(G Q^2 \left(186445 S^{10} \kappa ^{10}+12849564 S^8 \kappa ^8+120924000 S^6 \kappa ^6+223673472 S^4 \kappa ^4\right.\right.\\&\left.\left.-23065344 S^2 \kappa ^2-23307264\right)-4 \left(181125 S^{10} \kappa ^{10}+1563100 S^8 \kappa ^8+4311200 S^6 \kappa ^6+1815936 S^4 \kappa ^4\right.\right.\\&\left.\left.+10121472 S^2 \kappa ^2+2018304\right)\right)-8 \pi ^3 S^3 \left(G Q^2 \left(222425 S^{10} \kappa ^{10}+11501220 S^8 \kappa ^8+114647328 S^6 \kappa ^6\right.\right.\\&\left.\left.+323971200 S^4 \kappa ^4+335418624 S^2 \kappa ^2-172108800\right)-12 \left(9375 S^{10} \kappa ^{10}+93500 S^8 \kappa ^8+322400 S^6 \kappa ^6\right.\right.\\&\left.\left.+196480 S^4 \kappa ^4+768768 S^2 \kappa ^2+986112\right)\right)\end{aligned}\right)\\
		&M=\left(\begin{aligned}&\left(S^{10} \kappa ^{10}-172 S^8 \kappa ^8+1504 S^6 \kappa ^6+67968 S^4 \kappa ^4+131328 S^2 \kappa ^2-27648\right) Q^6+\\&48 \pi ^5 S \left(335 S^{10} \kappa ^{10}-1276 S^8 \kappa ^8+203232 S^6 \kappa ^6+2987136 S^4 \kappa ^4+3748608 S^2 \kappa ^2-248832\right) Q^4+\\&4 \pi ^4 S^2 \left(9 G Q^2 \left(1211 S^{10} \kappa ^{10}-1492 S^8 \kappa ^8+381600 S^6 \kappa ^6+5352576 S^4 \kappa ^4+8301312 S^2 \kappa ^2-5059584\right)\right.\\&\left.-4 \left(2575 S^{10} \kappa ^{10}+93180 S^8 \kappa ^8+1329696 S^6 \kappa ^6+4811904 S^4 \kappa ^4+6504192 S^2 \kappa ^2+1244160\right)\right) Q^2\\&+216 G^3 S^6 \left(1715 S^{10} \kappa ^{10}+16709 S^8 \kappa ^8+27552 S^6 \kappa ^6-50112 S^4 \kappa ^4+76032 S^2 \kappa ^2-20736\right)\\&+36 G^2 \pi  S^5 \left(19845 S^{10} \kappa ^{10}+194180 S^8 \kappa ^8+519552 S^6 \kappa ^6-147456 S^4 \kappa ^4+836352 S^2 \kappa ^2-82944\right)\end{aligned}\right)\\
		&N=\left(\begin{aligned}&\left(16 \left(9625 S^{10} \kappa ^{10}+96380 S^8 \kappa ^8+343680 S^6 \kappa ^6+129024 S^4 \kappa ^4+753408 S^2 \kappa ^2+82944\right)\right.\\&\left.-9 G Q^2 \left(2597 S^{10} \kappa ^{10}+147812 S^8 \kappa ^8+1883360 S^6 \kappa ^6+3608448 S^4 \kappa ^4-99072 S^2 \kappa ^2-138240\right)\right)\\&-12 \pi ^3 S^3 \left(G Q^2 \left(11165 S^{10} \kappa ^{10}+406044 S^8 \kappa ^8+5315232 S^6 \kappa ^6+14550912 S^4 \kappa ^4+18544896 S^2 \kappa ^2-7050240\right)\right.\\&\left.-64 \left(125 S^{10} \kappa ^{10}+1350 S^8 \kappa ^8+6160 S^6 \kappa ^6+4416 S^4 \kappa ^4+13824 S^2 \kappa ^2+13824\right)\right)\end{aligned}\right)\\
		&O=\left(\begin{aligned}&\left(S^6 \kappa ^6-12 S^4 \kappa ^4-336 S^2 \kappa ^2-576\right) Q^6+384 \pi ^5 \left(25 S^{10} \kappa ^{10}+36 S^8 \kappa ^8+14016 S^6 \kappa ^6+134400 S^4 \kappa ^4\right.\\&\left.+172800 S^2 \kappa ^2+27648\right) Q^4+48 \pi ^4 S \left(3 G Q^2 \left(175 S^{10} \kappa ^{10}+692 S^8 \kappa ^8+52192 S^6 \kappa ^6+452736 S^4 \kappa ^4\right.\right.\\&\left.\left.+762624 S^2 \kappa ^2-359424\right)-32 \left(25 S^{10} \kappa ^{10}+540 S^8 \kappa ^8+10088 S^6 \kappa ^6+25440 S^4 \kappa ^4+42624 S^2 \kappa ^2+13824\right)\right) Q^2\\&+9 G^3 S^5 \left(32585 S^{10} \kappa ^{10}+470988 S^8 \kappa ^8+1989792 S^6 \kappa ^6-3259008 S^4 \kappa ^4+3587328 S^2 \kappa ^2-746496\right)\\&+12 G^2 \pi  S^4 \left(48265 S^{10} \kappa ^{10}+669732 S^8 \kappa ^8+3424032 S^6 \kappa ^6-1662336 S^4 \kappa ^4+4416768 S^2 \kappa ^2-248832\right)\end{aligned}\right)\\
		&P=\left(\begin{aligned}&\left(9 G Q^2 \left(245 S^8 \kappa ^8+7700 S^6 \kappa ^6+123504 S^4 \kappa ^4+137664 S^2 \kappa ^2+55296\right)\right.\\&\left.-4 \left(2975 S^8 \kappa ^8+40200 S^6 \kappa ^6+243936 S^4 \kappa ^4+41472 S^2 \kappa ^2+518400\right)\right)\\&-16 \pi ^3 S^2 \left(3 G Q^2 \left(2555 S^{10} \kappa ^{10}+58268 S^8 \kappa ^8+950304 S^6 \kappa ^6+1650816 S^4 \kappa ^4+3158784 S^2 \kappa ^2-746496\right)\right.\\&\left.-8 \left(625 S^{10} \kappa ^{10}+8500 S^8 \kappa ^8+62880 S^6 \kappa ^6+42624 S^4 \kappa ^4+158976 S^2 \kappa ^2+82944\right)\right)\end{aligned}\right)\\
		&	R=\left(\begin{aligned}&\left(8 \pi ^4 Q^2 \left(S^2 \kappa ^2+28\right) b^4+4 G \pi ^3 Q^2 S \left(S^2 \kappa ^2-60\right) b^2+3 G^2 S^2 \left(4 \pi ^2 \left(S^2 \kappa ^2+12\right) Q^2\right.\right.\\&\left.\left.-4 \pi  S \left(5 S^2 \kappa ^2+4\right)+G S^2 \left(12-35 S^2 \kappa ^2\right)\right)\right)^2 \left(8 \pi ^4 Q^2 \left(S^2 \kappa ^2+12\right) b^4+12 G \pi ^3 Q^2 S \left(S^2 \kappa ^2-12\right) b^2\right.\\&\left.+3 G^2 S^2 \left(-12 \pi ^2 \left(S^2 \kappa ^2-4\right) Q^2+4 \pi  S \left(5 S^2 \kappa ^2-12\right)+3 G S^2 \left(7 S^2 \kappa ^2-12\right)\right)\right)^2\end{aligned}\right)\\
		&T=\left(\begin{aligned}&\left(25 S^6 \kappa ^6+516 S^4 \kappa ^4-1872 S^2 \kappa ^2+1728\right) b^{20}+64 G \pi ^{15} Q^6 \left(25 S^{10} \kappa ^{10}+2100 S^8 \kappa ^8+38880 S^6 \kappa ^6\right.\\&\left.+897408 S^4 \kappa ^4-145152 S^2 \kappa ^2-248832\right) b^{18}+48 G^2 \pi ^{12} Q^4 S \left(-2 \pi ^2 \left(325 S^{10} \kappa ^{10}+29820 S^8 \kappa ^8\right.\right.\\&\left.\left.+657888 S^6 \kappa ^6+4149888 S^4 \kappa ^4+2518272 S^2 \kappa ^2-1631232\right) Q^2+12 \pi  S \left(125 S^{10} \kappa ^{10}+8100 S^8 \kappa ^8\right.\right.\\&\left.\left.+146320 S^6 \kappa ^6+333120 S^4 \kappa ^4+142848 S^2 \kappa ^2+718848\right)+G S^2 \left(11725 S^{10} \kappa ^{10}+657180 S^8 \kappa ^8\right.\right.\\&\left.\left.+11183904 S^6 \kappa ^6+25370496 S^4 \kappa ^4+1347840 S^2 \kappa ^2+10202112\right)\right) b^{16}\\&-24 G^3 \pi ^{11} Q^4 S^2 \left(-6 \pi ^2 \left(775 S^{10} \kappa ^{10}+80460 S^8 \kappa ^8+2137632 S^6 \kappa ^6+12573312 S^4 \kappa ^4\right.\right.\\&\left.\left.+11904768 S^2 \kappa ^2-4174848\right) Q^2+\pi  S \left(11375 S^{10} \kappa ^{10}+1002900 S^8 \kappa ^8+21592800 S^6 \kappa ^6\right.\right.\\&\left.\left.+59393664 S^4 \kappa ^4+14549760 S^2 \kappa ^2+91653120\right)+3 G S^2 \left(23975 S^{10} \kappa ^{10}+2128620 S^8 \kappa ^8\right.\right.\\&\left.\left.+44788896 S^6 \kappa ^6+126117504 S^4 \kappa ^4+13803264 S^2 \kappa ^2+14183424\right)\right)\end{aligned}\right)\\
		&U=\left(\begin{aligned}&18 \pi ^4 \left(275 S^{10} \kappa ^{10}+38180 S^8 \kappa ^8+1307296 S^6 \kappa ^6+8815488 S^4 \kappa ^4+9962240 S^2 \kappa ^2-1072128\right) Q^4\\&-3 \pi ^3 S \left(6625 S^{10} \kappa ^{10}+694100 S^8 \kappa ^8+17063840 S^6 \kappa ^6+66961536 S^4 \kappa ^4+23291136 S^2 \kappa ^2+46310400\right) Q^2\\&+12 G^2 S^4 \left(42875 S^{10} \kappa ^{10}+1717380 S^8 \kappa ^8+3130848 S^6 \kappa ^6+3687552 S^4 \kappa ^4+10098432 S^2 \kappa ^2-4230144\right)\\&+2 G \pi  S^3 \left(58625 S^{10} \kappa ^{10}+2837700 S^8 \kappa ^8+7308000 S^6 \kappa ^6+22232448 S^4 \kappa ^4+35520768 S^2 \kappa ^2-3234816\right)\\&-3 \pi ^2 S^2 \left(G Q^2 \left(36575 S^{10} \kappa ^{10}+4271780 S^8 \kappa ^8+103819744 S^6 \kappa ^6+417627264 S^4 \kappa ^4+144458496 S^2 \kappa ^2\right.\right.\\&\left.\left.-127429632\right)-4 \left(625 S^{10} \kappa ^{10}+33500 S^8 \kappa ^8+107200 S^6 \kappa ^6+456960 S^4 \kappa ^4+744192 S^2 \kappa ^2+635904\right)\right)\end{aligned}\right)
		\end{split}
\end{equation}
\begin{equation}
	\begin{split}
		&V=\left(\begin{aligned}&-8 \pi ^4 \left(125 S^{10} \kappa ^{10}-155040 S^8 \kappa ^8-9940176 S^6 \kappa ^6-82216512 S^4 \kappa ^4-110343168 S^2 \kappa ^2+24966144\right) Q^4\\&-3 \pi ^3 S \left(10125 S^{10} \kappa ^{10}+1996300 S^8 \kappa ^8+70974560 S^6 \kappa ^6+368767104 S^4 \kappa ^4+192478464 S^2 \kappa ^2\right.\\&\left.+97790976\right) Q^2+54 G^2 S^4 \left(84525 S^{10} \kappa ^{10}+3700900 S^8 \kappa ^8+10162144 S^6 \kappa ^6+7202688 S^4 \kappa ^4\right.\\&\left.+21300480 S^2 \kappa ^2-9427968\right)+6 G \pi  S^3 \left(216125 S^{10} \kappa ^{10}+9450300 S^8 \kappa ^8+32879520 S^6 \kappa ^6\right.\\&\left.+71561088 S^4 \kappa ^4+122390784 S^2 \kappa ^2-20984832\right)+\pi ^2 S^2 \left(98750 S^{10} \kappa ^{10}+4179000 S^8 \kappa ^8\right.\\&\left.+17524800 S^6 \kappa ^6+53694720 S^4 \kappa ^4+84450816 S^2 \kappa ^2-3 G Q^2 \left(43575 S^{10} \kappa ^{10}+12099100 S^8 \kappa ^8+\right.\right.\\&\left.\left.423351648 S^6 \kappa ^6+2249453952 S^4 \kappa ^4+1349948160 S^2 \kappa ^2-1276425216\right)+67516416\right)\end{aligned}\right)\\
		&W=\left(\begin{aligned}&4 \pi ^6 \left(1225 S^{10} \kappa ^{10}-1380 S^8 \kappa ^8-15176736 S^6 \kappa ^6-159933312 S^4 \kappa ^4-266340096 S^2 \kappa ^2+103652352\right) Q^6\\&+\pi ^5 S \left(35625 S^{10} \kappa ^{10}+3335300 S^8 \kappa ^8+205961760 S^6 \kappa ^6+1380249216 S^4 \kappa ^4+1221297408 S^2 \kappa ^2\right.\\&\left.-207498240\right) Q^4+2 \pi ^4 S^2 \left(-105625 S^{10} \kappa ^{10}-3907500 S^8 \kappa ^8-26536800 S^6 \kappa ^6-69033600 S^4 \kappa ^4\right.\\&\left.-111165696 S^2 \kappa ^2+G Q^2 \left(103775 S^{10} \kappa ^{10}+10018260 S^8 \kappa ^8+605066688 S^6 \kappa ^6+4090823424 S^4 \kappa ^4\right.\right.\\&\left.\left.+4199337216 S^2 \kappa ^2-3692749824\right)-42052608\right) Q^2+36 G^3 S^6 \left(471625 S^{10} \kappa ^{10}+1190700 S^8 \kappa ^8\right.\\&\left.+876960 S^6 \kappa ^6-2706048 S^4 \kappa ^4+7236864 S^2 \kappa ^2-2737152\right)+12 G^2 \pi  S^5 \left(667625 S^{10} \kappa ^{10}+\right.\\&\left.2295300 S^8 \kappa ^8+3970080 S^6 \kappa ^6-3487104 S^4 \kappa ^4+14038272 S^2 \kappa ^2-2239488\right)\\&-2 \pi ^3 S^3 \left(3 G Q^2 \left(445375 S^{10} \kappa ^{10}+17414700 S^8 \kappa ^8+102854240 S^6 \kappa ^6+186039168 S^4 \kappa ^4\right.\right.\\&\left.\left.+319322880 S^2 \kappa ^2-115762176\right)-4 \left(9375 S^{10} \kappa ^{10}+47500 S^8 \kappa ^8+108000 S^6 \kappa ^6-28800 S^4 \kappa ^4\right.\right.\\&\left.\left.+449280 S^2 \kappa ^2+580608\right)\right)-2 G \pi ^2 S^4 \left(9 G Q^2 \left(471625 S^{10} \kappa ^{10}+20205500 S^8 \kappa ^8+103013728 S^6 \kappa ^6\right.\right.\\&\left.\left.+67128960 S^4 \kappa ^4+124915968 S^2 \kappa ^2-71525376\right)-8 \left(83125 S^{10} \kappa ^{10}+346500 S^8 \kappa ^8+756000 S^6 \kappa ^6\right.\right.\\&\left.\left.-224640 S^4 \kappa ^4+2757888 S^2 \kappa ^2+746496\right)\right)\end{aligned}\right)\\
		&X=\left(\begin{aligned}&12 \pi ^6 \left(105 S^{10} \kappa ^{10}+16852 S^8 \kappa ^8-695328 S^6 \kappa ^6-10456704 S^4 \kappa ^4-20051712 S^2 \kappa ^2+8377344\right) Q^6\\&+\pi ^5 S \left(8825 S^{10} \kappa ^{10}+77580 S^8 \kappa ^8+36898272 S^6 \kappa ^6+328708224 S^4 \kappa ^4+365140224 S^2 \kappa ^2-95966208\right) Q^4\\&+3 \pi ^4 S^2 \left(G Q^2 \left(27125 S^{10} \kappa ^{10}+114948 S^8 \kappa ^8+71384736 S^6 \kappa ^6+629665920 S^4 \kappa ^4+830013696 S^2 \kappa ^2\right.\right.\\&\left.\left.-662473728\right)-16 \left(1125 S^{10} \kappa ^{10}+51000 S^8 \kappa ^8+453320 S^6 \kappa ^6+1268448 S^4 \kappa ^4+1761408 S^2 \kappa ^2\right.\right.\\&\left.\left.+483840\right)\right) Q^2+432 G^3 S^6 \left(25725 S^{10} \kappa ^{10}+90160 S^8 \kappa ^8+86688 S^6 \kappa ^6-203904 S^4 \kappa ^4+463104 S^2 \kappa ^2\right.\\&\left.-165888\right)+72 G^2 \pi  S^5 \left(72275 S^{10} \kappa ^{10}+309260 S^8 \kappa ^8+653184 S^6 \kappa ^6-405504 S^4 \kappa ^4+1651968 S^2 \kappa ^2\right.\\&\left.-248832\right)-12 G \pi ^2 S^4 \left(3 G Q^2 \left(53655 S^{10} \kappa ^{10}+3056060 S^8 \kappa ^8+22143408 S^6 \kappa ^6+21196224 S^4 \kappa ^4\right.\right.\\&\left.\left.+14031360 S^2 \kappa ^2-10119168\right)-16 \left(4375 S^{10} \kappa ^{10}+21975 S^8 \kappa ^8+61020 S^6 \kappa ^6-4752 S^4 \kappa ^4\right.\right.\\&\left.\left.+164160 S^2 \kappa ^2+41472\right)\right)+3 \pi ^3 S^3 \left(8 \left(1875 S^{10} \kappa ^{10}+11500 S^8 \kappa ^8+36000 S^6 \kappa ^6+1920 S^4 \kappa ^4\right.\right.\\&\left.\left.+99072 S^2 \kappa ^2+138240\right)-G Q^2 \left(226275 S^{10} \kappa ^{10}+10731700 S^8 \kappa ^8+85631136 S^6 \kappa ^6\right.\right.\\&\left.\left.+170854272 S^4 \kappa ^4+243475200 S^2 \kappa ^2-101772288\right)\right)\end{aligned}\right)\\
		&Y=\left(\begin{aligned}&12 \pi ^6 \left(5 S^{10} \kappa ^{10}+5260 S^8 \kappa ^8-100128 S^6 \kappa ^6-2413440 S^4 \kappa ^4-4817664 S^2 \kappa ^2+1797120\right) Q^6\\&+\pi ^5 S \left(4925 S^{10} \kappa ^{10}-13020 S^8 \kappa ^8+7751712 S^6 \kappa ^6+92506752 S^4 \kappa ^4+111829248 S^2 \kappa ^2-27675648\right) Q^4\\&+\pi ^4 S^2 \left(-16875 S^{10} \kappa ^{10}-884500 S^8 \kappa ^8-9588000 S^6 \kappa ^6-35402112 S^4 \kappa ^4-42603264 S^2 \kappa ^2\right.\\&\left.+3 G Q^2 \left(17885 S^{10} \kappa ^{10}-36340 S^8 \kappa ^8+14857632 S^6 \kappa ^6+171531648 S^4 \kappa ^4+245221632 S^2 \kappa ^2-178578432\right)\right.\\&\left.-2626560\right) Q^2+96 G^2 \pi  S^5 \left(35525 S^{10} \kappa ^{10}+281085 S^8 \kappa ^8+554376 S^6 \kappa ^6-96192 S^4 \kappa ^4\right.\\&\left.+1085184 S^2 \kappa ^2-145152\right)+18 G^3 S^6 \left(403025 S^{10} \kappa ^{10}+2966460 S^8 \kappa ^8+2573088 S^6 \kappa ^6-5175936 S^4 \kappa ^4\right.\\&\left.+11162880 S^2 \kappa ^2-3732480\right)+3 G \pi ^2 S^4 \left(181125 S^{10} \kappa ^{10}+1563100 S^8 \kappa ^8+4311200 S^6 \kappa ^6+\right.\\&\left.1815936 S^4 \kappa ^4+10121472 S^2 \kappa ^2-G Q^2 \left(186445 S^{10} \kappa ^{10}+12849564 S^8 \kappa ^8+120924000 S^6 \kappa ^6\right.\right.\\&\left.\left.+223673472 S^4 \kappa ^4-23065344 S^2 \kappa ^2-23307264\right)+2018304\right)+\pi ^3 S^3 \left(3 \left(9375 S^{10} \kappa ^{10}\right.\right.\\&\left.\left.+93500 S^8 \kappa ^8+322400 S^6 \kappa ^6+196480 S^4 \kappa ^4+768768 S^2 \kappa ^2+986112\right)-G Q^2 \left(222425 S^{10} \kappa ^{10}\right.\right.\\&\left.\left.+11501220 S^8 \kappa ^8+114647328 S^6 \kappa ^6+323971200 S^4 \kappa ^4+335418624 S^2 \kappa ^2-172108800\right)\right)\end{aligned}\right)
		\end{split}
\end{equation}
\begin{equation}
	\begin{split}
		&Z=\left(\begin{aligned}&-36 \pi ^6 \left(S^{10} \kappa ^{10}-172 S^8 \kappa ^8+1504 S^6 \kappa ^6+67968 S^4 \kappa ^4+131328 S^2 \kappa ^2-27648\right) Q^6\\&+3 \pi ^5 S \left(335 S^{10} \kappa ^{10}-1276 S^8 \kappa ^8+203232 S^6 \kappa ^6+2987136 S^4 \kappa ^4+3748608 S^2 \kappa ^2-248832\right) Q^4\\&+\pi ^4 S^2 \left(-2575 S^{10} \kappa ^{10}-93180 S^8 \kappa ^8-1329696 S^6 \kappa ^6-4811904 S^4 \kappa ^4-6504192 S^2 \kappa ^2\right.\\&\left.+9 G Q^2 \left(1211 S^{10} \kappa ^{10}-1492 S^8 \kappa ^8+381600 S^6 \kappa ^6+5352576 S^4 \kappa ^4+8301312 S^2 \kappa ^2-5059584\right)\right.\\&\left.-1244160\right) Q^2+864 G^3 S^6 \left(1715 S^{10} \kappa ^{10}+16709 S^8 \kappa ^8+27552 S^6 \kappa ^6-50112 S^4 \kappa ^4+76032 S^2 \kappa ^2-20736\right)\\&+36 G^2 \pi  S^5 \left(19845 S^{10} \kappa ^{10}+194180 S^8 \kappa ^8+519552 S^6 \kappa ^6-147456 S^4 \kappa ^4+836352 S^2 \kappa ^2-82944\right)\\&+3 G \pi ^2 S^4 \left(4 \left(9625 S^{10} \kappa ^{10}+96380 S^8 \kappa ^8+343680 S^6 \kappa ^6+129024 S^4 \kappa ^4+753408 S^2 \kappa ^2+82944\right)\right.\\&\left.-9 G Q^2 \left(2597 S^{10} \kappa ^{10}+147812 S^8 \kappa ^8+1883360 S^6 \kappa ^6+3608448 S^4 \kappa ^4-99072 S^2 \kappa ^2-138240\right)\right)\\&-3 \pi ^3 S^3 \left(G Q^2 \left(11165 S^{10} \kappa ^{10}+406044 S^8 \kappa ^8+5315232 S^6 \kappa ^6+14550912 S^4 \kappa ^4+18544896 S^2 \kappa ^2\right.\right.\\&\left.\left.-7050240\right)-16 \left(125 S^{10} \kappa ^{10}+1350 S^8 \kappa ^8+6160 S^6 \kappa ^6+4416 S^4 \kappa ^4+13824 S^2 \kappa ^2+13824\right)\right)\end{aligned}\right)\\
		&A1=\left(\begin{aligned}&576 \pi ^6 S \kappa ^2 \left(S^6 \kappa ^6-12 S^4 \kappa ^4-336 S^2 \kappa ^2-576\right) Q^6+6 \pi ^5 \left(25 S^{10} \kappa ^{10}+36 S^8 \kappa ^8+14016 S^6 \kappa ^6\right.\\&\left.+134400 S^4 \kappa ^4+172800 S^2 \kappa ^2+27648\right) Q^4+3 \pi ^4 S \left(3 G Q^2 \left(175 S^{10} \kappa ^{10}+692 S^8 \kappa ^8+52192 S^6 \kappa ^6\right.\right.\\&\left.\left.+452736 S^4 \kappa ^4+762624 S^2 \kappa ^2-359424\right)-8 \left(25 S^{10} \kappa ^{10}+540 S^8 \kappa ^8+10088 S^6 \kappa ^6+25440 S^4 \kappa ^4\right.\right.\\&\left.\left.+42624 S^2 \kappa ^2+13824\right)\right) Q^2+9 G^3 S^5 \left(32585 S^{10} \kappa ^{10}+470988 S^8 \kappa ^8+1989792 S^6 \kappa ^6-3259008 S^4 \kappa ^4\right.\\&\left.+3587328 S^2 \kappa ^2-746496\right)+3 G^2 \pi  S^4 \left(48265 S^{10} \kappa ^{10}+669732 S^8 \kappa ^8+3424032 S^6 \kappa ^6-1662336 S^4 \kappa ^4\right.\\&\left.+4416768 S^2 \kappa ^2-248832\right)-8 G \pi ^2 S^5 \kappa ^2 \left(-2975 S^8 \kappa ^8-40200 S^6 \kappa ^6-243936 S^4 \kappa ^4-41472 S^2 \kappa ^2\right.\\&\left.+9 G Q^2 \left(245 S^8 \kappa ^8+7700 S^6 \kappa ^6+123504 S^4 \kappa ^4+137664 S^2 \kappa ^2+55296\right)-518400\right)\\&+\pi ^3 S^2 \left(2 \left(625 S^{10} \kappa ^{10}+8500 S^8 \kappa ^8+62880 S^6 \kappa ^6+42624 S^4 \kappa ^4+158976 S^2 \kappa ^2+82944\right)\right.\\&\left.-3 G Q^2 \left(2555 S^{10} \kappa ^{10}+58268 S^8 \kappa ^8+950304 S^6 \kappa ^6+1650816 S^4 \kappa ^4+3158784 S^2 \kappa ^2-746496\right)\right)\end{aligned}\right)\\
		&A2=\left(\begin{aligned}&\left(2 \pi ^4 Q^2 \left(S^2 \kappa ^2+28\right) b^4+G \pi ^3 Q^2 S \left(S^2 \kappa ^2-60\right) b^2+3 G^2 S^2 \left(\pi ^2 \left(S^2 \kappa ^2+12\right) Q^2\right.\right.\\&\left.\left.-\pi  S \left(5 S^2 \kappa ^2+4\right)+G S^2 \left(12-35 S^2 \kappa ^2\right)\right)\right) \left(2 \pi ^4 Q^2 \left(S^2 \kappa ^2+12\right) b^4+3 G \pi ^3 Q^2 S \left(S^2 \kappa ^2-12\right) b^2\right.\\&\left.+3 G^2 S^2 \left(-3 \pi ^2 \left(S^2 \kappa ^2-4\right) Q^2+\pi  S \left(5 S^2 \kappa ^2-12\right)+3 G S^2 \left(7 S^2 \kappa ^2-12\right)\right)\right)^2\times \\& \left(2 l^2 \pi ^4 Q^2 \left(S^2 \kappa ^2+28\right) b^4+G l^2\right.\\&\left. \pi ^3 Q^2 S \left(S^2 \kappa ^2-60\right) b^2+3 G^2 S^2 \left(\pi  \left(\pi  Q^2 \left(S^2 \kappa ^2+12\right)-S \left(5 S^2 \kappa ^2+4\right)\right) l^2+G S^2 \left(12-35 S^2 \kappa ^2\right)\right)\right)\end{aligned}\right)\\
		&A3=\left(\begin{aligned}&\left(2 \pi ^2 \left(5 S^2 \kappa ^2+12\right) b^4-15 G \pi  S \left(S^2 \kappa ^2+4\right) b^2+15 G^2 S^2 \left(S^2 \kappa ^2+12\right)\right)^2\times\\& \left(2 l^2 \pi ^4 Q^2 \left(S^2 \kappa ^2+12\right) b^4+3 G l^2 \pi ^3 Q^2 S \left(S^2 \kappa ^2-12\right) b^2+3 G^2 S^2 \left(\pi  \left(S \left(5 S^2 \kappa ^2-12\right)-3 \pi  Q^2 \left(S^2 \kappa ^2-4\right)\right) l^2\right.\right.\\&\left.\left.+3 G S^2 \left(7 S^2 \kappa ^2-12\right)\right)\right)\end{aligned}\right)
	\end{split}
\end{equation}
In section \ref{sec:5}, the GTD scalar for the dual CFT of NED AdS BH in Bekenstein Hawking entropy is
\begin{equation}
R_{GTD}=-\frac{144 c \pi ^2 \mathcal{S}^6 \left(50331648 c^9 \pi ^{13} \mathcal{Q}^4 A \mathit{b}^6+9216 c^5 \pi ^5 \mathcal{S}^4 B-6480 c^2 \pi ^2 \mathcal{S}^7 I\right) \mathcal{V}}{D}
\label{ap:10}
\end{equation}
where A, B, I, D a given below
\begin{equation}
	\begin{split}
		&A=\left(\begin{aligned}&\left(\pi ^2 \mathit{b}^2 \mathcal{Q}^2-26 \mathcal{S}^2\right) \mathit{b}^{16}-524288 c^8 \pi ^{10} \mathcal{Q}^2 \mathcal{S} \left(236 \pi ^4 \mathit{b}^4 \mathcal{Q}^4+3 \pi ^2 \mathit{b}^2 \left(246 \mathit{b}^2-1105\right) \mathcal{S}^2 \mathcal{Q}^2-414 \mathcal{S}^4\right) \mathit{b}^{12}\\&+196608 c^7 \pi ^7 \mathcal{S}^2 \left(604 \pi ^6 \mathit{b}^6 \mathcal{Q}^6+3 \pi ^4 \mathit{b}^4 \left(342 \mathit{b}^2-1675\right) \mathcal{S}^2 \mathcal{Q}^4-3 \pi ^2 \mathit{b}^2 \left(78 \mathit{b}^2+407\right) \mathcal{S}^4 \mathcal{Q}^2-126 \mathcal{S}^6\right) \mathit{b}^8\\&-24576 c^6 \pi ^6 \mathcal{S}^3 \left(1396 \pi ^6 \mathit{b}^6 \mathcal{Q}^6-3 \pi ^4 \mathit{b}^4 \left(9218 \mathit{b}^2+3537\right) \mathcal{S}^2 \mathcal{Q}^4+9 \pi ^2 \mathit{b}^2 \left(408 \mathit{b}^4-506 \mathit{b}^2-507\right) \mathcal{S}^4 \mathcal{Q}^2\right.\\&\left.+36 \left(18 \mathit{b}^2-25\right) \mathcal{S}^6\right)\end{aligned}\right)\\
		&B=\left(\begin{aligned}&\left(4816 \pi ^6 \mathit{b}^6 \mathcal{Q}^6+\pi ^4 \mathit{b}^4 \left(7505-92334 \mathit{b}^2\right) \mathcal{S}^2 \mathcal{Q}^4+6 \pi ^2 \mathit{b}^2 \left(2046 \mathit{b}^4-4187 \mathit{b}^2-700\right) \mathcal{S}^4 \mathcal{Q}^2\right.\\&\left.+3 \left(324 \mathit{b}^4+480 \mathit{b}^2-535\right) \mathcal{S}^6\right) \mathit{b}^4-2304 c^4 \pi ^4 \mathcal{S}^5 \left(14996 \pi ^6 \mathit{b}^6 \mathcal{Q}^6-3 \pi ^4 \mathit{b}^4 \left(89042 \mathit{b}^2-5785\right) \mathcal{S}^2 \mathcal{Q}^4\right.\\&\left.+3 \pi ^2 \mathit{b}^2 \left(15522 \mathit{b}^4-18405 \mathit{b}^2-475\right) \mathcal{S}^4 \mathcal{Q}^2-9 \left(396 \mathit{b}^6-360 \mathit{b}^4-365 \mathit{b}^2+200\right) \mathcal{S}^6\right) \mathit{b}^2\\&+18225 \mathcal{S}^{11} \left(13 \pi ^4 \mathcal{Q}^4+27 \mathcal{S}^4\right)+24300 c \pi  \mathcal{S}^8 \left(4 \pi ^6 \mathit{b}^2 \mathcal{Q}^6-\pi ^4 \left(183 \mathit{b}^2+2\right) \mathcal{S}^2 \mathcal{Q}^4\right.\\&\left.+3 \pi ^2 \left(5 \mathit{b}^2-9\right) \mathcal{S}^4 \mathcal{Q}^2+9 \left(1-8 \mathit{b}^2\right) \mathcal{S}^6\right)\end{aligned}\right)\\
		&I=\left(\begin{aligned}&\left(260 \pi ^6 \mathit{b}^4 \mathcal{Q}^6+3 \pi ^4 \mathit{b}^2 \left(15-2153 \mathit{b}^2\right) \mathcal{S}^2 \mathcal{Q}^4+3 \pi ^2 \left(281 \mathit{b}^4-425 \mathit{b}^2-20\right) \mathcal{S}^4 \mathcal{Q}^2+90 \mathit{b}^2 \left(2-9 \mathit{b}^2\right) \mathcal{S}^6\right)\\&+8640 c^3 \pi ^3 \mathcal{S}^6 \left(1212 \pi ^6 \mathit{b}^6 \mathcal{Q}^6-7 \pi ^4 \mathit{b}^4 \left(3423 \mathit{b}^2-143\right) \mathcal{S}^2 \mathcal{Q}^4+3 \pi ^2 \mathit{b}^2 \left(1464 \mathit{b}^4-2075 \mathit{b}^2-75\right) \mathcal{S}^4 \mathcal{Q}^2\right.\\&\left.+18 \left(-48 \mathit{b}^6+28 \mathit{b}^4+10 \mathit{b}^2-5\right) \mathcal{S}^6\right)\end{aligned}\right)\\
		&D=\left(\begin{aligned}&\left(32 c^2 \pi ^2 \mathit{b}^4-20 c \pi  \mathcal{S} \mathit{b}^2+15 \mathcal{S}^2\right)^2 \left(32 c^2 \pi ^4 \mathcal{Q}^2 \mathit{b}^4-9 \mathcal{S}^4+3 \pi ^2 \mathcal{Q}^2 \mathcal{S}^2-12 c \pi  \mathcal{S} \left(\pi ^2 \mathit{b}^2 \mathcal{Q}^2+\mathcal{S}^2\right)\right)^3\\& \left(224 c^2 \pi ^4 \mathcal{Q}^2 \mathit{b}^4-12 c \pi  \mathcal{S} \left(5 \pi ^2 \mathit{b}^2 \mathcal{Q}^2+\mathcal{S}^2\right)+9 \left(\mathcal{S}^4+\pi ^2 \mathcal{Q}^2 \mathcal{S}^2\right)\right)^2\end{aligned}\right)
	\end{split}
\end{equation}
The GTD scalar for the dual CFT of NED AdS black holes for R\'enyi entropy is given as
\begin{equation}
	R_{GTD}=-\frac{2304 c \pi ^2 \mathcal{S}^6 \mathcal{V} \left(\begin{aligned}&536870912 c^{10} \pi ^{16} \mathcal{Q}^6 \lambda  A \mathit{b}^{16}+524288 c^8 \pi ^{10} \mathcal{Q}^2 \mathcal{S}B \mathit{b}^{12}-\\&196608 c^7 \pi ^7 \mathcal{S}^2 \left(4 \pi ^6 \mathit{b}^6 I \mathcal{Q}^4+3 \pi ^2 \mathit{b}^2 \mathcal{S}^4 D\right) \mathit{b}^8+24576 c^6 \pi ^6 \mathcal{S}^3 \left(4 \pi ^6 \mathit{b}^6 E \mathcal{Q}^4-9 \pi ^2 \mathit{b}^2 \mathcal{S}^4 F\right) \mathit{b}^6\\&-9216 c^5 \pi ^5 \mathcal{S}^4 \left(16 \pi ^6 \mathit{b}^6 H \mathcal{Q}^4+6 \pi ^2 \mathit{b}^2 \mathcal{S}^4 J \mathcal{Q}^2+3 \mathcal{S}^6 K\right) \mathit{b}^4\\&+2304 c^4 \pi ^4 \mathcal{S}^5 \left(4 \pi ^6 \mathit{b}^6 L \mathcal{Q}^4+3 \pi ^2 \mathit{b}^2 \mathcal{S}^4 M \mathcal{Q}^2-9 \mathcal{S}^6 N\right) \mathit{b}^2-6075 \mathcal{S}^{10}O+\\&8100 c \pi  \mathcal{S}^8 \left(4 \pi ^6 \mathit{b}^2 P \mathcal{Q}^4-9 \pi ^2 \mathcal{S}^4 R\right)-8640 c^3 \pi ^3 \mathcal{S}^6 \left(4 \pi ^6 \mathit{b}^6 T \mathcal{Q}^2-18 \mathcal{S}^6 U\right)\\&-2160 c^2 \pi ^2 \mathcal{S}^7 \left(4 \pi ^6 \mathit{b}^4 V \mathcal{Q}^2+54 \mathcal{S}^6 W\right)\end{aligned}\right)}{X}
	\label{ap:11}
\end{equation}
where A, B, I, D, E, F, H, J, K, L, M, N, O, P, R, T, U, V, W, X are given below
\begin{equation}
	\begin{split}
			&A=\left(\begin{aligned}&\left(225 \mathcal{S}^3 \lambda ^3-660 \mathcal{S}^2 \lambda ^2+560 \mathcal{S} \lambda -192\right) \mathit{b}^{20}+8388608 c^9 \pi ^{13} \mathcal{Q}^4 \left(2 \pi ^2 \mathit{b}^2 \left(675 \mathcal{S}^5 \lambda ^5-11160 \mathcal{S}^4 \lambda ^4\right.\right.\\&\left.\left.+35952 \mathcal{S}^3 \lambda ^3-40896 \mathcal{S}^2 \lambda ^2+20992 \mathcal{S} \lambda -3072\right) \mathcal{Q}^2+3 \mathcal{S}^2 \left(-2025 \mathcal{S}^5 \lambda ^5+24480 \mathcal{S}^4 \lambda ^4-77664 \mathcal{S}^3 \lambda ^3\right/\right.\\&\left.\left.+125568 \mathcal{S}^2 \lambda ^2-125696 \mathcal{S} \lambda +53248\right)\right)\end{aligned}\right)\\
		&B= \left(\begin{aligned}&4 \pi ^4 \mathit{b}^4 \left(3375 \mathcal{S}^5 \lambda ^5+13500 \mathcal{S}^4 \lambda ^4-163872 \mathcal{S}^3 \lambda ^3+320640 \mathcal{S}^2 \lambda ^2-223488 \mathcal{S} \lambda +60416\right) \mathcal{Q}^4\\&-3 \pi ^2 \mathit{b}^2 \mathcal{S}^2 \left(-141075 \mathcal{S}^5 \lambda ^5+846180 \mathcal{S}^4 \lambda ^4-1858848 \mathcal{S}^3 \lambda ^3+2254464 \mathcal{S}^2 \lambda ^2-2215680 \mathcal{S} \lambda\right.\\&\left. +6 \mathit{b}^2 \left(55125 \mathcal{S}^5 \lambda ^5-280500 \mathcal{S}^4 \lambda ^4+457120 \mathcal{S}^3 \lambda ^3-260736 \mathcal{S}^2 \lambda ^2+82176 \mathcal{S} \lambda -41984\right)+1131520\right) \mathcal{Q}^2\\&-36 \mathcal{S}^4 \left(2025 \mathcal{S}^5 \lambda ^5-6120 \mathcal{S}^4 \lambda ^4-4584 \mathcal{S}^3 \lambda ^3+28128 \mathcal{S}^2 \lambda ^2-28544 \mathcal{S} \lambda +11776\right)\end{aligned}\right)\\
		&I=\left(\begin{aligned}&\left(25425 \mathcal{S}^5 \lambda ^5-139740 \mathcal{S}^4 \lambda ^4+131328 \mathcal{S}^3 \lambda ^3+243456 \mathcal{S}^2 \lambda ^2-384256 \mathcal{S} \lambda +154624\right) \mathcal{Q}^6\\&-3 \pi ^4 \mathit{b}^4 \mathcal{S}^2 \left(-388575 \mathcal{S}^5 \lambda ^5+2521500 \mathcal{S}^4 \lambda ^4-5675616 \mathcal{S}^3 \lambda ^3+5383808 \mathcal{S}^2 \lambda ^2-3149568 \mathcal{S} \lambda\right.\\&\left. +2 \mathit{b}^2 \left(624375 \mathcal{S}^5 \lambda ^5-3531900 \mathcal{S}^4 \lambda ^4+6351840 \mathcal{S}^3 \lambda ^3-3675776 \mathcal{S}^2 \lambda ^2+399104 \mathcal{S} \lambda -175104\right)+1715200\right)\end{aligned}\right)
	\end{split}
\end{equation}
\begin{equation}
	\begin{split}
			&D=\left(\begin{aligned}&\left(-33075 \mathcal{S}^5 \lambda ^5+93060 \mathcal{S}^4 \lambda ^4+214176 \mathcal{S}^3 \lambda ^3-840576 \mathcal{S}^2 \lambda ^2+890112 \mathcal{S} \lambda +6 \mathit{b}^2 \left(46125 \mathcal{S}^5 \lambda ^5-60300 \mathcal{S}^4 \lambda ^4\right.\right.\\&\left.\left.-192320 \mathcal{S}^3 \lambda ^3+402944 \mathcal{S}^2 \lambda ^2-143616 \mathcal{S} \lambda -13312\right)-416768\right) \mathcal{Q}^2+18 \mathcal{S}^6 \left(2025 \mathcal{S}^5 \lambda ^5-2520 \mathcal{S}^4 \lambda ^4\right.\\&\left.-4944 \mathcal{S}^3 \lambda ^3+4672 \mathcal{S}^2 \lambda ^2+8192 \mathcal{S} \lambda -7168\right)\end{aligned}\right)\\
		&E=\left(\begin{aligned}&\left(146025 \mathcal{S}^5 \lambda ^5-1160640 \mathcal{S}^4 \lambda ^4+2843376 \mathcal{S}^3 \lambda ^3-1972032 \mathcal{S}^2 \lambda ^2-390144 \mathcal{S} \lambda +357376\right) \mathcal{Q}^6\\&-3 \pi ^4 \mathit{b}^4 \mathcal{S}^2 \left(-1149075 \mathcal{S}^5 \lambda ^5+10041540 \mathcal{S}^4 \lambda ^4-27167904 \mathcal{S}^3 \lambda ^3+26592384 \mathcal{S}^2 \lambda ^2-8898304 \mathcal{S} \lambda\right.\\&\left. +2 \mathit{b}^2 \left(3700125 \mathcal{S}^5 \lambda ^5-26058300 \mathcal{S}^4 \lambda ^4+58201440 \mathcal{S}^3 \lambda ^3-41997696 \mathcal{S}^2 \lambda ^2-945920 \mathcal{S} \lambda +4719616\right)+3621888\right)\end{aligned}\right)\\
		&F=\left(\begin{aligned}&\left(20475 \mathcal{S}^5 \lambda ^5-8940 \mathcal{S}^4 \lambda ^4-652992 \mathcal{S}^3 \lambda ^3+1789952 \mathcal{S}^2 \lambda ^2-1572608 \mathcal{S} \lambda +24 \mathit{b}^4 \left(5625 \mathcal{S}^5 \lambda ^5-3500 \mathcal{S}^4 \lambda ^4\right.\right.\\&\left.\left.-32000 \mathcal{S}^3 \lambda ^3+103680 \mathcal{S}^2 \lambda ^2-35072 \mathcal{S} \lambda -17408\right)+\mathit{b}^2 \left(-249750 \mathcal{S}^5 \lambda ^5+305400 \mathcal{S}^4 \lambda ^4+3206400 \mathcal{S}^3 \lambda ^3\right.\right.\\&\left.\left.-7121408 \mathcal{S}^2 \lambda ^2+2864640 \mathcal{S} \lambda +518144\right)+519168\right) \mathcal{Q}^2-72 \mathcal{S}^6 \left(3 \mathit{b}^2 \left(7875 \mathcal{S}^5 \lambda ^5-3900 \mathcal{S}^4 \lambda ^4-11120 \mathcal{S}^3 \lambda ^3\right.\right.\\&\left.\left.-1088 \mathcal{S}^2 \lambda ^2+13056 \mathcal{S} \lambda -3072\right)-5 \left(405 \mathcal{S}^5 \lambda ^5-945 \mathcal{S}^4 \lambda ^4-540 \mathcal{S}^3 \lambda ^3+1680 \mathcal{S}^2 \lambda ^2+1856 \mathcal{S} \lambda -2560\right)\right)\end{aligned}\right)\\
		&H=\left(\begin{aligned}&\left(24375 \mathcal{S}^5 \lambda ^5-275265 \mathcal{S}^4 \lambda ^4+904812 \mathcal{S}^3 \lambda ^3-881392 \mathcal{S}^2 \lambda ^2-170432 \mathcal{S} \lambda +308224\right) \mathcal{Q}^6+\\&\pi ^4 \mathit{b}^4 \mathcal{S}^2 \left(1693125 \mathcal{S}^5 \lambda ^5-24046380 \mathcal{S}^4 \lambda ^4+86945568 \mathcal{S}^3 \lambda ^3-104446848 \mathcal{S}^2 \lambda ^2+24559872 \mathcal{S} \lambda\right.\\&\left. -6 \mathit{b}^2 \left(3363375 \mathcal{S}^5 \lambda ^5-31655100 \mathcal{S}^4 \lambda ^4+90183520 \mathcal{S}^3 \lambda ^3-83978368 \mathcal{S}^2 \lambda ^2+1259264 \mathcal{S} \lambda +15758336\right)+7685120\right)\end{aligned}\right)\\
		&J=\left(\begin{aligned}&6 \left(196875 \mathcal{S}^5 \lambda ^5-544500 \mathcal{S}^4 \lambda ^4+635200 \mathcal{S}^3 \lambda ^3-1520640 \mathcal{S}^2 \lambda ^2+760064 \mathcal{S} \lambda +349184\right) \mathit{b}^4\\&-\left(248625 \mathcal{S}^5 \lambda ^5-1243500 \mathcal{S}^4 \lambda ^4+11029440 \mathcal{S}^3 \lambda ^3-23867904 \mathcal{S}^2 \lambda ^2+10865408 \mathcal{S} \lambda +4287488\right) \mathit{b}^2+\\&10 \left(1755 \mathcal{S}^5 \lambda ^5-18576 \mathcal{S}^4 \lambda ^4+128256 \mathcal{S}^3 \lambda ^3-292224 \mathcal{S}^2 \lambda ^2+235776 \mathcal{S} \lambda -71680\right)\end{aligned}\right)\\
		&K=\left(\begin{aligned}&36 \left(73125 \mathcal{S}^5 \lambda ^5-42500 \mathcal{S}^4 \lambda ^4-86400 \mathcal{S}^3 \lambda ^3-6400 \mathcal{S}^2 \lambda ^2+71424 \mathcal{S} \lambda +9216\right) \mathit{b}^4-\\&240 \left(3375 \mathcal{S}^5 \lambda ^5-5085 \mathcal{S}^4 \lambda ^4-4140 \mathcal{S}^3 \lambda ^3+1072 \mathcal{S}^2 \lambda ^2+8512 \mathcal{S} \lambda -2048\right) \mathit{b}^2+\\&5 \left(6885 \mathcal{S}^5 \lambda ^5-39852 \mathcal{S}^4 \lambda ^4+15840 \mathcal{S}^3 \lambda ^3+57216 \mathcal{S}^2 \lambda ^2+42240 \mathcal{S} \lambda -109568\right)\end{aligned}\right)\\
		&L=\left(\begin{aligned}&\left(46725 \mathcal{S}^5 \lambda ^5-973680 \mathcal{S}^4 \lambda ^4+4501488 \mathcal{S}^3 \lambda ^3-6076992 \mathcal{S}^2 \lambda ^2-1024000 \mathcal{S} \lambda +3838976\right) \mathcal{Q}^6\\&-3 \pi ^4 \mathit{b}^4 \mathcal{S}^2 \left(4 \mathit{b}^2 \left(1225875 \mathcal{S}^5 \lambda ^5-17811450 \mathcal{S}^4 \lambda ^4+68946000 \mathcal{S}^3 \lambda ^3-86702144 \mathcal{S}^2 \lambda ^2+11210496 \mathcal{S} \lambda\right.\right.\\&\left.\left. +22794752\right)-5 \left(19845 \mathcal{S}^5 \lambda ^5-854316 \mathcal{S}^4 \lambda ^4+4875744 \mathcal{S}^3 \lambda ^3-7801216 \mathcal{S}^2 \lambda ^2+2357504 \mathcal{S} \lambda +1184768\right)\right)\end{aligned}\right)\\
		&M=\left(\begin{aligned}&6 \left(1888125 \mathcal{S}^5 \lambda ^5-7078500 \mathcal{S}^4 \lambda ^4+8421600 \mathcal{S}^3 \lambda ^3-7141760 \mathcal{S}^2 \lambda ^2+4999424 \mathcal{S} \lambda +2649088\right) \mathit{b}^4-\\&15 \left(131175 \mathcal{S}^5 \lambda ^5-721020 \mathcal{S}^4 \lambda ^4+2564096 \mathcal{S}^3 \lambda ^3-5375232 \mathcal{S}^2 \lambda ^2+2965760 \mathcal{S} \lambda +1256448\right) \mathit{b}^2+\\&5 \left(11205 \mathcal{S}^5 \lambda ^5-119268 \mathcal{S}^4 \lambda ^4+719808 \mathcal{S}^3 \lambda ^3-1605120 \mathcal{S}^2 \lambda ^2+1097472 \mathcal{S} \lambda -97280\right)\end{aligned}\right)\\
		&N=\left(\begin{aligned}&36 \left(65625 \mathcal{S}^5 \lambda ^5-22500 \mathcal{S}^4 \lambda ^4-76000 \mathcal{S}^3 \lambda ^3-3200 \mathcal{S}^2 \lambda ^2+42240 \mathcal{S} \lambda +11264\right) \mathit{b}^6-\\&120 \left(16875 \mathcal{S}^5 \lambda ^5-25650 \mathcal{S}^4 \lambda ^4-18040 \mathcal{S}^3 \lambda ^3+8480 \mathcal{S}^2 \lambda ^2+25984 \mathcal{S} \lambda +3072\right) \mathit{b}^4\\&+5 \left(66825 \mathcal{S}^5 \lambda ^5-289260 \mathcal{S}^4 \lambda ^4-42048 \mathcal{S}^3 \lambda ^3+61440 \mathcal{S}^2 \lambda ^2+356096 \mathcal{S} \lambda -74752\right) \mathit{b}^2\\&-100 \left(81 \mathcal{S}^5 \lambda ^5-594 \mathcal{S}^4 \lambda ^4+792 \mathcal{S}^3 \lambda ^3+480 \mathcal{S}^2 \lambda ^2+128 \mathcal{S} \lambda -2048\right)\end{aligned}\right)\\
		&O= \left(\begin{aligned}&-16 \pi ^6 \lambda  \left(\mathcal{S}^3 \lambda ^3-4 \mathcal{S}^2 \lambda ^2-80 \mathcal{S} \lambda +64\right) \mathcal{Q}^6+3 \pi ^4 \mathcal{S} \left(75 \mathcal{S}^5 \lambda ^5+836 \mathcal{S}^4 \lambda ^4-32 \mathcal{S}^3 \lambda ^3-29056 \mathcal{S}^2 \lambda ^2\right.\\&\left.+32512 \mathcal{S} \lambda +13312\right) \mathcal{Q}^4+36 \pi ^2 \mathcal{S}^4 \lambda  \left(75 \mathcal{S}^4 \lambda ^4+1040 \mathcal{S}^3 \lambda ^3-7072 \mathcal{S}^2 \lambda ^2+3072 \mathcal{S} \lambda -256\right) \mathcal{Q}^2+\\&27 \mathcal{S}^5 \left(2625 \mathcal{S}^5 \lambda ^5-14900 \mathcal{S}^4 \lambda ^4+16480 \mathcal{S}^3 \lambda ^3+35968 \mathcal{S}^2 \lambda ^2+20736 \mathcal{S} \lambda +3072\right)\end{aligned}\right)\\
		&P=\left(\begin{aligned}&\left(3 \mathcal{S}^5 \lambda ^5-72 \mathcal{S}^4 \lambda ^4-720 \mathcal{S}^3 \lambda ^3+5440 \mathcal{S}^2 \lambda ^2-4608 \mathcal{S} \lambda -3072\right) \mathcal{Q}^6-3 \pi ^4 \mathcal{S}^2 \left(\left(615 \mathcal{S}^5 \lambda ^5-6124 \mathcal{S}^4 \lambda ^4\right.\right.\\&\left.\left.-89248 \mathcal{S}^3 \lambda ^3+459648 \mathcal{S}^2 \lambda ^2-340224 \mathcal{S} \lambda -187392\right) \mathit{b}^2+4 \left(3 \mathcal{S}^5 \lambda ^5+70 \mathcal{S}^4 \lambda ^4+64 \mathcal{S}^3 \lambda ^3\right.\right.\\&\left.\left.-2688 \mathcal{S}^2 \lambda ^2+1024 \mathcal{S} \lambda -512\right)\right)\end{aligned}\right)
	\end{split}
\end{equation}
\begin{equation}
	\begin{split}
			&R=\left(\begin{aligned}&\left(45 \mathcal{S}^5 \lambda ^5+1988 \mathcal{S}^4 \lambda ^4-12480 \mathcal{S}^3 \lambda ^3+18432 \mathcal{S}^2 \lambda ^2-39168 \mathcal{S} \lambda +\mathit{b}^2 \left(3075 \mathcal{S}^5 \lambda ^5-71540 \mathcal{S}^4 \lambda ^4\right.\right.\\&\left.\left.+218176 \mathcal{S}^3 \lambda ^3-135168 \mathcal{S}^2 \lambda ^2-20224 \mathcal{S} \lambda +5120\right)-9216\right) \mathcal{Q}^2+27 \mathcal{S}^6 \left(-2325 \mathcal{S}^5 \lambda ^5+\right.\\&\left.13780 \mathcal{S}^4 \lambda ^4-24128 \mathcal{S}^3 \lambda ^3-24576 \mathcal{S}^2 \lambda ^2-17152 \mathcal{S} \lambda +4 \mathit{b}^2 \left(3375 \mathcal{S}^5 \lambda ^5-13350 \mathcal{S}^4 \lambda ^4\right.\right.\\&\left.\left.-1320 \mathcal{S}^3 \lambda ^3+14112 \mathcal{S}^2 \lambda ^2+11392 \mathcal{S} \lambda +2048\right)-1024\right)\end{aligned}\right)\\
		&T=\left(\begin{aligned}&\left(225 \mathcal{S}^5 \lambda ^5-26476 \mathcal{S}^4 \lambda ^4+205504 \mathcal{S}^3 \lambda ^3-412672 \mathcal{S}^2 \lambda ^2+40704 \mathcal{S} \lambda +310272\right) \mathcal{Q}^6-\\&\pi ^4 \mathit{b}^4 \mathcal{S}^2 \left(15525 \mathcal{S}^5 \lambda ^5+141612 \mathcal{S}^4 \lambda ^4-2393568 \mathcal{S}^3 \lambda ^3+5581440 \mathcal{S}^2 \lambda ^2-2288384 \mathcal{S} \lambda\right.\\&\left. +3 \mathit{b}^2 \left(59925 \mathcal{S}^5 \lambda ^5-2479940 \mathcal{S}^4 \lambda ^4+14938272 \mathcal{S}^3 \lambda ^3-26167168 \mathcal{S}^2 \lambda ^2+7443712 \mathcal{S} \lambda +8178688\right)\right.\\&\left.-1025024\right) \mathcal{Q}^4+3 \pi ^2 \mathit{b}^2 \mathcal{S}^4 \left(6 \left(192375 \mathcal{S}^5 \lambda ^5-1060300 \mathcal{S}^4 \lambda ^4+1476080 \mathcal{S}^3 \lambda ^3-666048 \mathcal{S}^2 \lambda ^2\right.\right.\\&\left.\left.+420864 \mathcal{S} \lambda +249856\right) \mathit{b}^4-\left(125325 \mathcal{S}^5 \lambda ^5-1090380 \mathcal{S}^4 \lambda ^4+3465792 \mathcal{S}^3 \lambda ^3-7205888 \mathcal{S}^2 \lambda ^2\right.\right.\\&\left.\left.+4384512 \mathcal{S} \lambda +2124800\right) \mathit{b}^2+5 \left(189 \mathcal{S}^5 \lambda ^5-3420 \mathcal{S}^4 \lambda ^4+20640 \mathcal{S}^3 \lambda ^3-59264 \mathcal{S}^2 \lambda ^2+42240 \mathcal{S} \lambda -15360\right)\right)\end{aligned}\right)\\
		&U=\left(\begin{aligned}&12 \left(16875 \mathcal{S}^5 \lambda ^5-17250 \mathcal{S}^4 \lambda ^4-22200 \mathcal{S}^3 \lambda ^3+6880 \mathcal{S}^2 \lambda ^2+17280 \mathcal{S} \lambda +4096\right) \mathit{b}^6-\\&4 \left(23625 \mathcal{S}^5 \lambda ^5-92550 \mathcal{S}^4 \lambda ^4-12960 \mathcal{S}^3 \lambda ^3+41152 \mathcal{S}^2 \lambda ^2+68480 \mathcal{S} \lambda +7168\right) \mathit{b}^4+\\&10 \left(945 \mathcal{S}^5 \lambda ^5-4986 \mathcal{S}^4 \lambda ^4+2904 \mathcal{S}^3 \lambda ^3+2592 \mathcal{S}^2 \lambda ^2+7808 \mathcal{S} \lambda -1024\right) \mathit{b}^2+\\&5 \left(-27 \mathcal{S}^5 \lambda ^5+216 \mathcal{S}^4 \lambda ^4-720 \mathcal{S}^3 \lambda ^3+64 \mathcal{S}^2 \lambda ^2+1024\right)\end{aligned}\right)\\
		&V=\left(\begin{aligned}&\left(405 \mathcal{S}^5 \lambda ^5+1620 \mathcal{S}^4 \lambda ^4-79264 \mathcal{S}^3 \lambda ^3+256640 \mathcal{S}^2 \lambda ^2-116480 \mathcal{S} \lambda -199680\right) \mathcal{Q}^6-\\&3 \pi ^4 \mathit{b}^2 \mathcal{S}^2 \left(\left(44625 \mathcal{S}^5 \lambda ^5+414900 \mathcal{S}^4 \lambda ^4-6968864 \mathcal{S}^3 \lambda ^3+18363008 \mathcal{S}^2 \lambda ^2-8641280 \mathcal{S} \lambda -6614016\right) \mathit{b}^2\right.\\&\left.+5 \left(-207 \mathcal{S}^5 \lambda ^5+1860 \mathcal{S}^4 \lambda ^4+32480 \mathcal{S}^3 \lambda ^3-145792 \mathcal{S}^2 \lambda ^2+72960 \mathcal{S} \lambda +9216\right)\right) \mathcal{Q}^4-\\&9 \pi ^2 \mathcal{S}^4 \left(\left(288375 \mathcal{S}^5 \lambda ^5-2646300 \mathcal{S}^4 \lambda ^4+5337440 \mathcal{S}^3 \lambda ^3-2675072 \mathcal{S}^2 \lambda ^2-399616 \mathcal{S} \lambda +287744\right) \mathit{b}^4-\right.\\&\left.5 \left(1845 \mathcal{S}^5 \lambda ^5-30756 \mathcal{S}^4 \lambda ^4+110464 \mathcal{S}^3 \lambda ^3-250112 \mathcal{S}^2 \lambda ^2+248576 \mathcal{S} \lambda +87040\right) \mathit{b}^2-\right.\\&\left.10 \left(9 \mathcal{S}^5 \lambda ^5+84 \mathcal{S}^4 \lambda ^4-976 \mathcal{S}^3 \lambda ^3+3264 \mathcal{S}^2 \lambda ^2-1536 \mathcal{S} \lambda +2048\right)\right)\end{aligned}\right)\\
		&W=\left(\begin{aligned}&\left(\left(106875 \mathcal{S}^5 \lambda ^5-315000 \mathcal{S}^4 \lambda ^4-198000 \mathcal{S}^3 \lambda ^3+164160 \mathcal{S}^2 \lambda ^2+217088 \mathcal{S} \lambda +46080\right) \mathit{b}^4\right.\\&\left.-10 \left(2925 \mathcal{S}^5 \lambda ^5-13830 \mathcal{S}^4 \lambda ^4+6392 \mathcal{S}^3 \lambda ^3+11872 \mathcal{S}^2 \lambda ^2+11904 \mathcal{S} \lambda +1024\right) \mathit{b}^2\right.\\&\left.+40 \mathcal{S} \lambda  \left(45 \mathcal{S}^4 \lambda ^4-279 \mathcal{S}^3 \lambda ^3+592 \mathcal{S}^2 \lambda ^2+416 \mathcal{S} \lambda +832\right)\right)\end{aligned}\right)\\
		&X=\begin{aligned}&\left(32 c^2 \pi ^2 (5 \mathcal{S} \lambda -4) \mathit{b}^4+20 c \pi  \mathcal{S} (4-3 \mathcal{S} \lambda ) \mathit{b}^2+15 \mathcal{S}^2 (\mathcal{S} \lambda -4)\right)^2\\& \left(32 c^2 \pi ^4 \mathcal{Q}^2 (3 \mathcal{S} \lambda -4) \mathit{b}^4-12 c \pi ^3 \mathcal{Q}^2 \mathcal{S} (\mathcal{S} \lambda -4) \mathit{b}^2-3 \pi ^2 \mathcal{Q}^2 \mathcal{S}^2 (\mathcal{S} \lambda +4)+12 c \pi  \mathcal{S}^3 (3 \mathcal{S} \lambda +4)+9 \mathcal{S}^4 (5 \mathcal{S} \lambda +4)\right)^3\\& \left(32 c^2 \pi ^4 \mathcal{Q}^2 (28-15 \mathcal{S} \lambda ) \mathit{b}^4+3 \pi ^2 \mathcal{Q}^2 \mathcal{S}^2 (\mathcal{S} \lambda +12)+9 \mathcal{S}^4 (15 \mathcal{S} \lambda +4)+12 c \left(\pi  (3 \mathcal{S} \lambda -4) \mathcal{S}^3+\pi ^3 \mathit{b}^2 \mathcal{Q}^2 (3 \mathcal{S} \lambda -20) \mathcal{S}\right)\right)^2\end{aligned}
	\end{split}
\end{equation}
The GTD scalar for the dual CFT for NED AdS black hole for the Kaniadakis entropy is given as
\begin{equation}
R_{GTD}=\frac{6912 c \pi ^2 \mathcal{S}^6 \mathcal{V} \left(\begin{aligned}&2147483648 c^{10} \pi ^{16} \mathcal{Q}^6 \mathcal{S} \kappa ^2 A \mathit{b}^{16}-1572864 c^8 \pi ^{10} \mathcal{Q}^2 \mathcal{S} B \mathit{b}^{12}\\&+589824 c^7 \pi ^7 \mathcal{S}^2 D \mathit{b}^8-221184 c^6 \pi ^6 \mathcal{S}^3 E \mathit{b}^6-9216 c^5 \pi ^5 \mathcal{S}^4 \left(16 \pi ^6 \mathit{b}^6 F \mathcal{Q}^4-18 \pi ^2 \mathit{b}^2 \mathcal{S}^4 H\right) \mathit{b}^4\\&+6912 c^4 \pi ^4 \mathcal{S}^5 \left(4 \pi ^6 \mathit{b}^6 I \mathcal{Q}^2+9 \mathcal{S}^6J\right) \mathit{b}^2+54675 \mathcal{S}^{11} K+72900 c \pi  \mathcal{S}^8 L\\&-25920 c^3 \pi ^3 \mathcal{S}^6 \left(4 \pi ^6 \mathit{b}^6 M \mathcal{Q}^2+6 \mathcal{S}^6 N\right)+19440 c^2 \pi ^2 \mathcal{S}^7 \left(4 \pi ^6 \mathit{b}^4 O\mathcal{Q}^2+2 \mathcal{S}^6P\right)\end{aligned}\right)}{R}
\label{ap:12}
\end{equation}
where A, B, D, E, F, , H, I, J, K, L, M, N, O, P, R are given below
\begin{equation}
	\begin{split}
& A=\left(\begin{aligned}&\left(25 \mathcal{S}^6 \kappa ^6+516 \mathcal{S}^4 \kappa ^4-1872 \mathcal{S}^2 \kappa ^2+1728\right) \mathit{b}^{20}+16777216 c^9 \pi ^{13} \mathcal{Q}^4 \left(\pi ^2 \mathit{b}^2 \left(25 \mathcal{S}^{10} \kappa ^{10}+2100 \mathcal{S}^8 \kappa ^8\right.\right.\\&\left.\left.+38880 \mathcal{S}^6 \kappa ^6+897408 \mathcal{S}^4 \kappa ^4-145152 \mathcal{S}^2 \kappa ^2-248832\right) \mathcal{Q}^2+9 \mathcal{S}^2 \left(125 \mathcal{S}^{10} \kappa ^{10}+8100 \mathcal{S}^8 \kappa ^8\right.\right.\\&\left.\left.+146320 \mathcal{S}^6 \kappa ^6+333120 \mathcal{S}^4 \kappa ^4+142848 \mathcal{S}^2 \kappa ^2+718848\right)\right)\end{aligned}\right)\\
	&B=\left(\begin{aligned}&4 \pi ^4 \mathit{b}^4 \left(325 \mathcal{S}^{10} \kappa ^{10}+29820 \mathcal{S}^8 \kappa ^8+657888 \mathcal{S}^6 \kappa ^6+4149888 \mathcal{S}^4 \kappa ^4+2518272 \mathcal{S}^2 \kappa ^2-1631232\right) \mathcal{Q}^4\\&+\pi ^2 \mathit{b}^2 \mathcal{S}^2 \left(11375 \mathcal{S}^{10} \kappa ^{10}+1002900 \mathcal{S}^8 \kappa ^8+21592800 \mathcal{S}^6 \kappa ^6+59393664 \mathcal{S}^4 \kappa ^4+14549760 \mathcal{S}^2 \kappa ^2\right.\\&\left.-2 \mathit{b}^2 \left(11725 \mathcal{S}^{10} \kappa ^{10}+657180 \mathcal{S}^8 \kappa ^8+11183904 \mathcal{S}^6 \kappa ^6+25370496 \mathcal{S}^4 \kappa ^4+1347840 \mathcal{S}^2 \kappa ^2\right.\right.\\&\left.\left.+10202112\right)+91653120\right) \mathcal{Q}^2+18 \mathcal{S}^4 \left(625 \mathcal{S}^{10} \kappa ^{10}+33500 \mathcal{S}^8 \kappa ^8+107200 \mathcal{S}^6 \kappa ^6+456960 \mathcal{S}^4 \kappa ^4\right.\\&\left.+744192 \mathcal{S}^2 \kappa ^2+635904\right)\end{aligned}\right)\\
	&D=\left(\begin{aligned}&4 \pi ^6 \mathit{b}^6 \left(775 \mathcal{S}^{10} \kappa ^{10}+80460 \mathcal{S}^8 \kappa ^8+2137632 \mathcal{S}^6 \kappa ^6+12573312 \mathcal{S}^4 \kappa ^4+11904768 \mathcal{S}^2 \kappa ^2-4174848\right) \mathcal{Q}^6\\&+\pi ^4 \mathit{b}^4 \mathcal{S}^2 \left(3 \left(6625 \mathcal{S}^{10} \kappa ^{10}+694100 \mathcal{S}^8 \kappa ^8+17063840 \mathcal{S}^6 \kappa ^6+66961536 \mathcal{S}^4 \kappa ^4+23291136 \mathcal{S}^2 \kappa ^2\right.\right.\\&\left.\left.+46310400\right)-2 \mathit{b}^2 \left(23975 \mathcal{S}^{10} \kappa ^{10}+2128620 \mathcal{S}^8 \kappa ^8+44788896 \mathcal{S}^6 \kappa ^6+126117504 \mathcal{S}^4 \kappa ^4\right.\right.\\&\left.\left.+13803264 \mathcal{S}^2 \kappa ^2+14183424\right)\right) \mathcal{Q}^4+\pi ^2 \mathit{b}^2 \mathcal{S}^4 \left(49375 \mathcal{S}^{10} \kappa ^{10}+2089500 \mathcal{S}^8 \kappa ^8+8762400 \mathcal{S}^6 \kappa ^6\right.\\&\left.+26847360 \mathcal{S}^4 \kappa ^4+42225408 \mathcal{S}^2 \kappa ^2-2 \mathit{b}^2 \left(58625 \mathcal{S}^{10} \kappa ^{10}+2837700 \mathcal{S}^8 \kappa ^8+7308000 \mathcal{S}^6 \kappa ^6\right.\right.\\&\left.\left.+22232448 \mathcal{S}^4 \kappa ^4+35520768 \mathcal{S}^2 \kappa ^2-3234816\right)+33758208\right) \mathcal{Q}^2+6 \mathcal{S}^6 \left(9375 \mathcal{S}^{10} \kappa ^{10}+47500 \mathcal{S}^8 \kappa ^8\right.\\&\left.+108000 \mathcal{S}^6 \kappa ^6-28800 \mathcal{S}^4 \kappa ^4+449280 \mathcal{S}^2 \kappa ^2+580608\right)\end{aligned}\right)\\
	&E=\left(\begin{aligned}&12 \pi ^6 \mathit{b}^6 \left(275 \mathcal{S}^{10} \kappa ^{10}+38180 \mathcal{S}^8 \kappa ^8+1307296 \mathcal{S}^6 \kappa ^6+8815488 \mathcal{S}^4 \kappa ^4+9962240 \mathcal{S}^2 \kappa ^2-1072128\right) \mathcal{Q}^6\\&+\pi ^4 \mathit{b}^4 \mathcal{S}^2 \left(10125 \mathcal{S}^{10} \kappa ^{10}+1996300 \mathcal{S}^8 \kappa ^8+70974560 \mathcal{S}^6 \kappa ^6+368767104 \mathcal{S}^4 \kappa ^4+192478464 \mathcal{S}^2 \kappa ^2\right.\\&\left.-2 \mathit{b}^2 \left(36575 \mathcal{S}^{10} \kappa ^{10}+4271780 \mathcal{S}^8 \kappa ^8+103819744 \mathcal{S}^6 \kappa ^6+417627264 \mathcal{S}^4 \kappa ^4+144458496 \mathcal{S}^2 \kappa ^2-127429632\right)\right.\\&\left.+97790976\right) \mathcal{Q}^4+\pi ^2 \mathit{b}^2 \mathcal{S}^4 \left(105625 \mathcal{S}^{10} \kappa ^{10}+3907500 \mathcal{S}^8 \kappa ^8+26536800 \mathcal{S}^6 \kappa ^6+69033600 \mathcal{S}^4 \kappa ^4\right.\\&\left.+111165696 \mathcal{S}^2 \kappa ^2+8 \mathit{b}^4 \left(42875 \mathcal{S}^{10} \kappa ^{10}+1717380 \mathcal{S}^8 \kappa ^8+3130848 \mathcal{S}^6 \kappa ^6+3687552 \mathcal{S}^4 \kappa ^4\right.\right.\\&\left.\left.+10098432 \mathcal{S}^2 \kappa ^2-4230144\right)-2 \mathit{b}^2 \left(216125 \mathcal{S}^{10} \kappa ^{10}+9450300 \mathcal{S}^8 \kappa ^8+32879520 \mathcal{S}^6 \kappa ^6\right.\right.\\&\left.\left.+71561088 \mathcal{S}^4 \kappa ^4+122390784 \mathcal{S}^2 \kappa ^2-20984832\right)+42052608\right) \mathcal{Q}^2-4 \mathcal{S}^6 \left(2 \mathit{b}^2 \left(83125 \mathcal{S}^{10} \kappa ^{10}\right.\right.\\&\left.\left.+346500 \mathcal{S}^8 \kappa ^8+756000 \mathcal{S}^6 \kappa ^6-224640 \mathcal{S}^4 \kappa ^4+2757888 \mathcal{S}^2 \kappa ^2+746496\right)-15 \left(1875 \mathcal{S}^{10} \kappa ^{10}\right.\right.\\&\left.\left.+11500 \mathcal{S}^8 \kappa ^8+36000 \mathcal{S}^6 \kappa ^6+1920 \mathcal{S}^4 \kappa ^4+99072 \mathcal{S}^2 \kappa ^2+138240\right)\right)\end{aligned}\right)\\
	&F=\left(\begin{aligned}&\left(125 \mathcal{S}^{10} \kappa ^{10}-155040 \mathcal{S}^8 \kappa ^8-9940176 \mathcal{S}^6 \kappa ^6-82216512 \mathcal{S}^4 \kappa ^4-110343168 \mathcal{S}^2 \kappa ^2+24966144\right) \mathcal{Q}^6\\&+3 \pi ^4 \mathit{b}^4 \mathcal{S}^2 \left(-35625 \mathcal{S}^{10} \kappa ^{10}-3335300 \mathcal{S}^8 \kappa ^8-205961760 \mathcal{S}^6 \kappa ^6-1380249216 \mathcal{S}^4 \kappa ^4-1221297408 \mathcal{S}^2 \kappa ^2\right.\\&\left.+\mathit{b}^2 \left(87150 \mathcal{S}^{10} \kappa ^{10}+24198200 \mathcal{S}^8 \kappa ^8+846703296 \mathcal{S}^6 \kappa ^6+4498907904 \mathcal{S}^4 \kappa ^4+2699896320 \mathcal{S}^2 \kappa ^2\right.\right.\\&\left.\left.-2552850432\right)+207498240\right)\end{aligned}\right)\\
	&H=\left(\begin{aligned}&\left(6 \left(84525 \mathcal{S}^{10} \kappa ^{10}+3700900 \mathcal{S}^8 \kappa ^8+10162144 \mathcal{S}^6 \kappa ^6+7202688 \mathcal{S}^4 \kappa ^4+21300480 \mathcal{S}^2 \kappa ^2-9427968\right) \mathit{b}^4\right.\\&\left.-\left(445375 \mathcal{S}^{10} \kappa ^{10}+17414700 \mathcal{S}^8 \kappa ^8+102854240 \mathcal{S}^6 \kappa ^6+186039168 \mathcal{S}^4 \kappa ^4+319322880 \mathcal{S}^2 \kappa ^2-115762176\right) \mathit{b}^2\right.\\&\left.+40 \left(1125 \mathcal{S}^{10} \kappa ^{10}+51000 \mathcal{S}^8 \kappa ^8+453320 \mathcal{S}^6 \kappa ^6+1268448 \mathcal{S}^4 \kappa ^4+1761408 \mathcal{S}^2 \kappa ^2+483840\right)\right) \mathcal{Q}^2\\&-9 \mathcal{S}^6 \left(4 \left(667625 \mathcal{S}^{10} \kappa ^{10}+2295300 \mathcal{S}^8 \kappa ^8+3970080 \mathcal{S}^6 \kappa ^6-3487104 \mathcal{S}^4 \kappa ^4+14038272 \mathcal{S}^2 \kappa ^2-2239488\right) \mathit{b}^4\right.\\&\left.-320 \left(4375 \mathcal{S}^{10} \kappa ^{10}+21975 \mathcal{S}^8 \kappa ^8+61020 \mathcal{S}^6 \kappa ^6-4752 \mathcal{S}^4 \kappa ^4+164160 \mathcal{S}^2 \kappa ^2+41472\right) \mathit{b}^2\right.\\&\left.+15 \left(9375 \mathcal{S}^{10} \kappa ^{10}+93500 \mathcal{S}^8 \kappa ^8+322400 \mathcal{S}^6 \kappa ^6+196480 \mathcal{S}^4 \kappa ^4+768768 \mathcal{S}^2 \kappa ^2+986112\right)\right)\end{aligned}\right)\\
	&I=\left(\begin{aligned}&\left(1225 \mathcal{S}^{10} \kappa ^{10}-1380 \mathcal{S}^8 \kappa ^8-15176736 \mathcal{S}^6 \kappa ^6-159933312 \mathcal{S}^4 \kappa ^4-266340096 \mathcal{S}^2 \kappa ^2+103652352\right) \mathcal{Q}^6\\&+\pi ^4 \mathit{b}^4 \mathcal{S}^2 \left(2 \mathit{b}^2 \left(103775 \mathcal{S}^{10} \kappa ^{10}+10018260 \mathcal{S}^8 \kappa ^8+605066688 \mathcal{S}^6 \kappa ^6+4090823424 \mathcal{S}^4 \kappa ^4+4199337216 \mathcal{S}^2 \kappa ^2\right.\right.\\&\left.\left.-3692749824\right)-5 \left(8825 \mathcal{S}^{10} \kappa ^{10}+77580 \mathcal{S}^8 \kappa ^8+36898272 \mathcal{S}^6 \kappa ^6+328708224 \mathcal{S}^4 \kappa ^4+365140224 \mathcal{S}^2 \kappa ^2\right.\right.\\&\left.\left.-95966208\right)\right) \mathcal{Q}^4-3 \pi ^2 \mathit{b}^2 \mathcal{S}^4 \left(6 \left(471625 \mathcal{S}^{10} \kappa ^{10}+20205500 \mathcal{S}^8 \kappa ^8+103013728 \mathcal{S}^6 \kappa ^6+67128960 \mathcal{S}^4 \kappa ^4\right.\right.\\&\left.\left.+124915968 \mathcal{S}^2 \kappa ^2-71525376\right) \mathit{b}^4-5 \left(226275 \mathcal{S}^{10} \kappa ^{10}+10731700 \mathcal{S}^8 \kappa ^8+85631136 \mathcal{S}^6 \kappa ^6\right.\right.\\&\left.\left.+170854272 \mathcal{S}^4 \kappa ^4+243475200 \mathcal{S}^2 \kappa ^2-101772288\right) \mathit{b}^2+5 \left(16875 \mathcal{S}^{10} \kappa ^{10}+884500 \mathcal{S}^8 \kappa ^8\right.\right.\\&\left.\left.+9588000 \mathcal{S}^6 \kappa ^6+35402112 \mathcal{S}^4 \kappa ^4+42603264 \mathcal{S}^2 \kappa ^2+2626560\right)\right)\end{aligned}\right)\\
	&J= \left(\begin{aligned}&4 \left(471625 \mathcal{S}^{10} \kappa ^{10}+1190700 \mathcal{S}^8 \kappa ^8+876960 \mathcal{S}^6 \kappa ^6-2706048 \mathcal{S}^4 \kappa ^4+7236864 \mathcal{S}^2 \kappa ^2-2737152\right) \mathit{b}^6\\&-40 \left(72275 \mathcal{S}^{10} \kappa ^{10}+309260 \mathcal{S}^8 \kappa ^8+653184 \mathcal{S}^6 \kappa ^6-405504 \mathcal{S}^4 \kappa ^4+1651968 \mathcal{S}^2 \kappa ^2-248832\right) \mathit{b}^4\\&+5 \left(181125 \mathcal{S}^{10} \kappa ^{10}+1563100 \mathcal{S}^8 \kappa ^8+4311200 \mathcal{S}^6 \kappa ^6+1815936 \mathcal{S}^4 \kappa ^4+10121472 \mathcal{S}^2 \kappa ^2+2018304\right) \mathit{b}^2\\&-400 \left(125 \mathcal{S}^{10} \kappa ^{10}+1350 \mathcal{S}^8 \kappa ^8+6160 \mathcal{S}^6 \kappa ^6+4416 \mathcal{S}^4 \kappa ^4+13824 \mathcal{S}^2 \kappa ^2+13824\right)\end{aligned}\right)
	\end{split}
\end{equation}
\begin{equation}
	\begin{split}
			&K=\left(\begin{aligned}&64 \pi ^6 \kappa ^2 \left(\mathcal{S}^6 \kappa ^6-12 \mathcal{S}^4 \kappa ^4-336 \mathcal{S}^2 \kappa ^2-576\right) \mathcal{Q}^6+\pi ^4 \left(175 \mathcal{S}^{10} \kappa ^{10}+692 \mathcal{S}^8 \kappa ^8+52192 \mathcal{S}^6 \kappa ^6\right.\\&\left.+452736 \mathcal{S}^4 \kappa ^4+762624 \mathcal{S}^2 \kappa ^2-359424\right) \mathcal{Q}^4-8 \pi ^2 \mathcal{S}^4 \kappa ^2 \left(245 \mathcal{S}^8 \kappa ^8+7700 \mathcal{S}^6 \kappa ^6+123504 \mathcal{S}^4 \kappa ^4\right.\\&\left.+137664 \mathcal{S}^2 \kappa ^2+55296\right) \mathcal{Q}^2+\mathcal{S}^4 \left(32585 \mathcal{S}^{10} \kappa ^{10}+470988 \mathcal{S}^8 \kappa ^8+1989792 \mathcal{S}^6 \kappa ^6-3259008 \mathcal{S}^4 \kappa ^4\right.\\&\left.+3587328 \mathcal{S}^2 \kappa ^2-746496\right)\end{aligned}\right)\\
		&L=\left(\begin{aligned}&4 \pi ^6 \mathit{b}^2 \left(\mathcal{S}^{10} \kappa ^{10}-172 \mathcal{S}^8 \kappa ^8+1504 \mathcal{S}^6 \kappa ^6+67968 \mathcal{S}^4 \kappa ^4+131328 \mathcal{S}^2 \kappa ^2-27648\right) \mathcal{Q}^6\\&+\pi ^4 \mathcal{S}^2 \left(50 \mathcal{S}^{10} \kappa ^{10}+72 \mathcal{S}^8 \kappa ^8+28032 \mathcal{S}^6 \kappa ^6+268800 \mathcal{S}^4 \kappa ^4+345600 \mathcal{S}^2 \kappa ^2\right.\\&\left.-\mathit{b}^2 \left(1211 \mathcal{S}^{10} \kappa ^{10}-1492 \mathcal{S}^8 \kappa ^8+381600 \mathcal{S}^6 \kappa ^6+5352576 \mathcal{S}^4 \kappa ^4+8301312 \mathcal{S}^2 \kappa ^2-5059584\right)+55296\right) \mathcal{Q}^4\\&+\pi ^2 \mathcal{S}^4 \left(-2555 \mathcal{S}^{10} \kappa ^{10}-58268 \mathcal{S}^8 \kappa ^8-950304 \mathcal{S}^6 \kappa ^6-1650816 \mathcal{S}^4 \kappa ^4-3158784 \mathcal{S}^2 \kappa ^2\right.\\&\left.+3 \mathit{b}^2 \left(2597 \mathcal{S}^{10} \kappa ^{10}+147812 \mathcal{S}^8 \kappa ^8+1883360 \mathcal{S}^6 \kappa ^6+3608448 \mathcal{S}^4 \kappa ^4-99072 \mathcal{S}^2 \kappa ^2-138240\right)+746496\right) \mathcal{Q}^2\\&+\mathcal{S}^6 \left(48265 \mathcal{S}^{10} \kappa ^{10}+669732 \mathcal{S}^8 \kappa ^8+3424032 \mathcal{S}^6 \kappa ^6-1662336 \mathcal{S}^4 \kappa ^4+4416768 \mathcal{S}^2 \kappa ^2\right.\\&\left.-96 \mathit{b}^2 \left(1715 \mathcal{S}^{10} \kappa ^{10}+16709 \mathcal{S}^8 \kappa ^8+27552 \mathcal{S}^6 \kappa ^6-50112 \mathcal{S}^4 \kappa ^4+76032 \mathcal{S}^2 \kappa ^2-20736\right)-248832\right)\end{aligned}\right)\\
		&M=\left(\begin{aligned}&\left(105 \mathcal{S}^{10} \kappa ^{10}+16852 \mathcal{S}^8 \kappa ^8-695328 \mathcal{S}^6 \kappa ^6-10456704 \mathcal{S}^4 \kappa ^4-20051712 \mathcal{S}^2 \kappa ^2+8377344\right) \mathcal{Q}^6\\&+\pi ^4 \mathit{b}^4 \mathcal{S}^2 \left(-4925 \mathcal{S}^{10} \kappa ^{10}+13020 \mathcal{S}^8 \kappa ^8-7751712 \mathcal{S}^6 \kappa ^6-92506752 \mathcal{S}^4 \kappa ^4-111829248 \mathcal{S}^2 \kappa ^2\right.\\&\left.+\mathit{b}^2 \left(27125 \mathcal{S}^{10} \kappa ^{10}+114948 \mathcal{S}^8 \kappa ^8+71384736 \mathcal{S}^6 \kappa ^6+629665920 \mathcal{S}^4 \kappa ^4+830013696 \mathcal{S}^2 \kappa ^2-662473728\right)\right.\\&\left.+27675648\right) \mathcal{Q}^4-\pi ^2 \mathit{b}^2 \mathcal{S}^4 \left(12 \left(53655 \mathcal{S}^{10} \kappa ^{10}+3056060 \mathcal{S}^8 \kappa ^8+22143408 \mathcal{S}^6 \kappa ^6+21196224 \mathcal{S}^4 \kappa ^4\right.\right.\\&\left.\left.+14031360 \mathcal{S}^2 \kappa ^2-10119168\right) \mathit{b}^4-\left(222425 \mathcal{S}^{10} \kappa ^{10}+11501220 \mathcal{S}^8 \kappa ^8+114647328 \mathcal{S}^6 \kappa ^6\right.\right.\\&\left.\left.+323971200 \mathcal{S}^4 \kappa ^4+335418624 \mathcal{S}^2 \kappa ^2-172108800\right) \mathit{b}^2+5 \left(2575 \mathcal{S}^{10} \kappa ^{10}+93180 \mathcal{S}^8 \kappa ^8\right.\right.\\&\left.\left.+1329696 \mathcal{S}^6 \kappa ^6+4811904 \mathcal{S}^4 \kappa ^4+6504192 \mathcal{S}^2 \kappa ^2+1244160\right)\right)\end{aligned}\right)\\
		&N=\left(\begin{aligned}&24 \left(25725 \mathcal{S}^{10} \kappa ^{10}+90160 \mathcal{S}^8 \kappa ^8+86688 \mathcal{S}^6 \kappa ^6-203904 \mathcal{S}^4 \kappa ^4+463104 \mathcal{S}^2 \kappa ^2-165888\right) \mathit{b}^6\\&-16 \left(35525 \mathcal{S}^{10} \kappa ^{10}+281085 \mathcal{S}^8 \kappa ^8+554376 \mathcal{S}^6 \kappa ^6-96192 \mathcal{S}^4 \kappa ^4+1085184 \mathcal{S}^2 \kappa ^2-145152\right) \mathit{b}^4\\&+10 \left(9625 \mathcal{S}^{10} \kappa ^{10}+96380 \mathcal{S}^8 \kappa ^8+343680 \mathcal{S}^6 \kappa ^6+129024 \mathcal{S}^4 \kappa ^4+753408 \mathcal{S}^2 \kappa ^2+82944\right) \mathit{b}^2\\&-5 \left(625 \mathcal{S}^{10} \kappa ^{10}+8500 \mathcal{S}^8 \kappa ^8+62880 \mathcal{S}^6 \kappa ^6+42624 \mathcal{S}^4 \kappa ^4+158976 \mathcal{S}^2 \kappa ^2+82944\right)\end{aligned}\right)\\
		&O=\left(\begin{aligned}&\left(5 \mathcal{S}^{10} \kappa ^{10}+5260 \mathcal{S}^8 \kappa ^8-100128 \mathcal{S}^6 \kappa ^6-2413440 \mathcal{S}^4 \kappa ^4-4817664 \mathcal{S}^2 \kappa ^2+1797120\right) \mathcal{Q}^6\\&+\pi ^4 \mathit{b}^2 \mathcal{S}^2 \left(\mathit{b}^2 \left(17885 \mathcal{S}^{10} \kappa ^{10}-36340 \mathcal{S}^8 \kappa ^8+14857632 \mathcal{S}^6 \kappa ^6+171531648 \mathcal{S}^4 \kappa ^4+245221632 \mathcal{S}^2 \kappa ^2\right.\right.\\&\left.\left.-178578432\right)-5 \left(335 \mathcal{S}^{10} \kappa ^{10}-1276 \mathcal{S}^8 \kappa ^8+203232 \mathcal{S}^6 \kappa ^6+2987136 \mathcal{S}^4 \kappa ^4+3748608 \mathcal{S}^2 \kappa ^2-248832\right)\right) \mathcal{Q}^4\\&-\pi ^2 \mathcal{S}^4 \left(\left(186445 \mathcal{S}^{10} \kappa ^{10}+12849564 \mathcal{S}^8 \kappa ^8+120924000 \mathcal{S}^6 \kappa ^6+223673472 \mathcal{S}^4 \kappa ^4-23065344 \mathcal{S}^2 \kappa ^2\right.\right.\\&\left.\left.-23307264\right) \mathit{b}^4-5 \left(11165 \mathcal{S}^{10} \kappa ^{10}+406044 \mathcal{S}^8 \kappa ^8+5315232 \mathcal{S}^6 \kappa ^6+14550912 \mathcal{S}^4 \kappa ^4+18544896 \mathcal{S}^2 \kappa ^2\right.\right.\\&\left.\left.-7050240\right) \mathit{b}^2+120 \left(25 \mathcal{S}^{10} \kappa ^{10}+540 \mathcal{S}^8 \kappa ^8+10088 \mathcal{S}^6 \kappa ^6+25440 \mathcal{S}^4 \kappa ^4+42624 \mathcal{S}^2 \kappa ^2+13824\right)\right)\end{aligned}\right) \\
		&P= \left(\begin{aligned}&3 \left(403025 \mathcal{S}^{10} \kappa ^{10}+2966460 \mathcal{S}^8 \kappa ^8+2573088 \mathcal{S}^6 \kappa ^6-5175936 \mathcal{S}^4 \kappa ^4+11162880 \mathcal{S}^2 \kappa ^2-3732480\right) \mathit{b}^4\\&-30 \left(19845 \mathcal{S}^{10} \kappa ^{10}+194180 \mathcal{S}^8 \kappa ^8+519552 \mathcal{S}^6 \kappa ^6-147456 \mathcal{S}^4 \kappa ^4+836352 \mathcal{S}^2 \kappa ^2-82944\right) \mathit{b}^2\\&+20 \mathcal{S}^2 \kappa ^2 \left(2975 \mathcal{S}^8 \kappa ^8+40200 \mathcal{S}^6 \kappa ^6+243936 \mathcal{S}^4 \kappa ^4+41472 \mathcal{S}^2 \kappa ^2+518400\right)\end{aligned}\right)\\
		&R=\left(\begin{aligned}&\left(-105 \kappa ^2 \mathcal{S}^6+36 \mathcal{S}^4-12 c \pi  \left(5 \mathcal{S}^2 \kappa ^2+4\right) \mathcal{S}^3+3 \pi ^2 \mathcal{Q}^2 \left(\mathcal{S}^2 \kappa ^2+12\right) \mathcal{S}^2+4 c \pi ^3 \mathit{b}^2 \mathcal{Q}^2 \left(\mathcal{S}^2 \kappa ^2-60\right) \mathcal{S}\right.\\&\left.+32 c^2 \pi ^4 \mathit{b}^4 \mathcal{Q}^2 \left(\mathcal{S}^2 \kappa ^2+28\right)\right)^2 \left(32 c^2 \pi ^2 \left(5 \mathcal{S}^2 \kappa ^2+12\right) \mathit{b}^4-60 c \pi  \mathcal{S} \left(\mathcal{S}^2 \kappa ^2+4\right) \mathit{b}^2+15 \mathcal{S}^2 \left(\mathcal{S}^2 \kappa ^2+12\right)\right)^2 \times\\&\left(32 c^2 \pi ^4 \mathcal{Q}^2 \left(\mathcal{S}^2 \kappa ^2+12\right) \mathit{b}^4-9 \left(-7 \kappa ^2 \mathcal{S}^6+12 \mathcal{S}^4+\pi ^2 \mathcal{Q}^2 \left(\mathcal{S}^2 \kappa ^2-4\right) \mathcal{S}^2\right)\right.\\&\left.+12 c \left(\pi  \left(5 \mathcal{S}^2 \kappa ^2-12\right) \mathcal{S}^3+\pi ^3 \mathit{b}^2 \mathcal{Q}^2 \left(\mathcal{S}^2 \kappa ^2-12\right) \mathcal{S}\right)\right)^3\end{aligned}\right)
	\end{split}
\end{equation}
In section \ref{sec:6}, the GTD scalar for the Euler Heisenberg AdS black hole in Bekenstein Hawking entropy is given as
\begin{equation}
R_{GTD}=\frac{256 G^4 l^2 \pi ^3 S^7 \left(\frac{\begin{aligned}&a^4 \pi ^{12} A Q^{12}+a^3 G \pi ^8 S^2 B Q^8+\\&a^2 G^2 \pi ^4 S^4 I Q^4-5 a G^3 \pi ^2 S^6 D Q^2\\&+50 G^4 S^9 E\end{aligned}}{F^2 \left(a \pi ^4 Q^4+G S^2 \left(-4 \pi ^2 Q^2+3 G S^2+4 \pi  S\right)\right)^2}+\frac{l^2 \left(\begin{aligned}&a^4 \pi ^{12} \left(307 \pi ^2 Q^2+2196 G S^2-402 \pi  S\right) Q^{12}\\&+2 a^3 G \pi ^8 S^2 G Q^8+8 a^2 G^2 \pi ^4 S^4H Q^4\\&-160 a G^3 \pi ^2 S^6 J Q^2+3200 G^4 S^9 K\end{aligned}\right)}{\begin{aligned}&\left(7 a \pi ^4 Q^4+4 G S^2 \left(-3 \pi ^2 Q^2-3 G S^2+\pi  S\right)\right)\times\\& \left(a \pi ^4 Q^4+4 G S^2 \left(-\pi ^2 Q^2+3 G S^2+\pi  S\right)\right)^2\times\\& \left(7 a l^2 \pi ^4 Q^4+4 G S^2 \left(l^2 \pi  \left(S-3 \pi  Q^2\right)-3 G S^2\right)\right)\end{aligned}}\right)}{\left(3 a \pi ^2 Q^2-10 G S^2\right)^2 \left(a l^2 \pi ^4 Q^4+4 G S^2 \left(\pi  \left(S-\pi  Q^2\right) l^2+3 G S^2\right)\right)}
\label{ap:13}
\end{equation}
where A, B, I, D, E, F, G, H, j, K are given below
\begin{equation}
	\begin{split}
		&A=\left(307 \pi ^2 Q^2+549 G S^2-402 \pi  S\right)\\
		&B=\left(-1146 \pi ^4 Q^4+3176 \pi ^3 S Q^2+162 G^2 S^4+792 G \pi  S^3-3 \pi ^2 \left(2569 G Q^2+424\right) S^2\right)\\
		&I=\left(\begin{aligned}&-1104 \pi ^6 Q^6-5960 \pi ^5 S Q^4+2 \pi ^4 \left(18003 G Q^2+4664\right) S^2 Q^2+729 G^3 S^6+378 G^2 \pi  S^5\\&-9 G \pi ^2 \left(211 G Q^2+168\right) S^4-24 \pi ^3 \left(383 G Q^2+88\right) S^3\end{aligned}\right)\\
		&D=\left(\begin{aligned}&-928 \pi ^6 Q^6-672 \pi ^5 S Q^4+48 \pi ^4 \left(259 G Q^2+76\right) S^2 Q^2+729 G^3 S^6+432 G^2 \pi  S^5\\&-18 G \pi ^2 \left(45 G Q^2+56\right) S^4-192 \pi ^3 \left(31 G Q^2+10\right) S^3\end{aligned}\right)\\
		&E=\left(81 G^3 S^5+36 G^2 \pi  S^4-16 \pi ^3 \left(27 G Q^2+8\right) S^2+16 \pi ^4 Q^2 \left(39 G Q^2+16\right) S-128 \pi ^5 Q^4\right)\\
		&F=\left(7 a \pi ^4 Q^4+G S^2 \left(-12 \pi ^2 Q^2-3 G S^2+4 \pi  S\right)\right)\\
		&G=\left(-573 \pi ^4 Q^4+1588 \pi ^3 S Q^2+1296 G^2 S^4+1584 G \pi  S^3-6 \pi ^2 \left(2569 G Q^2+106\right) S^2\right)\\
		&H= \left(\begin{aligned}&-138 \pi ^6 Q^6-745 \pi ^5 S Q^4+\pi ^4 \left(18003 G Q^2+1166\right) S^2 Q^2+5832 G^3 S^6+756 G^2 \pi  S^5\\&-18 G \pi ^2 \left(211 G Q^2+42\right) S^4-12 \pi ^3 \left(383 G Q^2+22\right) S^3\end{aligned}\right)\\
		&J=\left(\begin{aligned}&-29 \pi ^6 Q^6-21 \pi ^5 S Q^4+6 \pi ^4 \left(259 G Q^2+19\right) S^2 Q^2+1458 G^3 S^6+216 G^2 \pi  S^5\\&-9 G \pi ^2 \left(45 G Q^2+14\right) S^4-12 \pi ^3 \left(62 G Q^2+5\right) S^3\end{aligned}\right)\\
		&K=\left(81 G^3 S^5+9 G^2 \pi  S^4-\pi ^3 \left(27 G Q^2+2\right) S^2+\pi ^4 Q^2 \left(39 G Q^2+4\right) S-2 \pi ^5 Q^4\right)
	\end{split}
\end{equation}
The GTD scalar for the Euler Heisenberg AdS black hole in R\'enyi entropy is given as
\begin{equation}
R_{GTD}=\frac{4096 G^3 l^2 \pi ^3 S^6 \left(\frac{l^2 \left(\begin{aligned}&72 a^5 \pi ^{16} \lambda  A Q^{16}+a^4 G \pi ^{12} S B Q^{12}+\\&2 a^3 G^2 \pi ^8 S^3 I Q^8+8 a^2 G^3 \pi ^4 S^5 D Q^4\\&-160 a G^4 \pi ^2 S^7 E Q^2+3200 G^5 S^{10}F\end{aligned}\right)}{\begin{aligned}&H^2 J \left(a l^2 \pi ^4 (15 S \lambda -28) Q^4+\right.\\&\left.4 G S^2 \left(\pi  \left(\pi  (S \lambda +12) Q^2\right.\right.\right.\\&\left.\left.\left.+S (3 S \lambda -4)\right) l^2+3 G S^2 (15 S \lambda +4)\right)\right)\end{aligned}}+\frac{\begin{aligned}&72 a^5 \pi ^{16} \lambda  K Q^{16}+a^4 G \pi ^{12} S L Q^{12}\\&+a^3 G^2 \pi ^8 S^3 M Q^8+a^2 G^3 \pi ^4 S^5\times\\& \left(-16 \pi ^6 N Q^6-8 \pi ^5 S O Q^4-2 \pi ^4 S^2 P Q^2\right.\\&\left.+81 G^3 S^6 R+54 G^2 \pi  S^5 T-9 G \pi ^2 S^4 U\right) Q^4\\&-5 a G^4 \pi ^2 S^7 \left(32 \pi ^6 V Q^4-16 \pi ^4 S^2 W Q^2\right.\\&\left.+216 G^2 \pi  S^5 X-18 G \pi ^2 S^4 Y\right) Q^2+\\&50 G^5 S^{10} \left(-1024 \pi ^6 \lambda  Z Q^2+36 G^2 \pi  S^4 A1\right.\\&\left.+16 \pi ^3 S^2 A2\right)\end{aligned}}{A3^2 \left(\begin{aligned} & a \pi ^4 (15 S \lambda -28) Q^4\\&+G S^2 \left(4 \pi ^2 (S \lambda +12) Q^2\right.\\&\left.+4 \pi  S (3 S \lambda -4)+3 G S^2 (15 S \lambda +4)\right)\end{aligned}\right)^2}\right)}{\begin{aligned}&\left(10 G S^2 (S \lambda -4)-3 a \pi ^2 Q^2 (5 S \lambda -4)\right)^2\times\\& \left(a l^2 \pi ^4 (4-3 S \lambda ) Q^4+4 G S^2 \left(\pi  \left(S (3 S \lambda +4)-\pi  Q^2 (S \lambda +4)\right) l^2+3 G S^2 (5 S \lambda +4)\right)\right)\end{aligned}}
\label{ap:14}
\end{equation}
where A, B, I, D, E, F, H, J, K, L, M, N, O, P, R, T, U, V, W, X, Y, Z, A1, A2, A3 are given below
\begin{equation}
	\begin{split}
		&A=\left(225 S^3 \lambda ^3-660 S^2 \lambda ^2+560 S \lambda -192\right)\\
		&B=\left(\begin{aligned}&\pi ^2 \left(-176175 S^5 \lambda ^5+1138860 S^4 \lambda ^4-2244384 S^3 \lambda ^3+1784448 S^2 \lambda ^2-800512 S \lambda +314368\right) Q^2+\\&6 \pi  S \left(62775 S^5 \lambda ^5-376380 S^4 \lambda ^4+703584 S^3 \lambda ^3-486528 S^2 \lambda ^2+171776 S \lambda -68608\right)\\&+36 G S^2 \left(124875 S^5 \lambda ^5-634500 S^4 \lambda ^4+1015680 S^3 \lambda ^3-450304 S^2 \lambda ^2-105216 S \lambda +62464\right)\end{aligned}\right)\\
		&I=\left(\begin{aligned}&\pi ^4 \left(56025 S^5 \lambda ^5-1175220 S^4 \lambda ^4+4636512 S^3 \lambda ^3-5045120 S^2 \lambda ^2+1493248 S \lambda -586752\right) Q^4\\&-8 \pi ^3 S \left(78975 S^5 \lambda ^5-628560 S^4 \lambda ^4+1738872 S^3 \lambda ^3-1648224 S^2 \lambda ^2+500096 S \lambda -203264\right) Q^2\\&+144 G \pi  S^3 \left(10125 S^5 \lambda ^5-90000 S^4 \lambda ^4+195600 S^3 \lambda ^3-170048 S^2 \lambda ^2+18432 S \lambda +11264\right)\\&+432 G^2 S^4 \left(50625 S^5 \lambda ^5-168000 S^4 \lambda ^4+114000 S^3 \lambda ^3-18880 S^2 \lambda ^2-5632 S \lambda +3072\right)\\&-6 \pi ^2 S^2 \left(-4050 S^5 \lambda ^5+68040 S^4 \lambda ^4-305856 S^3 \lambda ^3+452352 S^2 \lambda ^2-252416 S \lambda\right.\\&\left. +G Q^2 \left(934875 S^5 \lambda ^5-8761500 S^4 \lambda ^4+22037280 S^3 \lambda ^3-16470656 S^2 \lambda ^2-1340672 S \lambda +2630656\right)+108544\right)\end{aligned}\right)\\
		&D=\left(\begin{aligned}&-2 \pi ^6 \left(225 S^5 \lambda ^5-69900 S^4 \lambda ^4+641248 S^3 \lambda ^3-1311872 S^2 \lambda ^2+272640 S \lambda +70656\right) Q^6\\&-\pi ^5 S \left(7425 S^5 \lambda ^5+1513260 S^4 \lambda ^4-6719424 S^3 \lambda ^3+9677824 S^2 \lambda ^2-3051776 S \lambda +762880\right) Q^4\\&+\pi ^4 S^2 \left(-44550 S^5 \lambda ^5+489240 S^4 \lambda ^4-1653696 S^3 \lambda ^3+3269376 S^2 \lambda ^2-2276864 S \lambda\right.\\&\left. -3 G Q^2 \left(55125 S^5 \lambda ^5+4295100 S^4 \lambda ^4-22795040 S^3 \lambda ^3+29770368 S^2 \lambda ^2-3058432 S \lambda\right.\right.\\&\left.\left. -6145024\right)+1193984\right) Q^2+648 G^3 S^6 \left(84375 S^5 \lambda ^5-22500 S^4 \lambda ^4-56000 S^3 \lambda ^3+32000 S \lambda +9216\right)\\&+108 G^2 \pi  S^5 \left(163125 S^5 \lambda ^5-118500 S^4 \lambda ^4-20000 S^3 \lambda ^3+17280 S^2 \lambda ^2+76032 S \lambda +7168\right)\\&-12 \pi ^3 S^3 \left(G Q^2 \left(151875 S^5 \lambda ^5-774900 S^4 \lambda ^4+2596320 S^3 \lambda ^3-3419776 S^2 \lambda ^2+964352 S \lambda +392192\right)\right.\\&\left.-4 \left(2025 S^5 \lambda ^5-2970 S^4 \lambda ^4-2016 S^3 \lambda ^3+2688 S^2 \lambda ^2+5888 S \lambda -5632\right)\right)\\&-18 G \pi ^2 S^4 \left(G \left(770625 S^5 \lambda ^5-4036500 S^4 \lambda ^4+5151200 S^3 \lambda ^3-1059200 S^2 \lambda ^2+35072 S \lambda +216064\right) Q^2\right.\\&\left.+6 \left(-21375 S^5 \lambda ^5+15900 S^4 \lambda ^4+160 S^3 \lambda ^3+2432 S^2 \lambda ^2-20224 S \lambda +7168\right)\right)\end{aligned}\right)\\
		&E=\left(\begin{aligned}&\pi ^6 \left(285 S^5 \lambda ^5-324 S^4 \lambda ^4-21984 S^3 \lambda ^3+99712 S^2 \lambda ^2-50944 S \lambda -29696\right) Q^6+\\&\pi ^5 S \left(-2115 S^5 \lambda ^5+2148 S^4 \lambda ^4+229120 S^3 \lambda ^3-516864 S^2 \lambda ^2+240384 S \lambda -21504\right) Q^4\\&-2 \pi ^4 S^2 \left(1215 S^5 \lambda ^5-12132 S^4 \lambda ^4+62976 S^3 \lambda ^3-107008 S^2 \lambda ^2+97024 S \lambda \right.\\&\left.+3 G Q^2 \left(1875 S^5 \lambda ^5-10860 S^4 \lambda ^4-321632 S^3 \lambda ^3+917888 S^2 \lambda ^2-374016 S \lambda -265216\right)-58368\right) Q^2\\&+108 G^2 \pi  S^5 \left(7125 S^5 \lambda ^5-43300 S^4 \lambda ^4+1680 S^3 \lambda ^3+15808 S^2 \lambda ^2+23040 S \lambda +2048\right)\\&+162 G^3 S^6 \left(16875 S^5 \lambda ^5-81500 S^4 \lambda ^4-44000 S^3 \lambda ^3+35200 S^2 \lambda ^2+43776 S \lambda +9216\right)\\&-12 \pi ^3 S^3 \left(G Q^2 \left(5625 S^5 \lambda ^5-50280 S^4 \lambda ^4+141344 S^3 \lambda ^3-292736 S^2 \lambda ^2+193280 S \lambda +63488\right)\right.\\&\left.-2 \left(135 S^5 \lambda ^5-1008 S^4 \lambda ^4+696 S^3 \lambda ^3+1056 S^2 \lambda ^2+896 S \lambda -2560\right)\right)\\&-9 G \pi ^2 S^4 \left(G Q^2 \left(30375 S^5 \lambda ^5-487500 S^4 \lambda ^4+981920 S^3 \lambda ^3-319616 S^2 \lambda ^2+34560 S \lambda +46080\right)\right.\\&\left.-2 \left(5175 S^5 \lambda ^5-31260 S^4 \lambda ^4+4512 S^3 \lambda ^3+1408 S^2 \lambda ^2+30464 S \lambda -7168\right)\right)\end{aligned}\right)\\
	&	F= \left(\begin{aligned}&-16 \pi ^6 \lambda  \left(S^3 \lambda ^3-4 S^2 \lambda ^2-80 S \lambda +64\right) Q^6+4 \pi ^5 \left(3 S^5 \lambda ^5+70 S^4 \lambda ^4+64 S^3 \lambda ^3-2688 S^2 \lambda ^2\right.\\&\left.+1024 S \lambda -512\right) Q^4+\pi ^4 S \left(3 G \left(75 S^5 \lambda ^5+836 S^4 \lambda ^4-32 S^3 \lambda ^3-29056 S^2 \lambda ^2+32512 S \lambda +13312\right) Q^2\right.\\&\left.+2 \left(9 S^5 \lambda ^5+84 S^4 \lambda ^4-976 S^3 \lambda ^3+3264 S^2 \lambda ^2-1536 S \lambda +2048\right)\right) Q^2+\\&9 G^2 \pi  S^4 \left(2325 S^5 \lambda ^5-13780 S^4 \lambda ^4+24128 S^3 \lambda ^3+24576 S^2 \lambda ^2+17152 S \lambda +1024\right)+\\&27 G^3 S^5 \left(2625 S^5 \lambda ^5-14900 S^4 \lambda ^4+16480 S^3 \lambda ^3+35968 S^2 \lambda ^2+20736 S \lambda +3072\right)\\&+12 G \pi ^2 S^4 \lambda  \left(3 G \left(75 S^4 \lambda ^4+1040 S^3 \lambda ^3-7072 S^2 \lambda ^2+3072 S \lambda -256\right) Q^2\right.\\&\left.+4 \left(45 S^4 \lambda ^4-279 S^3 \lambda ^3+592 S^2 \lambda ^2+416 S \lambda +832\right)\right)+\pi ^3 S^2 \left(3 G \left(45 S^5 \lambda ^5+1988 S^4 \lambda ^4\right.\right.\\&\left.\left.-12480 S^3 \lambda ^3+18432 S^2 \lambda ^2-39168 S \lambda -9216\right) Q^2+2 \left(27 S^5 \lambda ^5-216 S^4 \lambda ^4+720 S^3 \lambda ^3-64 S^2 \lambda ^2-1024\right)\right)\end{aligned}\right)\\
		&H=\left(a \pi ^4 Q^4 (3 S \lambda -4)-4 G S^2 \left(-\pi ^2 (S \lambda +4) Q^2+\pi  S (3 S \lambda +4)+3 G S^2 (5 S \lambda +4)\right)\right)\\
		&J=\left(a \pi ^4 (15 S \lambda -28) Q^4+4 G S^2 \left(\pi ^2 (S \lambda +12) Q^2+\pi  S (3 S \lambda -4)+3 G S^2 (15 S \lambda +4)\right)\right)\\
		&K=\left(225 S^3 \lambda ^3-660 S^2 \lambda ^2+560 S \lambda -192\right)\\
		&L=\left(\begin{aligned}&\pi ^2 \left(-176175 S^5 \lambda ^5+1138860 S^4 \lambda ^4-2244384 S^3 \lambda ^3+1784448 S^2 \lambda ^2-800512 S \lambda +314368\right) Q^2\\&+6 \pi  S \left(62775 S^5 \lambda ^5-376380 S^4 \lambda ^4+703584 S^3 \lambda ^3-486528 S^2 \lambda ^2+171776 S \lambda -68608\right)\\&+9 G S^2 \left(124875 S^5 \lambda ^5-634500 S^4 \lambda ^4+1015680 S^3 \lambda ^3-450304 S^2 \lambda ^2-105216 S \lambda +62464\right)\end{aligned}\right)
	\end{split}
	\end{equation}
	\begin{equation}
		\begin{split}
			&M=\left(\begin{aligned}&2 \pi ^4 \left(56025 S^5 \lambda ^5-1175220 S^4 \lambda ^4+4636512 S^3 \lambda ^3-5045120 S^2 \lambda ^2+1493248 S \lambda -586752\right) Q^4\\&-16 \pi ^3 S \left(78975 S^5 \lambda ^5-628560 S^4 \lambda ^4+1738872 S^3 \lambda ^3-1648224 S^2 \lambda ^2+500096 S \lambda -203264\right) Q^2\\&+72 G \pi  S^3 \left(10125 S^5 \lambda ^5-90000 S^4 \lambda ^4+195600 S^3 \lambda ^3-170048 S^2 \lambda ^2+18432 S \lambda +11264\right)\\&+54 G^2 S^4 \left(50625 S^5 \lambda ^5-168000 S^4 \lambda ^4+114000 S^3 \lambda ^3-18880 S^2 \lambda ^2-5632 S \lambda +3072\right)\\&-3 \pi ^2 S^2 \left(G Q^2 \left(934875 S^5 \lambda ^5-8761500 S^4 \lambda ^4+22037280 S^3 \lambda ^3-16470656 S^2 \lambda ^2-1340672 S \lambda +2630656\right)\right.\\&\left.-8 \left(2025 S^5 \lambda ^5-34020 S^4 \lambda ^4+152928 S^3 \lambda ^3-226176 S^2 \lambda ^2+126208 S \lambda -54272\right)\right)\end{aligned}\right)\\
		&N=\left(225 S^5 \lambda ^5-69900 S^4 \lambda ^4+641248 S^3 \lambda ^3-1311872 S^2 \lambda ^2+272640 S \lambda +70656\right)\\
		&O=\left(7425 S^5 \lambda ^5+1513260 S^4 \lambda ^4-6719424 S^3 \lambda ^3+9677824 S^2 \lambda ^2-3051776 S \lambda +762880\right)\\
		&P=\left(\begin{aligned}&3 G \left(55125 S^5 \lambda ^5+4295100 S^4 \lambda ^4-22795040 S^3 \lambda ^3+29770368 S^2 \lambda ^2-3058432 S \lambda -6145024\right) Q^2\\&+8 \left(22275 S^5 \lambda ^5-244620 S^4 \lambda ^4+826848 S^3 \lambda ^3-1634688 S^2 \lambda ^2+1138432 S \lambda -596992\right)\end{aligned}\right)\\
		&R=\left(84375 S^5 \lambda ^5-22500 S^4 \lambda ^4-56000 S^3 \lambda ^3+32000 S \lambda +9216\right)\\
		&T=\left(\begin{aligned}&\left(163125 S^5 \lambda ^5-118500 S^4 \lambda ^4-20000 S^3 \lambda ^3+17280 S^2 \lambda ^2+76032 S \lambda +7168\right)\\&-24 \pi ^3 S^3 \left(G Q^2 \left(151875 S^5 \lambda ^5-774900 S^4 \lambda ^4+2596320 S^3 \lambda ^3-3419776 S^2 \lambda ^2+964352 S \lambda +392192\right)\right.\\&\left.-16 \left(2025 S^5 \lambda ^5-2970 S^4 \lambda ^4-2016 S^3 \lambda ^3+2688 S^2 \lambda ^2+5888 S \lambda -5632\right)\right)\end{aligned}\right)\\
		&U=\left(\begin{aligned}& G Q^2 \left(770625 S^5 \lambda ^5-4036500 S^4 \lambda ^4+5151200 S^3 \lambda ^3-1059200 S^2 \lambda ^2+35072 S \lambda +216064\right)\\&-24 \left(21375 S^5 \lambda ^5-15900 S^4 \lambda ^4-160 S^3 \lambda ^3-2432 S^2 \lambda ^2+20224 S \lambda -7168\right)\end{aligned}\right)\\
		&V=\begin{aligned}&\left(285 S^5 \lambda ^5-324 S^4 \lambda ^4-21984 S^3 \lambda ^3+99712 S^2 \lambda ^2-50944 S \lambda -29696\right) Q^6\\&-32 \pi ^5 S \left(2115 S^5 \lambda ^5-2148 S^4 \lambda ^4-229120 S^3 \lambda ^3+516864 S^2 \lambda ^2-240384 S \lambda +21504\right)\end{aligned}\\
		&W=\left(\begin{aligned}&3 G \left(1875 S^5 \lambda ^5-10860 S^4 \lambda ^4-321632 S^3 \lambda ^3+917888 S^2 \lambda ^2-374016 S \lambda -265216\right) Q^2\\&+4 \left(1215 S^5 \lambda ^5-12132 S^4 \lambda ^4+62976 S^3 \lambda ^3-107008 S^2 \lambda ^2+97024 S \lambda -58368\right)\end{aligned}\right)\\
		&X=\left(\begin{aligned}&\left(7125 S^5 \lambda ^5-43300 S^4 \lambda ^4+1680 S^3 \lambda ^3+15808 S^2 \lambda ^2+23040 S \lambda +2048\right)\\&+81 G^3 S^6 \left(16875 S^5 \lambda ^5-81500 S^4 \lambda ^4-44000 S^3 \lambda ^3+35200 S^2 \lambda ^2+43776 S \lambda +9216\right)\\&-96 \pi ^3 S^3 \left(G Q^2 \left(5625 S^5 \lambda ^5-50280 S^4 \lambda ^4+141344 S^3 \lambda ^3-292736 S^2 \lambda ^2+193280 S \lambda +63488\right)\right.\\&\left.-8 \left(135 S^5 \lambda ^5-1008 S^4 \lambda ^4+696 S^3 \lambda ^3+1056 S^2 \lambda ^2+896 S \lambda -2560\right)\right)\end{aligned}\right)\\
		&Y=\left(\begin{aligned}&G Q^2 \left(30375 S^5 \lambda ^5-487500 S^4 \lambda ^4+981920 S^3 \lambda ^3-319616 S^2 \lambda ^2+34560 S \lambda +46080\right)\\&-8 \left(5175 S^5 \lambda ^5-31260 S^4 \lambda ^4+4512 S^3 \lambda ^3+1408 S^2 \lambda ^2+30464 S \lambda -7168\right)\end{aligned}\right)\\
		&Z=\left(\begin{aligned}&\left(S^3 \lambda ^3-4 S^2 \lambda ^2-80 S \lambda +64\right) Q^6+256 \pi ^5 \left(3 S^5 \lambda ^5+70 S^4 \lambda ^4+64 S^3 \lambda ^3-2688 S^2 \lambda ^2+1024 S \lambda -512\right) Q^4\\&+16 \pi ^4 S \left(3 G \left(75 S^5 \lambda ^5+836 S^4 \lambda ^4-32 S^3 \lambda ^3-29056 S^2 \lambda ^2+32512 S \lambda +13312\right) Q^2\right.\\&\left.+8 \left(9 S^5 \lambda ^5+84 S^4 \lambda ^4-976 S^3 \lambda ^3+3264 S^2 \lambda ^2-1536 S \lambda +2048\right)\right)\end{aligned}\right)\\
		&A1=\left(\begin{aligned}&\left(2325 S^5 \lambda ^5-13780 S^4 \lambda ^4+24128 S^3 \lambda ^3+24576 S^2 \lambda ^2+17152 S \lambda +1024\right)\\&+27 G^3 S^5 \left(2625 S^5 \lambda ^5-14900 S^4 \lambda ^4+16480 S^3 \lambda ^3+35968 S^2 \lambda ^2+20736 S \lambda +3072\right)\\&+48 G \pi ^2 S^4 \lambda  \left(3 G \left(75 S^4 \lambda ^4+1040 S^3 \lambda ^3-7072 S^2 \lambda ^2+3072 S \lambda -256\right) Q^2\right.\\&\left.+16 \left(45 S^4 \lambda ^4-279 S^3 \lambda ^3+592 S^2 \lambda ^2+416 S \lambda +832\right)\right)\end{aligned}\right)\\
		&A2=\left(\begin{aligned}&3 G \left(45 S^5 \lambda ^5+1988 S^4 \lambda ^4-12480 S^3 \lambda ^3+18432 S^2 \lambda ^2-39168 S \lambda -9216\right) Q^2\\&+8 \left(27 S^5 \lambda ^5-216 S^4 \lambda ^4+720 S^3 \lambda ^3-64 S^2 \lambda ^2-1024\right)\end{aligned}\right)\\
		&A3=\left(a \pi ^4 Q^4 (3 S \lambda -4)-G S^2 \left(-4 \pi ^2 (S \lambda +4) Q^2+4 \pi  S (3 S \lambda +4)+3 G S^2 (5 S \lambda +4)\right)\right)\\
		&
	\end{split}
\end{equation}
The GTD scalar for the Euler Heisenberg AdS black hole in Kaniadakis entropy is given as
\begin{equation}
R_{GTD}=\frac{12288 G^3 l^2 \pi ^3 S^7}{V} \left(-\frac{\begin{aligned}&-288 a^5 \pi ^{16} \kappa ^2 A Q^{12}+a^3 G^2 \pi ^8 S^2 B Q^8\\&+a^2 G^3 \pi ^4 S^4 I Q^4-5 a G^4 \pi ^2 S^6\times\\& \left(96 \pi ^6 D Q^2+288 G^2 \pi  S^5 E\right) Q^2\\&+50 G^5 S^9 \left(36864 \pi ^6 S \kappa ^2 F-\right.\\&\left.32 G \pi ^2 S^5 \kappa ^2 H\right)\end{aligned}}{J^2}-\frac{l^2 \left(\begin{aligned}&-288 a^5 \pi ^{16} \kappa ^2 K Q^{12}\\&+2 a^3 G^2 \pi ^8 S^2 L Q^8+8 a^2 G^3 \pi ^4 S^4\times\\& \left(-6 \pi ^6 M Q^2+216 G^3 S^6 N\right) Q^4\\&-160 a G^4 \pi ^2 S^6 \left(3 \pi ^6 O+54 G^3 S^6 P\right) Q^2\\&+3200 G^5 S^9 \left(576 \pi ^6 S \kappa ^2 R+3 G^2 \pi  S^4 T\right)\end{aligned}\right)}{U}\right)
\label{ap:15}
\end{equation}
where A, B, I, D, E, F, H, J, K, L, M, N, O, P, R, T, U, V are given below
\begin{equation}
	\begin{split}
			&	A=\left(\begin{aligned}&\left(25 S^6 \kappa ^6+516 S^4 \kappa ^4-1872 S^2 \kappa ^2+1728\right) Q^{16}+3 a^4 G \pi ^{12} \left(-\pi ^2 \left(2875 S^{10} \kappa ^{10}+242420 S^8 \kappa ^8\right.\right.\\&\left.\left.+4807264 S^6 \kappa ^6+17493120 S^4 \kappa ^4+5066496 S^2 \kappa ^2+8487936\right) Q^2+6 \pi  S \left(1625 S^{10} \kappa ^{10}+105700 S^8 \kappa ^8\right.\right.\\&\left.\left.+1937120 S^6 \kappa ^6+5124480 S^4 \kappa ^4+665856 S^2 \kappa ^2+1852416\right)+3 G S^2 \left(6125 S^{10} \kappa ^{10}+346620 S^8 \kappa ^8\right.\right.\\&\left.\left.+5985312 S^6 \kappa ^6+15343488 S^4 \kappa ^4+794880 S^2 \kappa ^2-5059584\right)\right)\end{aligned}\right)\\
		&B=\left(\begin{aligned}&2 \pi ^4 \left(32875 S^{10} \kappa ^{10}+2155460 S^8 \kappa ^8+42126816 S^6 \kappa ^6+216097920 S^4 \kappa ^4+94037760 S^2 \kappa ^2+47526912\right) Q^4\\&-24 \pi ^3 S \left(12625 S^{10} \kappa ^{10}+832100 S^8 \kappa ^8+11155840 S^6 \kappa ^6+41822208 S^4 \kappa ^4+12697344 S^2 \kappa ^2+10976256\right) Q^2\\&+72 G \pi  S^3 \left(20125 S^{10} \kappa ^{10}+915300 S^8 \kappa ^8+3568800 S^6 \kappa ^6+5668992 S^4 \kappa ^4+1734912 S^2 \kappa ^2-912384\right)\\&+18 G^2 S^4 \left(79625 S^{10} \kappa ^{10}+3137820 S^8 \kappa ^8+9365472 S^6 \kappa ^6+5678208 S^4 \kappa ^4-1804032 S^2 \kappa ^2-746496\right)\\&-9 \pi ^2 S^2 \left(G Q^2 \left(46375 S^{10} \kappa ^{10}+3518180 S^8 \kappa ^8+44851168 S^6 \kappa ^6+158007936 S^4 \kappa ^4+50400000 S^2 \kappa ^2\right.\right.\\&\left.\left.-71027712\right)-8 \left(5625 S^{10} \kappa ^{10}+273500 S^8 \kappa ^8+1285600 S^6 \kappa ^6+3008640 S^4 \kappa ^4+1366272 S^2 \kappa ^2+1465344\right)\right)\end{aligned}\right)\\
		&I=\left(\begin{aligned}&-48 \pi ^6 \left(2375 S^{10} \kappa ^{10}+58820 S^8 \kappa ^8+3679136 S^6 \kappa ^6+25832832 S^4 \kappa ^4+19996416 S^2 \kappa ^2-1907712\right) Q^6\\&+8 \pi ^5 S \left(230125 S^{10} \kappa ^{10}+3789300 S^8 \kappa ^8+95542560 S^6 \kappa ^6+442733184 S^4 \kappa ^4+258031872 S^2 \kappa ^2+61793280\right) Q^4\\&+2 \pi ^4 S^2 \left(G Q^2 \left(1463875 S^{10} \kappa ^{10}+22132740 S^8 \kappa ^8+566110944 S^6 \kappa ^6+2419569792 S^4 \kappa ^4\right.\right.\\&\left.\left.+1655209728 S^2 \kappa ^2-1493240832\right)-24 \left(134375 S^{10} \kappa ^{10}+1620500 S^8 \kappa ^8+13175200 S^6 \kappa ^6\right.\right.\\&\left.\left.+29923200 S^4 \kappa ^4+21765888 S^2 \kappa ^2+16118784\right)\right) Q^2+27 G^3 S^6 \left(728875 S^{10} \kappa ^{10}+1572900 S^8 \kappa ^8\right.\\&\left.+2469600 S^6 \kappa ^6-2616192 S^4 \kappa ^4+4790016 S^2 \kappa ^2-2239488\right)+18 G^2 \pi  S^5 \left(2051875 S^{10} \kappa ^{10}\right.\\&\left.+5810700 S^8 \kappa ^8+17845920 S^6 \kappa ^6-3169152 S^4 \kappa ^4+11010816 S^2 \kappa ^2-1741824\right)\\&-24 \pi ^3 S^3 \left(G Q^2 \left(814625 S^{10} \kappa ^{10}+10281300 S^8 \kappa ^8+74781600 S^6 \kappa ^6+120698496 S^4 \kappa ^4\right.\right.\\&\left.\left.+79176960 S^2 \kappa ^2-31767552\right)-24 \left(9375 S^{10} \kappa ^{10}+37500 S^8 \kappa ^8+160000 S^6 \kappa ^6+76800 S^4 \kappa ^4s\right.\right.\\&\left.\left.+149760 S^2 \kappa ^2+304128\right)\right)-3 G \pi ^2 S^4 \left(G Q^2 \left(4808125 S^{10} \kappa ^{10}+70480620 S^8 \kappa ^8+423889056 S^6 \kappa ^6\right.\right.\\&\left.\left.+309260160 S^4 \kappa ^4-17397504 S^2 \kappa ^2-52503552\right)-24 \left(336875 S^{10} \kappa ^{10}+1120500 S^8 \kappa ^8\right.\right.\\&\left.\left.+4500000 S^6 \kappa ^6+1123200 S^4 \kappa ^4+3048192 S^2 \kappa ^2+1741824\right)\right)\end{aligned}\right)
	\end{split}
\end{equation}
\begin{equation}
	\begin{split}
	&D=\left(\begin{aligned}&\left(425 S^{10} \kappa ^{10}+5388 S^8 \kappa ^8-277728 S^6 \kappa ^6-3063936 S^4 \kappa ^4-3877632 S^2 \kappa ^2+801792\right) Q^6\\&-96 \pi ^5 S \left(1525 S^{10} \kappa ^{10}-44540 S^8 \kappa ^8-1677472 S^6 \kappa ^6-10727040 S^4 \kappa ^4-9960192 S^2 \kappa ^2-580608\right) Q^4\\&-16 \pi ^4 S^2 \left(3 G \left(1225 S^{10} \kappa ^{10}-145028 S^8 \kappa ^8-4701792 S^6 \kappa ^6-28819584 S^4 \kappa ^4-31470336 S^2 \kappa ^2\right.\right.\\&\left.\left.+21482496\right) Q^2+4 \left(-625 S^{10} \kappa ^{10}+436500 S^8 \kappa ^8+3713760 S^6 \kappa ^6+9728640 S^4 \kappa ^4+10513152 S^2 \kappa ^2+4727808\right)\right)\end{aligned}\right)\\
	&E=\left(\begin{aligned}&\left(23275 S^{10} \kappa ^{10}+432600 S^8 \kappa ^8+861624 S^6 \kappa ^6-61344 S^4 \kappa ^4+1067904 S^2 \kappa ^2-124416\right)\\&+27 G^3 S^6 \left(145775 S^{10} \kappa ^{10}+2349060 S^8 \kappa ^8+2050272 S^6 \kappa ^6-3680640 S^4 \kappa ^4+6697728 S^2 \kappa ^2-2239488\right)\\&-128 \pi ^3 S^3 \left(4 G Q^2 \left(175 S^{10} \kappa ^{10}+175710 S^8 \kappa ^8+1288818 S^6 \kappa ^6+2643624 S^4 \kappa ^4+2724192 S^2 \kappa ^2-964224\right)\right.\\&\left.-9 \left(625 S^{10} \kappa ^{10}+16500 S^8 \kappa ^8+62400 S^6 \kappa ^6+49920 S^4 \kappa ^4+99072 S^2 \kappa ^2+138240\right)\right)\\&+6 G \pi ^2 S^4 \left(G \left(30625 S^{10} \kappa ^{10}-12405876 S^8 \kappa ^8-80933472 S^6 \kappa ^6-99363456 S^4 \kappa ^4-5204736 S^2 \kappa ^2\right.\right.\\&\left.\left.+11197440\right) Q^2+24 \left(27125 S^{10} \kappa ^{10}+579100 S^8 \kappa ^8+1676640 S^6 \kappa ^6+913536 S^4 \kappa ^4+2536704 S^2 \kappa ^2+580608\right)\right)\end{aligned}\right)\\
	&F=\left(\begin{aligned}&\left(S^6 \kappa ^6-12 S^4 \kappa ^4-336 S^2 \kappa ^2-576\right) Q^6+384 \pi ^5 \left(25 S^{10} \kappa ^{10}+36 S^8 \kappa ^8+14016 S^6 \kappa ^6+134400 S^4 \kappa ^4\right.\\&\left.+172800 S^2 \kappa ^2+27648\right) Q^4+48 \pi ^4 S \left(3 G Q^2 \left(175 S^{10} \kappa ^{10}+692 S^8 \kappa ^8+52192 S^6 \kappa ^6+452736 S^4 \kappa ^4\right.\right.\\&\left.\left.+762624 S^2 \kappa ^2-359424\right)-32 \left(25 S^{10} \kappa ^{10}+540 S^8 \kappa ^8+10088 S^6 \kappa ^6+25440 S^4 \kappa ^4+42624 S^2 \kappa ^2+13824\right)\right) Q^2\\&+9 G^3 S^5 \left(32585 S^{10} \kappa ^{10}+470988 S^8 \kappa ^8+1989792 S^6 \kappa ^6-3259008 S^4 \kappa ^4+3587328 S^2 \kappa ^2-746496\right)\\&+12 G^2 \pi  S^4 \left(48265 S^{10} \kappa ^{10}+669732 S^8 \kappa ^8+3424032 S^6 \kappa ^6-1662336 S^4 \kappa ^4+4416768 S^2 \kappa ^2-248832\right)\end{aligned}\right)\\
	&H=\left(\begin{aligned}&\left(9 G Q^2 \left(245 S^8 \kappa ^8+7700 S^6 \kappa ^6+123504 S^4 \kappa ^4+137664 S^2 \kappa ^2+55296\right)\right.\\&\left.-4 \left(2975 S^8 \kappa ^8+40200 S^6 \kappa ^6+243936 S^4 \kappa ^4+41472 S^2 \kappa ^2+518400\right)\right)\\&-16 \pi ^3 S^2 \left(3 G Q^2 \left(2555 S^{10} \kappa ^{10}+58268 S^8 \kappa ^8+950304 S^6 \kappa ^6+1650816 S^4 \kappa ^4+3158784 S^2 \kappa ^2-746496\right)\right.\\&\left.-8 \left(625 S^{10} \kappa ^{10}+8500 S^8 \kappa ^8+62880 S^6 \kappa ^6+42624 S^4 \kappa ^4+158976 S^2 \kappa ^2+82944\right)\right)\end{aligned}\right)\\
	&J=\left(\begin{aligned}&\left(a \pi ^4 \left(S^2 \kappa ^2+28\right) Q^4+G S^2 \left(-4 \pi ^2 \left(S^2 \kappa ^2+12\right) Q^2+4 \pi  S \left(5 S^2 \kappa ^2+4\right)+G S^2 \left(35 S^2 \kappa ^2-12\right)\right)\right)^2\\& \left(a \pi ^4 \left(S^2 \kappa ^2+12\right) Q^4+G S^2 \left(12 \pi ^2 \left(S^2 \kappa ^2-4\right) Q^2+3 G S^2 \left(12-7 S^2 \kappa ^2\right)+\pi  \left(48 S-20 S^3 \kappa ^2\right)\right)\right)\end{aligned}\right)\\
	&K=\left(\begin{aligned}&\left(25 S^6 \kappa ^6+516 S^4 \kappa ^4-1872 S^2 \kappa ^2+1728\right) Q^{16}+3 a^4 G \pi ^{12} \left(-\pi ^2 \left(2875 S^{10} \kappa ^{10}+242420 S^8 \kappa ^8\right.\right.\\&\left.\left.+4807264 S^6 \kappa ^6+17493120 S^4 \kappa ^4+5066496 S^2 \kappa ^2+8487936\right) Q^2+6 \pi  S \left(1625 S^{10} \kappa ^{10}+105700 S^8 \kappa ^8\right.\right.\\&\left.\left.+1937120 S^6 \kappa ^6+5124480 S^4 \kappa ^4+665856 S^2 \kappa ^2+1852416\right)+12 G S^2 \left(6125 S^{10} \kappa ^{10}+346620 S^8 \kappa ^8\right.\right.\\&\left.\left.+5985312 S^6 \kappa ^6+15343488 S^4 \kappa ^4+794880 S^2 \kappa ^2-5059584\right)\right)\end{aligned}\right)
	\end{split}
\end{equation}
\begin{equation}
\begin{split}
&L=\left(\begin{aligned}&\pi ^4 \left(32875 S^{10} \kappa ^{10}+2155460 S^8 \kappa ^8+42126816 S^6 \kappa ^6+216097920 S^4 \kappa ^4+94037760 S^2 \kappa ^2+47526912\right) Q^4\\&-12 \pi ^3 S \left(12625 S^{10} \kappa ^{10}+832100 S^8 \kappa ^8+11155840 S^6 \kappa ^6+41822208 S^4 \kappa ^4+12697344 S^2 \kappa ^2+10976256\right) Q^2\\&+144 G \pi  S^3 \left(20125 S^{10} \kappa ^{10}+915300 S^8 \kappa ^8+3568800 S^6 \kappa ^6+5668992 S^4 \kappa ^4+1734912 S^2 \kappa ^2-912384\right)\\&+144 G^2 S^4 \left(79625 S^{10} \kappa ^{10}+3137820 S^8 \kappa ^8+9365472 S^6 \kappa ^6+5678208 S^4 \kappa ^4-1804032 S^2 \kappa ^2-746496\right)\\&-18 \pi ^2 S^2 \left(G Q^2 \left(46375 S^{10} \kappa ^{10}+3518180 S^8 \kappa ^8+44851168 S^6 \kappa ^6+158007936 S^4 \kappa ^4+50400000 S^2 \kappa ^2\right.\right.\\&\left.\left.-71027712\right)-2 \left(5625 S^{10} \kappa ^{10}+273500 S^8 \kappa ^8+1285600 S^6 \kappa ^6+3008640 S^4 \kappa ^4+1366272 S^2 \kappa ^2+1465344\right)\right)\end{aligned}\right)\\
&M=\left(\begin{aligned}&\left(2375 S^{10} \kappa ^{10}+58820 S^8 \kappa ^8+3679136 S^6 \kappa ^6+25832832 S^4 \kappa ^4+19996416 S^2 \kappa ^2-1907712\right) Q^6+\\&\pi ^5 S \left(230125 S^{10} \kappa ^{10}+3789300 S^8 \kappa ^8+95542560 S^6 \kappa ^6+442733184 S^4 \kappa ^4+258031872 S^2 \kappa ^2+61793280\right) Q^4\\&+\pi ^4 S^2 \left(G Q^2 \left(1463875 S^{10} \kappa ^{10}+22132740 S^8 \kappa ^8+566110944 S^6 \kappa ^6+2419569792 S^4 \kappa ^4+1655209728 S^2 \kappa ^2\right.\right.\\&\left.\left.-1493240832\right)-6 \left(134375 S^{10} \kappa ^{10}+1620500 S^8 \kappa ^8+13175200 S^6 \kappa ^6+29923200 S^4 \kappa ^4\right.\right.\\&\left.\left.+21765888 S^2 \kappa ^2+16118784\right)\right)\end{aligned}\right)\\
&N=\left(\begin{aligned}&\left(728875 S^{10} \kappa ^{10}+1572900 S^8 \kappa ^8+2469600 S^6 \kappa ^6-2616192 S^4 \kappa ^4+4790016 S^2 \kappa ^2-2239488\right)\\&+36 G^2 \pi  S^5 \left(2051875 S^{10} \kappa ^{10}+5810700 S^8 \kappa ^8+17845920 S^6 \kappa ^6-3169152 S^4 \kappa ^4+11010816 S^2 \kappa ^2-1741824\right)\\&-12 \pi ^3 S^3 \left(G Q^2 \left(814625 S^{10} \kappa ^{10}+10281300 S^8 \kappa ^8+74781600 S^6 \kappa ^6+120698496 S^4 \kappa ^4+79176960 S^2 \kappa ^2\right.\right.\\&\left.\left.-31767552\right)-6 \left(9375 S^{10} \kappa ^{10}+37500 S^8 \kappa ^8+160000 S^6 \kappa ^6+76800 S^4 \kappa ^4+149760 S^2 \kappa ^2+304128\right)\right)\\&-6 G \pi ^2 S^4 \left(G Q^2 \left(4808125 S^{10} \kappa ^{10}+70480620 S^8 \kappa ^8+423889056 S^6 \kappa ^6+309260160 S^4 \kappa ^4-17397504 S^2 \kappa ^2\right.\right.\\&\left.\left.-52503552\right)-6 \left(336875 S^{10} \kappa ^{10}+1120500 S^8 \kappa ^8+4500000 S^6 \kappa ^6+1123200 S^4 \kappa ^4+3048192 S^2 \kappa ^2+1741824\right)\right)\end{aligned}\right)\\
&O=\left(\begin{aligned}&\left(425 S^{10} \kappa ^{10}+5388 S^8 \kappa ^8-277728 S^6 \kappa ^6-3063936 S^4 \kappa ^4-3877632 S^2 \kappa ^2+801792\right) Q^6\\&+3 \pi ^5 S \left(-1525 S^{10} \kappa ^{10}+44540 S^8 \kappa ^8+1677472 S^6 \kappa ^6+10727040 S^4 \kappa ^4+9960192 S^2 \kappa ^2+580608\right) Q^4\\&-2 \pi ^4 S^2 \left(-625 S^{10} \kappa ^{10}+436500 S^8 \kappa ^8+3713760 S^6 \kappa ^6+9728640 S^4 \kappa ^4+10513152 S^2 \kappa ^2\right.\\&\left.+3 G Q^2 \left(1225 S^{10} \kappa ^{10}-145028 S^8 \kappa ^8-4701792 S^6 \kappa ^6-28819584 S^4 \kappa ^4-31470336 S^2 \kappa ^2+21482496\right)\right.\\&\left.+4727808\right) Q^2+144 G^2 \pi  S^5 \left(23275 S^{10} \kappa ^{10}+432600 S^8 \kappa ^8+861624 S^6 \kappa ^6-61344 S^4 \kappa ^4+1067904 S^2 \kappa ^2\right.\\&\left.-124416\right)\end{aligned}\right)\\
&P=\left(\begin{aligned}&\left(145775 S^{10} \kappa ^{10}+2349060 S^8 \kappa ^8+2050272 S^6 \kappa ^6-3680640 S^4 \kappa ^4+6697728 S^2 \kappa ^2-2239488\right)\\&-4 \pi ^3 S^3 \left(16 G Q^2 \left(175 S^{10} \kappa ^{10}+175710 S^8 \kappa ^8+1288818 S^6 \kappa ^6+2643624 S^4 \kappa ^4+2724192 S^2 \kappa ^2-964224\right)\right.\\&\left.-9 \left(625 S^{10} \kappa ^{10}+16500 S^8 \kappa ^8+62400 S^6 \kappa ^6+49920 S^4 \kappa ^4+99072 S^2 \kappa ^2+138240\right)\right)\\&+3 G \pi ^2 S^4 \left(G \left(30625 S^{10} \kappa ^{10}-12405876 S^8 \kappa ^8-80933472 S^6 \kappa ^6-99363456 S^4 \kappa ^4-5204736 S^2 \kappa ^2\right.\right.\\&\left.\left.+11197440\right) Q^2+6 \left(27125 S^{10} \kappa ^{10}+579100 S^8 \kappa ^8+1676640 S^6 \kappa ^6+913536 S^4 \kappa ^4\right.\right.\\&\left.\left.+2536704 S^2 \kappa ^2+580608\right)\right)\end{aligned}\right)\\
&R=\left(\begin{aligned}&\left(S^6 \kappa ^6-12 S^4 \kappa ^4-336 S^2 \kappa ^2-576\right) Q^6+6 \pi ^5 \left(25 S^{10} \kappa ^{10}+36 S^8 \kappa ^8+14016 S^6 \kappa ^6+134400 S^4 \kappa ^4\right.\\&\left.+172800 S^2 \kappa ^2+27648\right) Q^4+3 \pi ^4 S \left(3 G Q^2 \left(175 S^{10} \kappa ^{10}+692 S^8 \kappa ^8+52192 S^6 \kappa ^6+452736 S^4 \kappa ^4\right.\right.\\&\left.\left.+762624 S^2 \kappa ^2-359424\right)-8 \left(25 S^{10} \kappa ^{10}+540 S^8 \kappa ^8+10088 S^6 \kappa ^6+25440 S^4 \kappa ^4+42624 S^2 \kappa ^2+13824\right)\right) Q^2\\&+9 G^3 S^5 \left(32585 S^{10} \kappa ^{10}+470988 S^8 \kappa ^8+1989792 S^6 \kappa ^6-3259008 S^4 \kappa ^4+3587328 S^2 \kappa ^2-746496\right)\end{aligned}\right)\\
&T=\left(\begin{aligned}&\left(48265 S^{10} \kappa ^{10}+669732 S^8 \kappa ^8+3424032 S^6 \kappa ^6-1662336 S^4 \kappa ^4+4416768 S^2 \kappa ^2-248832\right)\\&-8 G \pi ^2 S^5 \kappa ^2 \left(-2975 S^8 \kappa ^8-40200 S^6 \kappa ^6-243936 S^4 \kappa ^4-41472 S^2 \kappa ^2+9 G Q^2 \left(245 S^8 \kappa ^8\right.\right.\\&\left.\left.+7700 S^6 \kappa ^6+123504 S^4 \kappa ^4+137664 S^2 \kappa ^2+55296\right)-518400\right)+\pi ^3 S^2 \left(2 \left(625 S^{10} \kappa ^{10}+8500 S^8 \kappa ^8\right.\right.\\&\left.\left.+62880 S^6 \kappa ^6+42624 S^4 \kappa ^4+158976 S^2 \kappa ^2+82944\right)-3 G Q^2 \left(2555 S^{10} \kappa ^{10}+58268 S^8 \kappa ^8+950304 S^6 \kappa ^6\right.\right.\\&\left.\left.+1650816 S^4 \kappa ^4+3158784 S^2 \kappa ^2-746496\right)\right)\end{aligned}\right)\\
&U=\left(\begin{aligned}&\left(a \pi ^4 \left(S^2 \kappa ^2+28\right) Q^4+4 G S^2 \left(-\pi ^2 \left(S^2 \kappa ^2+12\right) Q^2+\pi  S \left(5 S^2 \kappa ^2+4\right)+G S^2 \left(35 S^2 \kappa ^2-12\right)\right)\right)\times\\& \left(a \pi ^4 \left(S^2 \kappa ^2+12\right) Q^4+4 G S^2 \left(3 \pi ^2 \left(S^2 \kappa ^2-4\right) Q^2+3 G S^2 \left(12-7 S^2 \kappa ^2\right)+\pi  \left(12 S-5 S^3 \kappa ^2\right)\right)\right)^2\times\\& \left(a l^2 \pi ^4 \left(S^2 \kappa ^2+28\right) Q^4+4 G S^2 \left(\pi  \left(S \left(5 S^2 \kappa ^2+4\right)-\pi  Q^2 \left(S^2 \kappa ^2+12\right)\right) l^2+G S^2 \left(35 S^2 \kappa ^2-12\right)\right)\right)\end{aligned}\right)\\
&V=\begin{aligned}&\left(10 G S^2 \left(S^2 \kappa ^2+12\right)-3 a \pi ^2 Q^2 \left(5 S^2 \kappa ^2+12\right)\right)^2\times\\& \left(a l^2 \pi ^4 \left(S^2 \kappa ^2+12\right) Q^4+4 G S^2 \left(\pi  \left(-5 \kappa ^2 S^3+12 S+3 \pi  Q^2 \left(S^2 \kappa ^2-4\right)\right) l^2+3 G S^2 \left(12-7 S^2 \kappa ^2\right)\right)\right)\end{aligned}
\end{split}
\end{equation}
In section \ref{sec:7}, the GTD scalar for the dual CFT of Euler Heisenberg AdS black hole in Bekenstein Hawking entropy is given as
\begin{equation}
R_{GTD}=-\frac{512 \pi ^2 c \mathcal{S}^7 \mathcal{V} \left(\begin{aligned}&4096 \pi ^3 c^3 \mathcal{S}^7 A+128 \pi ^4 c^2 \mathcal{Q}^2 \mathcal{S}^4 B+8 c I-307 \pi ^{14} \alpha ^4 \mathcal{Q}^{14}-6 \pi ^{12} \alpha ^3 (366 \alpha -191) \mathcal{Q}^{12} \mathcal{S}^2\\&+12 \pi ^{10} \alpha ^2 (2569 \alpha +92) \mathcal{Q}^{10} \mathcal{S}^4-8 \pi ^8 \alpha  \left(324 \alpha ^2+18003 \alpha +580\right) \mathcal{Q}^8 \mathcal{S}^6\\&+48 \pi ^6 \alpha  (633 \alpha +5180) \mathcal{Q}^6 \mathcal{S}^8-96 \pi ^4 \left(486 \alpha ^2+675 \alpha +1300\right) \mathcal{Q}^4 \mathcal{S}^{10}\\&+233280 \pi ^2 \alpha  \mathcal{Q}^2 \mathcal{S}^{12}-259200 \mathcal{S}^{14}\end{aligned}\right)}{\left(10 \mathcal{S}^2-3 \pi ^2 \alpha  \mathcal{Q}^2\right)^2 D^2 \left(16 \pi  c \mathcal{S}^3+\pi ^4 \alpha  \mathcal{Q}^4-4 \pi ^2 \mathcal{Q}^2 \mathcal{S}^2+12 \mathcal{S}^4\right)^3}
\label{ap:16}
\end{equation}
where A, B, I, D are given below
\begin{equation}
	\begin{split}
&A=\left(33 \pi ^4 \alpha ^2 \mathcal{Q}^4-150 \pi ^2 \alpha  \mathcal{Q}^2 \mathcal{S}^2+100 \mathcal{S}^4\right)\\
&B=\left(159 \pi ^6 \alpha ^3 \mathcal{Q}^6-1166 \pi ^4 \alpha ^2 \mathcal{Q}^4 \mathcal{S}^2+12 \pi ^2 \alpha  (63 \alpha +190) \mathcal{Q}^2 \mathcal{S}^4-40 (63 \alpha +40) \mathcal{S}^6\right)\\
&I=\left(201 \pi ^{13} \alpha ^4 \mathcal{Q}^{12} \mathcal{S}-1588 \pi ^{11} \alpha ^3 \mathcal{Q}^{10} \mathcal{S}^3-4 \pi ^9 \alpha ^2 (396 \alpha -745) \mathcal{Q}^8 \mathcal{S}^5+48 \pi ^7 \alpha  (383 \alpha -35) \mathcal{Q}^6 \mathcal{S}^7-16 \pi ^5 \left(189 \alpha ^2+3720 \alpha -200\right) \mathcal{Q}^4 \mathcal{S}^9+8640 \pi ^3 (2 \alpha +5) \mathcal{Q}^2 \mathcal{S}^{11}-14400 \pi  \mathcal{S}^{13}\right)\\
&D=\left(-16 \pi  c \mathcal{S}^3-7 \pi ^4 \alpha  \mathcal{Q}^4+12 \pi ^2 \mathcal{Q}^2 \mathcal{S}^2+12 \mathcal{S}^4\right)
	\end{split}
\end{equation}
The GTD scalar for the dual CFT of Euler Heisenberg AdS black hole in R\'enyi entropy is given as
\begin{equation}
R_{GTD}=\frac{\begin{aligned}&8192 c \pi ^2 \mathcal{S}^6 \mathcal{V} \left(72 \pi ^{16} \alpha ^5 \lambda  A \mathcal{Q}^{16}+\pi ^{14} \mathcal{S} \alpha ^4 B \mathcal{Q}^{14}+2 \pi ^{12} \mathcal{S}^3 \alpha ^3 I \mathcal{Q}^{12}-4 \pi ^{10} \mathcal{S}^5 \alpha ^2 D \mathcal{Q}^{10}+8 \pi ^8 \mathcal{S}^7 \alpha  E \mathcal{Q}^8\right.\\&\left.-16 \pi ^6 \mathcal{S}^9 F \mathcal{Q}^6+96 \pi ^4 \mathcal{S}^{11}G \mathcal{Q}^4-2880 \pi ^2 \mathcal{S}^{13} H \mathcal{Q}^2+86400 \mathcal{S}^{15} J+8192 c^3 \pi ^3 \mathcal{S}^8 K+128 c^2 \pi ^2 \mathcal{S}^5 L\right.\\&\left.+8 c \pi  \mathcal{S}^2 \left(3 \pi ^{12} \alpha ^4 M \mathcal{Q}^{12}-8 \pi ^{10} \mathcal{S}^2 \alpha ^3 N \mathcal{Q}^{10}+4 \pi ^8 \mathcal{S}^4 \alpha ^2 O \mathcal{Q}^8-16 \pi ^6 \mathcal{S}^6 \alpha P \mathcal{Q}^6+16 \pi ^4 \mathcal{S}^8 Q \mathcal{Q}^4\right.\right.\\&\left.\left.-960 \pi ^2 \mathcal{S}^{10} R\mathcal{Q}^2+14400 \mathcal{S}^{12} S\right)\right)\end{aligned}}{\begin{aligned}&T^2 \left(\pi ^4 \alpha  (4-3 \mathcal{S} \lambda ) \mathcal{Q}^4-4 \pi ^2 \mathcal{S}^2 (\mathcal{S} \lambda +4) \mathcal{Q}^2+16 c \pi  \mathcal{S}^3 (3 \mathcal{S} \lambda +4)+12 \mathcal{S}^4 (5 \mathcal{S} \lambda +4)\right)^3\times \\& \left(\pi ^4 \alpha  (15 \mathcal{S} \lambda -28) \mathcal{Q}^4+4 \pi ^2 \mathcal{S}^2 (\mathcal{S} \lambda +12) \mathcal{Q}^2+16 c \pi  \mathcal{S}^3 (3 \mathcal{S} \lambda -4)+12 \mathcal{S}^4 (15 \mathcal{S} \lambda +4)\right)^2\end{aligned}}
\label{ap:17}
\end{equation}
where A, B, I, D, E, F, G, H, J, K, L, M, N, O, P, Q, R, S, T are given below
\begin{equation}
\begin{split}
&A=\left(225 \mathcal{S}^3 \lambda ^3-660 \mathcal{S}^2 \lambda ^2+560 \mathcal{S} \lambda -192\right)\\
&B=\left(-176175 \mathcal{S}^5 \lambda ^5+1138860 \mathcal{S}^4 \lambda ^4-2244384 \mathcal{S}^3 \lambda ^3+1784448 \mathcal{S}^2 \lambda ^2-800512 \mathcal{S} \lambda +314368\right)\\
&I=\left(\begin{aligned}&56025 \mathcal{S}^5 \lambda ^5-1175220 \mathcal{S}^4 \lambda ^4+4636512 \mathcal{S}^3 \lambda ^3-5045120 \mathcal{S}^2 \lambda ^2+1493248 \mathcal{S} \lambda \\&+18 \alpha  \left(124875 \mathcal{S}^5 \lambda ^5-634500 \mathcal{S}^4 \lambda ^4+1015680 \mathcal{S}^3 \lambda ^3-450304 \mathcal{S}^2 \lambda ^2-105216 \mathcal{S} \lambda +62464\right)-586752\end{aligned}\right)\\
&D=\left(\begin{aligned}&4 \left(225 \mathcal{S}^5 \lambda ^5-69900 \mathcal{S}^4 \lambda ^4+641248 \mathcal{S}^3 \lambda ^3-1311872 \mathcal{S}^2 \lambda ^2+272640 \mathcal{S} \lambda +70656\right)+\\&3 \alpha  \left(934875 \mathcal{S}^5 \lambda ^5-8761500 \mathcal{S}^4 \lambda ^4+22037280 \mathcal{S}^3 \lambda ^3-16470656 \mathcal{S}^2 \lambda ^2-1340672 \mathcal{S} \lambda +2630656\right)\end{aligned}\right)\\
&E=\left(\begin{aligned}&108 \left(50625 \mathcal{S}^5 \lambda ^5-168000 \mathcal{S}^4 \lambda ^4+114000 \mathcal{S}^3 \lambda ^3-18880 \mathcal{S}^2 \lambda ^2-5632 \mathcal{S} \lambda +3072\right) \alpha ^2\\&-3 \left(55125 \mathcal{S}^5 \lambda ^5+4295100 \mathcal{S}^4 \lambda ^4-22795040 \mathcal{S}^3 \lambda ^3+29770368 \mathcal{S}^2 \lambda ^2-3058432 \mathcal{S} \lambda -6145024\right) \alpha\\& -20 \left(285 \mathcal{S}^5 \lambda ^5-324 \mathcal{S}^4 \lambda ^4-21984 \mathcal{S}^3 \lambda ^3+99712 \mathcal{S}^2 \lambda ^2-50944 \mathcal{S} \lambda -29696\right)\end{aligned}\right)\\
&F=\left(\begin{aligned}&9 \left(770625 \mathcal{S}^5 \lambda ^5-4036500 \mathcal{S}^4 \lambda ^4+5151200 \mathcal{S}^3 \lambda ^3-1059200 \mathcal{S}^2 \lambda ^2+35072 \mathcal{S} \lambda +216064\right) \alpha ^2\\&-60 \left(1875 \mathcal{S}^5 \lambda ^5-10860 \mathcal{S}^4 \lambda ^4-321632 \mathcal{S}^3 \lambda ^3+917888 \mathcal{S}^2 \lambda ^2-374016 \mathcal{S} \lambda -265216\right) \alpha \\&+3200 \mathcal{S} \lambda  \left(\mathcal{S}^3 \lambda ^3-4 \mathcal{S}^2 \lambda ^2-80 \mathcal{S} \lambda +64\right)\end{aligned}\right)\\
&G= \left(\begin{aligned}&54 \left(84375 \mathcal{S}^5 \lambda ^5-22500 \mathcal{S}^4 \lambda ^4-56000 \mathcal{S}^3 \lambda ^3+32000 \mathcal{S} \lambda +9216\right) \alpha ^2\\&+15 \left(30375 \mathcal{S}^5 \lambda ^5-487500 \mathcal{S}^4 \lambda ^4+981920 \mathcal{S}^3 \lambda ^3-319616 \mathcal{S}^2 \lambda ^2+34560 \mathcal{S} \lambda +46080\right) \alpha +\\&100 \left(75 \mathcal{S}^5 \lambda ^5+836 \mathcal{S}^4 \lambda ^4-32 \mathcal{S}^3 \lambda ^3-29056 \mathcal{S}^2 \lambda ^2+32512 \mathcal{S} \lambda +13312\right)\end{aligned}\right)\\
&H=\left(\begin{aligned}&9 \alpha  \left(16875 \mathcal{S}^5 \lambda ^5-81500 \mathcal{S}^4 \lambda ^4-44000 \mathcal{S}^3 \lambda ^3+35200 \mathcal{S}^2 \lambda ^2+43776 \mathcal{S} \lambda +9216\right)\\&-40 \mathcal{S} \lambda  \left(75 \mathcal{S}^4 \lambda ^4+1040 \mathcal{S}^3 \lambda ^3-7072 \mathcal{S}^2 \lambda ^2+3072 \mathcal{S} \lambda -256\right)\end{aligned}\right)\\
&J=\left(2625 \mathcal{S}^5 \lambda ^5-14900 \mathcal{S}^4 \lambda ^4+16480 \mathcal{S}^3 \lambda ^3+35968 \mathcal{S}^2 \lambda ^2+20736 \mathcal{S} \lambda +3072\right)\\
&K=\left(\begin{aligned}&1350 \lambda ^5 \mathcal{S}^9-10800 \lambda ^4 \mathcal{S}^8-450 \lambda ^3 \left(9 \pi ^2 \mathcal{Q}^2 \alpha  \lambda ^2-80\right) \mathcal{S}^7+160 \lambda ^2 \left(189 \pi ^2 \mathcal{Q}^2 \alpha  \lambda ^2-20\right) \mathcal{S}^6+\\&45 \pi ^2 \mathcal{Q}^2 \alpha  \lambda ^3 \left(135 \pi ^2 \mathcal{Q}^2 \alpha  \lambda ^2-464\right) \mathcal{S}^5-10 \left(891 \pi ^4 \mathcal{Q}^4 \alpha ^2 \lambda ^4+3168 \pi ^2 \mathcal{Q}^2 \alpha  \lambda ^2+5120\right) \mathcal{S}^4-\\&672 \pi ^2 \mathcal{Q}^2 \alpha  \lambda  \left(9 \pi ^2 \mathcal{Q}^2 \alpha  \lambda ^2+40\right) \mathcal{S}^3+384 \pi ^2 \mathcal{Q}^2 \alpha  \left(21 \pi ^2 \mathcal{Q}^2 \alpha  \lambda ^2+200\right) \mathcal{S}^2+17664 \pi ^4 \mathcal{Q}^4 \alpha ^2 \lambda  \mathcal{S}\\&-16896 \pi ^4 \mathcal{Q}^4 \alpha ^2\end{aligned}\right)\\
&L=\left(\begin{aligned}&19200 \lambda  \left(45 \mathcal{S}^4 \lambda ^4-279 \mathcal{S}^3 \lambda ^3+592 \mathcal{S}^2 \lambda ^2+416 \mathcal{S} \lambda +832\right) \mathcal{S}^9-40 \pi ^2 \mathcal{Q}^2 \left(9 \alpha  \left(5175 \mathcal{S}^5 \lambda ^5\right.\right.\\&\left.\left.-31260 \mathcal{S}^4 \lambda ^4+4512 \mathcal{S}^3 \lambda ^3+1408 \mathcal{S}^2 \lambda ^2+30464 \mathcal{S} \lambda -7168\right)-20 \left(9 \mathcal{S}^5 \lambda ^5+84 \mathcal{S}^4 \lambda ^4\right.\right.\\&\left.\left.-976 \mathcal{S}^3 \lambda ^3+3264 \mathcal{S}^2 \lambda ^2-1536 \mathcal{S} \lambda +2048\right)\right) \mathcal{S}^6+4 \pi ^4 \mathcal{Q}^4 \alpha  \left(10 \left(1215 \mathcal{S}^5 \lambda ^5-12132 \mathcal{S}^4 \lambda ^4\right.\right.\\&\left.\left.+62976 \mathcal{S}^3 \lambda ^3-107008 \mathcal{S}^2 \lambda ^2+97024 \mathcal{S} \lambda -58368\right)+27 \alpha  \left(21375 \mathcal{S}^5 \lambda ^5-15900 \mathcal{S}^4 \lambda ^4\right.\right.\\&\left.\left.-160 \mathcal{S}^3 \lambda ^3-2432 \mathcal{S}^2 \lambda ^2+20224 \mathcal{S} \lambda -7168\right)\right) \mathcal{S}^4-2 \pi ^6 \mathcal{Q}^6 \alpha ^2 \left(22275 \mathcal{S}^5 \lambda ^5-244620 \mathcal{S}^4 \lambda ^4\right.\\&\left.+826848 \mathcal{S}^3 \lambda ^3-1634688 \mathcal{S}^2 \lambda ^2+1138432 \mathcal{S} \lambda -596992\right) \mathcal{S}^2+3 \pi ^8 \mathcal{Q}^8 \alpha ^3 \left(2025 \mathcal{S}^5 \lambda ^5\right.\\&\left.-34020 \mathcal{S}^4 \lambda ^4+152928 \mathcal{S}^3 \lambda ^3-226176 \mathcal{S}^2 \lambda ^2+126208 \mathcal{S} \lambda -54272\right)\end{aligned}\right)\\
&M=\left(62775 \mathcal{S}^5 \lambda ^5-376380 \mathcal{S}^4 \lambda ^4+703584 \mathcal{S}^3 \lambda ^3-486528 \mathcal{S}^2 \lambda ^2+171776 \mathcal{S} \lambda -68608\right)\\
&N=\left(78975 \mathcal{S}^5 \lambda ^5-628560 \mathcal{S}^4 \lambda ^4+1738872 \mathcal{S}^3 \lambda ^3-1648224 \mathcal{S}^2 \lambda ^2+500096 \mathcal{S} \lambda -203264\right)\\
&O=\left(\begin{aligned}&-7425 \mathcal{S}^5 \lambda ^5-1513260 \mathcal{S}^4 \lambda ^4+6719424 \mathcal{S}^3 \lambda ^3-9677824 \mathcal{S}^2 \lambda ^2+3051776 \mathcal{S} \lambda +\\&36 \alpha  \left(10125 \mathcal{S}^5 \lambda ^5-90000 \mathcal{S}^4 \lambda ^4+195600 \mathcal{S}^3 \lambda ^3-170048 \mathcal{S}^2 \lambda ^2+18432 \mathcal{S} \lambda +11264\right)-762880\end{aligned}\right)\\
&P= \left(\begin{aligned}&3 \alpha  \left(151875 \mathcal{S}^5 \lambda ^5-774900 \mathcal{S}^4 \lambda ^4+2596320 \mathcal{S}^3 \lambda ^3-3419776 \mathcal{S}^2 \lambda ^2+964352 \mathcal{S} \lambda +392192\right)\\&-5 \left(2115 \mathcal{S}^5 \lambda ^5-2148 \mathcal{S}^4 \lambda ^4-229120 \mathcal{S}^3 \lambda ^3+516864 \mathcal{S}^2 \lambda ^2-240384 \mathcal{S} \lambda +21504\right)\end{aligned}\right)\\
&Q=\left(\begin{aligned}&27 \left(163125 \mathcal{S}^5 \lambda ^5-118500 \mathcal{S}^4 \lambda ^4-20000 \mathcal{S}^3 \lambda ^3+17280 \mathcal{S}^2 \lambda ^2+76032 \mathcal{S} \lambda +7168\right) \alpha ^2\\&+60 \left(5625 \mathcal{S}^5 \lambda ^5-50280 \mathcal{S}^4 \lambda ^4+141344 \mathcal{S}^3 \lambda ^3-292736 \mathcal{S}^2 \lambda ^2+193280 \mathcal{S} \lambda +63488\right) \alpha\\& +400 \left(3 \mathcal{S}^5 \lambda ^5+70 \mathcal{S}^4 \lambda ^4+64 \mathcal{S}^3 \lambda ^3-2688 \mathcal{S}^2 \lambda ^2+1024 \mathcal{S} \lambda -512\right)\end{aligned}\right)\\
&R=\left(\begin{aligned}&9 \alpha  \left(7125 \mathcal{S}^5 \lambda ^5-43300 \mathcal{S}^4 \lambda ^4+1680 \mathcal{S}^3 \lambda ^3+15808 \mathcal{S}^2 \lambda ^2+23040 \mathcal{S} \lambda +2048\right)\\&+5 \left(-45 \mathcal{S}^5 \lambda ^5-1988 \mathcal{S}^4 \lambda ^4+12480 \mathcal{S}^3 \lambda ^3-18432 \mathcal{S}^2 \lambda ^2+39168 \mathcal{S} \lambda +9216\right)\end{aligned}\right) \\
&S=\left(2325 \mathcal{S}^5 \lambda ^5-13780 \mathcal{S}^4 \lambda ^4+24128 \mathcal{S}^3 \lambda ^3+24576 \mathcal{S}^2 \lambda ^2+17152 \mathcal{S} \lambda +1024\right)\\
&T=\left(-10 \lambda  \mathcal{S}^3+40 \mathcal{S}^2+15 \pi ^2 \mathcal{Q}^2 \alpha  \lambda  \mathcal{S}-12 \pi ^2 \mathcal{Q}^2 \alpha \right)
\end{split}
\end{equation}
The GTD scalar for the dual CFT for Euler Heisenberg AdS black hole in Kaniadakis entropy is given as
\begin{equation}
R_{GTD}=-\frac{24576 c \pi ^2 \mathcal{S}^7 \mathcal{V} \left(\begin{aligned}&-288 \pi ^{16} \alpha ^5 \kappa ^2 A \mathcal{Q}^{16}-3 \pi ^{14} \alpha ^4 B \mathcal{Q}^{14}+2 \pi ^{12} \mathcal{S}^2 \alpha ^3 I\mathcal{Q}^{12}-12 \pi ^{10} \mathcal{S}^4 \alpha ^2 D \mathcal{Q}^{10}\\&+8 \pi ^8 \mathcal{S}^6 \alpha E \mathcal{Q}^8-48 \pi ^6 \mathcal{S}^8 F \mathcal{Q}^6+96 \pi ^4 \mathcal{S}^{10} G \mathcal{Q}^4-2880 \pi ^2 \mathcal{S}^{12} H \mathcal{Q}^2+28800 \mathcal{S}^{14} J\\&+4096 c^3 \pi ^3 \mathcal{S}^7 K+128 c^2 \pi ^2 \mathcal{S}^4 L+8 c \left(4800 \pi M \mathcal{S}^{13}-960 \pi ^3 \mathcal{Q}^2 N \mathcal{S}^{11}+16 \pi ^5 \mathcal{Q}^4 O\mathcal{S}^9\right.\\&\left.-48 \pi ^7 \mathcal{Q}^6 \alpha P \mathcal{S}^7+4 \pi ^9 \mathcal{Q}^8 \alpha ^2 Q \mathcal{S}^5-12 \pi ^{11} \mathcal{Q}^{10} \alpha ^3 R \mathcal{S}^3+9 \pi ^{13} \mathcal{Q}^{12} \alpha ^4 S \mathcal{S}\right)\end{aligned}\right)}{U^3V^2 \left(-10 \kappa ^2 \mathcal{S}^4+15 \left(\pi ^2 \mathcal{Q}^2 \alpha  \kappa ^2-8\right) \mathcal{S}^2+36 \pi ^2 \mathcal{Q}^2 \alpha \right)^2}
\label{ap:18}
\end{equation}
where A, B, I, D, E, F, G, H, J, K, L, M, N, O, P, Q, R, S, U, V are given below
\begin{equation}
	\begin{split}
		&A=\left(25 \mathcal{S}^6 \kappa ^6+516 \mathcal{S}^4 \kappa ^4-1872 \mathcal{S}^2 \kappa ^2+1728\right)\\
		&B=\left(2875 \mathcal{S}^{10} \kappa ^{10}+242420 \mathcal{S}^8 \kappa ^8+4807264 \mathcal{S}^6 \kappa ^6+17493120 \mathcal{S}^4 \kappa ^4+5066496 \mathcal{S}^2 \kappa ^2+8487936\right)\\
		&I=\left(\begin{aligned}&32875 \mathcal{S}^{10} \kappa ^{10}+2155460 \mathcal{S}^8 \kappa ^8+42126816 \mathcal{S}^6 \kappa ^6+216097920 \mathcal{S}^4 \kappa ^4+94037760 \mathcal{S}^2 \kappa ^2\\&+18 \alpha  \left(6125 \mathcal{S}^{10} \kappa ^{10}+346620 \mathcal{S}^8 \kappa ^8+5985312 \mathcal{S}^6 \kappa ^6+15343488 \mathcal{S}^4 \kappa ^4+794880 \mathcal{S}^2 \kappa ^2-5059584\right)+47526912\end{aligned}\right) \\
		&D=\left(\begin{aligned}&4 \left(2375 \mathcal{S}^{10} \kappa ^{10}+58820 \mathcal{S}^8 \kappa ^8+3679136 \mathcal{S}^6 \kappa ^6+25832832 \mathcal{S}^4 \kappa ^4+19996416 \mathcal{S}^2 \kappa ^2-1907712\right)\\&+3 \alpha  \left(46375 \mathcal{S}^{10} \kappa ^{10}+3518180 \mathcal{S}^8 \kappa ^8+44851168 \mathcal{S}^6 \kappa ^6+158007936 \mathcal{S}^4 \kappa ^4+50400000 \mathcal{S}^2 \kappa ^2-71027712\right)\end{aligned}\right)\\
		&E= \left(\begin{aligned}&36 \left(79625 \mathcal{S}^{10} \kappa ^{10}+3137820 \mathcal{S}^8 \kappa ^8+9365472 \mathcal{S}^6 \kappa ^6+5678208 \mathcal{S}^4 \kappa ^4-1804032 \mathcal{S}^2 \kappa ^2-746496\right) \alpha ^2\\&+\left(1463875 \mathcal{S}^{10} \kappa ^{10}+22132740 \mathcal{S}^8 \kappa ^8+566110944 \mathcal{S}^6 \kappa ^6+2419569792 \mathcal{S}^4 \kappa ^4+1655209728 \mathcal{S}^2 \kappa ^2\right.\\&\left.-1493240832\right) \alpha -60 \left(425 \mathcal{S}^{10} \kappa ^{10}+5388 \mathcal{S}^8 \kappa ^8-277728 \mathcal{S}^6 \kappa ^6-3063936 \mathcal{S}^4 \kappa ^4-3877632 \mathcal{S}^2 \kappa ^2+801792\right)\end{aligned}\right)\\
		&F=\left(\begin{aligned}&\left(4808125 \mathcal{S}^{10} \kappa ^{10}+70480620 \mathcal{S}^8 \kappa ^8+423889056 \mathcal{S}^6 \kappa ^6+309260160 \mathcal{S}^4 \kappa ^4-17397504 \mathcal{S}^2 \kappa ^2-52503552\right) \alpha ^2\\&-20 \left(1225 \mathcal{S}^{10} \kappa ^{10}-145028 \mathcal{S}^8 \kappa ^8-4701792 \mathcal{S}^6 \kappa ^6-28819584 \mathcal{S}^4 \kappa ^4-31470336 \mathcal{S}^2 \kappa ^2+21482496\right) \alpha\\& -38400 \mathcal{S}^2 \kappa ^2 \left(\mathcal{S}^6 \kappa ^6-12 \mathcal{S}^4 \kappa ^4-336 \mathcal{S}^2 \kappa ^2-576\right)\end{aligned}\right)\\
		&G= \left(\begin{aligned}&18 \left(728875 \mathcal{S}^{10} \kappa ^{10}+1572900 \mathcal{S}^8 \kappa ^8+2469600 \mathcal{S}^6 \kappa ^6-2616192 \mathcal{S}^4 \kappa ^4+4790016 \mathcal{S}^2 \kappa ^2-2239488\right) \alpha ^2\\&-5 \left(30625 \mathcal{S}^{10} \kappa ^{10}-12405876 \mathcal{S}^8 \kappa ^8-80933472 \mathcal{S}^6 \kappa ^6-99363456 \mathcal{S}^4 \kappa ^4-5204736 \mathcal{S}^2 \kappa ^2+11197440\right) \alpha\\& +300 \left(175 \mathcal{S}^{10} \kappa ^{10}+692 \mathcal{S}^8 \kappa ^8+52192 \mathcal{S}^6 \kappa ^6+452736 \mathcal{S}^4 \kappa ^4+762624 \mathcal{S}^2 \kappa ^2-359424\right)\end{aligned}\right)\\
		&H=\left(\begin{aligned}&80 \mathcal{S}^2 \left(245 \mathcal{S}^8 \kappa ^8+7700 \mathcal{S}^6 \kappa ^6+123504 \mathcal{S}^4 \kappa ^4+137664 \mathcal{S}^2 \kappa ^2+55296\right) \kappa ^2+\\&3 \alpha  \left(145775 \mathcal{S}^{10} \kappa ^{10}+2349060 \mathcal{S}^8 \kappa ^8+2050272 \mathcal{S}^6 \kappa ^6-3680640 \mathcal{S}^4 \kappa ^4+6697728 \mathcal{S}^2 \kappa ^2-2239488\right)\end{aligned}\right)\\
		&J=\left(32585 \mathcal{S}^{10} \kappa ^{10}+470988 \mathcal{S}^8 \kappa ^8+1989792 \mathcal{S}^6 \kappa ^6-3259008 \mathcal{S}^4 \kappa ^4+3587328 \mathcal{S}^2 \kappa ^2-746496\right)
	\end{split}
\end{equation}
\begin{equation}
	\begin{split}
&K=\left(\begin{aligned}&62500 \kappa ^{10} \mathcal{S}^{14}-6250 \kappa ^8 \left(9 \pi ^2 \mathcal{Q}^2 \alpha  \kappa ^2-136\right) \mathcal{S}^{12}+375 \kappa ^6 \left(225 \pi ^4 \mathcal{Q}^4 \alpha ^2 \kappa ^4-3960 \pi ^2 \mathcal{Q}^2 \alpha  \kappa ^2+16768\right) \mathcal{S}^{10}\\&+900 \kappa ^4 \left(375 \pi ^4 \mathcal{Q}^4 \alpha ^2 \kappa ^4-6240 \pi ^2 \mathcal{Q}^2 \alpha  \kappa ^2+4736\right) \mathcal{S}^8+57600 \kappa ^2 \left(25 \pi ^4 \mathcal{Q}^4 \alpha ^2 \kappa ^4-78 \pi ^2 \mathcal{Q}^2 \alpha  \kappa ^2+276\right) \mathcal{S}^6\\&+69120 \left(10 \pi ^4 \mathcal{Q}^4 \alpha ^2 \kappa ^4-129 \pi ^2 \mathcal{Q}^2 \alpha  \kappa ^2+120\right) \mathcal{S}^4+103680 \pi ^2 \mathcal{Q}^2 \alpha  \left(13 \pi ^2 \mathcal{Q}^2 \alpha  \kappa ^2-120\right) \mathcal{S}^2+2737152 \pi ^4 \mathcal{Q}^4 \alpha ^2\end{aligned}\right)\\
&L=\left(\begin{aligned}&3200 \kappa ^2 \left(2975 \mathcal{S}^8 \kappa ^8+40200 \mathcal{S}^6 \kappa ^6+243936 \mathcal{S}^4 \kappa ^4+41472 \mathcal{S}^2 \kappa ^2+518400\right) \mathcal{S}^{10}-\\&120 \pi ^2 \mathcal{Q}^2 \left(80 \left(25 \mathcal{S}^{10} \kappa ^{10}+540 \mathcal{S}^8 \kappa ^8+10088 \mathcal{S}^6 \kappa ^6+25440 \mathcal{S}^4 \kappa ^4+42624 \mathcal{S}^2 \kappa ^2+13824\right)\right.\\&\left.+3 \alpha  \left(27125 \mathcal{S}^{10} \kappa ^{10}+579100 \mathcal{S}^8 \kappa ^8+1676640 \mathcal{S}^6 \kappa ^6+913536 \mathcal{S}^4 \kappa ^4+2536704 \mathcal{S}^2 \kappa ^2+580608\right)\right) \mathcal{S}^6\\&+4 \pi ^4 \mathcal{Q}^4 \alpha  \left(9 \alpha  \left(336875 \mathcal{S}^{10} \kappa ^{10}+1120500 \mathcal{S}^8 \kappa ^8+4500000 \mathcal{S}^6 \kappa ^6+1123200 \mathcal{S}^4 \kappa ^4\right.\right.\\&\left.\left.+3048192 \mathcal{S}^2 \kappa ^2+1741824\right)+10 \left(-625 \mathcal{S}^{10} \kappa ^{10}+436500 \mathcal{S}^8 \kappa ^8+3713760 \mathcal{S}^6 \kappa ^6+9728640 \mathcal{S}^4 \kappa ^4\right.\right.\\&\left.\left.+10513152 \mathcal{S}^2 \kappa ^2+4727808\right)\right) \mathcal{S}^4-6 \pi ^6 \mathcal{Q}^6 \alpha ^2 \left(134375 \mathcal{S}^{10} \kappa ^{10}+1620500 \mathcal{S}^8 \kappa ^8+13175200 \mathcal{S}^6 \kappa ^6\right.\\&\left.+29923200 \mathcal{S}^4 \kappa ^4+21765888 \mathcal{S}^2 \kappa ^2+16118784\right) \mathcal{S}^2+9 \pi ^8 \mathcal{Q}^8 \alpha ^3 \left(5625 \mathcal{S}^{10} \kappa ^{10}+273500 \mathcal{S}^8 \kappa ^8\right.\\&\left.+1285600 \mathcal{S}^6 \kappa ^6+3008640 \mathcal{S}^4 \kappa ^4+1366272 \mathcal{S}^2 \kappa ^2+1465344\right)\end{aligned}\right)\\
&M= \left(48265 \mathcal{S}^{10} \kappa ^{10}+669732 \mathcal{S}^8 \kappa ^8+3424032 \mathcal{S}^6 \kappa ^6-1662336 \mathcal{S}^4 \kappa ^4+4416768 \mathcal{S}^2 \kappa ^2-248832\right)\\
&N=\left(\begin{aligned}&5 \left(2555 \mathcal{S}^{10} \kappa ^{10}+58268 \mathcal{S}^8 \kappa ^8+950304 \mathcal{S}^6 \kappa ^6+1650816 \mathcal{S}^4 \kappa ^4+3158784 \mathcal{S}^2 \kappa ^2-746496\right)\\&+12 \alpha  \left(23275 \mathcal{S}^{10} \kappa ^{10}+432600 \mathcal{S}^8 \kappa ^8+861624 \mathcal{S}^6 \kappa ^6-61344 \mathcal{S}^4 \kappa ^4+1067904 \mathcal{S}^2 \kappa ^2-124416\right)\end{aligned}\right)\\
&O=\left(\begin{aligned}&9 \left(2051875 \mathcal{S}^{10} \kappa ^{10}+5810700 \mathcal{S}^8 \kappa ^8+17845920 \mathcal{S}^6 \kappa ^6-3169152 \mathcal{S}^4 \kappa ^4+11010816 \mathcal{S}^2 \kappa ^2-1741824\right) \alpha ^2\\&+320 \left(175 \mathcal{S}^{10} \kappa ^{10}+175710 \mathcal{S}^8 \kappa ^8+1288818 \mathcal{S}^6 \kappa ^6+2643624 \mathcal{S}^4 \kappa ^4+2724192 \mathcal{S}^2 \kappa ^2-964224\right) \alpha\\& +600 \left(25 \mathcal{S}^{10} \kappa ^{10}+36 \mathcal{S}^8 \kappa ^8+14016 \mathcal{S}^6 \kappa ^6+134400 \mathcal{S}^4 \kappa ^4+172800 \mathcal{S}^2 \kappa ^2+27648\right)\end{aligned}\right) \\
&P= \left(\begin{aligned}&\alpha  \left(814625 \mathcal{S}^{10} \kappa ^{10}+10281300 \mathcal{S}^8 \kappa ^8+74781600 \mathcal{S}^6 \kappa ^6+120698496 \mathcal{S}^4 \kappa ^4+79176960 \mathcal{S}^2 \kappa ^2-31767552\right)\\&+5 \left(-1525 \mathcal{S}^{10} \kappa ^{10}+44540 \mathcal{S}^8 \kappa ^8+1677472 \mathcal{S}^6 \kappa ^6+10727040 \mathcal{S}^4 \kappa ^4+9960192 \mathcal{S}^2 \kappa ^2+580608\right)\end{aligned}\right)\\
&Q=\left(\begin{aligned}&230125 \mathcal{S}^{10} \kappa ^{10}+3789300 \mathcal{S}^8 \kappa ^8+95542560 \mathcal{S}^6 \kappa ^6+442733184 \mathcal{S}^4 \kappa ^4+258031872 \mathcal{S}^2 \kappa ^2\\&+36 \alpha  \left(20125 \mathcal{S}^{10} \kappa ^{10}+915300 \mathcal{S}^8 \kappa ^8+3568800 \mathcal{S}^6 \kappa ^6+5668992 \mathcal{S}^4 \kappa ^4\right.\\&\left.+1734912 \mathcal{S}^2 \kappa ^2-912384\right)+61793280\end{aligned}\right)\\
&R=\left(12625 \mathcal{S}^{10} \kappa ^{10}+832100 \mathcal{S}^8 \kappa ^8+11155840 \mathcal{S}^6 \kappa ^6+41822208 \mathcal{S}^4 \kappa ^4+12697344 \mathcal{S}^2 \kappa ^2+10976256\right)\\
&S=\left(1625 \mathcal{S}^{10} \kappa ^{10}+105700 \mathcal{S}^8 \kappa ^8+1937120 \mathcal{S}^6 \kappa ^6+5124480 \mathcal{S}^4 \kappa ^4+665856 \mathcal{S}^2 \kappa ^2+1852416\right)\\
&U=\left(-84 \kappa ^2 \mathcal{S}^6+144 \mathcal{S}^4+16 c \pi  \left(12-5 \mathcal{S}^2 \kappa ^2\right) \mathcal{S}^3+12 \pi ^2 \mathcal{Q}^2 \left(\mathcal{S}^2 \kappa ^2-4\right) \mathcal{S}^2+\pi ^4 \mathcal{Q}^4 \alpha  \left(\mathcal{S}^2 \kappa ^2+12\right)\right)\\
&V= \left(\pi ^4 \alpha  \left(\mathcal{S}^2 \kappa ^2+28\right) \mathcal{Q}^4-4 \pi ^2 \mathcal{S}^2 \left(\mathcal{S}^2 \kappa ^2+12\right) \mathcal{Q}^2+16 c \pi  \mathcal{S}^3 \left(5 \mathcal{S}^2 \kappa ^2+4\right)+4 \mathcal{S}^4 \left(35 \mathcal{S}^2 \kappa ^2-12\right)\right)
\end{split}
\end{equation}

\end{document}